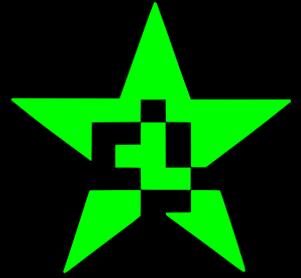

# Computed fingertip touch for the instrumental control of musical sound

with an excursion on the
computed retinal afterimage

Staas de Jong

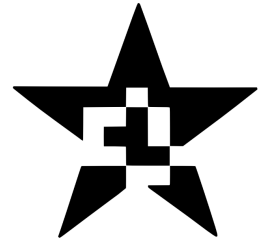

Computed fingertip touch
for the instrumental control
of musical sound

with an excursion on the
computed retinal afterimage

Staas de Jong

**Computed fingertip touch for the instrumental control of musical sound with an excursion on the computed retinal afterimage**

Proefschrift

ter verkrijging van
de graad van Doctor aan de Universiteit Leiden,
op gezag van Rector Magnificus prof. mr. C.J.J.M. Stolker,
volgens besluit van het College voor Promoties
te verdedigen op woensdag 4 november 2015
klokke 13:45 uur

door

Anastasius Prosper Anselmus de Jong

geboren te Roosendaal en Nispen
in 1980

Promotor:

Prof. dr. S. Haring

Promotiecommissie:

Prof. dr. A.M.L. Kappers (Vrije Universiteit, Amsterdam)
Prof. S. O'Modhrain, Ph.D. (University of Michigan, Ann Arbor, VS)
Prof. dr. J.N. Kok
Prof. F.C. de Ruiter

*"I determined, however, like a man who walks alone in the darkness, to go forward so slowly, and with so much circumspection at every step, that, even if I made very little progress, I should at least avoid a fall."*

– René Descartes, *Discourse on the Method*, 1637.

*"Dulces ante omnia musae"* *("Sweet above all to me are the muses")*

– Janus Dousa, closely involved in the founding of Leiden University in 1575. Motto carved in the wall of the courtyard of the Academiegebouw.

*"...und nicht seines Nuzens wegen allein, sonst würde er gleich fertig seyn."*

– Wolfgang Mozart, about the piano builder Johann Stein, 1777.

Dedicated in gratitude to my country of birth, the Netherlands.

# MAIN CONTENTS

# Introduction: Thesis outline and summary

This thesis is about the unique and central role computation has to play in the development of how humans make music.

Like other natural phenomena, human music making can be described as a causal chain, consisting of components that are each subject to measurement and empirical investigation. Therefore, as a working definition, in Section 1.1 we propose *instrumental control of musical sound:* the phenomenon where human actions make changes to a sound-generating process, resulting in heard sound which induces musical experiences within the brain.

To create a technique for making music, then, is to set up a causal relationship, between aspects of human action and changes in heard musical sound. The development of different techniques may then be motivated by a single, overarching question: How can the instrumental control of musical sound be improved? As is discussed in Section 1.5, answering this question requires also answering another, fundamental question: *What forms of instrumental control of musical sound are possible to implement?*

Parts of the answer to this question are already present in the many techniques for making music that have been developed from prehistory onward. These techniques initially involved the human body only, or made use of mechanical technologies. More recently, electromechanical technologies have also come into development and use (see Section 1.2). However, it is only when *computational* technologies are given the central role of causally linking human action to heard musical sound, that a unique advantage appears: Unlike earlier technologies, Turing-complete automata combined with transducers explicitly minimize the constraints on implementable causations (see Sections 1.4.4 and 1.5).

There is evidence for this theoretical advantage resulting in the practical implementation of new forms of instrumental control of musical sound: The combination of electronic digital computer and electric loudspeaker, introduced in the 1950s, has enabled the development of a wide variety of new types of sound-generating processes, which have come into wide use. In Section 1.4, an overview of these types is given, based on how instrumental control moved away from direct manipulation of the wave table.

Considering this evidence, the following question in particular seems potentially rewarding: How can we systematically *extend* the scope of computational technologies, from the sound-generating process, to the other components of the causal chain? In Section 1.4.4, we first formulate a *computed sound* model, to describe the fundamentals which enabled the wide variety of implementable sound-generating processes. Here, a Turing-complete automaton must become causally linked, via a transducer, to human auditory perception. Then, in Section 1.5.1.1, we generalize this to the notion of *completely computed instrumental control of musical sound*. Here, the Turing-complete automaton can track, represent and induce all relevant aspects of human action and perception. This seems to approach a general capability for implementing *all* perceivably different causal relationships between human actions and changes in heard musical sound.

Realizing a system capable of completely computed instrumental control seems hard. We can work toward this goal, however, by progressively developing transducers combined with Turing-complete automata. We call this process the *computational liberation* of instrumental control, as it will *gradually minimize constraints* on implementable causal relationships (see Section 1.5.1.2). We view computational liberation then as a de facto ongoing historical process, exemplified in the development of computed sound.

What area to consider next for computational liberation? Fingertip use is extremely important to the instrumental control of musical sound (see Section 1.2.5). Therefore, it makes sense to give priority to the area of fingertip touch.

To describe the prerequisites for computational liberation that are specific to fingertip use, we formulate a model for *computed fingertip touch* (see Section 1.5.2.1). Here, a Turing-complete automaton may be causally linked, via transducers, both to human somatosensory perception and to human motor activity involving the fingertips. The model is explicitly rooted in existing knowledge on fingertip movement and touch, including its anatomy, physiology, and neural processes (see Section 1.3 and Appendix A). This allows us to explicitly distinguish between different fundamental subtypes of touch. In Chapter 4, this results in working definitions for *computed passive touch*, *computed active touch*, and *computed manipulation*. These working definitions differentiate based on the presence and dynamics of a perceptually induced exterospecific component.

Within the area of fingertip use, a specific type of movement deserves additional priority. In Section 1.2, after considering fingertip use throughout time and across musical instruments, we identify *unidirectional fingertip movement orthogonal to a surface* as a widespread, common component. Here, fingertip movement approximates a single path of movement, at right angles with a surface, and extending across at most a few centimeters. Technologies implementing the computed fingertip touch model should support this type of fingertip movement with priority, because it offers advantages in terms of control precision, effort, speed, and multiplication (see Section 1.2.5). This motivates a number of choices directing transducer development: to have the human fingerpad as transducer source and target; to create transducers based on flat, closed, and rigid contact surfaces; and to provide orthogonal as well as parallel force output to the fingerpad (see Section 1.6.2).

Finally, for reasons that are discussed in Sections 1.4.4 and 1.6.7, it is crucial that technologies implementing the computed fingertip touch model are developed in such a way as to be mass-producible, cheap, and powerful.

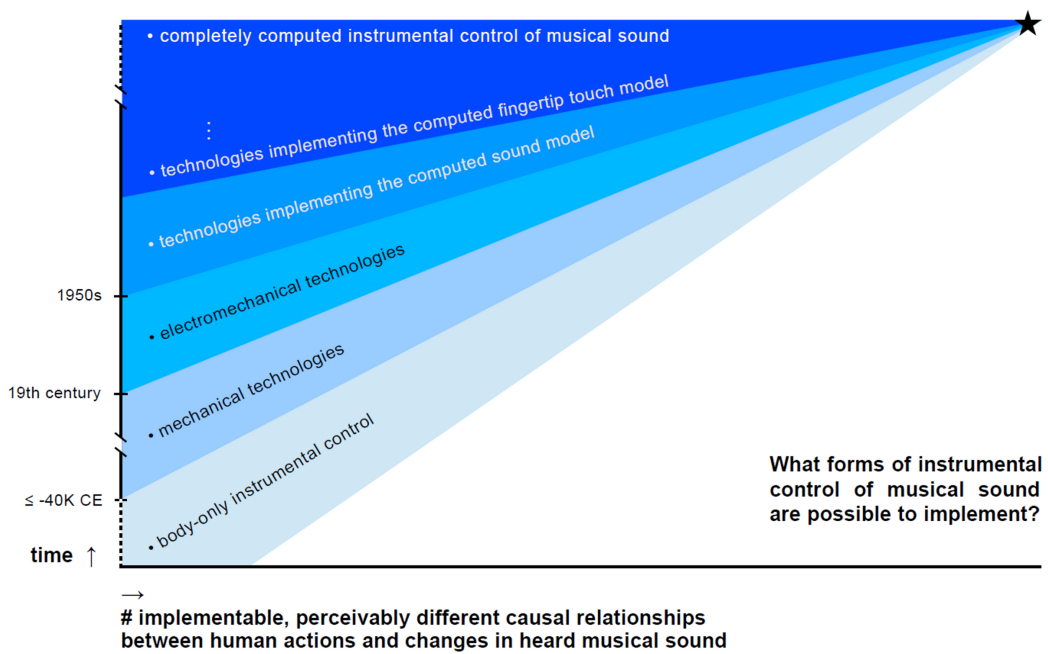

**Figure I** *Overview. What forms of instrumental control of musical sound are possible to implement?*

• *Along the vertical axis: time. Along the horizontal axis: the number of implementable and perceivably different causal relationships between human actions and changes in heard musical sound.*

• *The rising edges then indicate the increasing contributions, over time, by body-only instrumental control and by different types of technology. The gradual shift in background hue at each edge is meant to indicate that these different contributions combine, to enable the overall total of implementable causal relationships.*

• *As is shown, body-only instrumental control has contributed from the earliest times. Mechanical technologies have contributed from at least 40 millennia BCE; while electromechanical technologies have done so, roughly, from the 19th century onward.*

• *From the 1950s onward, the labels are shown in white, to highlight the turn to a process of computational liberation: Here, the use of computational technologies enables an explicit minimization of constraints on implementable causal relationships.*

• *Zooming in, contributing technologies here include those implementing the computed sound model; the computed fingertip touch model; and other, as yet unspecified models.*

• *The star in the upper right of the graph marks a hypothetical point in the future where this development arrives at a theoretical endpoint: the completely computed instrumental control of musical sound.*

## Hypothesis

Based on the above, we formulate the following *Hypothesis:*

> **H**
>
> Developing technologies that newly fit the computed fingertip touch model enables the implementation of new forms of instrumental control of musical sound.

Here, the implementation of new forms of instrumental control should result after some *limited* amount of time and effort. This makes it possible to design experiments disproving the hypothesis, ensuring its falsifiability.

## Testable predictions

In Chapters 2 and 3, we develop the *cyclotactor (CT)* system, which provides fingerpad-orthogonal force output while tracking surface-orthogonal fingertip movement. This system will newly implement the computed fingertip touch model. Then, by deduction, the Hypothesis yields the following *Prediction 1:*

> **P 1**
>
> The CT system developed in Chapters 2 and 3 will enable the implementation of new forms of instrumental control of musical sound.

In Chapters 2 and 3, we also develop the *kinetic surface friction transducer (KSFT)* system, which provides fingerpad-parallel force output while tracking surface-parallel fingertip movement. This system, too, will newly implement the computed fingertip touch model. Then, by deduction, the Hypothesis yields the following *Prediction 2*, as well:

> **P 2**
>
> The KSFT system developed in Chapters 2 and 3 will enable the implementation of new forms of instrumental control of musical sound.

## Experimental results

The first part of the experimental results is about technology: How do the CT and KSFT systems each indeed newly implement the computed fingertip touch model? Tables IIa and IIb, below, first describe the main qualitative and quantitative aspects of both technologies. Their resulting capabilities, in terms of human action and perception, are then summarized in Table IIc. Here, the phrase "complete integration with computed sound", used in Table IIa, means that input from human motor activity, output to somatosensory perception, and output to auditory perception are easily combined within a single written algorithm. Table III then concludes this part, by listing novel aspects of each technology. An important property here is *to avoid the use of connected mechanical parts moving relative to the target anatomical site*. We emphasize this as a general principle for transducer construction, with the potential benefit of enabling more precise output to somatosensory perception.

| qualitative aspect | system | |
|---|---|---|
| based on the use of a flat, closed, and rigid contact surface | CT | KSFT |
| provides *fingerpad-orthogonal* force output, tracks *surface-orthogonal* fingertip movement | CT | |
| provides *fingerpad-parallel* force output and tracks *surface-parallel* fingertip movement | | KSFT |
| I/O made programmable in physical units; via SuperCollider classes | CT | KSFT |
| complete integration with computed sound | CT | KSFT |
| precisely adjustable, personal fit, for more accurate & comfortable I/O | CT | |
| cheaply mass-producible | CT | KSFT |

**Table IIa**  *Qualitative aspects of the technologies developed in Chapters 2 and 3.*

| quantitative aspect | | CT system | KSFT system |
|---|---|---|---|
| input | spatial resolution | 0.2 mm | 0.02 mm |
| | spatial range | 35.0 mm | hundreds of cm$^2$ |
| | temporal resolution | 4000 Hz | 125 Hz, average |
| output | force resolution | ± 0.003 N | N/A |
| | force range | bipolar, varying over distance: see Chapter 3, Figure 3.2 | 0.14-1.43 N, kinetic friction |
| | temporal resolution | accurate wave output up to 1000 Hz | features between 1-10 ms |
| I/O | latency | 4.0 ms | 20.5 ms, average |

**Table IIb**  *Quantitative aspects of the technologies developed in Chapters 2 and 3.*

| resulting capability | system | |
|---|---|---|
| force output can co-determine the movement of fingertip control actions | CT | KSFT |
| I/O can induce aspects of haptic perception | CT | |
| accurate mechanical wave output across the frequency ranges involved in fingertip vibration perception | CT | |
| excellent support for real-time instrumental control of musical sound | CT | |
| I/O can induce high-resolution aspects of fingertip surface texture perception during active touch | | KSFT |

**Table IIc**  *Resulting capabilities, in terms of human action and perception, of the technologies developed in Chapters 2 and 3.*

| novel aspect | system |
|---|---|
| explicit and specific support for unidirectional fingertip movement orthogonal to a surface | CT |
| transducer *completely avoids* the use of connected mechanical parts moving relative to the target anatomical site | CT |
| I/O specific to those flexing movements of the human finger that are independent, precise, and directly controlled by the motor cortex | CT |
| transducer *partially avoids* the use of connected mechanical parts moving relative to the target anatomical site | KSFT |
| inducing high-resolution aspects of fingertip surface texture perception using cheap, off-the-shelf optical mouse sensor input | KSFT |

**Table III** *Novel aspects of the technologies developed in Chapters 2 and 3.*

The final part of the experimental results is about forms of instrumental control: What new forms have been implemented using both systems? This is summarized in Table IV. In its first examples, computed fingertip touch was used to display the state of the sound-generating process – at a higher level of detail than provided by existing technologies. This to better inform, and thereby alter, fingertip control actions. In the final examples, new forms of instrumental control were pursued more directly: Here, computed touch was used to implement new types of fingertip control action.

| see | type | system | key points |
|---|---|---|---|
| 4.2.2 | passive touch display | CT | • display of granular synthesis at the level of individual grains<br>• via presence, duration, amplitude, and vibrational content of fingerpad-orthogonal force pulses<br>• using a timescale identical to that of sonic grains |
| 4.3.2 | active touch display | KSFT | • during actions similar to turntable scratching: display more specific to the stored sound fragment<br>• via fingerpad-parallel friction, millisecond resolution |
| 4.3.3 | active touch display | CT | • during surface-orthogonal percussive fingertip movements<br>• touch display expanded outside moment of impact |
| 4.5.2 | control action | KSFT | • pushing against a virtual surface bump<br>• using horizontally applied output forces, during horizontally directed fingertip movements |
| 4.5.3 | control action | CT | • fingertip tensing during force wave output<br>• can be used simultaneously with control based on surface-orthogonal fingertip movement |

**Table IV** *New forms of fingertip instrumental control presented in Chapter 4.*

# Discussion

Below, we will first discuss the results of two research excursions (for their motivation: see the Acknowledgments). We then conclude this summary by reflecting on the main experimental results, presented above. In the final chapter, Section 7.6 will follow up on this, with some brief speculation on future developments based on the concepts of *computed manipulandum* and *computed instrument*.

The first research excursion followed the phenomenon of unidirectional fingertip movement orthogonal to a surface elsewhere. Chapter 5 presents *one-press control*, a fingertip input technique for pressure-sensitive computer keyboards, based on the detection and classification of pressing movements on the already held-down key. We show how this new technique can be seamlessly integrated with existing practices on ordinary computer keyboards, and how it can be used to simplify existing user interactions by replacing modifier key combinations with single key presses. In general, the proposed technique can be used to navigate GUI interaction options, to get full previews of potential outcomes, and then to either commit to one outcome or abort altogether – all in the course of one key press/release cycle. The results of user testing indicate that effective one-press control can be learned within about a quarter of an hour of hands-on operating practice time.

The second research excursion followed the idea of using computation to induce aspects of human perception elsewhere. In Chapter 6, we first consider how the use of techniques incorporating stages of automated computation offers visual artists a control over perceived visual complexity that is otherwise unattainable. This motivates the question whether computed output also can induce 2D shape perception in the *retinal afterimage:* the familiar effect in the human visual system where the ongoing perception of light is influenced by the preceding exposure to it. A fundamental problem here is how to exclude the possibility that shape recognition is caused by normal viewing of the stimuli, which may occur simultaneously. To solve this, we develop a rasterization method, a model of the afterimage intensities it induces, and then a series of candidate formal strategies for concrete rule sets computing stimuli. The Electronic Appendix to the chapter provides video examples, the image sequences used in a pilot experiment, and also software implementing the approach, in source format. The results of the pilot experiment testing the approach confirm shape display specific to the retinal afterimage. This result also demonstrates feasibility of the *computed retinal afterimage* in general.

Turning to the main results, the contents of Tables IIa to IV confirm both Prediction 1 and Prediction 2. This supports the Hypothesis: Developing technologies that newly fit the computed fingertip touch model enables the implementation of new forms of instrumental control of musical sound. At the outset of this summary, we presented an articulated, empirical view on what human music making is, and on how this relates to computation. The experimental evidence which we obtained seems to indicate that this view can be used as a *tool* for systematically generating models, hypotheses and new technologies that enable an ever more complete answer to the fundamental question as to what forms of instrumental control of musical sound are possible to implement.

# Acknowledgments

This thesis is the outcome of an "*autonomous Ph.D.*" trajectory, a probably unusual experiment in graduate research.

Here, as elsewhere, the candidate is recruited from the ranks of M.Sc. students, and then employed by the University. However, unlike elsewhere, the candidate is also given carte blanche as to what to research – a very rare opportunity indeed. Furthermore, the candidate does not become part of a research group (for a definition, see [Gosling and Noordam 2006]); does not receive supervision by a scientist publishing in the field of choice; and is encouraged to work on multiple subjects in different fields.

I thank Clara Takken for being a rock solid true friend throughout the ups and downs this entailed.

Aside from the research excursions of Chapters 5 and 6, it turned out that the main subject itself already required entering multiple different areas of expertise. These included computer science; new musical interfaces; the history and prehistory of musical instruments; related fundamentals and specifics of human anatomy and neural processes; the construction of high-performance electronics; and haptics.

Of these areas, my formal background included computer science only. Many thanks, therefore, to the Electronics Department (ELD) at the Leiden Institute of Physics, to the Fijnmechanische Dienst (FMD) at the Faculty of Science, and to the Elektronische Werkplaats (EWP) at the Royal Conservatoire – more specifically, to Rene Overgauw, Arno van Amersfoort, and Lex van den Broek – for generously providing their indispensable advice and practical support, regarding certain electronic components, during various stages of prototype development (see Chapters 2 and 3).

Also thanks to Adriana Zekveld, Edwin van der Heide, Gerrit van der Veer, Laura Wisse, Prosper de Jong, and Stephen Sinclair, for kindly providing references, comments, pointers, and support. Thanks to Art Bos and Irene Nooren for the institutional support when filing the patent.

Thanks to all test subjects for kindly volunteering in the absence of payment.

One consequence of not being part of a research group was receiving first feedback during peer review. This provided a non-optimal learning environment. I thank Walter Kosters for creating an exception in this regard, by kindly providing beforehand feedback on the formal notation used in Chapter 6. The research excursion in that chapter was inspired by the Lilac Chaser illusion by Jeremy Hinton [Hinton 2005], and by its online explanation by Michael Bach. In general, I would like to thank the anonymous reviewers who provided constructive criticisms and helpful suggestions.

Another consequence of not being part of a research group was sole authorship. The happy exception in this regard was setting up an informal group with the M.Sc.

students Jeroen Jillissen, Dünya Kirkali, Alwin de Rooij, Hanna Schraffenberger, and Arnout Terpstra, for the research excursion in Chapter 5. Many thanks to them, for their sustained efforts participating in this project. Also, many thanks to the Media Technology programme at Leiden University, for funding the presentation by one of the M.Sc. students at the UIST conference in Canada. And of course, many thanks to Paul Dietz and the Microsoft Applied Sciences Group, for providing us with pressure-sensitive keyboard prototypes.

Going to all conferences alone (in Europe, including Paris, Genova, and the workshop in Stockholm), it was very important to meet nice people, from various continents. I thank them all, and especially Chris Kiefer, Robin Price, Peter Bennett, and Andy Dolphin, for the interesting discussions and good company.

I have powerful memories of the three separate visits to the US to present my work at conferences. These were held at Carnegie Mellon in Pittsburgh, at MIT in greater Boston, and at Georgia Tech in Atlanta. Thanks to the US National Science Foundation (NSF) for funding one of these visits. Also, many thanks to Mark D. Gross, Ellen Yi-Luen Do, Ivan Poupyrev, and Ian Oakley, who guided the fantastic TEI Graduate Student Consortium I attended, and to Ofer M. Shir for the tour of Princeton.

Another consequence of not being part of a research group was the urgent need to try and find senior scientists publishing in the field through conference visits. I was happy to meet Sile O'Modhrain during my demonstration session at MIT, which then turned into an exciting ad hoc experimentation session, testing the KSFT system of Chapter 3 to its limits. I received her advice and also a kind invitation to visit the Sonic Arts Research Centre (SARC) at Queen's University Belfast, where I also met some of her Ph.D. students and other visitors.

I would also like to thank the people I met at the EuroHaptics conference in Amsterdam. Given the many participants and my lack of background in haptics, winning the Best Demonstration Award with the CT system of Chapter 3 was something I had not dared to hope for. It was a very special moment. I am also very grateful to Cécile Pacoret, who subsequently invited me and the CT system over to the Institut des Systèmes Intelligents et de Robotique (ISIR) in Paris. I was glad to be there, to see everything and meet everyone, including Vincent Hayward, to whom I am grateful for his feedback and advice.

Last, but certainly not least – considering the very good memories of the many and happily busy demonstration sessions – thanks to all the people who have tried my demos over the years, including those shown on the opposite page. Thank you for your feedback, and thank you for your meaningful encouragement.

Staas de Jong, Leiden, 2014.

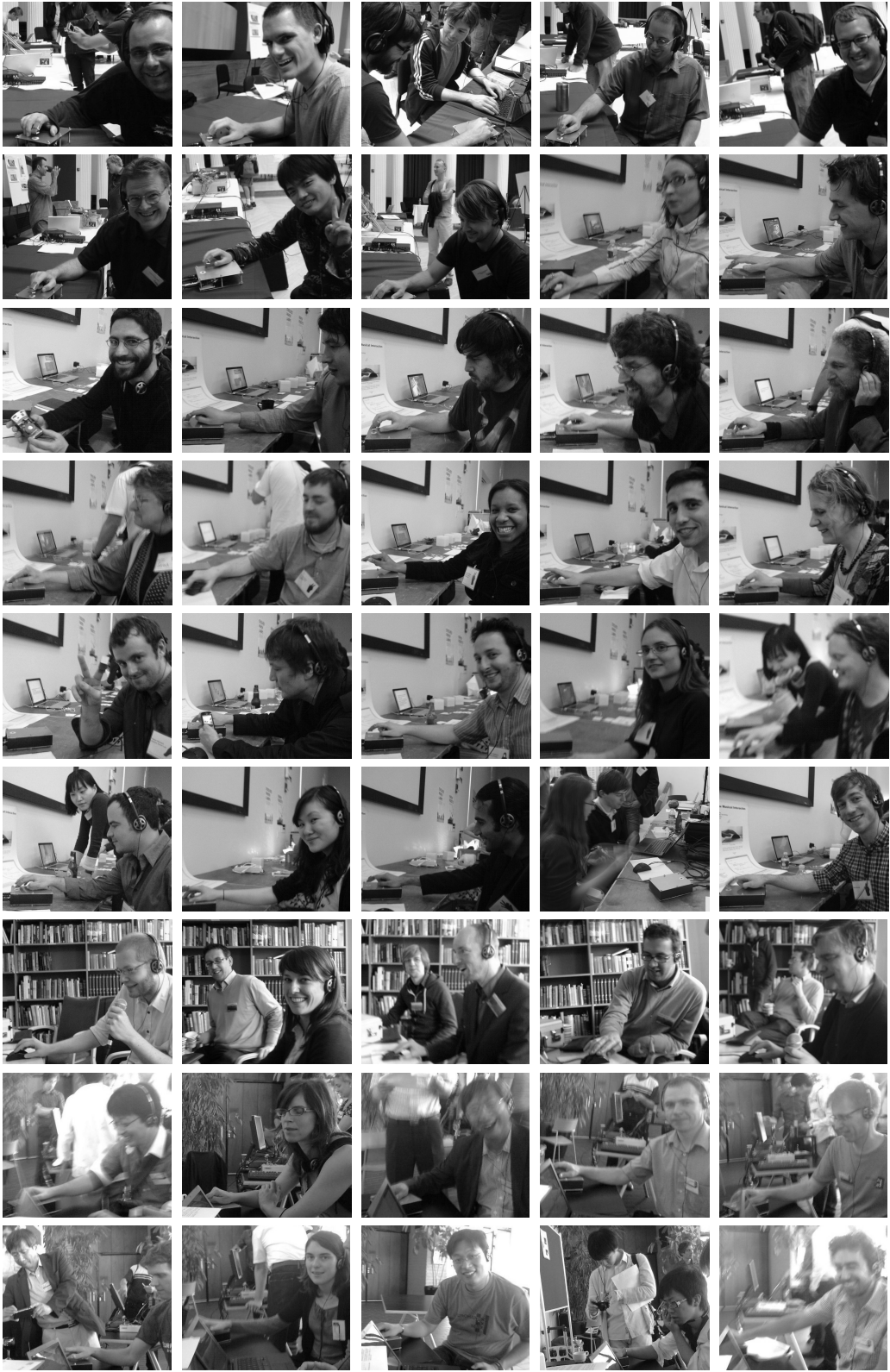

# 1. The computational liberation of music making

This text will be concerned with the human activity of making music. In Section 1.1, this is defined more precisely as *instrumental control of musical sound:* the phenomenon where human actions result in sound which induces musical experiences. This involves the *instrumental control* of human actions making changes to some *sound-generating process*, resulting in *heard sound* which induces *musical experiences* within the brain. *Musical instruments* are understood separately, as tools, external to the human body, which may be used in this overall process.

In Section 1.2, fingertip use is discussed chronologically, from ancient prehistory to the beginning of the 21st century, and across a wide range of widely used musical instrument types, incorporating acoustic, electromechanical, and electronic sound-generating processes. A widespread, common component is identified: *unidirectional fingertip movement orthogonal to a surface*.

To articulate an understanding of the human fundamentals of fingertip instrumental control, in Section 1.3 the functional anatomy, physiology, and neural processes of fingertip movement and touch are discussed. Central turn out to be *consciously* and *unconsciously* occurring processes of *passive touch*, *active touch*, and *manipulation*. During these, the activation of *learned motor programs* guided by *somatosensory transduction only* gives crucial advantages for instrumental control of musical sound.

In Section 1.4, the introduction of *computed sound* is discussed, and how it led to a wide variety of new and widely used types of sound-generating processes. Of these, an overview is given, based on how instrumental control moved away from direct manipulation of the wave table. Fundamental reasons for this historical development are discussed, including Turing-completeness of the *computed sound model*, and the development of cheaply mass-producible, powerful implementations of this model.

In Section 1.5, computed sound is regarded as an example of a larger historical development: the *computational liberation* of instrumental control. Here, in contrast with earlier technologies, Turing-complete automata combined with transducers explicitly minimize constraints on implementable causal relationships between human actions and changes in heard musical sound. This makes computational liberation itself a fundamental goal when generally aiming to improve instrumental control of musical sound. We will pursue this goal, based on a *computed fingertip touch model*.

This is made more precise in Section 1.6, resulting in a series of interrelated, concrete goals for the body of this thesis. These are divided into obtaining *new transducer technology* (Chapter 2); for *new systems for computed fingertip touch* (Chapter 3); to then enable *new forms of instrumental control* (Chapter 4).

# 1.1 Making music: Instrumental control of musical sound

This text [1] will be concerned with the human activity of making music. We will aim to approach this subject using clearly defined concepts based on empirically observable physical processes. Our main subject will be *instrumental control of musical sound,* which we define as the phenomenon where human actions result in sound which induces musical experiences. This means regarding *musical sound* as sound which induces musical experiences, and it may be appropriate to first further clarify this, by briefly addressing three basic questions: What do we mean by sound; how does it lead to musical experiences; and how can these be characterized? By sound, we mean the propagation of pressure waves through some medium, usually the air, without this necessarily being subject to detection by the human senses. Such sound explicitly becomes heard sound if it leads to movement of the human eardrums, which is then transformed into nerve impulses going into the brain. There, the auditory cortex area, indicated in Figure 1.1, acts as a gateway for heard sound entering human consciousness [Mathews 2001].

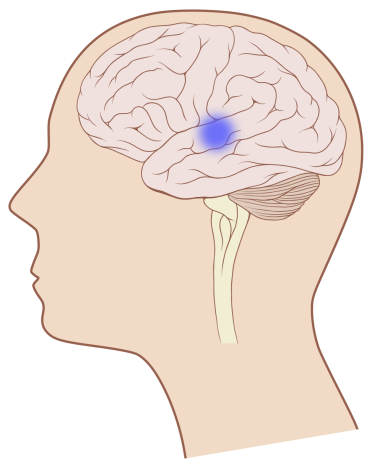

**Figure 1.1**  *The auditory cortex of the human brain (location indicated). Activity in many other regions of the brain is involved in musical experiences, but this area of the left and right hemispheres acts as a gateway for heard sound entering consciousness [Wolters and Groenewegen 2004]. (Image based on a diagram of the human skull and brain by Patrick J. Lynch and C. Carl Jaffe, 2006.)*

Musical experiences in the brain, and how heard sound leads to them, are the subject of ongoing research. It has been made possible to detect the presence of various types of musical activity in the brain by measurement. The perception of musical aspects of heard sound, such as pitch, rhythm, timbre, and harmony, are established subjects of neuroscientific and neuropsychological research. A popular introduction to this, with many scientific references covering the subjects of musical activity in the human brain as well as the evolutionary basis of music, is [Levitin 2006]. In general, it can be said that musical experiences represent a complex and much-valued aspect of human experience, omnipresent across cultures and over time.

---

1   This chapter, while also introductory in nature, should be read *after* the Introduction, which explicitly motivates all of its parts. These parts here follow a very different progression, however, intended to be more natural to the non-specialist reader.

Consideration of music as a phenomenon of the brain underscores that musical experiences may also be induced by other types of brain activity than hearing, such as memory recall, imaginative mental processes, dreams, and hallucinations. Activity in other parts of the body may be involved as well, for example during body movement such as dancing. We will assume situations where it is primarily the sound being heard which induces musical experiences.

Where there is sound, there always is a sound-generating process. Human actions can result in sound by making changes to this sound-generating process, whatever it may be. Here, by human actions, we will mean observable bodily actions, as opposed to purely mental acts such as thoughts. A basic example of how such actions may make changes to a sound-generating process is shown in Figure 1.2, which depicts a single person hand-clapping rhythmically.

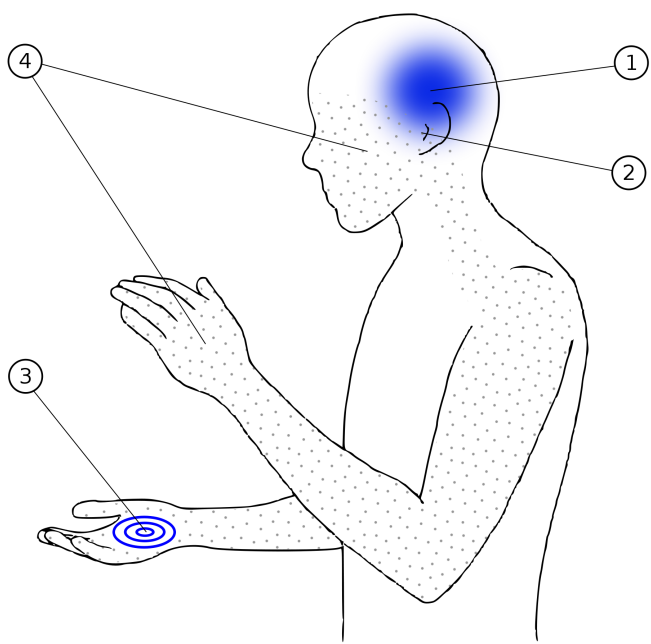

**Figure 1.2** *Symbolic representation of a basic example of the instrumental control of musical sound: a single person hand-clapping rhythmically. Four causally related aspects are highlighted by the symbols in the figure. Clockwise, from the top right: (1) The induced musical experience within the brain. (2) Heard sound, the pressure waves in the air and surroundings that are heard as they resonate with the human eardrums. (3) The sound-generating process. In this case, the contacts made over time between the surfaces of the hands. (4) Instrumental control of this process. Here, the rhythmical clapping actions that extend from the brain to the limbs, performed with feedback from the senses, such as touch, audition, and vision.*

The example of Figure 1.2 emphasizes that for making musical sound, musical instruments external to the human body are not at all required. Other examples that could have been used to make this point are whistling, and using the voice to sing. The various body parts involved in such examples are sometimes also put under the term "musical instrument". And in recent classification systems for musical instruments, sometimes those parts of the body engaged with external objects to make musical sound were also considered elements in the overall "instrument" (see [Kvifte 2007]). We will not do this, and reserve the term "musical instrument" exclusively for objects external to the human body. This in order to support a clear distinction in regarding musical instruments as external tools, on the one side, expanding the possibilities of bodily actions, on the other side, for the control of musical sound. The influential organologist Hornbostel defined "musical instrument" as everything with which sound can be produced intentionally [Kvifte 2007]. In the introductory text of [Gillespie 2001], any device under the control of a human player that produces sound which carries meaning is considered a "musical instrument". Here, we would consider such instruments musical only if the sound produced induces musical experiences.

We will regard *instrumental control* as the way in which observable bodily actions make changes to a sound-generating process, via tools or purely via the use of parts of the body. In this way, instrumental control of musical sound is related to a wide range of other forms of tool use and bodily behavior that have evolved in humans and other organisms. Figure 1.2 further illustrates how we understand the instrumental control of musical sound to be composed of the causally related concepts of *instrumental control*, *sound-generating process*, *heard sound* and *induced musical experience*.

This definition of instrumental control is wide-ranging also because it covers any bodily action resulting in musical sound. For example, in the context of Western classical music, this means everything from a composer acting to note down a musical idea, to a musician playing a musical instrument, to an engineer operating sound recording technology, to a consumer activating "play" on his or her listening device. The sound-generating processes operated on by human actions can then be thought of as long, branching chains which connect many interrelated people and instruments, possibly stretching across an extended period of time. Indeed, by the time the musical experience is induced, the composer may be long dead. Instrumental control in this way not being tied to individual musical instruments is one important difference with a concept such as "instrument control" as defined in [Kvifte 2007]. Charting the many possible types of chains of instrumental control would reflect the many types of cooperation on and co-authorship of musical works that may exist over time. Here, however, we will typically consider much shorter timescales, as in the situation illustrated by Figure 1.2. This means that instrumental control happens in real time, simultaneously with the resulting heard sound.

In general, not all musical sound will be the result of instrumental control as we have defined it. For example, a person might go outside, listen to naturally occurring birdsong and as a result have musical experiences. Our definition involving human actions making changes to a sound-generating process would be stretched too far, if it were considered to also cover such a situation. Another important point is that imaginative mental processes, which of course also determine how musical sound is made, may very well occur quite separately from any bodily actions making changes to a sound-generating process.

Still, the concrete ways in which humans actually make musical sound are crucial in determining what musical ideas can be made heard and how this happens. Instrumental control can be a direct source for musical imagination, in the exploration of sonic possibilities and resulting musical experiences that it allows. Musical experiences had in the past, in general, will have been shaped in part by the specific types of musical sound heard resulting from specific types of instrumental control. This too, more indirectly, will influence imaginative mental processes, however detached they may seem from observable bodily actions. For such reasons, the study and development of instrumental control of musical sound is important to the boundaries and possibilities of musical sound, and thereby musical experience, in general.

## 1.2 Use of the fingertips: The prevalence of unidirectional fingertip movement orthogonal to a surface

**1.2.1 Ancient origins**    A fundamental reason for the instrumental control of musical sound to naturally move from the tool-like use of parts of the body to the use of all kinds of external musical instruments, is to gain access to more types of sounds, and thereby musical experiences. External instruments may become used e.g. via the mouth, hands, and feet. The fingertips, due to this transition, may at once become much more effective as means for instrumental control: In general, their use will bring all the potential advantages for fine manipulation and control of tangible objects that are inherent in being the anatomical endpoints of the hands and fingers. However, where sound is made using only the body, these advantages may be overshadowed by equally inherent limitations. For example, body-only percussion by snapping the fingers may produce musical sound as clearly perceivable as when clapping the hands; but for many other types of percussive movement, this will not be the case, simply because of the smaller sizes and applied pressures involved when using the fingertips. The use of appropriate external sounding objects, either found or made, can eliminate this as a problem. Transitioning to these, the spatial and temporal precision in manipulation that is associated with the fingertips can be put to use, with each finger adding more possibilities for simultaneous change to the sound-generating process being controlled.

A prime and ancient example of this is the flute, a musical instrument that can typically be described as a pipe perforated with holes. The mouth is used to blow onto one of the openings, creating a moving volume of air across it. The fingertips can then be used to cover and uncover other holes, causing resonations within the air column inside the instrument to occur at differing frequencies – which may then be heard as musical sound. The flute is the oldest type of musical instrument to have been recovered so far. At the time of writing, the most ancient undisputed finds have been of bird bone and mammoth ivory flutes, made at least 42 millennia ago [Higham et al. 2012] [Conard et al. 2009]. For a relatively recent example, see Figure 1.3.

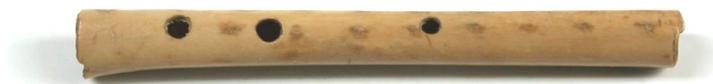

**Figure 1.3** *Ancient bone flute with holes for the fingertips. A 20 centuries old provincial Roman flute, made from a bird bone (15-30 CE, Rijksmuseum voor Oudheden). The oldest flute that has been found dates further back to much earlier, prehistoric times. Made at least 420 centuries ago, it is the oldest musical instrument known to exist.*

**1.2.2  Historical times**    From later, historical times, more evidence is available of other types of widely used musical instruments based on use of the fingertips. This includes drums, another ancient category of instruments. In the top row of Figure 1.4, the terracotta figure of a woman playing a frame drum, a type of hand-held drum, is shown. With this instrument, the fingertips can be used to make various types of rapid percussive movements, on the membrane of the drum, resulting in heard rhythmic sound. In the same row of Figure 1.4, a 4.6 millennia old picture of a man playing a lyre is shown. Here, the fingertips are used to pluck or strum separate strings, which when moved produce sound at differing frequencies.

The middle row of Figure 1.4 gives a snapshot of stringed instruments more than 4000 years later, showing examples from the Western classical tradition that are still in wide use today. Of these, the guitar, too, is an instrument where fingertips are used to pluck and strum strings, possibly using a small tool called a plectrum. Here however, the fingertips are also used to press against strings, in order to change their effective length and thereby the produced sound frequencies. With the violin, the plucking and strumming of one hand is largely replaced by bowing. The bow is moved across the strings mainly due to hand and arm movement, but is still held between the fingertips. Finally, in the keyboard instruments that have culminated via Bartolomeo Cristofori's fortepiano hammer action [Harding 1933] in the modern piano, there no longer is direct contact between fingers and strings. The fingertips are used to press down on a row of keys, activating hammers that hit the strings producing the sound. The general usefulness of keyboard control is illustrated by its presence in other widely used instruments, such as the pipe organ (not shown). There, pressing keys controls whether air moves through pipes, which then produce sound in a way similar to that of flutes. Another similarity to flutes, regarding their fingertip operation, exists in other instruments, including many types of wind instrument. In woodwind instruments such as the clarinet, the oboe and Adolphe Sax's saxophone, pressing the fingertips is again used to close holes, either directly or via some mechanism. In brass instruments such as the trumpet, the horn and Wilhelm Wieprecht and father and son Moritz's tuba, fingertip presses operate valves in order to change the vibrating air column.

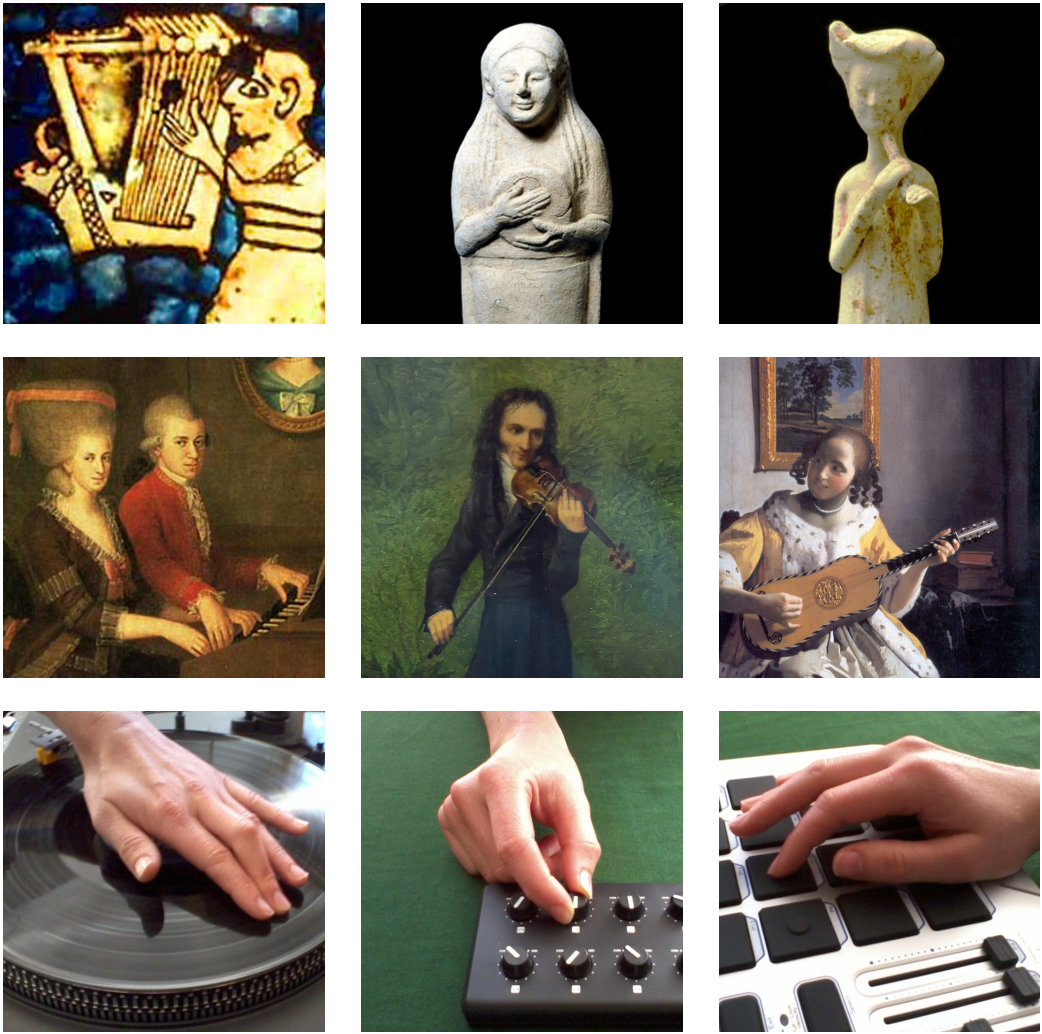

**Figure 1.4** *Fingertips on musical instruments. Top row, left: man playing a lyre, a stringed instrument of ancient Ur (detail from a mosaic, 27th-26th century BCE, British Museum). Top row, middle: woman playing a Phoenician frame drum (detail of a terracotta figure from Tharros, Sardinia, 7th-6th century BCE, British Museum). Top row, right: woman holding a flute (detail from a group of musician statues, China, Sui dynasty, 581-618, Museum Rietberg). Middle row, left: Mozart and his sister playing an early piano (detail from a contemporary family portrait oil-painted by Della Croce, 1780-1781). Middle row, middle: Paganini playing a violin (detail from a portrait oil-painted by Kersting, 1830-1831). Middle row, right: girl playing a baroque guitar (detail from an oil painting by Vermeer, 1670-1672). Bottom row: photographs showing fingertips on three musical instrument types from the late 20th century. From left to right: fingertips on a vinyl record on a turntable; fingertips on a plastic rotary knob; fingertip touching a pressure-sensitive rubber pad.*

**1.2.3 The 20<sup>th</sup> century**    The instruments mentioned so far have all been acoustic, meaning that their sound-generating processes are based on mechanical waves only. During the 19<sup>th</sup> century and the two centuries before it, many discoveries and inventions relating to electricity were made. During the 20<sup>th</sup> century, this enabled the introduction of many new instruments based on electromechanical and electronic sound-generating processes. A widely popular example of this is the solid-body electric guitar, in which the mechanical vibrations of strings are picked up by an electric coil wound around a permanent magnet, generating an electric signal. This signal can then control separate amplifier equipment, producing sound at louder levels than those of an acoustic guitar, and thereby suitable for larger audiences. Also, the signal can be electronically altered, resulting in sounds dramatically different from those produced by acoustic guitars. Still, the ways in which the fingers are used to play the instrument remain similar, with the fingertips of one hand pressing down on strings that are plucked and strummed via the fingertips of the other hand.

In another type of electricity-based instrument, the analog synthesizer, heard sound originates from a purely electronic process. Circuits connecting electronic oscillators with many other types of components are used to generate signals that then control sound-producing electromechanical devices. Circuit properties altering the resulting musical sound can be made subject to manual control. Sometimes, this happens using controller devices specifically designed for musical application, such as pitch-bend and modulation wheels operated by the fingertips. However, there also has been a major trend to adopt general-purpose, fingertip-operated input devices for the instrumental control of musical sound. This includes switches that can be flipped, buttons that can be pushed, linear sliders that can be moved, and rotary knobs or dials that can be turned. (See the bottom row of Figure 1.4 for an example.) All of these devices can also be found in the control panels of other types of electronic equipment. However, to activate variably-pitched sounds, the favorite controller remained a specifically musical, earlier device: the piano-type keyboard.

Another, again electromechanical instrument for producing musical sound is the gramophone record player, or turntable. Here, a needle traces a spiralling groove on an otherwise flat spinning disc, converting the mechanical vibrations due to small variations along the groove to electric signals. After initial activation, playback happens automatically, as this method of sound reproduction was not developed for makers of, but for listeners to musical sound. Still, by the 1970s, people following the example of DJ Kool Herc were manually changing turntable playback, so as to alter and prolong fragments of rhythmic music being played to dancing crowds [Chang 2005]. This was done by electronically switching between the signals from two separate turntables using an electronic mixer device, while manipulating the two rotating records using the fingertips. This shows that it is appropriate to consider even the very end stages, intended only for the final conversion of stored signal to heard sound, as part of a chain of instrumental control of musical sound: As long as manipulative interventions are possible, they may occur, extending the chain while drastically altering, or even largely determining, the resulting musical sound. The further development of turntablism brought widely used techniques that went beyond recognizably reproducing and recombining stored fragments of sound. In scratching, for example, the fingertips are pressed on the record and then moved parallel to the turntable surface using swift hand and finger movements. This drags the record groove back and forth under the needle (see Figure 1.4, bottom row). This distinct type of fingertip operation results in distinct types of musical sound, which may no longer be recognizable as variations of the specific stored sound being used.

Around the same time, another new category of instruments was developed that was also based on the playback of recorded sound fragments. In the digital sampler, however, sound signals were now stored purely electronically. The control of their timed playback often occurred via the same types of fingertip-operated devices already seen in analog synthesizers. However, in drum sound samplers, such as the pioneering and widely used instruments designed by Roger Linn [Rundgren 2011], specialized pressure-sensitive pads were also included. The fingertips contact these pads as the result of downward percussive hand and finger movements, not unlike those made onto the membranes of acoustic drums. This triggers playback of the selected sound fragment, often with its loudness dependent on the force of impact. Varying the contact pressure after the initial impact may further modulate aspects of playback (see Figure 1.4, bottom row).

Of course, the appearance of the sampler was part of a much broader development of electromechanical and electronic digital technologies during the second half of the 20th century. As part of this, starting from the 1980s devices for personal computing became widely used. Among their many, progressively developed uses were also the creation, recording and processing of musical sound. This meant a further continuation of the major trend to adopt general-purpose, fingertip-operated input devices for the instrumental control of musical sound. Beginning with the desktop personal computer, such devices included the computer keyboard and mouse.

Comparing the computer keyboard to the piano-type keyboard, there are similarities as well as fundamental differences in manual operation. For their general layout, both types use rows of keys arranged next to eachother in the same general direction in which the fingers occur next to eachother on each hand. A basic use of the computer keyboard is typing, the activation of graphical symbols serially combined into words and sentences. On observation, such keyboard use will be characterized by strings of separately occurring fingertip presses, as mostly, the activated graphical symbols are not intended to be used simultaneously. Historically, such use has become supported by layouts using multiple rows of keys covering a relatively compact area, much like the melody-side keyboard of a chromatic button accordion. The basic application of the piano-type keyboard, on the other hand, is the activation of pitched sounds. For this, it has turned out that simultaneous activation is very much desired, e.g. to create musical sound based on chords and simultaneous melodies. Where in typing words, simultaneously occurring keypresses are the exception, here they became the norm. Historically, this has become supported by a standard layout where the same number of keys will be arranged across a much wider area, in one row of wider white keys which are also deeper, so as to become interspersed by a second, raised row of black keys. For a visual comparison of the keyboard layouts, see Figure 1.5.

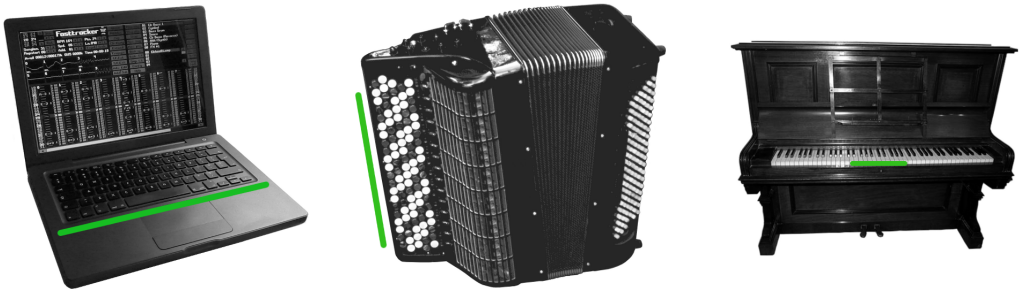

**Figure 1.5** *Comparing different keyboard layouts. From left to right, a computer keyboard, a button accordion's melody-side keyboard, and a piano keyboard. The green line added to each photograph indicates a distance of approximately 30 cm.*

Within the context of these different layouts, the individual keys are manipulated by the fingertips in press/release cycles. Initially, a fingertip may rest on the surface of a particular key, or contact it while moving. Increasingly pressing down on a key's surface will overcome its initial resistance, leading into a phase of downward travel. Travel distance extends to a certain level, beyond which increasing upward counterpressure effectively blocks further movement. Upward fingertip movement back to the initial surface level can again free the finger from counterpressure, ending the cycle so that it can be repeated again. A crucial difference within this cycle is that with a computer key, after overcoming the initial counterpressure of the key's surface, further subtleties in movement and applied pressure are not picked up or used for control, something which does happen with piano-type keys.

Still, computer keys, too, were increasingly used in the instrumental control of musical sound. This included typing the code for software producing musical sound, as well as issuing commands to such software as it executed. As Graphical User Interfaces (GUIs) became the dominant way to operate desktop personal computers, issuing such commands was also done via a mouse. (With a GUI, a computer could now dynamically draw images on an electronic screen, with one of these showing a pointer. Using one hand, the mouse, usually placed next to the keyboard, could then be moved across the underlying flat surface, with the computer changing the screen position of the pointer image correspondingly, in two directions. Changes to the information represented visually around the pointer could then be made, by pressing a mouse button located under one of the fingertips of the hand moving the mouse.) When compared to piano-type keys, mouse buttons share the similarities and difference mentioned above for computer keys.

Parallel to the general development of the desktop personal computer, many special-purpose add-on devices were developed for the manual control of software producing musical sound. These would usually incorporate existing device types such as piano-type keyboards, pitch wheels, and knobs and sliders, as these remained the favored choices for real-time instrumental control of electronic sound-generating processes.

**1.2.4 The beginning of the 21$^{st}$ century**  Into the 21$^{st}$ century, these developments continued, as the desktop computer was followed by smaller form factors for personal computing. Fingertip operation of such devices was extended with the use of rigid, flat surface areas registering the two-dimensional position of fingertip contact, during continuous movement as well as for separate taps. In the laptop personal computer, this took the form of a touch pad usually integrated next to the keyboard, in the same housing. Now, fingertip movements could be used to move the GUI pointer. In handheld personal computers that became widely used, such as smartphones and tablets, such surfaces were made to become transparent, and to coincide with the visual display screen. Electronic images could now be pointed at and changed directly at their perceived location on the screen, dispensing with the need for a separate pointer image. Moreover, multi-touch technology was introduced, simultaneously tracking the positions of such actions for multiple fingertips.

In order to get an impression of the forms of fingertip use present in recent research into new musical instruments, we will now refer to work reported at the annual international conference on New Interfaces for Musical Expression (NIME). We will do this for the five-year period from 2006 to 2010, and based on a number of trends that can be identified.

A first trend can be defined as *fingertip use based on the manipulation of existing musical instruments*. This happens in systems where an existing instrument is augmented by additional technology which expands the possibilities of playing, without fundamentally changing the existing manipulations involved. In some systems, this is done by recording and analyzing sound signals coming from the existing instrument in real time, and producing an automatically computed sonic response. Examples of this expand on the saxophone [Hsu 2006], the frame drum [Aimi 2007], and the electric guitar [Lähdeoja 2008]. In other systems, sensors are used to track the movement of parts of the existing instrument during manipulation, so as to identify and respond to various types of gestures. Examples for this expand on bowing the cello [Young et al. 2006] and violin [Bevilacqua et al. 2006], and on moving the slide on a trombone [Farwell 2006]. In another group of augmented instruments, new and separate manipulations are added to the existing ones. For the fingertips, this often means presses or percussive hits, applied to specific parts of the instrument where detection occurs via added sensors and general-purpose input devices. Examples of this expand on the electric guitar using acoustic sensing [Lähdeoja 2009] and buttons [Bouillot et al. 2008]; on the cello [Freed et al. 2006] and guitar plectrum [Vanegas 2007] using pressure sensors; on the saxophone [Schiesser and Traube 2006], flute [Palacio-Quintin 2008] and trumpet [Leeuw 2009] using buttons and pressure sensors; on the piano using key displacement sensing [McPherson and Kim 2010]; and on the turntable-with-mixer using a range of general-purpose input devices [Lippit 2006]. In yet another group of systems, the operation via the fingertips can be characterized as presenting a manipulative likeness of an existing instrument. Examples of this mimick the violin [Poepel and Overholt 2006], [Carrillo and Bonada 2010]; the piano [Takegawa et al. 2007, 2008]; the acoustic [Zoran and Maes 2008] and electric [Maruyama et al. 2010] guitar; and the ocarina [Wang 2009] and turntable [Yerkes et al. 2010].

A second trend, continuing the existing major trend already seen above, can be defined as *fingertip use adopting general-purpose input devices* for the instrumental control of musical sound. In the context of the laptop personal computer, this sometimes includes explicitly new uses for the fingertip operations on computer keyboards and touch pads, as in [Fiebrink et al. 2007]; sometimes, these fingertip operations are implicitly present, as in the screen-based instruments of [Magnusson 2006] and [Zadel and Scavone 2006]. Add-on input devices for personal computers, too, are sometimes adopted, such as the pen-like stylus held between the fingertips in [Zbyszyński et al. 2007]. Where handheld personal computers are the starting point, projects have implicitly incorporated the types of fingertip operation associated with touch screens [Geiger 2006] and multi-touch screens [Bryan et al. 2010], [Oh et al. 2010], [Essl and Müller 2010]. Other work explicitly focuses on the use of multi-touch fingertip operations, considering larger display form factors [Davidson and Han 2006] and new manipulations generalizable from the laptop touch pad [Schlei 2010].

A third trend, related to the second, can be defined as *fingertip use implicitly present in Tangible User Interfaces (TUIs)* applied to the instrumental control of musical sound. Here, the TUIs [Ishii and Ullmer 1997] usually are based on the automated tracking of untethered objects placed manually on a flat tabletop surface. Input results from the presence or absence of specific objects on the surface, from their locations, their rotational angles in the surface plane, and from the choice of down-facing side. The fingertips hold and release objects as they are picked up, laid down, and moved across or turned around on the surface. Often, as in the influential reacTable platform [Jordà et al. 2007], this is extended with the multi-touch detection of fingertip presses made directly onto the surface, and also with back projection on the table surface for dynamic visual display. Examples of projects of this type include [Jordà and Alonso 2006], [Dimitrov et al. 2008], [Mann et al. 2009], [Hochenbaum et al. 2010], and [Heinz and O'Modhrain 2010]. Some projects also allow and sense deformation of the objects being manipulated, including this possibility [Taylor and Hook 2010] or exclusively focusing on it [Kiefer 2010]. Here, the fingertips are pressing into the objects as they are manually deformed.

This overlaps with a fourth trend, which can be defined as *fingertip use via new input technologies developed specifically for the instrumental control of musical sound*. As in touch pads, the aim here often is to track the two-dimensional position on a flat surface of contact made using the fingertips. In one example of this [Nishibori and Iwai 2006], a low-cost LED-based sensing grid [Hudson 2004] is used to this end. Usually, an additional aim is to also capture and use variations in the pressure applied by the fingertips during contact. In [Wessel et al. 2007] this is done using touch pads arranged in a grid; in [Crevoisier and Kellum 2008], using infrared reflective fingertip position tracking with acoustic tap detection; in [Freed 2009] using resistive sensing techniques; and in [Jones et al. 2009] using a low-cost capacitive sensing grid. As an example of another type of input technology, in [Räisänen 2008] each non-thumb finger can operate a separate, custom rotary wooden knob.

### 1.2.5  Reflection: Unidirectional fingertip movement orthogonal to a surface    In Section 1.2, up to this point, we have mentioned many widely used musical

instruments with which use of the fingertips is crucial. The examples have included aerophones, where the sound-generating process is based primarily on mechanical vibration in volumes of air, as in the flute, oboe, clarinet, ocarina, horn, pipe organ, trumpet, tuba and saxophone. Examples have also included membranophones, where sound generation is based primarily on the mechanical vibration of membranes, as in the frame drum. Also included were chordophones, based on the mechanical vibration of strings, such as the acoustic guitar, solid-body electric guitar, violin, cello, and (forte)piano. Newer examples, of electronic instruments, included the analog synthesizer and digital sampler, as well as personal computers in their various form factors. Overall, the unomittable presence in a wide range of widely used musical instruments shows that, from prehistory to the present, *fingertip use is extremely important to the instrumental control of musical sound*.

Also, the fingertip use *in all of the examples recapitulated in the previous paragraph* contains a common component. Whether the fingertip is used to open and close holes and valves on aerophones; or to strike pads and membranes; or to tap and press on sensor surfaces; or to press strings against instrument bodies; or to perform press/release cycles on push-buttons, computer keys or the keys of piano-type keyboards: there is *unidirectional fingertip movement orthogonal to a surface*.

The conditions of its presence will vary: with different instruments, fingertip movement will result in different tangible responses. Also, other fingertip movement may be occurring simultaneously, such as the sideways movement on pressed flute holes and strings which is also used to change the sound produced. And of course, any fingertip movement may occur as part of a wide range of dynamically changing hand, arm and body postures. Still, in all of the above cases, the fingertip movement itself can be characterized as approximating a single path of movement, at right angles with a surface, extending across at most a few centimeters. This is illustrated in Figure 1.6.

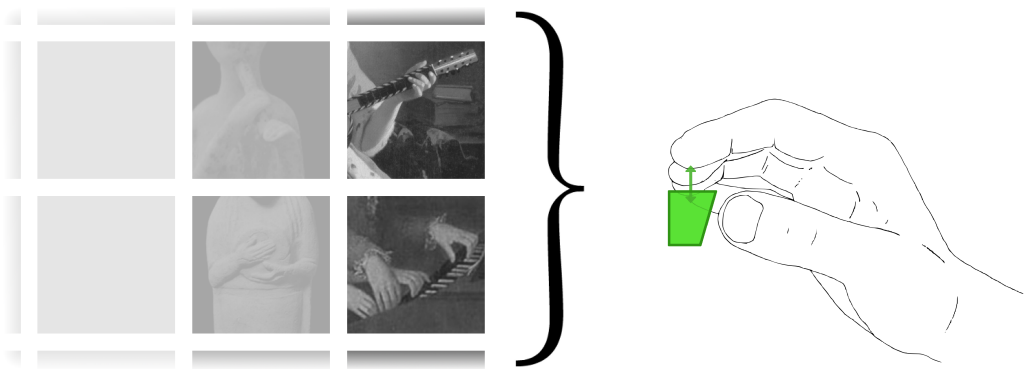

**Figure 1.6** *Unidirectional fingertip movement orthogonal to a surface: To the left, details from Figure 1.4 showing fingertip use on a flute, frame drum, guitar, and piano. This as part of a wider range of musical instruments which share a common component in their fingertip use: unidirectional fingertip movement orthogonal to a surface. This is depicted to the right, using an intermediate hand posture, and with an example surface, as wide as the fingers, seen from below.*

For the person performing it, this type of movement is associated with relatively precise control costing relatively little effort. It can also be performed at relatively high speeds, and by all fingers separately and simultaneously. This allows making more changes to a sound-generating process over a given time period. All of these positive properties for control may help explain why unidirectional fingertip movement orthogonal to a surface has been widely used in the instrumental control of musical sound, across many cultures, and for a long time, going back from the present at least hundreds of centuries.

## 1.3 The anatomy, physiology and neural processes of fingertip use

In the previous section, we have recalled many examples of fingertip use in the instrumental control of musical sound. These examples served as evidence for the great importance of fingertip use to the instrumental control of musical sound. How, in general, across all of these cases, are the changes to sound-generating processes being made via the fingertips? A first answer could be: by human actions attempting to move or hold still the fingertip. Then, to more completely answer the question, it becomes necessary to consider the implicit principles underlying this.

To do so, we will now provide an overview of some important, basic facts. These will be, first, about the anatomy of those human body parts the combined motion of which results in fingertip movement. Then, the physiology of fingertip movement and touch will be discussed. Finally, this extends to the level of neural processes. Together, these subject areas are of fundamental importance for the understanding of human fingertip use in general. Explicitly considering them may clarify how existing forms of fingertip use, and their advantages for instrumental control, are realized. Also, explicit consideration will point to fundamental boundaries of what is possible, and relevant, when fingertip instrumental control is to be improved or extended in any way.

**1.3.1 The functional anatomy of fingertip movement**    **1.3.1.1** *Scope*    Where anatomy studies the structural parts of the human body, functional anatomy studies how these parts interact in dynamical processes. For the process of fingertip movement, the relevant anatomy clearly also includes the larger body parts that incorporate the fingertips. For example, even as a pianist gently presses a single key only, the whole finger will move. More forceful downward pressing movements producing louder sounds will involve displacement of the rest of the hand, as well as the forearm. When a cellist adds vibrato to a tone, by moving a fingertip which is already being pressed down on a sounding string, the required body movement may include the upper arm, too [Fiste 2007]. These examples illustrate that, for the functional anatomy of fingertip instrumental control, it will be relevant to discuss the moving parts of the human body, starting from the fingertip, at least up to the shoulder.

Of course, other examples of instrumental control could be given to illustrate that the human body beyond the shoulder also is quite relevant here. Strictly speaking, any fingertip movement relative to the surroundings will be the combined result of movements of the entire body. It is not hard to imagine a concrete situation where, using a sensor attached to the fingertip, its relative position within a large room is made to directly control the generation of musical sound in some way. Then, movements of the entire body may be made to clearly contribute to instrumental control via the fingertip, from the smallest finger movements, to jumping up and down while running around.

However, we may also consider more ordinary situations, where people fully concentrate on the instrumental control of musical sound while using the fingertips with traditional musical instruments. It seems fair to say that the focus of attention in such situations probably is more on the more precise movements taking place from shoulder to fingertip, than on those which contribute from the torso and the other extremities. The latter movements may often be well described as enabling the other, more precise movements, by providing stable placement of the body within the surroundings, a suitably adapting dynamic body posture, and overall balance. Therefore, below, we will focus on the functional anatomy of fingertip movement relative to the torso.

**1.3.1.2** *Fundamentals of body movement*   In general, movement of the human body is produced by muscles. A muscle consists of bundles of fibers which, when activated via connecting nerve fibers, attempt to contract along a specific direction. This changes *muscle tension*, the force applied by the muscle to something else it is attached to. However, muscle tension is not only the result of the muscle contracting when activated via a nerve. In rest, too, the muscle will resist being stretched via its attachments, with counterforces increasing as it is being stretched further. When this being stretched suddenly ends for some reason, the backlash may temporarily result in the muscle being contracted. Overall muscle tension, resulting from these factors, may or may not produce *muscle excursion*: a displacement of the connecting end of the muscle. This will depend on the presence of other forces being applied simultaneously.

Often, a muscle will be attached to two separate bones, so that its contraction can move the pair relative to eachother. The muscle is then said to *originate* on the one bone, and *insert* on the other. The term *articulation* then refers to the area where such bones are closest together, and where their surfaces may be shaped to enable and guide their relative movement. Additional structures made of other living materials, including various types of cartilage, may be present to form an overall joint, which connects the bones and holds them together. The term "articulation" also refers to the process in which the bones actually move relative to eachother. Here, if a bone articulates in such a way as to rotate toward another, adjacent bone, this is often called *flexion*. Articulation in the opposite direction is then called *extension*.

It is not just so that at least two separate muscles must be present for this, so that the contraction of the one may produce flexion, the other extension. Such pairs of muscles in fact will contract simultaneously, in both cases, with movement in either direction resulting from the specific combination of muscle tensions being applied. Movement at the joints of the human body always happens based on this principle of antagonist contraction [Wolters and Groenewegen 2004]. This also means that muscles are not controlled individually, already for the most basic cases of body movement.

Especially for fingertip movement, many of the relevant muscles do not insert on bones directly, but via tendons. These bands of tissue, fitting through narrow trajectories within the body, project muscle tension and excursion at locations away from the muscle itself. Within the hand, tendons glide through sheaths containing synovial fluid, a nurturing slime [Tubiana et al. 1996]. The synovial sheaths in turn are surrounded by fibrous sheaths, which attach and position a tendon relative to the bones it passes along and inserts on. The fibrous sheaths are part of a wider, *fibrous skeleton* of the hand. Its components, unlike those of the *bony (osseous) skeleton*, are supple, and can have their shapes changed during movement. The fibrous skeleton crucially restricts the ways in which the bones of the hand can be moved relative to eachother. Also, some of its components are shaped to act like pulleys for the tendons of the hand, while others are shaped as movable hoods, dynamically applying muscle tension within the finger during its movement. In general, the attachments by ligaments of the fibrous skeleton contribute, together with the muscles and tendons, to the combinations of applied forces which determine the movement at each articulation of the hand.

**1.3.1.3** *Joint movements causing fingertip movement*      From the shoulder to the fingertip, numerous bones are involved in fingertip movement. From nearest in attachment to the torso, or most *proximal*, to further away, or more *distal*, these are: the *humerus*; the *ulna* and *radius*; the eight *carpal bones*; the five *metacarpal bones*; and the *proximal*, *middle*, and *distal phalanges* of the five digits of the hand. These bones are discussed in detail in appendix Section A-1, and shown in Figure A-1.

How the bones listed are movable relative to eachother determines fingertip movement during the instrumental control of musical sound. Specific anatomical terms can be used to indicate specific types of externally observable movement resulting from the various possible articulations. Here, we will pair such terms with the names of the joints involved, to unambiguously indicate main types of movement which may occur. Such joint movements, contributing from shoulder to fingertip, then include: *shoulder anteversion/retroversion*, *shoulder adduction/abduction*, and *shoulder endorotation/exorotation*; *elbow flexion/extension*; *forearm pronation/supination*; *wrist radial/ulnar deviation* and *wrist palmar/dorsal flexion*; *little finger opposition*; *MP joint flexion/extension* and *MP joint radial/ulnar deviation*; *PIP joint flexion/extension*; and *DIP joint flexion/extension*. These joint movements are discussed in detail in appendix Section A-2, and illustrated below in Figure 1.7.

An overview of fingertip movement as the result of combined joint movements may seem to suggest a model not unlike that for, say, a segmented robot arm. However, such an impression could easily lead to a number of implicit assumptions which in fact would not be true. To explicitly avoid this, we point out here that joint movements often do not produce straight movement trajectories; that they often do not happen around fixed axes of rotation; that they often are not replicated identically across joints; and that they often do not happen fully orthogonal to eachother – which also includes, near the fingertip, the mirroring of PIP and DIP joint flexion/extension. These points are discussed in detail in appendix Section A-3.

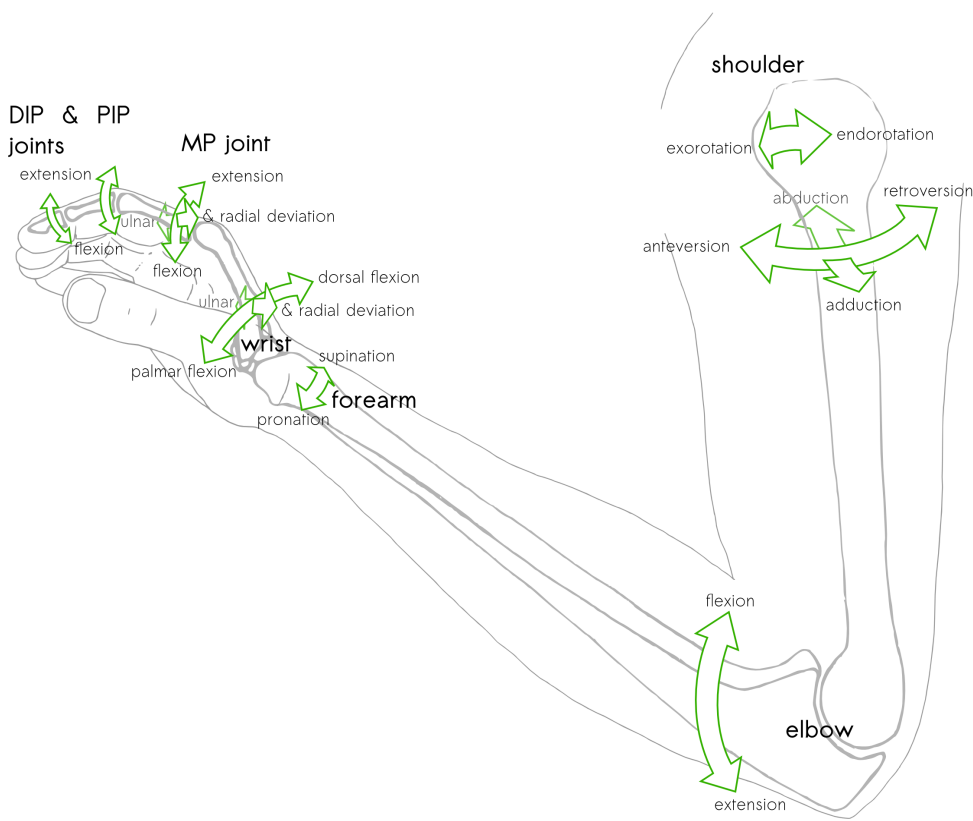

**Figure 1.7** *Joint movements causing index fingertip movement. Here, shown for the right hand and arm.*

**1.3.1.4** *Further anatomy of the moving fingertip*    Other anatomical structures, too, are involved in movement of the fingertip, especially as it comes into contact with other objects. Here, we will discuss such structures, more local to the fingertip itself, going from its palmar to its dorsal side.

The palmar side of the fingertip is fully covered by an outer layer of skin. This is *palmar skin*, a type of skin which is hairless (glabrous), and which also covers the palmar side of the rest of the hand. When looking at the palmar skin of a fingertip, apart from its overall curvature toward the dorsal side, the finer structures of three types of grooves and ridges in its surface stand out. First, there is a crease across the fingertip, reflecting how the skin is folded during DIP joint flexion, and delimiting the proximal end of the fingertip. Second, there are finer, curved ridges, running in parallel across the entire surface of this side of the fingertip. Finally, crossing the fine curved ridges run other, typically straighter lines, which are the result of tension within the skin [Tubiana et al. 1996].

The outer sublayer of palmar skin, just described, is called its *horny layer*. Via this layer, contact between the fingertip and other objects is made, takes its course, and is lost again. During contact with an external object, the skin will often mechanically resist movement of the object relative to the fingertip. For example, the palmar skin will resist relative movements in a dorsal direction, i.e. where objects press into the fingertip. This resistance is supported by the relative thickness of the horny layer of palmar skin of the hand [Tubiana et al. 1996]. The palmar skin, also, will resist relative movement in directions parallel to its surface. This type of resistance is supported by the presence of the fine curved ridges, mentioned above. These *friction ridges* are especially well-developed in areas of the hand where the skin contacts objects during grasping, and the ridges reduce sliding during grip [Tubiana et al. 1996].

The friction ridges also reflect the deeper-lying boundary between the skin's *epidermis* and *dermis* layers. The epidermis lies on top of the dermis, and has the horny layer as its outermost sublayer [Wolters and Groenewegen 2004]. The horny layer is made up of dead skin cells, but the other sublayers below it are alive, and contain relatively many receptor nerve endings responding to mechanical deformation. These *mechanoreceptors* signal the occurrence of mechanical events at the skin to the brain.

(This makes them examples of *afferent nerves*, nerves that send signals to the brain. Receptors present in the tissue of muscles and tendons are other examples of this. *Efferent nerves*, on the other hand, are nerves that send signals from the brain to other parts of the body, e.g. activating muscle contraction.)

The mechanoreceptors in the palmar skin of the fingertip will respond during contact, as mechanical deformation occurs, for example during resistance against the types of relative movement mentioned above. For relative movement parallel to the skin's surface, it is relevant that some mechanoreceptors are grouped and embedded specifically along the deeper boundary structure which is reflected in the friction ridges at the surface. Also, at the fingertips, the concentric layout of the friction ridges ensures that some will mechanically resist and deform during gliding contact, whatever the direction of relative movement.

As elsewhere in the body, the blood pumped around through the blood vessels of the hand supplies the cells of its tissues with nutrients and oxygen, and removes their waste products. However, most of the blood flow through the hand is for *thermoregulation* [Tubiana et al. 1996]. In this process, variably increased or restricted blood flow produces regulated heat loss via the skin, helping to keep overall body temperatures within safe and optimal bounds. Here, sweat released by glands in the skin makes the skin cool faster. In palmar skin, there typically are around 400 sweat glands per $cm^2$. Their sweating also hydrates the various sublayers of the skin, keeping them supple. This is supported by a lipid surface film on top of the horny layer [Tubiana et al. 1996]. The adhesion and mechanical resistance of the skin during contact, as well as the mechanical deformations to which its mechanoreceptors respond, partly depend on these processes.

Underneath the skin (Latin: cutis) of the palmar side of the fingertip, there is a compartmented layer of *subcutaneous fat*. This layer is also present elsewhere in the hand, forming fat pads at various areas of the palm, and also between the flexion creases of each finger. The area of the fingertip where its fat pad is underneath the palmar skin is called the *pulp*. Due to the fat pad, the pulp of the fingertip can have its shape changed during contact, in response to how pressure is being applied across the skin.

Continuing dorsally, the pulp is fixed against the distal phalanx bone, and beyond it, against the nail. The presence and shape of both nail and bone support the pulp during mechanical contacts. The nail of course only partially covers the back of the fingertip, with the remaining area covered by *dorsal skin*. This type of skin is much thinner than palmar skin, and lacks friction ridges reflecting the boundary between epidermis and dermis.

**1.3.2   The physiology of fingertip movement and touch**   Where anatomy studies the structural parts of the human body, physiology studies the processes that occur at smaller scales within and between these parts. For fingertip movement, a discussion of such processes obviously must include muscle contraction, activated via efferent nerves. Various types of receptor signaling, by afferent nerves, should be included as well, however. Such signaling conveys various types of events which occur simultaneous to and as a result of the muscle contractions. As will be clarified in Section 1.3.3, this then directly influences the course of movement.

**1.3.2.1**   *Muscle contraction, the neuromuscular transition, and motor units*   The *sarcomere* is considered the basic unit of muscle fiber tissue. It is a molecular structure, the length of which can change as the actin and myosin molecules it is composed of shift relative to eachother. Sarcomeres link up in series, which in turn line up in parallel, thereby forming a contractable muscle. The more sarcomeres placed in series, the longer the potential muscle excursion. The more series placed in parallel, the more forceful the potential muscle tension. Depending on the balance between these two structural principles, different muscle tissue with the same number of sarcomeres may present different trade-offs in potential muscle tension versus excursion [Tubiana et al. 1996]. Most of the muscle contractions happening within the human body will change both tension and excursion of the muscles involved [Wolters and Groenewegen 2004].

The efferent nerves which activate the muscles consist of cells called *motor neurons*. Given a muscle fiber, it will be attached, at specific locations on its membrane, to a motor neuron. This motor neuron may *fire*, meaning that there is a brief rise and fall, an *action potential*, in the electric voltage across the interior and exterior of its cell membrane. This action potential initiates an electrochemical chain reaction, in which a subsequent release of neurotransmitters triggers another action potential at the muscle cells, causing them to release $Ca^{2+}$ ions. Based on the presence of other chemicals, including adenosinetriphosphate (ATP), these $Ca^{2+}$ ions then enable the interaction between actin and myosin which causes the shift within the sarcomeres, producing contraction. The higher the frequency of motor neuron action potentials, the higher the concentration of released $Ca^{2+}$, and the more powerful the contraction. When action potentials cease to occur, the $Ca^{2+}$ concentration decreases, and relaxation follows contraction. After this, depending on the available amount of ATP, and of $Ca^{2+}$ being reabsorbed by the muscle cells, another cycle of contraction and relaxation can occur [Wolters and Groenewegen 2004].

A *motor unit* consists of one motor neuron and the muscle fibers it attaches to. The motor units of the interosseous and lumbrical muscles inside the hand contain some tens of muscle fibers. In contrast, elsewhere in the body, e.g. in the leg muscles, there may be hundreds or thousands of muscle fibers per motor neuron. In general, most muscles of the human body are divided into a range of motor units of different sizes. Based on this, the overall force they apply during contraction is regulated according to the *size principle*: over time, muscle force increases as the result of the subsequent activation of motor units ordered from small to large [Wolters and Groenewegen 2004].

Muscle contraction may also be activated by an externally applied voltage, for example via electrodes attached to the skin.

**1.3.2.2** *Somatosensory receptor nerve endings*  In Section 1.3.1.4, we mentioned the presence of mechanoreceptors in the skin, responding locally to its deformation. These receptors are part of a more general class of *somatosensory receptors*, which respond to events occurring nearby and inside the body. Of the sensory receptors, the somatosensory receptors detect the most specific information on ongoing fingertip movement, in the sense of responding to events that are local to the moving body parts involved.

Each somatosensory receptor transduces the physical measurables it responds to into an electrical voltage, the *receptor potential* [Wolters and Groenewegen 2004]. This receptor potential is then transformed into action potentials on the connecting afferent neuron. For a number of receptor types, this transduction happens via specialized structures, which encapsulate the nerve ending. Other receptor types do not have such *end organs*, and are called *free nerve endings*. Free nerve endings are present throughout almost all bodily tissues [Kahle 2001].

Another difference between receptor types is how action potentials are triggered in response to changes in the state of the physical measurables. If the change to a new state results in a brief train of action potentials, the receptor is said to be *rapidly adapting (RA)*. If, on the other hand, the train of action potentials is sustained for the duration of a new state, without decreasing much in rate, the receptor is said to be *slowly adapting (SA)*. Within the skin, the adaptation of different types of mechanoreceptors is determined by the mechanical behavior of their end organs [Goldstein 2002].

Since the skin mechanoreceptors respond to applied pressure based on changes in skin deformation, removal of an object may also result in sensations. For example, when a hat is taken off the head, the pattern of its rim may become felt, as the skin of the head springs back in the areas where the hat rim previously kept it depressed [Carlson 1998]. However, during prolonged wearing, adaptation may have occurred as well. This may cause a more counterintuitive experience, as the hat seems to become felt only after it has been removed.

There exists a concept analogous to that of motor unit for the somatosensory receptors: the *sensory unit*, consisting of one somatosensory neuron and the receptors it attaches to [Wolters and Groenewegen 2004]. For the skin, the resulting *receptive field* of a somatosensory neuron is defined as the area of skin which, when stimulated, influences the neuron's firing rate [Goldstein 2002].

In some cases, signaling by somatosensory receptors may also be the result of voltages being applied externally, for example via electrodes attached to the skin.

**1.3.2.3** *Somatosensory receptors for fingertip movement*　There exist various types of somatosensory receptors that are essential to normal human fingertip movement, distributed across various anatomical areas. These receptors include: *muscle spindles*, *Golgi tendon organs*, *receptors similar to Golgi tendon organs*, *Ruffini mechanoreceptors*, *Vater-Pacini mechanoreceptors*, *Merkel (SA1) mechanoreceptors*, *Meissner (RA1) mechanoreceptors*, *nociceptors*, and *thermoreceptors*. These receptor types are discussed in detail in appendix Section A-4. The contents of this section have been summarized in Table 1.8.

| type | anatomical location | transduces |
|---|---|---|
| *muscle spindles* | • within the muscles | • muscle length<br>• speed of muscle length changes |
| *Golgi tendon organs* | • between muscles and tendons | • muscle tension<br>• changes in muscle tension |
| *receptors similar to Golgi tendon organs* | • in tissues within and around the joints | • events related to joint movement |
| *Ruffini mechanoreceptors* | • in tissues within and around the joints<br>• in the skin below the epidermis | • mechanical deformation<br>• mechanical vibration (15-400 Hz) |
| *Vater-Pacini mechanoreceptors* | • in tissues within and around the joints<br>• in the skin below the epidermis<br>• at the surface of tendons and fascia<br>• in the walls of blood vessels<br>• in the periosteum | • changes in mechanical deformation<br>• mechanical vibration (10-500 Hz) |
| *Merkel (SA1) mechanoreceptors* | • in the skin, near the epidermis-dermis boundary | • mechanical deformation<br>• mechanical vibration (0.3-3 Hz) |
| *Meissner (RA1) mechanoreceptors* | • in the skin, near the epidermis-dermis boundary | • (light) mechanical deformation<br>• mechanical vibration (3-40 Hz) |
| *nociceptors* | • throughout various tissues, including the skin | • mechanical, thermal and chemical events associated with tissue damage |
| *thermoreceptors* | • in the skin | • changes in temperature |

**Table 1.8** *Somatosensory receptor types essential to normal fingertip movement and touch. (Words have been colored brightly to facilitate viewing the receptor types grouped by their anatomical locations. Precise references are given in Section A-4.)*

**1.3.3  Neural processes of fingertip movement and touch**    Beyond the motor units and the sensory units directly producing and responding to fingertip movement, lie attached the billions of interconnected nerve cells that make up the human *central nervous system*. This also includes the neurons of the brain. We will not discuss the structure of individual neurons, nor how they are attached to eachother, nor how they receive, process and transmit electrochemical signals amongst eachother. For an introduction to these subjects, see for example [Wolters and Groenewegen 2004]. Instead, we will move directly to discussing neural processes that are the result of the combined activity of groups of neurons, and that are fundamental to human actions attempting to move or hold still the fingertip.

**1.3.3.1  *Basic processes in the motor system beyond the motor units***    Beyond the motor units, the *motor system*, as a part of the wider central nervous system, controls and coordinates muscle contractions such as those producing fingertip movement. The motor system extends across parts of the spinal cord, brainstem and cerebral cortex (the outer layer of the upper part of the brain) [Wolters and Groenewegen 2004]. The motor neurons of motor units – and also of the muscle spindles – that are involved in fingertip movement stretch out from within the spinal cord, and there, and within the brainstem, the *contraction of individual muscles* is controlled.

Within the primary motor cortex, rather than the control of individual muscles, the *activation of muscle groups, in dynamically changing combinations* occurs. Also involved in this are the premotor cortex and the supplementary motor cortex. These areas, like the primary motor cortex, are part of the cortex surface of the frontal lobe of the cerebrum. The primary motor cortex, in this way, controls the direction, speed, and force of body movements. Its connection to the neurons that control individual muscle contractions typically is indirect, via intermediate circuits of neurons which are also under the influence of other areas of the brain [Wolters and Groenewegen 2004]. The big exception is for movements involving the muscles of the forearm and hand. For these, the motor cortex connects directly to the motor neurons within the spinal cord. In this way, independent, fine finger movement is more directly under the control of the motor cortex.

From different anatomical parts of the primary motor cortex, movement in different anatomical parts of the overall body is controlled. In this regard, too, the fingers are exceptional: the size of their cortical representations is relatively much larger than their sizes as parts of anatomy. This relationship, between anatomy and cortical representation, is often visualized for the whole of the body as the *motor homunculus*, a deformed little man spread out across the surface of the primary motor cortex.

The *more abstract activation and coordination of sequential and simultaneous body movements* occurs within areas of the frontal lobe, basal ganglia and cerebellum. The basal ganglia are a group of clustered neurons located at the base of the cerebrum. The cerebellum (Latin: little brain) lies attached underneath the cerebrum. The results of this process are signaled to the premotor and primary motor cortex.

**1.3.3.2** *Basic processes in the somatosensory system beyond the sensory units*
Beyond the sensory units, the *somatosensory system*, as a part of the wider central nervous system, processes and integrates the results of transduction by the somatosensory receptors, including those related to fingertip movement. The somatosensory receptors first connect to other neurons within the spinal cord or the brainstem [Wolters and Groenewegen 2004], thereby *signaling individual somatosensory events*.

Further anatomical parts of the brain that are also involved include the thalamus, which lies directly above the brainstem, and the primary and secondary somatosensory cortex areas. In the thalamus of certain monkeys, neurons have been found which respond to the specific directional placement of edged objects on the skin. Other neurons have been found to respond to specific directional movement of such objects. In the human cortex, too, there are neurons which respond to the orientation and movement direction of objects contacting the skin [Goldstein 2002]. Such tactile feature detector neurons are examples of neurons *signaling integrated somatosensory events*.

Like for the primary motor cortex, the links between the anatomy of the primary somatosensory cortex and that of the overall body can be represented by a homunculus. In this *sensory homunculus*, too, the fingers, and especially the skin of the fingertips, occupy large areas quite disproportionate to their relative anatomical sizes. Because of this large number of neurons in the cortex, linked to receptors covering small receptive fields of fingertip skin, there exists excellent detail perception via the fingertips [Goldstein 2002].

The structure which the sensory homunculus represents is not static and permanent: it changes in response to experiences. More generally, this phenomenon is called *brain plasticity*. A standard example of it is that musicians playing stringed instruments, over time, obtain an increased cortical representation specifically for the fingers of their left hand – i.e. for those fingers which directly contact the strings [Elbert et al. 1995].

**1.3.3.3** *Basic processes of sensorimotor integration*  Clearly, within the wider nervous system, the motor system and somatosensory system do not exist in isolation from eachother. We will now briefly discuss some basic processes of *sensorimotor integration*, relevant to fingertip movement, and organized here according to a characterization of the overall effect of sensory transduction.

First, the reflexes are a basic type of sensorimotor integration, where the transduction of a specific sensory event rapidly causes a typical, usually fixed motor response, in absence of a consciously experienced decision process. Reflexes occur within the spinal cord and brainstem, but also beyond these within other parts of the central nervous system. The simplest reflexes are activated by pairs of neurons, where one afferent sensory nerve is connected to one efferent motor nerve. Most reflexes, however, are activated by more than two neurons [Wolters and Groenewegen 2004]. Reflexes may involve fingertip movement, for example when the thermoreceptors and nociceptors of the skin trigger a withdrawal reflex, avoiding potential damage. In this way, reflexes are an example of *sensory transduction directly triggering movement*.

Other movement will not be triggered directly by the transduction of sensory events, but rather by other parts of the nervous system. Still, the results of sensory transduction will remain very important here, too. For example, before the actual coordinated activation of muscle groups by the motor system happens, a process that prepares this will have occurred. Its outcome depends partly on the results of sensory transduction, including those results which reflect the current state of the muscles, tendons and joints [Wolters and Groenewegen 2004]. This is an example of *sensory transduction informing the preparation of movement*.

Furthermore, once a movement is being executed by the motor system, ongoing sensory transduction may still lead to adjustments. The cerebellum is hypothesized to be the site of a process of comparison between a neural representation of the movement that has been prepared, and information on its ongoing execution, which continuously comes in from the somatosensory, visual, vestibular, and other sensory receptors. Moreover, this process also will trigger corrective changes to ongoing execution, so as to reduce detected discrepancies between prepared and executing movement [Wolters and Groenewegen 2004]. This is an example of *sensory transduction providing feedback during movement*.

Also hypothesized to occur within the cerebellum is the process of *motor learning*. A given movement may be repeated over time, but with some change made to its preparation, which during execution reduces the occurrence of sensory feedback signaling the need for correction. This is an example of *feedforward* motor activity. If the movement is then repeatedly executed over time, according to the modified preparation, its representation becomes stored, through brain plasticity of the cortex of the cerebellum [Wolters and Groenewegen 2004]. Such a stored representation is often called a *motor program*, and its retrieval and use in the execution of movement can realize a number of advantages. One is faster execution, through the reduction of feedback-induced corrective motor activity. Execution may also become more effective by other measures [Wolters and Groenewegen 2004]. Furthermore, execution may increasingly rely on somatosensory feedback, requiring less feedback from other senses. In fact, historically, in engineering psychology it has been regarded as a defining characteristic of motor programs that their activation altogether *removes* the need for guidance and correction based on visual feedback [Wickens 1992].

The activation of motor programs typically also means that movement execution will occupy *attention* and *consciousness* less, or even not at all. This is reflected in the everyday experience of executing movements which, once initiated, disappear from consciousness, freeing it for other activity. A standard example of such reliance on motor programs, acquired through prolonged motor learning involving the cerebellum, is skilled piano playing [Wolters and Groenewegen 2004]. Skilled piano playing also is a standard example for visual feedback not being used during execution anymore [Wickens 1992]. In general, the formation of motor programs through motor learning can be seen as an example of *sensory transduction underlying feedforward during movement*.

**1.3.3.4** *Passive touch via the fingertip*    In the previous section, while discussing the effects of motor programs, our scope again widened from more basic neural processes to consciousness. In this section and the next, we will focus on motor and somatosensory activity within the central nervous system which may be experienced consciously. Our brief introduction to this vast subject will be based on the fundamental distinction between passive and active touch (see, for example, [Carlson 1998] [Goldstein 2002]). Here, we will understand *passive touch* via the fingertip as some overall process involving the central nervous system, in which sensations result from transduction by the somatosensory receptors, occurring in the absence of motor activity causing fingertip movement.

Now, given our basic interest in fingertip movement as a medium for changes made to sound-generating processes during the instrumental control of musical sound, to discuss a phenomenon where fingertip movement, or at least its human activation, is absent, could seem to have little priority. This is not the case, however: we have already mentioned, above, how sensory transduction informs the preparation of movement during processes of sensorimotor integration. Similarly, episodes of passive touch may very well have an effect on future fingertip movement. For example, the sensations resulting from how a fingertip is resting on the surface of a piano key may influence how exactly it is subsequently pressed to produce a specific sound.

Various types of sensation may occur in such a way as to fall under the above definition of passive touch. This includes sensations that can roughly be indicated as *skin pressure*, *skin stretch*, and *skin vibration*; *vibration sensations originating from other tissues*; *proprioceptive sensations*; *kinesthetic sensations*; *pain sensations*; and *temperature sensations*. Which of the somatosensory receptor types previously discussed in Section 1.3.2.3 causally underlie each of these types of sensation is discussed in appendix Section A-5. There, also, the *two-point threshold* for skin pressure sensations is briefly discussed, which, across the human body, is smallest at the fingertips.

**1.3.3.5** *Active touch and manipulation via the fingertip*    Here, we will understand *active touch* via the fingertip as some overall process involving the central nervous system, in which sensations result from transduction by the somatosensory receptors, occurring in the presence of motor activity causing fingertip movement. As for passive touch, this may include sensations that are consciously experienced and described as skin pressure, skin stretch, skin and other tissue vibration, proprioception, kinesthesis, pain, or temperature. However, the presence of simultaneous motor activity typically results in a different kind of perceptual experience than that of passive touch [Goldstein 2002].

[Gibson 1962] was seminal in pointing out and clarifying this difference. In conditions of passive touch, bringing some rigid, protruding object into mechanical contact with the skin, then moving it across the skin, typically results in sensations reported in terms of an indentation moving across the skin. However, if the same relative movement between skin and object is produced in conditions of active touch, and is only due to motor activity by the observer, a different kind of sensation results, typically reported in terms of the surface properties of an external, non-moving object. For this external object to be perceived, as non-moving, a fundamental integration of both the pattern of skin indentation traveling across the skin and the simultaneous motor activity moving the fingertip must precede conscious perception. [Gibson 1962] points out that this is non-trivial, and as a neural process similar to visual perception resulting from stimulation of the retina during oculomotor activity by the observer. Here, that, which based on simultaneous motor activity becomes separated out to represent an external object, is called the *exterospecific* component of sensory stimulation. The resulting experiences of active touching and looking are then reported as corresponding to the environment, instead of to events at the sensory surface [Gibson 1962]. Based on this, we will understand the difference between passive touch and active touch also as a fundamental shift in experience, where a spatial aspect of the sensations being induced, tying them to locations on or within the body, changes to imply location *outside* of the body.

For other aspects than location outside of the body, various active touch sensations may be experienced as mutually different in nature. This includes types of experience reported as object shape, object hardness, object surface texture, and object slipperiness and stickyness [Carlson 1998]. In the case of object surface texture perception via the skin, a *spatial* experience of *roughness* for surfaces with details separated by distances of 0.125 mm or more, changes into a *temporal* experience of *vibration* for microgeometric surfaces, where the details are smaller and measured in μm. This shift is due to transduction by Vater-Pacini mechanoreceptors [LaMotte and Srinivasan 1991] [Rovan and Hayward 2000]. For surface texture, as well as other sensations of active touch, the hand movements involved have been identified as falling into a number of distinct types, called exploratory procedures [Lederman and Klatzky 1987, 1990] [Goldstein 2002]. Judging texture, for example, involves the exploratory procedure of lateral motion, and to a lesser extent of contour following, across the object.

[Gibson 1962] explicitly does not consider acts of manipulation, such as the relocation of objects, or tool use in general, to fall under active touch. Instead, active touch is primarily regarded as a mode of sensing properties of external objects, through their exterospecific contribution to sensory transduction during motor activity. Here, then, we will understand *manipulation* to be a type of process different from active touch in the effects of its motor activity: these must change the exterospecific components of subsequent active touch, resulting in a changed perception of object properties.

**1.3.4   Reflection: Human fundamentals for fingertip instrumental control of musical sound**   In the previous Sections 1.3.1 to 1.3.3, we have considered the functional anatomy, physiology, and neural processes that underlie human actions attempting to move or hold still the fingertip. This has included, from shoulder to fingertip, a discussion of the human osseous skeleton (see Figure A-1) and contributing joint movements (see Figure 1.7); also, a brief introductory overview of the physiology of muscle contraction and receptor signaling, followed by a discussion of specific somatosensory receptor types that are involved in fingertip movement and touch (see Table 1.8); and, finally, a discussion of neural processes involved in fingertip movement and touch, culminating in the concepts of passive touch, active touch, and manipulation.

From these areas, we will now highlight and discuss a number of points that are fundamental for instrumental control of musical sound via the fingertips. Naturally, this also covers those forms of control characterized by unidirectional fingertip movement orthogonal to a surface, as defined in Section 1.2.5.

**1.3.4.1**   *Anatomy*      An important anatomical fundamental is that *in fingertip instrumental control, the fingers are especially suited to flexing movements*. This follows from the general anatomy of the human hand being especially suited to the execution of gripping movements enclosing objects.

The overall placement and relative lengths of the various bones of the hand are precisely related in ways that specifically enable various gripping movements [Tubiana et al. 1996]. Other anatomical factors further support the suitable combined orientation of the phalanges during the execution of gripping movements. These factors correspond to the joint movements involved not being straight, not happening around fixed rotational axes, not being identical across joints, and not happening orthogonally; issues which more generally have been highlighted in Section 1.3.1.3.

For example, in many gripping movements, the fingers flex together toward the palm. Here, the relative lengths of the phalanx bones of each finger, decreasing from proximal to distal, allow each finger as a whole to progressively wrap around ever smaller objects. Then, DIP joint flexion, occurring together with, but slower than PIP joint flexion, locks grip at the end of movement [Tubiana et al. 1996].

The small phalanx bones may then apply large forces to the objects enclosed in flexion, due to the projection of muscle tension via tendons. Here, it is telling that the two flexor tendons inside each finger can apply considerably greater muscle tension than the one extensor tendon [Tubiana et al. 1996]. Extension is further limited by the fibrous skeleton, at the articulations of the digits, where extension approaching dorsal flexion ("hyperextension") is blocked.

The fingers are also better suited to flexion and gripping due to the type of skin on their palmar sides. Through its thickness, sweat glands, and friction ridges, palmar skin provides mechanical resistance against relative movements when making contact. It is further supported in this by the subcutaneous fat pads, also present on the palmar side of the hand, which cause the hand to mold around objects during their enclosure, thereby increasing grip surface [Tubiana et al. 1996]. As has been discussed in Section 1.3.1.4, these palmar fat pads also include the pulp of the fingertip, supported dorsally by the distal phalanx bone and fingernail.

**1.3.4.2** *Neural processes*     Focusing on neural processes, it can first be said that *fingertip instrumental control is based on neural processes of sensorimotor integration, and is especially suited to these*. This includes processes involving motor learning, recall and activation of motor programs, passive touch, active touch, and manipulation. Such neural processes are supported by detailed perception via the fingertips, simultaneous to precise and independent finger movements. The latter are possible because of the lumbrical and interosseous muscles inside the hand [Tubiana et al. 1996], which act under direct control of the motor cortex. There, the fingers have an exceptionally large representation in the motor homunculus. In the sensory homunculus, the fingers, and especially the skin of the fingertips, have a large cortical representation as well. As mentioned in Section 1.3.3.2, this representation may increase significantly over time as the result of instrumental control of musical sound. Detailed perception is further supported by the high density of small receptive fields in fingertip skin [Kahle 2001] [Goldstein 2002] [Wolters and Groenewegen 2004], which is reflected in the two-point threshold being lowest across the human body at the fingertips.

Then, of the neural processes of sensorimotor integration, those involving *motor programs are crucial to fingertip instrumental control*. As discussed in Section 1.3.3.3, spontaneously occurring neural processes of motor learning may result in fingertip instrumental control which largely relies on somatosensory transduction alone. This is associated with *crucial advantages for instrumental control*. One is faster movement execution through a reduced reliance on corrective sensory feedback. Another is the reduced or even absent claim on resources of attention and consciousness, once the execution of a movement has been initiated. In combination, these advantages may very well increase the number of changes that can be made intentionally and successfully to the sound-generating process over a given period of time. This is beneficial, in the sense of increasing the number of different possibilities for resulting musical sound. Also, the activation of motor programs may allow attention and consciousness to become more occupied by resulting musical experiences, which are the ultimate object of instrumental control. These advantages, realized, can be found e.g. in the study and practice of traditional acoustic instruments, which over time results in increasingly complex musical pieces becoming playable.

This also implies that *when studying fingertip instrumental control, it is a pitfall to consider somatosensory transduction only as a means of feedback*. As we have seen, somatosensory transduction informs the preparation of movement, consciously and unconsciously, and it underlies the feedforward of motor programs.

A second implication is that *when studying fingertip instrumental control, it is a pitfall to consider only the consciously experienced aspects of sensorimotor integration*. Consciously experienced processes of passive touch, active touch, and manipulation might intuitively seem the most significant, simply because they are noticed. However, as we have just seen e.g. for motor programs, it may be the very absence from consciousness that indicates successfully realized benefits to instrumental control. The importance of both conscious and unconscious aspects of instrumental control is emphasized in the schematic overview shown in Figure 1.9.

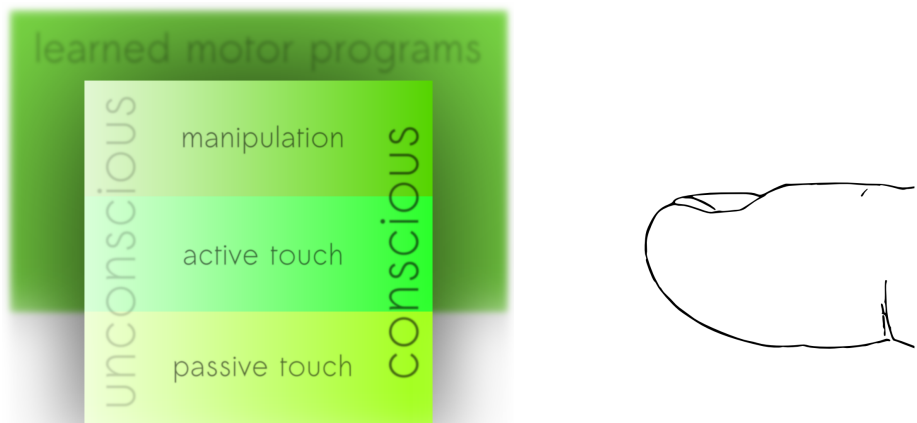

**Figure 1.9** *Important types of human sensorimotor activity underlying instrumental control of musical sound via the fingertips.*

Finally, from the previous points, it follows that *compared to other types of sensory transduction, somatosensory transduction is especially important to fingertip instrumental control*. The neural processes of sensorimotor integration underlying fingertip instrumental control, involving motor programs, active touch, and manipulation, rely on somatosensory transduction.

Still, during instrumental control, it may very well be other types of sensory transduction, perhaps visual or auditory, which effectively convey to the human that a change has been made to the sound-generating process. For example, a touch screen being pressed may light up, or trigger a clicking sound, to unambiguously confirm its activation. Or, the heard musical sound itself may be what confirms that a change has been made. Also, besides forms of feedback, visual transduction may more generally provide information on the current state of the sound-generating process, and this may continuously guide fingertip instrumental control. (Examples of this include the GUI displays of sound-generating algorithms.) Visually transduced information may also specifically relate to the process of making changes itself, e.g. as when a pianist, for orientation between key presses, visually tracks hand position during large and rapid movements across the keyboard.

But it is an increased reliance in fingertip instrumental control on the feedforward of motor programs, with further guidance by somatosensory transduction alone, that will realize the crucial advantages discussed near the beginning of this section.

**1.3.4.3** *Fingertip instrumental control: Active touch or manipulation?*     In instrumental control via the fingertips, changes to the sound-generating process may be caused by processes that involve active touch as well as manipulation. However, of the two, only manipulation (as defined in Section 1.3.3.5) will, consciously or unconsciously, confirm to the human that a change has been effected through the perception of changed object properties.

Given a concrete example of instrumental control, such as playing the piano, are the changes made via the fingertips to the sound-generating process the result of active touch, or of manipulation? As we have discussed in Section 1.2.3, there are similarities in the manual operation of piano-type keyboards and computer keyboards, and pressing computer keyboard keys is sometimes regarded as an everyday example of active touch [Goldstein 2002]. This would seem to suggest active touch as the answer.

However, perhaps the answer may be both, depending on the time scale, and on how the invariant exterospecific components probed by active touch are defined. If we consider the piano keyboard on a time scale corresponding to probing before and after a performance, we might say the changes we heard were made due to active touch: The piano has not changed, its keyboard still feels the same to us. However, if we choose to probe on a smaller time scale, before and during a key press, we might say the changes we hear are being made due to manipulation: The depressed key presents a change in exterospecific components, and it, and the keyboard in general, do feel different to us.

In this text, we will opt for the latter interpretation of what constitutes manipulation.

**1.3.4.4** *Applicability to unidirectional fingertip movement orthogonal to a surface*     In Section 1.2, we noted that unidirectional fingertip movement orthogonal to a surface can be regarded as a common component in many ancient and existing forms of fingertip instrumental control. We characterized it as fingertip motion which approximates a single path of movement, at right angles with a surface, and extending across at most a few centimeters. We noted some of the advantages of this for the control of musical sound.

The previous discussion of human fundamentals in Sections 1.3.4.1 to 1.3.4.3 now enables a fuller understanding of this type of movement. Summarizing, this is as a product of the human hand's predisposition toward gripping postures, while allowing independent and precise flexing movements by the fingers. The latter will be the result of conscious and unconscious processes of sensorimotor integration, possibly involving motor programs, passive touch, active touch, and manipulation. Naturally, each point discussed in the above sections relating to fingertip instrumental control in general will also apply in the cases where there is unidirectional fingertip movement orthogonal to a surface.

# 1.4  The importance of computed sound

**1.4.1  The seminal confluence of two technologies**    In the middle of the 20[th] century, two major historical developments in technology were combined: the electric loudspeaker, and the electronic digital computer. Early loudspeakers were developed during the second half of the 19[th] century as devices for the reproduction of sound. Mechanical types of loudspeaker were developed to reproduce sound, usually musical sound, from recordings made on newly developed phonograph cylinders and gramophone discs. The initial development of electric loudspeakers was closely tied to the development of the wired electric telephone. Its main application became the live transmission and reproduction of the human voice across distances much greater than feasible with purely acoustical means. This continued, at the beginning of the 20[th] century, with the development of radio technology for wireless electromagnetic transmission of human vocal sounds. Again, electric loudspeakers were necessary, with their development closely tied to that of the new technology. During the subsequent decades, electric loudspeakers increasingly became means for sound reproduction in general.

The *dynamic loudspeaker* became the most widely used type of general-purpose electric loudspeaker. Here, the electric voltage across the terminals of a wire is varied over time. A section of this wire is wound to a coil and attached to a membrane: the loudspeaker's cone. The coil itself is suspended between the poles of a permanent magnet. As the changing voltage induces changing levels of electric current through the coil, the interaction between its magnetic field and that of the permanent magnet causes the coil, and thereby the speaker cone, to move inward and outward along the suspended coil's axis. This generates acoustic waves, emanating from the moving speaker membrane into the surrounding volume of air, where the waves can be detected as changes in air pressure, or be heard as sound.

Regarding automatic computing machinery, by the middle of the 20[th] century proposed mechanical designs pioneered during the 19[th] century had given way to the first electronic and digital computers, implemented during and immediately after World War II. Early in the 1950s, such electronic digital computers, connected to electric loudspeakers, for the first time were used to generate musical sound [Doornbusch 2005]. However, new possibilities enabled by this development were not investigated or realized until later in the decade, with the pioneering work by Max Mathews and his colleagues. In 1957, this work involved executing programmed computations on a machine in New York City, transporting the output stored on magnetic tapes by car to Murray Hill, New Jersey, to then input it to a second machine capable of automatically converting the computed 12-bit number sequences to sound [Roads et al. 1996]. Soon, such computation and conversion could be implemented on a single machine. The principle of conversion remained the same, however, and has since. A sequence of numbers, called samples, stored in a memory buffer as the result of computation, over time is converted to amplitude changes in an electric voltage. These amplitude changes are proportional to the digital values, and conversion happens automatically, via a Digital-to-Analog Converter (DAC). The changing voltage then controls a loudspeaker, and thereby, acoustic waves in the air. The memory buffer containing the amplitude values is often called the *wave table* for this reason. This setup is described in [Mathews et al. 1969], where its capability for producing arbitrary heard sound is emphasized.

**1.4.2 Important subsequent developments: wide variety**     Over the second half of the 20[th] century, the beginnings of computed sound [2] were followed by a great expansion. This resulted in many newly developed and widely used types of processes for generating musical sound. Some of these mimicked or improved upon existing processes not involving digital computation, and were operated in ways similar to their predecessors. Others opened up possibilities for human action resulting in musical sound that previously simply did not exist. An overview of the many, simultaneously occurring developments could be given in many different ways. It might be done using a more or less chronological ordering, or perhaps based on the various types of non-digital predecessor technologies involved. Here, we will attempt a brief overview based on three organizing principles.

First, we will consider a limited number of widely used algorithmic types of sound generation, with each type corresponding to many implementations over the years, and indeed, to a separate established field of research. Second, since the wave table usually is fundamental to the various forms of computed sound, we will start from processes that allow its direct manipulation. Third, since we are concerned with the instrumental control of musical sound, we will divide the types into eight groups, according to common ways in which the instrumental control they enable moves away from direct manipulation of the wave table. In each case, the instrumental control itself may produce input for the sound-generating process via any of a wide range of devices. Many of these, involving the fingertips, have been discussed in Section 1.2.

**1.4.2.1** *Direct wave table manipulation*     The first group, then, contains types of direct wave table manipulation, or Wave Table Synthesis. Here, instrumental control can make changes to the wave table directly in terms of its positions and values. This includes the manual entry of amplitude-over-time values, as can be done using sound file editors. It also includes the use of programming systems to enter algorithms which directly modify the wave table. In this way, waveforms can be synthesized: precise and accurate approximations of arbitrary repeating or non-repeating pressure waves can be specified, and then made heard [Mathews et al. 1969].

**1.4.2.2** *Parametrized waveform synthesis*     A second group builds on this, and contains types of parametrized waveform synthesis. Here, instrumental control can make changes to the properties of automatically generated, repeating amplitude-over-time waveforms, which also may be combined in various ways. A fundamental example of such a waveform is the frequency-controlled sine wave, since heard sound passes through hair cells in the inner ear that resonate with, and respond to, sine wave pressure variations at specific frequencies. Multiple sine waves may computationally be added together, for example to simultaneously create a complex pitched sound. In general, this type of sound generation is called Additive Synthesis.

---

2    The terms *computed sound* and *computed touch* are used in this thesis to emphasize that sound and touch fit similarly within a larger theoretical context (see also: Introduction). Here, *computed sound* may also be regarded as shorthand for *computer-generated sound*.

Another way to combine elementary waveforms is to multiply one by another, in a process known as Amplitude Modulation (AM). Examples of this are often based on an oscillating carrier wave, which by itself would produce hearable sound, but is first multiplied by a second, modulator wave. If the modulator wave changes at a slower rate than the carrier, it may be used to control the development over time of the perceived loudness of the carrier wave. Classical examples of this are Attack-Decay-Sustain-Release (ADSR) envelope functions, and Low-Frequency Oscillators (LFOs). If the modulator wave oscillates at a rate comparable to that of the carrier, amplitude modulation may be used to create a new, more complex basic waveform. This happens in a computationally inexpensive way, e.g. considering how similar results would be arrived at via additive synthesis of sine waves.

In another type of sound generation, pioneered by John Chowning, one waveform combines with another by controlling its frequency. This is called Frequency Modulation (FM), and its development has provided another fundamental approach for the efficient generation of complex waveforms for musical use [Chowning 1973].

**1.4.2.3** *Sampling*    A third group contains types of what is often called sampling. Here, instrumental control can trigger playback of the amplitude-over-time sequences of arbitrary sound fragments. Especially recorded sound fragments are often used, which at some point may have been captured and converted to wave table values via an Analog-to-Digital Converter (ADC). In this way, a great and open-ended variety of sound fragments is available for use as components of musical sound. Usually, instrumental control also extends to additional parameters altering the triggered playback in various ways. Typical examples of this are overall loudness, playback rate (e.g. to raise or lower the perceived pitch of a given sound fragment), looped playback over parts of the sound fragment, and controlled interpolation between multiple sound fragments.

**1.4.2.4** *Spectrum-based operations*    A fourth group contains types of spectrum-based operations. Here, instrumental control makes changes to sound-generating processes involving the automatic separation into and integration of component waves at different frequencies. Typically, the pressure changes resulting in heard sound do not consist of a single, constantly repeating sine wave, but instead develop arbitrarily over time. The hair cells of the inner ear will each resonate to a different extent, which increases and decreases over time. In this way, a separation into simultaneously occurring waves of varying intensity, across a range (or spectrum) of frequencies, underlies the human perception of sound. This is one fundamental motivation to also use spectrum-based methods when generating musical sound.

In the case of computed sound, this will require methods for the automatic conversion to and from the amplitude-over-time representation of the wave table. Methods based on the Fourier transform convert to a representation describing how the signal can be constructed as the sum of a spectrum of sine waves. The amplitude and phase of these sine waves is tracked over time, based on an initial segmentation of the signal into a series of time windows, which usually have a short and uniform duration. In other methods, based on the wavelet transform, the time windows used are not uniform, but depend on the frequency analysed, so that a whole number of sine wave cycles may fit in each time window used.

Given a spectral representation, sound again may be generated based on direct manipulation, similar to that of the wave table. Building on this, many new types of sound generation have been realized, especially types for modifying recorded sound. For example, instrumental control may shift, scale or re-order the partials within the higher-frequency regions of arbitrary pitched sounds, while leaving the frequency region of their fundamentals intact. Other examples include various methods for having the spectral characteristics of one sound alter those of another, in order to generate musical sound.

**1.4.2.5** *Amplitude-based signal processing*   A fifth group contains types of amplitude-based signal processing. Here, instrumental control can change the properties of processes which automatically modify arbitrary amplitude-over-time input to again produce amplitude-over-time output. One example of this are delays implemented using memory buffers. Added to the original signal, and made part of a feedback loop in processing, these can be used to add echoes and otherwise emulate the effects of spatial reverberation of acoustic waves. Also, a difference in delay introduced between stereo channels can be used to vary the perceived spatial origin of a signal.

Another area is that of digital filters which, notwithstanding their amplitude-over-time input and output, are used especially for altering the spectral characteristics of arbitrary sound. Classical examples implement lowpass and highpass filters, and combinations based on these. In use, such filters will diminish the intensity of some of the spectral components relative to others present in a given sound. This has led to the term Subtractive Synthesis for this type of sound generation.

Yet another area is that of dynamic range processing. Here, algorithms track how the amplitude dynamics of an arbitrary sound signal change over time, to then variably amplify or attenuate the sound based on this. Classical applications of this include limiting, expanding and compressing parts of the dynamic amplitude range of a given sound.

**1.4.2.6** *Granular synthesis*   A sixth group contains types of granular synthesis. Here, instrumental control makes changes to sound-generating processes that involve the automatic separation into and integration of brief sound fragments. Such sound fragments, called "grains", are characterized by having a duration from 1 to 100 ms. During sound generation, their contents often derive from a second wave table not directly used for playback. This memory buffer may contain the representation of arbitrary sound, such as a synthesized waveform or a recording. At each point in time when the generation of a grain starts, its data is read from a given position within the buffer, in a given direction for playback. For the duration of the grain, the data are read out and converted according to the grain's individual playback rate. Over its duration, the grain's amplitude is modulated according to a given fixed envelope function, usually chosen to be continuous and smooth. The grain's overall amplitude again may be individual to it.

By adding large numbers of separate grains, new types of musical sound can be synthesized. As these are heard, the original source sounds may very well not be recognizable anymore. Aside from types of granular sound generation using amplitude-over-time representations, there are also other, spectrum-based approaches. Historically, various approaches have been developed and then used directly by composers, including Iannis Xenakis [Xenakis 1960], Curtis Roads [Roads 2004] and Barry Truax [Truax 1986]. Granular synthesis has especially found application in new experimental forms of music.

**1.4.2.7** *Physical modeling*     A seventh group contains types of physical modeling. Here, instrumental control makes changes to the properties of physical models simulating mechanical, acoustic and electronic sound-generating processes. As examples, we will briefly discuss three widely used types, which are based on imitating the excitation and propagation through space of mechanical waves. All of these naturally include mimicked acoustic waves directly resulting in writes to the wave table used for sound output. Generally, the more wide-ranging and detailed the simulation of the causal mechanisms producing the acoustic waves, the more preliminary steps of computation are required. We will order the examples according to this, since increasing amounts of preliminary computation also relate to the instrumental control moving further away from direct manipulation of the wave table.

In Karplus-Strong modeling [Karplus and Strong 1983], the contents of a relatively small wave table are played back repeatedly and continuously. Instrumental control can trigger the wave table to be filled with a series of random values. In a basic example, for each subsequent playback iteration, adjacent samples are then averaged in pairs, producing the new contents of the wave table. This results in the playback of a pitched tone with an initially bright-sounding spectrum, which over time reduces to a single sine wave. Classical applications of this type of sound generation are simulations of the plucked strings of acoustic instruments and, using extended versions, simulations of drum sounds.

To construct more detailed simulations of a wider range of acoustic instruments, Waveguide modeling can be applied to musical sound generation [Smith 1987]. A waveguide model simulates the result of the propagation through space of some pattern of excitation. To do this, a series of computations is specified, corresponding to an interconnected set of basic components. Mechanical excitations are mimicked by algorithms generating amplitude-over-time representations of the instances of impacts and oscillations. The propagation of these through specific types of material to locations at specific distances, as happens mechanically in strings, membranes and air columns, is then simulated by algorithms using memory buffers to implement corresponding time delays. How the incoming wave may scatter from a given location is then imitated by algorithms passing on portions of the wave to various further delays, which may represent its reflection back to its origin, or propagation to other locations. Algorithms implementing digital filters are added to this, to simulate the spectral effects on the wave of passing through resonating bodies with complex geometries.

Ultimately, one of the locations of the simulation will correspond to a transfer to acoustic waves, and its state of vibration will pass on to the wave table used for sound output. Given an instance of such a waveguide model, instrumental control can change the various properties affecting this heard wave, such as the pattern of excitation or the physical dimensions and material characteristics of the mechanical structure through which the pattern is simulated to propagate. In this way, waveguide models can be used to simulate the operation of many types of acoustic instruments. This includes not only many possible variations of striking a string or a drum surface, but also the playing of wind and bowed-string instruments, and imitations of the human voice.

For simulations at a still greater level of detail, Mass-Spring modeling can be used. This approach followed initial work in [Hiller and Ruiz 1971]. Here, computed sound is the result of the simulation of how, over time, mechanical excitations travel through some location within a set of masses interconnected by springs. A system of equations is used to describe how the masses and springs would interact mechanically as varying forces would be applied to them. Algorithms approximating a solution to these equations compute the representation of a resulting acoustic wave, which is then written to the wave table used for sound output. Instrumental control may change various properties of processes of this type, especially the pattern of excitation applied externally to the structure being simulated. Mass-spring modeling, in general, can be used to simulate the basic mechanisms of sound production for many types of vibrating mechanical objects. By extension, it can be used to simulate the mechanisms of excitation and resonation for a wide variety of acoustic instruments. This, however, typically requires fine structural detail in the simulation, and therefore, much computation.

**1.4.2.8**  *Sequencing*      An eighth group contains types of sequencing. Here, instrumental control makes changes to the relative temporal placement of instances of generated sound combined in playback. The instances themselves may be generated using any of the types of sound generation already discussed. Sequencing may not only determine the beginning and duration of each instance, but also any additional parameters specific to the type of sound generation used. Instrumental control may directly make such changes to individual instances, or it may alter the specification and execution of algorithms which in turn do this. This may also involve the automatic conversion to and from various types of musical notation. Classical applications of sequencing software and hardware include combining timbrally modulated, pitched sounds into harmonies and melodies, and assembling rhythms from percussive sounds.

**1.4.2.9**  *Overview*      The types of sound generation discussed above do not constitute a complete list, and more digital techniques for generating musical sound exist apart from those mentioned. For introductions to a wide range of methods for producing musical sound, based on electronic digital computation, see for example [Roads et al. 1996], [Wishart 1994] and [Puckette 2007]. Also, the types of sound generation discussed here certainly do not have the character of mutually exclusive categories. On the contrary: they may be combined in any way that is algorithmically possible. An overview, emphasizing this, is shown in Figure 1.10.

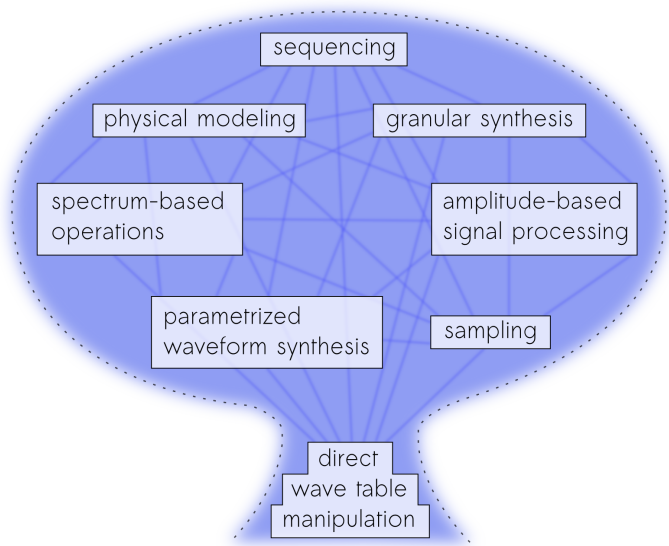

**Figure 1.10** *Some important developments in computed musical sound during the second half of the 20th century. The various groups in this overview of types of new sound-generating processes are based on different ways in which instrumental control has moved away from direct manipulation of the wave table. All groups are shown interconnected, since the various types of sound generation may be combined in any way that is algorithmically possible.*

**1.4.3 Important subsequent developments: wide use**   Above, we have discussed the wide range of sound-generating processes developed over the decades as part of computed sound. Below, we will discuss how the various types of sound generation identified became widely used in the instrumental control of musical sound. This happened based on many successive implementations of the basic pairing of electronic digital computer and electric loudspeaker.

The development of loudspeakers and related amplification technologies continued, leading to better-quality sound reproduction across many form factors. This ranged from sound systems for large crowds, via home stereo systems, to lightweight headphones used with mobile cassette players. Historically, these developments, begun before the introduction of electronic digital computers, continued to evolve independently. In practice, this meant that widespread use of specific forms of computed sound usually was fundamentally limited not by the availability of suitable loudspeaker technology, but by the state of ongoing developments in devices for electronic information processing.

These developments first of all included the digital encoding of information, itself. This provided a fundamental representation shared increasingly widely across many different types of devices. Also, implementations of this were of an increasingly high reliability, with devices retaining or changing their state reliably over many iterations of information processing. These iterations, in turn, were of ever shorter duration, for ever faster computing and information processing. Also, ever more information became subject to automatic processing, with increasing digital memory sizes giving ever greater storage per device. Between devices, communication over distance increasingly became possible, and at increasing transmission speeds. Devices implementing a given technical level of the previously listed characteristics became ever smaller in size and ever cheaper to obtain commercially. Due to the development and implementation of user-friendly human-machine interfaces, such as the GUIs discussed in Section 1.2.3, devices also became increasingly easy to operate.

As the state of these general developments in devices for electronic information processing allowed it, wide use in the instrumental control of musical sound became a reality for many of the new types of sound-generating processes developed as part of computed sound. For example, faster processing enabled the professional and live use of sampling, as a standard technique that became heavily used in the creation of popular music. Faster processing also enabled the live use of various types of parametrized waveform synthesis and physical modeling, with digital synthesizers based on frequency modulation and waveguide modeling becoming commercially available. Sequencing, too, became widely used professionally, especially in combination with sampling, e.g. for assembling percussion-like sound fragments into rhythmic musical sound. In general, it is safe to say that the combination of sequencing and sampling has had a great impact both on how much of popular music is created, and on how it sounds.

In the further processing of musical sound, many types of amplitude-based signal processing and spectrum-based operations became part of standard toolkits for filtering, mixing, equalizing and mastering sound signals. As practically usable memory sizes increased, this was combined with reliable and high-resolution digital implementations of professional multitrack sound recording and editing. Various forms of sequencing and score editing became integrated as much of music recording studio technology in general became digital. Since costs lowered continuously, most of these technologies also became available to, and widely used by, amateurs. Here, low cost often was the more decisive advantage of computed sound, for example where sampling became used to mimick the sound of existing acoustic instruments. Many digital keyboard devices were built and sold based on their ability to provide a cheap imitation of the instrumental control and musical sound of acoustic pianos.

Of course, computed sound also became widely used in the playback of musical sound. Here too, increasing processing speeds, storage sizes and reliability enabled the practical use of high-quality digital representations of sound. The physical media and playback technology of the Compact Disc (CD) consumer format became a very successful example of this. In general, digital technology eliminated sound degradation on repeated playback, copying and storage as a practical problem. The later MP3 format for digitally stored sound became a good example both of the flexibility resulting from the basic digital representation shared across devices, and of the increasing possibility of communication between devices, as many people began to use the format to obtain and distribute musical sound across a wide variety of devices, all connected over the internet. Into the 21$^{st}$ century, this also began to include small, cheap and easy-to-use handheld personal computers, wirelessly connected to the internet. Due to the wide availability of such devices, which give instant access to sound stored in large memories incorporated locally or at remote machines, many people in many parts of the world have quickly grown accustomed to the possibility of listening, anywhere in their daily surroundings, at any time, to the high-quality playback of any musical sound out of a very wide range of choices.

### 1.4.4 Reflection: Theoretical and practical fundamentals for computed sound

We have seen in Section 1.4.2 how the introduction of computed sound, over the decades, resulted in the development of a wide range of new sound-generating processes, including forms of direct wave table manipulation, parametrized waveform synthesis, sampling, spectrum-based operations, amplitude-based signal processing, granular synthesis, physical modeling, and sequencing. In Section 1.4.3, we have discussed how these became widely used for producing musical sound. Clearly, computed sound has become important to the instrumental control of musical sound, in general. But what are the reasons underlying its newness, wide variety, and wide use?

First of all, reasons can be found in the basic model of computed sound. The basic setup of electronic digital computer, wave table and electric loudspeaker described in Section 1.4.1 can be abstracted to a basic model illustrated in Figure 1.11a. Here, in the middle is the electric loudspeaker, as a sound-producing transducer. This transducer has a state, which can be defined by the current values for a set of measurable physical properties describing what of the transducer will vary over time. On the one hand, human sensory perception is exposed to this transducer state, as it changes over time, resulting in perceptual phenomena being induced (in this case, in sounds being heard). On the other hand, the transducer state has been made to causally depend on part of the state of an automaton (in this case, on the wave table of an electronic digital computer). This part of the automaton's state also is subject to the computations it can perform. Crucially, these are *Turing-complete*.

Turing Machines are the well-known models of automata that can be described by a tuple ($\Gamma$, $\Sigma$, $\square$, $Q$, $q_0$, $F$, $\delta$) (see for example [Linz 1997]). Here, $\Gamma$ is an alphabet for symbols read from and written to a hypothetical infinite tape, with $\Gamma$ including a subset $\Sigma$ for input placed initially on the tape, and also a separate blank symbol $\square$. As it operates on the infinite tape, the automaton has an internal symbolic state taken from the set $Q$, initially $q_0$. If this internal state at some point changes into one of the final states contained in the subset $F$, execution within the automaton halts. This point may or may not be reached, depending on the initial input and the transition function $\delta$, which describes how reading a symbol from the tape automatically results in a new internal state, a new replacement symbol written to the tape, and left/right movement to the next tape symbol to be processed. According to the famous Church-Turing conjecture, any algorithmic method for computing numbers can be modeled and executed in this way. A special case, related to this, is the Universal Turing Machine, which accepts the description of any Turing machine with any input as its input, and then models the execution of an exact simulation. In this way, a universal model of computation is provided [Turing 1936]. The computational steps of an actual, practically implemented automaton are said to be Turing-complete if and only if they have a similar capability for simulating any possible Turing machine.

In the context of computed sound, this is important not only for the computation of numbers in a narrow sense, but also for the related ability to simulate any machine that can be characterized by automatic causal transitions between a set of possible discrete states. Many different machines of such a general type may be developed, given a specific type of transducer, and causal dependence for controlling its state, in order to produce heard sound in various specific ways. Where the computed sound model is followed, given sufficient memory and computational speed, the Turing-complete automaton can be used to effectively implement any such conceivable machine. This is what fundamentally underlies the wide variety of sound-generating processes developed after the introduction of computed sound.

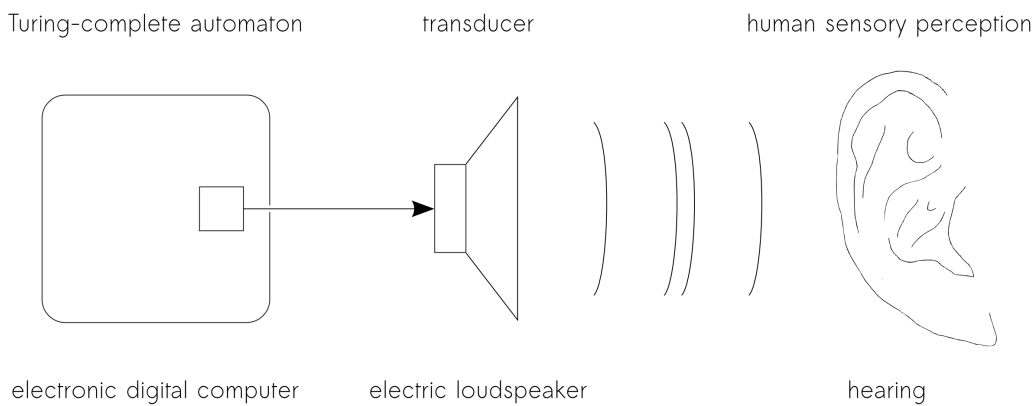

**Figure 1.11a** *A basic model of computed sound. Sound becomes heard as human perception is exposed to changes over time in the state of a sound-producing transducer. The transducer state causally depends on part of the state of a Turing-complete automaton.*

Second, more reasons for the wide variety, and also the wide use of computed sound can be found in *the existence of increasingly powerful implementations of the computed sound model*. These enabled actual development and use. Given some practically implemented Turing-complete automaton, the causal relationship between part of its state and the transducer state usually will be well-known (as for the initial combination of electric loudspeaker and electronic digital computer discussed in Section 1.4.1). But what heard sound this relationship ultimately can result in is subject to discovery, and may be specific to the implementation used. On the one hand, the transducing technologies used for this were made to be increasingly powerful, in the sense of transducer state changes over time being capable of inducing a widening range of perceptual phenomena. (This was often tied to a general goal of increased realism in sound reproduction, driven by demand from basic applications such as the playback of musical sound, film soundtracks, and various forms of voice communication.) On the other hand, the computing technologies used also were made to be increasingly powerful, in the sense of practically allowing the execution of algorithms of increasing complexity. The flawless execution, with infinite memory, in unspecified time of the Turing machine model was matched by the increasing reliability, larger finite memories and faster processing speeds of actual machines. These developments, already mentioned in Section 1.4.3, all resulted from ongoing miniaturization of reliable interconnected electronic switches, strongly driven by the commercial value of the resulting technology. This has been described by Moore's Law, the famous heuristic for integrated electronic circuits, which predicted the sustained and exponential increase, over the years, in the number of transistors per unit area [Moore 1965].

Third, more reasons for the wide variety and wide use of computed sound can be found in *the increasing availability of implementations of the computed sound model*. This enabled the use of computed sound by many people, and thereby accelerated its development. One way of characterizing this increasing availability could be to trace the presence of component technologies in affordable consumer electronics products. For transducing technologies, this would for example include the appearance of home stereo systems. For computing technologies, it would include the later introduction of home- and personal computers. In general, the increasing availability of computing technologies was possible because of the shrinking physical sizes and costs of increasingly user-friendly devices, already discussed in Section 1.4.3. The underlying reason for this, again, was the ongoing miniaturization described by Moore's Law. In this way, the existence and wide availability of powerful computing technology were closely intertwined, with existence depending on miniaturization – only possible because of the financial rewards of mass production and mass use. In the development of transducing technologies, e.g. for personal audio, expected mass use was similarly present as a fundamental factor.

Concluding, fundamental reasons for the newness, wide variety, and wide use of computed sound can be summarized as the development of cheaply mass-producible, powerful implementations of the computed sound model.

## 1.5  The need for computed fingertip touch

### 1.5.1  Computational liberation of instrumental control
Reflection on the discussion of computed sound presented in Section 1.4 may lead us to ask a broader question: What is the *general* role of computation in the development – since its earliest beginnings – of instrumental control of musical sound?

#### 1.5.1.1  *Completely computed instrumental control of musical sound*
As a thought experiment, suppose we would have a humanly programmable, Turing-complete automaton that, within the scope of instrumental control of musical sound, would somehow be capable, first, of monitoring and internally representing all aspects of human action; and second, of internally representing and inducing all aspects of human sensory perception. Suppose that the memory and processing speed of the automaton would suffice for the implementation of such a set of algorithmic relationships, involving the above discrete-state representations, that for any additionally definable, but non-implementable relationship, the set would contain some implementable counterpart not humanly discernible in its overall outcomes. We might call such a system capable of *completely computed instrumental control of musical sound*, with the word "completely" referring both to the system's coverage of aspects of human action and sensory perception, and to its Turing-completeness.

As before, we understand instrumental control of musical sound as the process where human actions make changes to a sound-generating process, resulting in heard sound inducing musical experiences (see Section 1.1). The aspects of sensory perception covered by the definition, above, therefore would include all possible heard musical sound. (Being made heard, sound might no longer be present as a mechanical wave, however: e.g. in implementations based on direct neural stimulation.)

Therefore, the above definition also implies a capability to represent and execute all algorithmically definable, perceivably different causal relationships involving discrete state representations tracking any possible human action and potentially inducing any change in heard musical sound. This seems to approach a general capability for implementing all perceivably different causal relationships between human actions and changes in heard musical sound. This seems relevant in the general context of the development, since prehistory, of technology for the instrumental control of musical sound. This development can be seen as a great search to extend the possibilities of causal relationships between human actions and heard musical sound. A recurring question here could be formulated simply as: *How can the instrumental control of musical sound be improved?* [3] Investigating this – *which* forms of instrumental control would be better to use – depends on also determining *what* forms are possible to implement, in the first place. Completely computed instrumental control would represent a general means to this end, maximizing the range for potential implementation based on its wide-ranging underlying criteria. Obtaining it, therefore, would support the search for better instrumental control in general.

**1.5.1.2** *Computational liberation* Actually realizing completely computed instrumental control may be hard, but gradually approaching it seems possible. Instead of full coverage, we might require only an *increasing* coverage, of *some* of the relevant aspects of human action and sensory perception, and their possible causal relationships. Indeed, defined like this, such a gradual approach can be recognized as a de facto ongoing historical process. The development of computed sound is a prominent example of this, especially increasing the coverage of heard musical sound.

This gradual approach, then, does not occur in a technological vacuum, and it may for example involve the addition of computational components to existing systems. These systems originally may have enabled causal relationships between human actions and heard musical sound purely via mechanical and acoustic components, or via analog electric and electronic ones. (Examples have been discussed in Section 1.2.) However, unlike means of Turing-complete computation, such components do not explicitly minimize constraints imposed on implementable causations. In practice, the more aspects of human action and sensory perception become subject to computation, the less constraints we may expect to remain on implementable forms of instrumental control. Because of this reduction in constraints, it seems appropriate to refer to this gradual approach as the *computational liberation* of instrumental control of musical sound.

---

3   *Improvement* here explicitly includes giving access to new sonic possibilities.

Increasingly making human action and sensory perception the subject of computation will require both the development of automaton output, i.e. realized causal dependences of aspects of sensory perception on automaton states; and automaton input, i.e. realized causal dependences of automaton states on aspects of human action. However the process of computational liberation may unfold, it must enable these types of causal dependences, and this will require the invention and development of transducers.

**1.5.2  Computed fingertip touch**    Because of its extreme importance to the instrumental control of musical sound (see Section 1.2.5), fingertip use will be an important area for computational liberation. To describe the prerequisites for computational liberation that are specific to fingertip use, we may, as for computed sound, outline a general model: one of *computed fingertip touch*.

**1.5.2.1** *A basic model of computed fingertip touch*    In general, computed fingertip touch will be based on the extension of a humanly programmable, Turing-complete automaton with transducers for output and input. Each transducer will have a state that can be defined by the current values for a set of measurable physical properties, describing what of the transducer will vary over time. For each output transducer, its state will causally depend on the state of the Turing-complete automaton. Changes over time in output transducer state will, in turn, induce aspects of human somatosensory perception involving the fingertips (see Sections 1.3.2 and 1.3.3). For each input transducer, its state will causally depend on aspects of human motor activity involving the fingertips (see Sections 1.3.1 to 1.3.3). Changes over time in input transducer state will, in turn, influence the state of the Turing-complete automaton. For computed fingertip touch to be meaningful, i.e. perceivable, at least the presence of output transducers will be mandatory. This basic model is illustrated in Figure 1.11b.

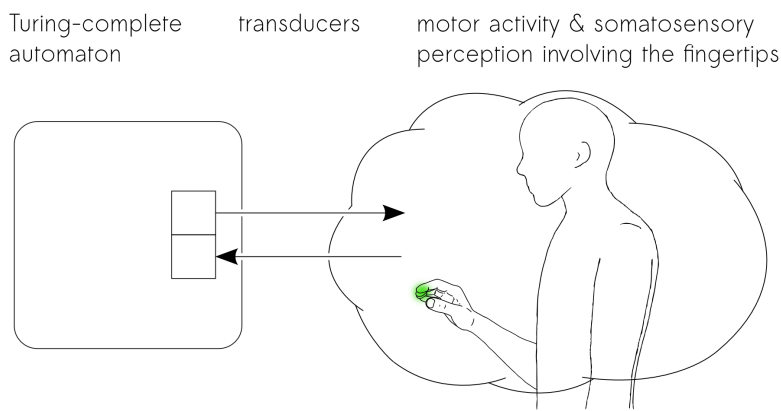

**Figure 1.11b**  *A basic model of computed fingertip touch, further supporting the computational liberation of instrumental control of musical sound.*

# 1.6  Pursuing computed fingertip touch

Our primary goal will be to further the computational liberation of instrumental control of musical sound, as defined in Section 1.5. Recapitulating, here, in contrast with earlier technologies, Turing-complete automata combined with transducers explicitly minimize constraints on implementable causal relationships between human actions and changes in heard musical sound.

To work toward this goal, our strategy will be to innovate those technologies that enable fingertip use in the instrumental control of musical sound. This choice to focus on fingertip use is motivated by its existing extreme importance to instrumental control (see Section 1.2.5). Innovating applied technologies specific to fingertip use will have to fit the computed fingertip touch model of Section 1.5, if it is indeed to enable further computational liberation.

In the subsections below, discussion of a number of issues will give further, more specific direction to our pursuit of computed fingertip touch. In the final subsection, this is summarized, in a series of interrelated goals for the body of this thesis.

**1.6.1  The need for new, specific transducer technology**     Clearly, the mechanical systems enabling fingertip use in existing, fully acoustic musical instruments do not fit the computed fingertip touch model: they do not implement a humanly programmable, Turing-complete automaton. Even where such an automaton is present, and combined with a mechanical system for fingertip use taken from an acoustic instrument, there still may be no fit with the model. For example, a piano-type keyboard, separate from any complete acoustic piano, may very well be used for the instrumental control of computed sound (see Section 1.4). But the keyboard's mechanical system will function only as a fingertip input transducer to the Turing-complete automaton, not providing it with transducer output to fingertip-related somatosensory perception. Therefore, there will still be no fit with the model.

For the same reason, the general-purpose, fingertip-operated input devices discussed in Sections 1.2.3 and 1.2.4 also do not fit the model. These devices included buttons, switches, linear sliders, rotary knobs, dials, mouse and keyboard keys, touch pads, and touch screens. In Section 1.2, we discussed how adopting general-purpose input devices such as these for the instrumental control of musical sound has been a major trend during the 20[th] century, continuing into the 21[st]. This trend quite possibly may continue into the future. But is it enough to just wait for the next iteration of consumer technology; adopt its fingertip input for the instrumental control of musical sound; and hope that one day, this will happen to fit the computed fingertip touch model? We do not think so. To obtain the prerequisites for furthering the computational liberation of instrumental control, what is needed, rather than adoption of the technologies mentioned, is the active pursuit of new transducer technologies which specifically fit the computed fingertip touch model.

The possible directions for such active pursuit seem quite varied: based on the discussion in Section 1.3 of the anatomy, physiology and neural processes of human fingertip use, many anatomical source and target locations for transducer causations are imaginable. Also, here, it might seem natural to first categorize human receptors as targets for transducer output; and human effectors as sources for transducer input. However, transducer output to human effectors, e.g. the electric activation of muscle tissue, might also be a means to induce aspects of somatosensory perception, e.g. via the muscle spindles. Similarly, transducer input from human receptors, e.g. the recording of mechanoreceptor firing patterns, might be a means to monitor aspects of motor activity, e.g. during mechanical contact with external objects. This further illustrates the wide range of apparent possibilities when pursuing new fingertip transducer technology.

**1.6.2 The need for support of unidirectional fingertip movement orthogonal to a surface**   In Section 1.2.5, we defined unidirectional fingertip movement orthogonal to a surface as fingertip movement that can be characterized as approximating a single path of movement, at right angles with a surface, and extending across at most a few centimeters. To be comprehensive, computed fingertip touch should also support the forms of fingertip instrumental control where this particular type of movement occurs.

Moreover, realizing this support has some priority over realizing support for other possible forms of fingertip instrumental control: unidirectional fingertip movement orthogonal to a surface has both a prevalence in, and specific advantages for fingertip instrumental control (see Section 1.2.5).

Concretely, this implies a need for such transducer technology as can monitor aspects of motor activity, and induce aspects of somatosensory perception, that are involved in this type of movement.

**1.6.2.1** *Consequent choice: the human fingerpad as transducer source and target* During unidirectional fingertip movement orthogonal to a surface, the *fingerpad*, or pulp of the fingertip (see Section 1.3.1.4) typically is the area making mechanical surface contact. Via this contact, changes are made to sound-generating processes during instrumental control. Therefore, we will pursue transducer technology which follows the same principle, with transducer causations having the fingerpad as their anatomical source and target location.

**1.6.2.2** *Consequent choice: transducers based on flat, closed, rigid contact surfaces* During the mechanical contact just mentioned, the fingerpad often has an approximately flat, closed, and rigid surface as its immediate counterpart outside the human body. Surely, this is not always the case: for example, the fingertip may open and close holes on aerophones, or press strings against instrument bodies. However, many existing forms of instrumental control do occur via this type of surface: for example, the fingertip may open and close valves on aerophones; strike pads and membranes; tap and press sensor surfaces; and perform press/release cycles on push-buttons, computer keys, and the keys of piano-type keyboards (see Section 1.2.5).

These existing forms of instrumental control can be grouped together as based on *unidirectional fingertip movement orthogonal to flat, closed, rigid surfaces*. Then, having the same type of surface making contact with the fingerpad could also be a useful principle for implementing transducer technology: this may allow support for many of the aspects of motor activity and somatosensory perception that are involved in these forms of instrumental control. Also, obtaining such support has some priority: because of the existing prevalence of this subtype of instrumental control, realizing support for it represents an important part of realizing support for unidirectional fingertip movement orthogonal to a surface, in general.

**1.6.2.3** *Consequent choice: orthogonal as well as parallel force output to the fingerpad*
The above transducers will be used to implement forms of instrumental control based on dynamic mechanical contact between the surface of the transducer and that of the human fingerpad. This will involve a mutual application of forces, where mandatory transducer output (see Section 1.5.2.1) produces forces applied to the fingerpad surface. Since the wider context here is providing support for unidirectional fingertip movement orthogonal to a surface, such force output at least must include forces orthogonal to the fingerpad surface.

These forces may then be resisted mechanically by the palmar side of the fingertip, causing the somatosensory perception of fingertip touch. However, as discussed in Section 1.3.1.4, via its skin, the fingerpad also offers another main type of mechanical resistance resulting in the perception of fingertip touch: the resistance against relative movements *parallel* to the fingerpad surface. To also cover this other main type of sensory transduction, force output parallel to the fingerpad surface is needed, too.

**1.6.3 The need for physical units in algorithms**     The previous discussion already points to the development of rather specific transducer technology. It is important to emphasize that any such development, even when completed, will provide only a starting point for further implementation. A general issue here is that even if a Turing-complete automaton and a concrete set of transducers have been obtained – which together enable computational liberation – this does not also imply the simultaneous realization of any direct correspondences between automaton states on the one hand, and aspects of human action and perception on the other. On the contrary: such correspondences quite probably will have to be built and discovered.

For example, in computed sound, the combination of digital wave table and electric loudspeaker has enabled computational liberation. But clearly, the first historical implementations of this combination only provided the starting points for a long and still ongoing process of further implementation, covering ever more aspects of heard musical sound (see Section 1.4). Here, not only have obstacles been overcome so as to match known aspects of heard musical sound with known automaton states. Continuing implementation has also reflected a process of more fundamental discovery: Transducer technology enabled the development and implementation of precisely defined and executed algorithmic relationships between measurable physical properties such as time and air pressure. This was then used to induce various aspects of heard musical sound in a reproducible manner. Based on this, a greater understanding has been obtained also of the very nature of aspects of heard musical sound. This increased fundamental understanding includes conceptual as well as experiential knowledge. A seminal example of this has been the generation of heard stable sine waves, made possible by the use of digital wave tables [Mathews et al. 1969].

In computed fingertip touch, we might expect an analogous process of construction and discovery. In any case, after initial transducer implementation, there will be a need for further implementation, so as to match automaton states to those aspects of motor activity and somatosensory perception deemed relevant to the further development of instrumental control of musical sound. To facilitate this, it is important to ensure from the outset that transducer state is algorithmically represented in terms of physical units. After all, any and all matching will be based, in the first place, on the specific changes over time in the physical state of the individual output or input transducers used. Therefore, these physical changes should be transparently accessible to someone writing algorithms for the transducer technology.

**1.6.4  The need for support of passive touch, active touch, and manipulation**   In Section 1.3.4 and before, we discussed how the human fundamentals of fingertip instrumental control include processes of passive touch, active touch, and manipulation. Therefore, support for these important types of human sensorimotor activity should be explicitly present in computed fingertip touch.

For the development of transducer technology, this means that aside from output transducers, the presence of input transducers will be mandatory as well: Aspects of active touch, and thereby of manipulation, will be possible to induce only if automaton states somehow represent concurrent human motor activity.

**1.6.5  The need for integration with computed sound**   Clearly, to further the computational liberation of instrumental control of musical sound, systems implementing computed fingertip touch should not exist in isolation from those implementing computed sound. On the contrary, computed fingertip touch should be completely integrated with computed sound, enabling instrumental control of all of the types of sound-generating processes discussed in Section 1.4.2.

**1.6.6  The need for support of real-time instrumental control**   In Section 1.1, we announced a focus on real-time forms of instrumental control of musical sound. Here, processes of instrumental control overlap in time with the resulting heard musical sound. Certainly, there exist important non real-time forms of instrumental control: the instrumental control of various forms of sequencing, for example (see Section 1.4.2.8, but also outside of computed sound). On the other hand, when considering widely used types of musical instrument, from the earliest examples available, via traditional acoustic instruments, up to contemporary instruments (see Section 1.2), it will be clear that real-time processes of instrumental control are important in determining heard musical sound. Therefore, integrated systems combining computed fingertip touch and computed sound should also support real-time instrumental control.

**1.6.7  The need for cheaply mass-producible, powerful implementations**   At the end of Section 1.4, we found that fundamental reasons for the newness, wide variety, and wide use of computed sound can be summarized as the development of cheaply mass-producible, powerful implementations of the computed sound model. Implementations of computed fingertip touch may be regarded as less or more powerful in similar terms, e.g. automaton processing speed, memory size, and transducer fidelity. Clearly, computational liberation may benefit from implementations that are more powerful in such terms, as long as this results in humanly perceivable differences.

A still greater benefit can be expected if here again, powerful implementations also are suitable for cheap mass production. Getting answers to the fundamental question what forms of instrumental control are possible (see Section 1.5.1) is better served by technology that can be made available cheaply, i.e. to many people. Such wide availability helped bring about the existing wide variety of forms of computed sound (see Section 1.4), and similarly, could help bring about a wide variety of forms of instrumental control based on computed fingertip touch.

**1.6.8 The need to demonstrate computational liberation**    Above, we have discussed a series of needs, each deserving to be addressed when implementing computed fingertip touch. To ensure that in doing this, we do not lose track of the overall goal of implementation itself, we explicitly state a final need: to actually demonstrate computational liberation.

**1.6.9 Recapitulation: goals for the body of this thesis**    The series of needs that has just been discussed gives more specific direction to our pursuit of computed fingertip touch. We will now concisely recapitulate this, in a series of interrelated, concrete goals for the body of this thesis. Each goal below is followed by one or more section numbers, pointing back to its motivation. The goals then are:

• Chapter 2: To obtain *new transducer technology*, which
   • provides novel I/O fitting the computed fingertip touch model (see 1.6.1 & 1.6.4);
   • supports unidirectional fingertip movement orthogonal to a surface (see 1.6.2);
   • has the human fingerpad as its source and target location (see 1.6.2.1);
   • is based on flat, closed, rigid contact surfaces (see 1.6.2.2);
   • provides orthogonal and parallel force output to the fingerpad surface (see 1.6.2.3).

• Chapter 3: To then expand this to *new systems for computed fingertip touch*, which
   • algorithmically represent transducer state using physical units (see 1.6.3);
   • integrate computed sound (see 1.6.5);
   • support real-time instrumental control of musical sound (see 1.6.6);
   • are powerful and cheaply mass-producible (see 1.6.7).

• Chapter 4: To then realize support for *new forms of instrumental control* of musical sound, which
   • covers passive touch, active touch, and manipulation (see 1.6.4);
   • demonstrates computational liberation (see 1.6.8).

Chapters 5 and 6 present the two research excursions already announced in the Thesis outline and summary.

Chapter 7 then presents the conclusion of this thesis, also revisiting the goals stated above.

# 2. New transducer technology

**CHAPTER SUMMARY**

In this chapter, after discussing the relationship between computed touch and haptics, we develop two transducer technologies implementing the computed fingertip touch model: the *cyclotactor (CT) device*, which provides fingerpad-orthogonal force output while tracking surface-orthogonal fingertip movement; and the *kinetic surface friction transducer (KSFT) device*, which provides fingerpad-parallel force output while tracking surface-parallel fingertip movement.

Both technologies are based on the use of a flat, closed, and rigid contact surface. Both, also, have a force output range large enough for automaton output to potentially co-determine the movement of fingertip control actions. This enables implementing a greater range of causal relationships between human actions and changes in heard musical sound.

Another aspect common to both technologies, and novel relative to earlier, related haptic transducers, is the idea to enable more precise output to somatosensory perception by *avoiding the use of connected mechanical parts moving relative to the target anatomical site*. In the KSFT device, this idea is achieved partially; in the CT device, completely.

Important properties that are specific to the CT device include support for both downward and upward force output to the fingertip; accurate mechanical wave output across the frequency ranges involved in fingertip vibration perception; and the capability to induce aspects of haptic perception.

Novel aspects specific to the CT device include support, specifically and explicitly, for unidirectional fingertip movement orthogonal to a surface; and I/O that is specific to those flexing movements of the human finger that are independent, precise, and directly controlled by the motor cortex. This is supported by a precisely adjustable, personal fit, enabling accurate yet comfortable I/O.

An important property that is specific to the KSFT device is support for inducing high-resolution aspects of fingertip surface texture perception during active touch. A novel aspect is that this is done using displacement input based on cheap, off-the-shelf optical mouse sensor technology.

De Jong S, 2006 A tactile closed-loop device for musical interaction. In *Proceedings of the 2006 international conference on New Interfaces for Musical Expression* (NIME06, June 4-8 2006, Paris, France) 79-80.

De Jong S, 2009 Developing the cyclotactor. In *Proceedings of the 2009 international conference on New Interfaces for Musical Expression* (NIME09, June 3-6 2009, Pittsburgh, PA, USA) 163-164.

De Jong A P A, 2010d Apparatus comprising a base and a finger attachment. *US patent application no. 12/792,432* (June 2 2010) 1-56.

# 2.1 Introduction

In this chapter, we will pursue the first set of goals identified in Section 1.6.9. The main goal is to obtain novel I/O fitting the computed fingertip touch model. To do this, we will build new transducer technologies. These will have the human fingerpad as their source and target location; and will be based on flat, closed, and rigid contact surfaces.

In Section 2.2, we will develop a technology of the above type that specifically supports unidirectional fingertip movement orthogonal to a surface, while providing force output orthogonal to the fingerpad surface. In Section 2.3, we will develop an alternative technology that provides force output parallel to the fingerpad surface.

Both of these sections will follow the same structure: First, the operating principles and existing technological context of the proposed transducer are discussed. Then, a detailed description follows of how the computed fingertip touch model was actually implemented, in a progressive series of prototypes. Finally, it is summarized how, in the resulting technology, the research goals of this chapter were attained.

However, before any of this, and especially before technological context can be discussed in detail, it is necessary to first clarify the relationship between computed touch and the field of Haptics.

**2.1.1 Haptics and computed touch**    To define the term *computed touch,* we can simply generalize from the concept of *computed fingertip touch* (see Section 1.5.2.1). Again, computed touch will be based on a humanly programmable, Turing-complete automaton, causally linked to the physical state of a set of transducers. These provide input or output, depending on the direction of causality. Again, for computed touch to be meaningful, at least the presence of output transducers will be mandatory. However, the difference is now that changes over time in output transducer state will induce aspects of human somatosensory perception *in general*; while similarly, input transducer state will causally depend on aspects of human motor activity, *in general*.

This means that the I/O of systems for computed touch may involve other parts of the limbs than the fingertips, as well as parts of the torso, neck, and head. In the context of instrumental control of musical sound, this especially includes output to the lips, mouth, tongue, and throat. Related input might then for example come from human motor activity associated with blowing, whistling, and making vocal sounds.

The term *haptics*, according to [Gibson 1962], was proposed for a hitherto unrecognized mode of human experience in [Révész 1950]. In his book, Révész defined haptics as "the impressions conveyed by the tactile and kinematic sense". These impressions are often understood to be spatial in nature, and the result of exploration by the hand [Goldstein 2002]. Processes of active touch and manipulation (as discussed in Section 1.3.3.5) must therefore play a central role in haptics.

Is all computed touch haptics? This will probably depend on the answers to a number of other questions. These include: whether haptics is deemed to include pain and temperature sensing; whether it is deemed to include passive touch (see Section 1.3.3.4); and whether it is deemed to also employ other body parts than the hands, such as the mouth and throat. [1]

In any case, not all haptics is computed touch. Of course, much technology for haptics from recent years has been microprocessor-based, so that haptic devices may provide computed touch, and vice versa. But haptics does not require the presence of a humanly programmable, Turing-complete automaton. In fact, just as the instrumental control of musical sound, it does not even require the presence of technology in general.

## 2.2 Fingerpad-orthogonal force output: The cyclotactor (CT)

**2.2.1 Operating principles**    A key principle underlying the first transducer technology presented in this chapter, is to avoid the use of any connected mechanical parts moving relative to the target anatomical site. Such components, by participating in mechanical phenomena such as friction and vibration, will introduce artefacts in transducer state over time. This negatively affects the precision of output to human somatosensory perception. One way to avoid this problem is to altogether avoid the use of motors, linkages, and connected, moving mechanical parts in general.

Since one of our goals is to support unidirectional fingertip movement orthogonal to a surface (see Section 1.2.5), mechanical movement by the fingertip itself will be a given. Another given is the rigid contact surface we have decided to use for applying orthogonal forces to the fingerpad. A first step, then, to avoid attaching any mechanical parts moving relative to the target anatomical site, is to rigidly attach this contact surface to the fingertip.

The aspects of human motor activity that we are interested in always involve orthogonal fingertip movement relative to some primary surface. Therefore, suitable transducer input may be obtained by tracking the vertical distance between the fingerpad contact surface and this primary surface. This may be done using some type of sensor that does not use connected moving mechanical parts, and that itself can be rigidly attached relative to e.g. the primary surface.

The aspects of human somatosensory perception that we are interested in involve forces applied orthogonally to the fingerpad. Therefore, suitable transducer output may be obtained by rigidly attaching a permanent magnet to the fingerpad contact surface, and applying a variable, orthogonally directed force to it via magnetism. This may be done, without introducing any moving mechanical parts, by using an electromagnet, rigidly attached to the primary surface.

---

1    Further discussion would take into account e.g. [Lederman and Klatzky 2009].

One property of the proposed setup is that transducer output can apply both upward and downward forces to the fingertip. This is guaranteed by the rigid attachment of the permanent magnet to the fingertip, and the dynamically reversible polarity of the electromagnet.

Another property of the proposed setup is that changes in output may directly cause changes in input: the force output to the fingertip may alter its distance trajectory. Also, as a general property of computed touch, changes in input can be made to directly cause changes in output: via the programmable Turing-complete automaton. Taken together, this means that the causal relationships that can be implemented between input and output may become *cyclical* in nature. The proposed transducer was therefore named *cyclotactor (CT)*, a term combining the Greek κύκλος, for ring or circle, with the Latin *tangere:* to grasp or to touch.

A final property of the proposed setup is that the transducer component attached to the fingertip will be central to how new programmable causations between input and output actually happen. It was therefore named *keystone*, a term from architecture for the central supporting stone that closes the top of an arch.

**2.2.2 Technological context**    In general, haptic devices may provide computed touch (see Section 2.1.1). It is therefore logical to consider their use also in the specific present context, where we seek to implement computed touch that supports unidirectional fingertip movement orthogonal to a surface.

This is further supported by the fact that the potential of using haptic devices, in the context of instrumental control of musical sound, has also been highlighted based on other lines of argumentation. An example of this is [O'Modhrain 2000], where the incorporation of haptic feedback into new musical instruments was studied. This was followed by efforts to facilitate and standardize such incorporation, using off-the-shelf force-feedback devices, in [Sinclair and Wanderley 2007] and [Berdahl et al. 2009].

The principle of avoiding connected mechanical parts moving relative to the target anatomical site has been previously implemented in haptic transducer technology: The 6 DOF magnetic levitation device of [Berkelman and Hollis 2000] [The Magnetic Levitation Haptics Consortium 2009] was designed for use via a hand-grasped joystick [Grieve et al. 2009], however, not as a fingerpad transducer during orthogonal movement.

Now, before further considering devices from the modern field of haptics, we will first consider certain mechanical devices invented before the 1950s, because some of their relevant fundamental characteristics have been implicitly carried over into more recent haptic devices that are suitable for computed fingertip touch.

Traditional mechanical devices such as pressed-down strings and piano-type keys (see Section 1.2.3) effectively apply only upward forces to the fingertip. Of course, at some point during a press/release cycle, the fingerpad might stick somewhat to the contact surface being used, resulting in downward forces being applied. But these forces are relatively small, they are not applied in a controlled manner, and they probably do very little to determine the course of fingertip control actions making changes to heard musical sound.

When using the proposed CT device, on the other hand, downward forces may be large enough to move the fingertip; are applied under the control of a Turing-complete automaton; and therefore, may be used to co-determine how fingertip control actions happen. In this way, more types of fingertip control action are supported.

This means that the controlled application of downward forces is a fundamental difference between the proposed CT device and the mechanical devices mentioned: Its availability implies that a greater range of causal relationships between human actions and changes in heard musical sound may be implemented (see Section 1.5.1.1).

The same fundamental difference also is present in haptic devices that have extended from the piano-type key, such as the actuated lever pressed down by the fingertip introduced with the Cordis system of [Cadoz et al. 1984]. In later work, multiple such actuated levers were mechanically connected to implement various types of actuated levers, joysticks, pliers, and other devices. However, this apparently did not include a device applying downward forces to the fingertip [Cadoz et al. 2003].

To actuate each lever, a motor was used. In [Cadoz et al. 1990], a custom motor type was introduced for this, based on a flat electromagnet coil, movable relative to a nearby fixed permanent magnet. This motor was then mechanically connected to its piano-type key, enabling actuation. Another important difference with the proposed CT device, therefore, is the use of connected mechanical parts moving relative to the target anatomical site (see Section 2.2.1).

The piano-type keys in [Gillespie 1996] and [Oboe and De Poli 2002], also actuated by mechanically connected motors, are subject to the same differences: no controlled application of downward forces, and the use of connected, moving mechanical parts.

A different haptics technology, not extending from the piano-type key, is the Phantom device of [Massie and Salisbury 1994]. This device provides a contact point that can be moved around within a volume of approximately 3 dm$^3$, along all three spatial directions. Along the same three directions, force output can occur (see also the overview in [Hayward et al. 2004]). The contact point is often operated using a stylus, but a single fingertip may also be attached. Clearly then, the Phantom device could also support unidirectional fingertip movement orthogonal to a surface, including downward force output to the fingertip. However, here too, force output is applied by motors via mechanical linkages (cables), whereas in the proposed CT device, there are no connected mechanical parts moving relative to the fingertip.

**2.2.3 Prototype 1    2.2.3.1** *Transducer components*    In the first prototype, the permanent magnet of the keystone was made of ceramic ferrite [Kato and Takei 1930], and was of a type normally used to keep doors closed [Allegro MicroSystems 1997].

The overall shape of the keystone was that of a small, flat block, held between the tips of the thumb and of the index and middle fingers (see Figure 2.1).

A light-dependent resistor (LDR) [Smith 1873] was used as the distance sensor. The LDR was mounted beneath an aperture, cut out from an otherwise closed plastic surface. This aperture was made similar in its rectangular shape and size to the down-facing side of the keystone. The LDR, facing upward, then tracked the increasing blockage of environment light caused by decreasing distance between the keystone and the aperture. For better spatial precision, black tape was applied to the bottom and side surfaces of the LDR, shielding it from incoming light from other directions. The LDR measurement voltage was converted to digital input using a Microlab voltage↔MIDI converter [Elektronische Werkplaats of the Royal Conservatoire 2013].

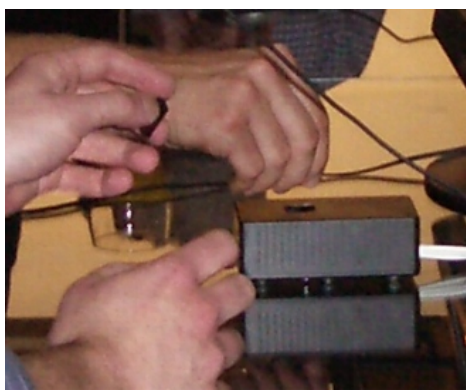

**Figure 2.1** *CT prototype 1.*

The solid-core electromagnet [Sturgeon 1825] that was used originally was part of an RXAB010 electromechanical relay, and was taken from defunct cinema equipment. Its two poles both ended at the same vertical distance below the plastic device surface. One pole, being the top surface of the cylindrical core of the electromagnet, had a roughly circular surface area about the size of the human fingerpad. The other pole, formed by a ferromagnetic bridge outside of the coil, had a rectangular top surface, much wider than deep, which was placed more distally relative to the fingertips (see Figure 2.3). As a whole, the electromagnet was mounted with its core directly beneath the distance sensor and aperture, so as to apply vertical forces to the keystone. These forces were based on a control voltage that was multiplied to a larger voltage, and then applied to the coil terminals of the electromagnet. The control voltage itself was produced from digital output using the same Microlab device.

**2.2.3.2** *Distance input*     In software, each incoming distance measurement was linearly scaled to a unitless, normalized range, corresponding to the actual range of integer values currently being produced by the LDR. Correspondingly, after switching on the device, this normalized distance input was calibrated by performing a full downward keystone movement toward the surface aperture. The minimum of the normalized range then would reliably correspond to the keystone being located at the surface aperture. The maximum would vary in its actual location, however, depending on the intensity and angle of incoming environment light. Distance resolution, the number of steps detected between the minimum and maximum, similarly would vary with lighting conditions.

Slow fluctuations in environment light (e.g. due to passing clouds) were problematic, especially in being detected as the keystone moving in and out of range. To counter this, the maximum 5% of the available sensor range was mapped to a value indicating "no object detected". Rapid fluctuations in environment light, on the other hand (e.g. due to 50 Hz light bulbs) would also introduce errors in the keystone distance measurement. To counter this, a low-sensitivity backup input was added in software.

Overall, when using daylight or stable electric light, distance input typically had a range of about 4 cm, a resolution of around 105 steps, and a sampling rate of approximately 100 Hz.

**2.2.3.3** *Force output*    It turned out that the electromagnet could be used to apply downward as well as upward forces to the keystone without the need for coil current reversal: At low current levels, magnetization of the electromagnet core by the nearby keystone permanent magnet could already generate considerable downward forces.

Maximum coil current was 2 A, corresponding to an upward force sufficient to make single-handedly touching the electromagnet top surface with the keystone nearly impossible. The force output range seemed to change when the electromagnet became hot, however. Changes in output force could be made felt within a range of up to 10 cm above the electromagnet. Force output had a resolution of 26 steps between maximum downward and maximum upward force, and a sample rate of 200 Hz.

**2.2.4  Prototype 2    2.2.4.1** *Transducer components*    In the second prototype, the keystone was rigidly attached to one finger using a velcro strap [De Mestral 1955]. This was done to obtain I/O specific to a single finger. The keystone's main contact surface now pressed against the palmar side of the distal phalanx, while also extending proximally to the middle phalanx (see Figure 2.2).

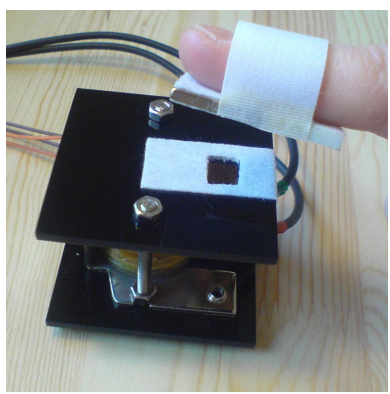

**Figure 2.2**  *CT prototype 2.*

Distance sensing was now done using the reflection of infrared (IR) light. From beneath the surface aperture, a light-emitting diode (LED) directed IR light of a fixed intensity [Losev 1927] [Braunstein 1955] [Biard and Pittman 1966] up toward the keystone. If present above the aperture, the keystone would reflect this IR light, via a layered surface on its down-facing side. The amount of light coming back through the aperture, which varied with reflector distance, was then converted into an electric signal by an IR detector [Shive 1953] [Nishizawa and Watanabe 1953].

The distance sensor and surface aperture no longer were placed straight above the electromagnet core, but rather, proximally beside it (see Figure 2.2). For this reason, the keystone reflective surface extended to under the middle phalanx. Digital I/O to and from the electronic sensing circuit was now done using an Ipsonlab voltage↔OSC converter, with custom firmware for faster I/O rates. This was kindly provided by Lex van den Broek, creator of the device [Elektronische Werkplaats of the Royal Conservatoire 2013].

To apply forces to the keystone, the same electromagnet as before was used. The keystone permanent magnet was changed, however, to a neodymium magnet [Sagawa et al. 1984] [Croat et al. 1984]. A thermistor circuit [Faraday 1833] [Grisdale 1941] was added to electronically track the temperature of the electromagnet. Digital I/O to and from the electromagnet and thermistor circuits was also implemented using the Ipsonlab device.

**2.2.4.2** *Distance input: toward a fixed range*    In prototype 1, the LDR-based distance sensing depended on environment light, and its range would fluctuate with changing lighting conditions. To obtain a fixed distance input range, a number of alternative sensing technologies were considered for use in prototype 2. These included: Hall-effect sensing [Hall 1879] [Maupin and Vorthmann 1971]; ultrasonic reflective sensing [Sokolov 1929] [Firestone 1942]; capacitive proximity sensing [Philipp 2002]; and IR reflective sensing. IR reflective sensing was chosen because it could provide the most suitable combination of spatial range, spatial resolution, and temporal resolution.

This choice meant that distance sensing again would be based on the intensity of incoming light, varying with the distance between keystone and device surface. However, the dependence on fluctuating environment light was now replaced by dependence on a known and stable light source: the IR LED. Also, distance tracking would be minimally affected by fluctuating visible light from the environment, since measurement was now based on a reflected beam occurring in the infrared part of the spectrum.

Four IR reflective sensor models of different brands were considered for use in prototype 2. Selection between them was based on the provision of explicit measures against incoming environment light, and on the amount of noise in the resulting electric signal. The effective range of the selected sensor only began some distance above its surface, however. Placing this sensor directly on top of the electromagnet core, like the LDR in prototype 1, would therefore have led to the loss of a significant part of the vertical area where force output could be strongest. To avoid this problem, the distance sensor was now placed next to the electromagnet, at its proximal side, and at a lower position relative to the device surface.

A number of light-reflecting materials were tested for use on the keystone. This included various types of plastics, metals, and mirror coatings. Specularly reflecting materials would return more IR light, also from further distances, and thereby potentially enabled a larger distance sensing range. However, their use also introduced a new obstacle to obtaining a fixed distance range.

Changes in not the distance, but the orientation of the fingertip (e.g., its roll) could now, via specular reflections, lead to spikes in the detected intensity of the reflected IR beam. These spikes were then indistinguishable from similar results caused by rapid changes in fingertip distance. Therefore, to safeguard accuracy, a layered combination of materials was used, selected for the suitable trade-off between more diffuse reflection and high reflection intensity that it was found to provide.

Another problem affecting the accuracy of distance input remained, however. It was found that sunlight falling on the device surface could still change the IR-based distance measurement significantly.

**2.2.4.3** *Distance input: obtaining a linear range*  Software objects were written in the Max/MSP programming environment [Puckette 2002] to obtain, monitor and log voltage measurements coming from the IR reflective sensor circuit. Here, the custom Ipsonlab device was used at its maximum 400 Hz input sampling rate. It was then determined how, when using the reflective surface discussed in Section 2.2.4.2, voltage measurements would vary with keystone vertical distance. This was done using a setup based on a digital caliper with a resolution of 0.01 mm (see Figure 2.3). The resulting measurements, converted to an interpolation table, were then used to implement the run-time conversion of voltage measurements to linear distance input.

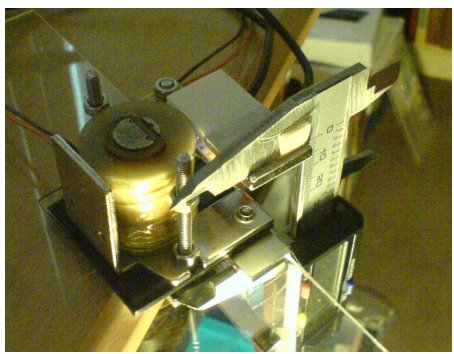 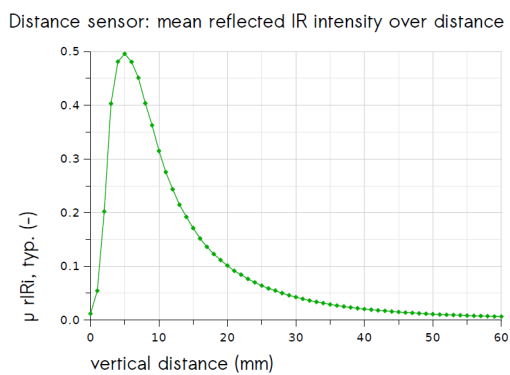

**Figure 2.3**  *Distance input: obtaining a linear range. Left: taking measurements. Right: results.*

For the actual construction of algorithms based on the new distance input, a remaining goal was then to make the input compatible with the Max/MSP digital signal processing (DSP) object library. To do this, additional conversion of the separately timed distance measurements into a continuous, fixed-interval signal vector representation was implemented and tested as well.

**2.2.4.4** *Distance input: obtaining sufficient spatial resolution for orthogonal fingertip movement*   In further measurements, the keystone reflective surface again was fixed at different distances above the sensor, but now to assess the effects of noise in the new linear distance input. Noise amplitude was found to be smallest at close range, and then to increase, more than linearly, over distance.

In order to have a distance input that would change with actual keystone movement, and not with noise alone, a noise suppression filter was created in software. This filter sampled and held linear distance input until the change introduced by a new input value would exceed a preset sensitivity threshold value. Then, the new value would be sampled and held, and the cycle would repeat, indefinitely.

Although a much smaller sensitivity threshold could be used at close range, giving greater input precision, it was decided to use the same threshold value across the entire distance range. This in order to prevent that tracking of the same, small-amplitude fingertip movement could become falsely suppressed when occurring at greater distances.

This also meant that distance input now offered a reconfigurable trade-off between sensitivity and range: Any desirably small sensitivity threshold value, at some distance, would be crossed by the increasing noise – making the distance range beyond that point unusable for input. In this way, lower sensitivity meant a greater range, and higher sensitivity, a smaller range.

In prototype 1, distance input typically had had a spatial resolution of around 0.4 mm. However, from everyday experience, it seemed possible to execute voluntary orthogonal fingertip movements with a smaller amplitude than this. Therefore, spatial sensitivity was doubled to 0.2 mm. This then allowed a distance range of 17.0 mm.

**2.2.4.5** *Force output: obtaining a fixed and linear magnetic field strength range*   Two electromagnet types, especially, were considered for use in prototype 2. The first type again was the RXAB010 electromagnet of prototype 1, which originally had been intended for alternating current (AC) operation. The second type was taken from an EKS 2184 solenoid, which also used a cylinder core, but originally had been intended for direct current (DC) operation. Metal housing was partially cut away by a remaining Leiden blacksmith, so that the second type could closely follow the arrangement of the first, with a core pole, a bridge pole, and an air gap inbetween.

The two types were then compared in their electric current, temperature, and magnetic field strength responses, over time, to applied coil voltages. Coil current was measured using a digital multimeter; coil temperature using a mercury-in-glass thermometer [Fahrenheit 1724]; and magnetic field strength using a ratiometric, linear Hall-effect sensor [Allegro MicroSystems 1997].

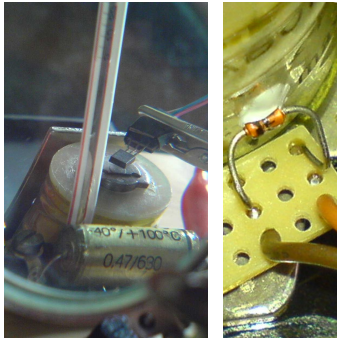
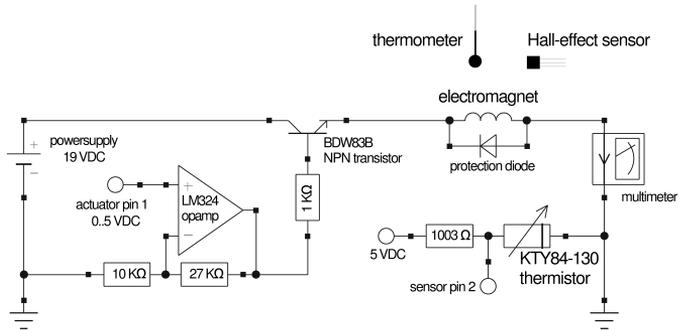

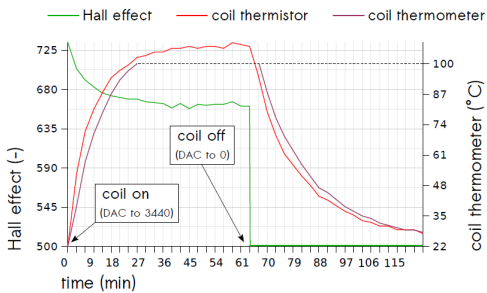

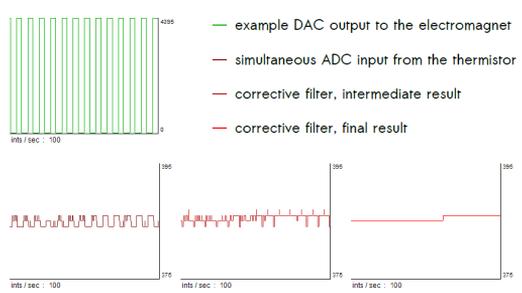

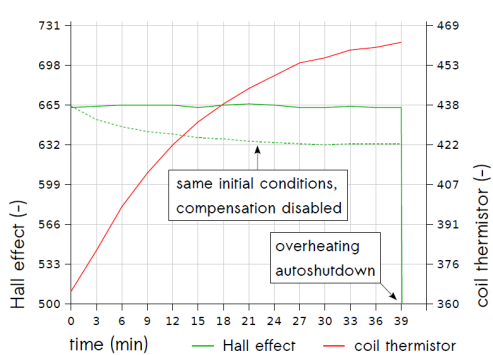

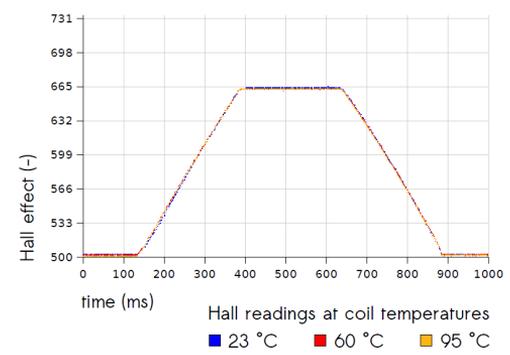

**Figure 2.4** *Force output: obtaining a fixed and linear magnetic field strength range.*
***Top***, *from left to right: the electromagnet with thermometer and Hall-effect sensor (placed above the core; fixed more rigidly for the measurements shown in the graphs). Then, the thermistor, attached to the electromagnet coil. Finally, the schematic of the electronic circuit controlling the electromagnet, including relevant sensors.*
***Middle***, *left: the magnetic field strength produced by a fixed coil voltage, over time, with coil temperature tracked by both thermometer and thermistor. Middle, right: demonstration of the filtering algorithm correcting thermistor input during a rise in coil temperature caused by block-pulsed example output.*
***Bottom***, *left: demonstration of fixed-level magnetic field strength output when automatically compensating coil voltage for coil temperature. Bottom, right: superimposed magnetic field strength recordings, taken across the operational temperature range, during output of the same one-second full-range linear pattern.*

The prototype 1 electromagnet turned out to be both the strongest, in terms of attainable magnetic field strength levels, and the fastest, in terms of the rise and fall times of controlled changes in magnetic field strength. Therefore, this type was chosen for re-use in prototype 2. It also became clear that, given a fixed applied coil voltage, coil electric resistance, coil current, and, indeed, the actual magnetic field strength would vary considerably with coil temperature.

A software object was written in the Max/MSP programming environment to now control coil voltage via the custom Ipsonlab device. Then, to digitally track coil temperature, a thermistor was attached to the coil using thermally conductive paste (see Figure 2.4). Subsequent measurements showed that, as a fixed coil voltage was applied, the thermistor input could be used to accurately track changes in magnetic field strength, up to a coil temperature of 100 °C (see also Figure 2.4).

However, in practice, a varying amount of error was added to the thermistor input, with probabilities that were found to vary with the applied coil voltage and with coil temperature. To counter this, a filtering algorithm was written which applied a series of empirically determined heuristics, based on the control voltage currently applied to the electromagnet; the average recent thermistor input level; the expected direction of coil temperature change, given recent output; and the last probably correct thermistor reading available. Figure 2.4 shows an example of this filter correcting thermistor input, during block-pulsed test output causing a rise in coil temperature.

Using the thermistor and Hall-effect measurements, a software object was written which, by encapsulating temperature-compensated coil voltage output, transparently provided control over magnetic field strength. This is demonstrated in Figure 2.4, where over time magnetic output is held at a new stable maximum level, which can be compared to results without temperature compensation. In the same graph, after 39 minutes, coil temperature rises above 100 °C. At this point, the software object can be seen to automatically switch off coil current, both to avoid providing incorrect output, and for better safety.

In final tests, a one-second output series was used which specified linear traversal of the full magnetic field strength range, in both directions. The results of its output were recorded at different points across the operational temperature range. This is shown last in Figure 2.4, with the Hall-effect recordings superimposed, to demonstrate near-identical output of the linear pattern.

In this way, a fixed and linear magnetic field strength range was obtained for force output to the keystone.

**2.2.4.6** *Force output: toward a range suitable for orthogonal fingertip movement*
Just as different types of distance sensor and electromagnet were considered for use in the second prototype, different types of permanent magnet were considered as well. These included rubber ferrite, ceramic ferrite, and neodymium magnets, of various grades. Some were flexible strips, but most were mechanically rigid and had either a block- or disc-shaped geometry, of varying dimensions. The main goal pursued was to determine which permanent magnet, out of 24 obtained types, could provide the largest output force range to the fingertip, when combined with the selected electromagnet (see Section 2.2.4.5). To find out, each magnet was tightly taped to the fingertip, in turn, so that a series of subjective pairwise comparisons could be made, and the magnet types could be sorted by the perceived difference between minimum and maximum output force. After this, one type was tentatively selected, and used as the permanent magnet for prototype 2.

However, a number of issues had come up during the measurements, which made clear that more criteria than perceived maximum force difference alone would have to be considered, before a permanent magnet type and its associated output force range could be settled upon.

One such issue was that some permanent magnets which enabled a large force range to the fingertip, unfortunately, also would painfully press onto other fingers having permanent magnets attached to them. This could be countered by constantly keeping a hand in question spread; but since this was tiring, it seemed that selecting certain magnets would also mean making the future use of multiple keystones on a single hand prohibitively impractical.

Another issue, moreover, was that some permanent magnets enabling a large force range would also tend to slip sideways during use, disrupting unidirectional fingertip movement orthogonal to the device surface.

**2.2.4.7** *Force output: obtaining sufficient amplitude resolution for fingertip vibration perception* Human somatosensory perception involving the fingertip clearly includes the perception of mechanical vibrations. To induce such aspects of perception, the CT device should allow the force applied to the fingertip to follow periodic wave functions of time. Moreover, so as to be capable of inducing as many perceptual phenomena as possible, the transducer should also allow these wave functions to make controlled transitions from not being noticeable, to being noticeable, and vice versa.

However, in prototype 1, the amplitude resolution of force output was not yet sufficient for this: single-step changes in force amplitude could be felt.

To improve amplitude resolution, different techniques for electronically controlling the coil current were implemented and tested. This included the use of digital-to-analog converters (DACs) and various forms of pulse width modulation (PWM), with frequencies up to 17 KHz. The end result of this was that in prototype 2, amplitude resolution was increased from 26 to 2195 steps.

This was deemed sufficient, since single-step changes in force amplitude no longer seemed noticeable.

**2.2.4.8** *Force output: toward sufficient temporal resolution for fingertip vibration perception* Also in order to potentially induce as many perceptual phenomena as possible, the frequencies of the wave functions for fingertip vibration perception (see

Section 2.2.4.7) should be freely variable within a certain range. This range will be determined, at least in part, by the properties of the different types of human mechanoreceptor that have been discussed in Section 1.3.2.

In prototype 1, the sample rate of force output to the fingertip had enabled a Nyquist frequency of 100 Hz (see Section 2.2.3.3). Based on statements in [Marshall and Wanderley 2006] and [Brewster and Brown 2004], on human cutaneous vibration perception and vibration perception via the fingers being most sensitive around 250 Hz, this sample rate was regarded as clearly insufficient.

For prototype 2, a software object was written in Max/MSP with the explicit goal of providing, now via the custom Ipsonlab device, control over magnetic field strength at the fastest possible update rate. Added to this was the capability of automatically downsampling arbitrary signal vector data to OSC-over-Ethernet updates, so that the Max/MSP library of DSP software objects could be used to generate wave functions. However, five problems were encountered.

First, raising the sample rate above 200 Hz turned out to be impossible. This was attempted on a laptop computer running the Windows XP operating system, with an Audigy 2 ZS Notebook sound card for audio output. On this machine, the possible combinations of Max/MSP signal vector sizes, hardware I/O vector sizes, scheduling options, and timer options were comprehensively explored. In all scenarios, however, update intervals shorter than 5 ms resulted in updates being dropped.

Second, wave output amplitude steeply dropped over frequency. Digital fixed-amplitude sine wave output to the stages beyond the Ipsonlab device showed a 67% drop in actual peak-to-peak magnetic field strength amplitude when frequency was increased from 0 to 100 Hz.

Third, sampling was too coarse to accurately produce most waves with frequencies above 25 Hz. Here, if the wave period did not happen to be a suitable multiple of the 5 ms sample period, repeating wave cycles would in fact differ significantly in their digital amplitude representations. These differences were found repeated in the resulting magnetic field strength output. Perceptually, when attempting to output a single, fixed-frequency sine wave, this seemed to result in the distinct impression of simultaneous, added fluctuations of lower frequency.

Fourth, the accurate generation of waves with frequencies of 30 Hz and above was also prevented by jitter. Already before transfer to the TCP/IP stack and Ethernet hardware, messaging within Max/MSP itself was found to introduce jitter with a range of at least -1.5 to +1.5 ms. This, too, caused the digital representation of repeated identical wave cycles to vary over time. Again, this was found to be probably repeated in the magnetic field strength output, and seemed to result in the perception of additional, lower-frequency fluctuations.

Fifth, the accurate generation of waves with frequencies of 27 Hz and above was also prevented by continuously occurring collapses in magnetic field strength output.

These collapses were characterized by a 20 ms drop to 5% amplitude, followed by a 20 ms rise back to normal output. In some cases, this could be avoided at the cost of significantly reducing digital wave output amplitude (e.g. by 50% at 100 Hz). The collapses were sharply felt.

In combination, these five problems made accurate wave output impossible outside the 0-25 Hz range. This meant that for the Merkel (SA1) mechanoreceptors, located in the skin near the epidermis-dermis boundary, wave output could be generated across the full 0.3-3 Hz range of mechanical vibration sensitivity (see Section 1.3.2.3, Table 1.8). However, the 3-40 Hz range of Meissner (RA1) mechanoreceptors, located in the same anatomical area, was covered only partially, while the 15-400 Hz and 10-500 Hz ranges of the deeper-lying Ruffini and Vater-Pacini mechanoreceptors remained largely out of reach. It seemed clear, therefore, that many aspects of vibration perception involving the fingertip could not yet be induced via the CT device.

**2.2.5 Prototype 3** **2.2.5.1** *Transducer components* In the third prototype, the device surface was made larger, so that the hand producing orthogonal fingertip movement could otherwise lie still, resting on this surface. For the user, this prevented fatigue, and made it easier to keep unidirectional fingertip movements directly above the distance sensor and electromagnet. Using a hand rest also enabled transduction via the CT device to be specific to finger flexing movements that are independent, precise, and directly controlled by the motor cortex: Muscle contractions outside of the forearm and hand – not under direct control of the motor cortex (see Sections 1.3.3.1, 1.3.4.2 and 1.3.4.4) – were no longer needed, and would no longer affect fingertip movement.

The main contact surface of the keystone was made smaller, and made to press against the fingertip only. Also, using additional lateral velcro strips, the contact surface was placed at an angle to the general direction of the distal phalanx. In this way, force output could be applied to a larger area of the fingerpad surface, in a direction more orthogonal to it (see Figure 2.5).

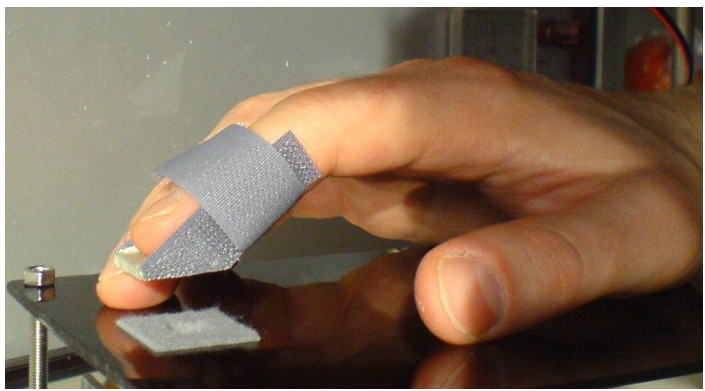

**Figure 2.5** *CT prototype 3.*

To more accurately track the distance above surface of the fingertip itself, the down-facing side of the keystone used for IR reflective sensing was now moved to directly below the fingerpad. This meant that the sensor also had to be moved, to a location again directly above the electromagnet core. This was accomplished with minimal loss of vertical distance above the core by extending the reflective sensor with a custom mirror assembly.

Another change to distance input was that reflective sensing now happened based on pulsed IR output. New voltage I/O for this was implemented using a commercially available Motu 828mkII audio signals interface.

Finally, a new electromagnet was self-built from scratch, using crucial advice given by Rene Overgauw of the Electronics Department at the Leiden Institute of Physics [ELD 2014]. A new electronic circuit for controlling coil current was kindly made available and fine-tuned by the same person.

**2.2.5.2** *Distance input: obtaining a fixed range*   When using IR reflective sensing based on pulsed output, there will be an "on-phase" during which the IR LED is lit, and measurement can occur as before. During a subsequent "off-phase" in which the IR LED is dimmed, however, the sensor response will depend on the environment light. This light may have changed since the on-phase measurement; however, the difference in amount will be small if the on/off phases alternate rapidly, compared to the rate of change in environment light. Then, the off-phase measurements can be used to effectively compensate the on-phase measurements for the influence of environment light.

As will be described in more detail in Section 2.2.5.3, in prototype 3, distance input was implemented based on this principle. As a result, it was found that sunlight falling on the CT device surface no longer affected distance measurement accuracy. In this way, a fixed range for distance input was obtained.

**2.2.5.3** *Distance input: obtaining sufficient spatial range for orthogonal fingertip movement*   Doing IR reflective sensing based on wave-modulated output is a standard technique, but the now much-higher 96 KHz sampling rate for voltage I/O, provided by the Motu 828mkII device, made it possible to carefully implement this using a custom algorithm. For higher precision, each on-phase and off-phase measurement (see Section 2.2.5.2) was averaged from at least 10 individual measurements. Also, the electronic amplification circuit between the IR detector and digital voltage input was replaced, giving a more linear response. The noise generated by the new circuit then was found to have an especially strong 10 KHz component. Suppression was attempted using various electronic means, but a significantly better result was obtained by carefully selecting the averaging window size (and thereby, pulse frequency) of the measuring algorithm. This still enabled the sampling rate of distance input to be improved by a factor of 10, to 4000 Hz.

Different materials and construction methods were tested for the mirror assembly above the electromagnet core. Available materials were tested especially for high IR reflectivity, while construction especially involved a trade-off in physical dimensions. Smaller sizes here meant losing less of the potential distance range of IR reflective sensing within the mirror trajectory. Smaller sizes also meant that the device surface – and thereby, the keystone during use – could come nearer to the electromagnet core, enabling a larger output force range. Larger sizes for the mirror assembly, on the other hand, meant losing less of the reflecting IR light, giving better distance input. A series of mirrors were manually cut, polished and tested to resolve this trade-off (see Figure 2.6 for an initial example).

After this, linear distance input was re-implemented using the new algorithm for pulsed IR reflective sensing in combination with the final version of the mirror assembly (see Figure 2.6).

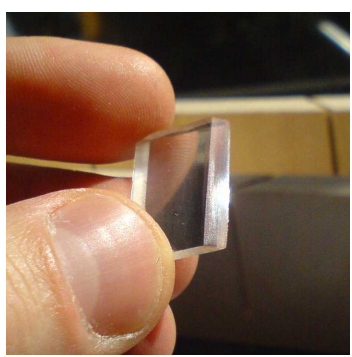 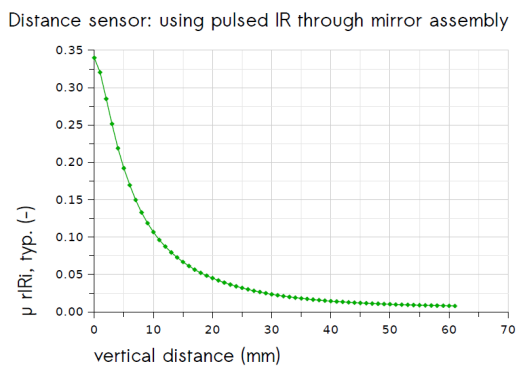

**Figure 2.6** *Distance input: re-implementing a linear range, based on pulsed IR reflective sensing through a mirror assembly. Left: one of the mirrors that were made and tested. Right: measurement results, over distance above device surface, using the final version of both mirror assembly and I/O algorithm.*

Then, the same trade-off between spatial sensitivity and spatial range was made as before (see Section 2.2.4.4). Now, however, this resulted in a sensitivity of 0.2 mm allowing a range of 35.0 mm. Compared to the ranges of (pre)historical examples of unidirectional fingertip movement orthogonal to a surface (see Section 1.2.5), the previous 17 mm range for distance input had been regarded as too small. Now, in the same context, the new 35 mm range was regarded as acceptable.

**2.2.5.4** *Force output: obtaining a range suitable for orthogonal fingertip movement*
One major goal for force output in prototype 3 was to enable faster changes in magnetic field strength, while retaining the overall amplitude range of prototype 2. This was made possible by re-implementing both the electronic circuit controlling coil current and the electromagnet itself. The new circuit incorporated coil current feedback, making the use of a separate mechanism for temperature compensation no longer necessary. Eleven electromagnet types were considered for use in prototype 3. Six of these were taken from electromechanical relays and solenoids, and five were built from scratch. After manually shaped plastic parts of the initial self-built electromagnet had melted during tests, the Department of Fine Mechanics at the Leiden Institute of Physics [FMD 2014] kindly provided heat-resistant versions of the affected component, cut to specification. One of the subsequently made candidate electromagnets provided a single magnetic pole at the device surface; the others provided two.

The criteria for comparing the candidate electromagnet types included their electric resistance and inductance, as well as the maximum magnetic field strength amplitudes of level and sine wave output.

For the keystone, 27 candidate types were considered: see Figure 2.7. These were made to test different permanent magnet materials and geometries, and different variations of attachment to the fingertip.

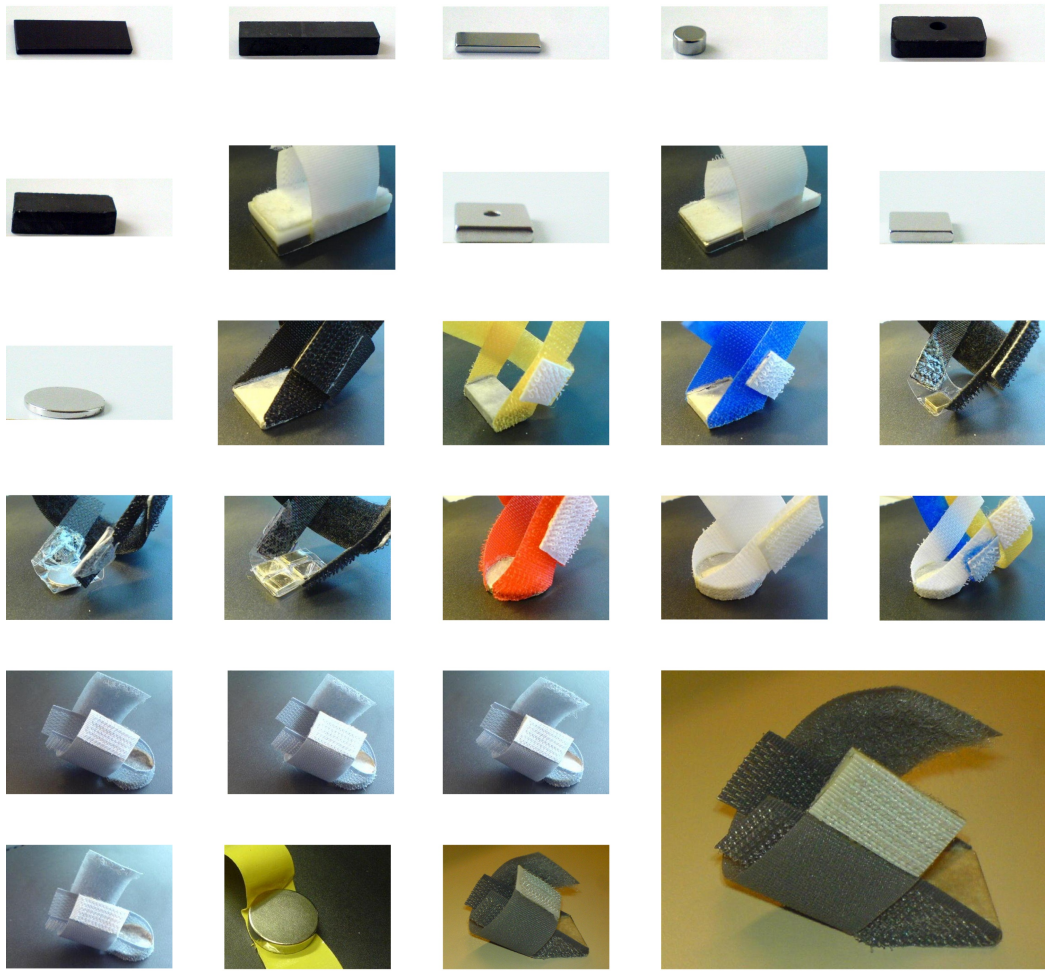

**Figure 2.7** *Force output: obtaining a range suitable for orthogonal fingertip movement. From left to right, top to bottom: candidate keystone types 1 to 26. (Keystones shown without attachments were taped to the fingertip during testing.) Keystone 26 existed in two versions: one using lateral velcro strips placed at a 30 ° angle to the fingerpad contact surface; and another, shown enlarged, bottom right, using a 35 ° angle.*

The criteria for comparing candidate keystone types included the weight of the permanent magnet; the rigidity of the attachment of contact surface to fingerpad; the comfort of the fingertip attachment during use (ideally, its lack of noticeability); and the practicality of potential use on multiple fingers (see Section 2.2.4.6).

The main criterion that was used to compare different keystone-electromagnet pairs related to their support for unidirectional fingertip movement orthogonal to the device surface. This movement could become disrupted by sideways slips, as was first encountered in prototype 2 (see Section 2.2.4.6). These slips would occur during fixed, maximum magnetic field strength output, when attempting to maintain a force equilibrium at decreasing vertical distances above the electromagnet core, pressing down the fingertip with increasing force. It appeared that small changes in the roll of the fingertip could initiate the brisk diagonal slips, of keystone and fingertip, down toward the device surface and away from the electromagnet core. Here, using different permanent magnet geometries seemed to give different results.

To take this issue into account, a measure labeled "maximum practicable static rejection" (MPSR) was used: the maximum upward force which, across the vertical distance range of a given keystone-electromagnet pair, could be countered, in stable equilibrium, by the index finger – without the keystone slipping sideways. To measure MPSR, a custom device was made which could track the vertical force acting between an electromagnet and a fingertip-attached keystone, with a precision of 0.01 N.

After measurements, based on a trade-off between MPSR, keystone weight, and potential multi-finger practicality, self-built electromagnet 4 and candidate keystone 26 were selected for use in prototype 3. To provide rigid but comfortable attachment to different fingers, keystone 26 was made in two versions: one with the lateral velcro strips at a 30 ° angle to the fingerpad contact surface, and another with this angle at 35 ° (see Figure 2.7).

In this way, force output with a suitable range for orthogonal fingertip movement was obtained.

**2.2.5.5** *Force output: obtaining sufficient temporal resolution for fingertip vibration perception*    In prototype 2, accurate wave output for fingertip vibration perception had been prevented, above 25 Hz, by five problems: a low sample rate; associated coarse sampling; jitter; collapses in magnetic field strength during wave output; and a drop in wave output amplitude over frequency (see Section 2.2.4.8).

In prototype 3, these problems were removed, by new voltage I/O, a new electronic circuit for controlling coil current, and the new custom electromagnet. Using the Motu 828mkII device, the sample rate of voltage output was increased from 200 Hz to 96 KHz. Waves generated with frequencies of 0-1000 Hz were verified to be reproduced in magnetic field strength output, improving on the previous 0-100 Hz range. Coarse sampling was verified not to be a practical issue anymore: The relative error in digital peak-to-peak amplitude of generated sine waves now was found to be less than 0.1% across the 1-1000 Hz range. Jitter, too, no longer was a practical issue: Some lower-frequency fluctuation was still seen in magnetic field strength output when generating waves at the higher end of the 1-1000 Hz range, but this occurred with very limited amplitudes.

Perceptually, during the output of single, fixed-frequency sine waves across the new range, the resulting subjective impressions changed to feeling a single vibration, without additional lower-frequency fluctuations. Also, this was no longer interrupted by any collapses in magnetic field strength output, which in prototype 2 probably had been due to the coil current control circuit.

Also, the previous 67% drop in magnetic field strength amplitude over 0-100 Hz was replaced by a dip of 5% across the 0-400 Hz range. Accurate wave output was now possible not only across the frequency range of the Merkel (SA1) mechanoreceptors, but also across the ranges of the Meissner (RA1), Ruffini, and Vater-Pacini mechanoreceptors (see Section 2.2.4.8). Therefore, force output by the CT device was now regarded as sufficient for inducing fingertip vibration perception.

**2.2.6  Prototype 4  2.2.6.1** *Transducer components*  In the fourth prototype, the two keystone velcro strips that had been lateral were made palmar and dorsal, respectively. This meant that the third, other velcro strip no longer functioned as an adjustable ring around the finger, only. By detaching and then re-attaching it at shifted distal positions on the now palmar and dorsal strips, the keystone's contact surface could be placed at an angle more parallel to the specific human fingerpad in question (see Figure 2.8). In this way, the new keystone improved on transducing fingerpad-orthogonal forces via a rigid yet comfortable attachment.

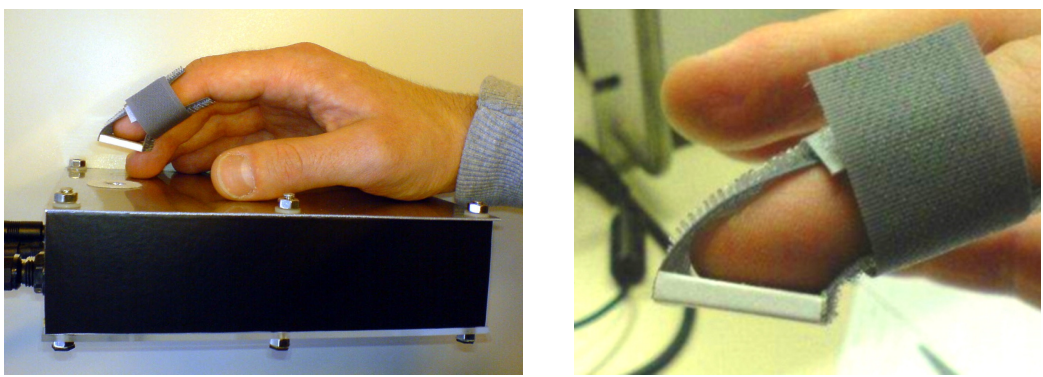

**Figure 2.8**  *CT prototype 4. Left: side view. Right: dorsal and palmar velcro strips, enabling precise adjustment of the keystone fit.*

Another change was that the DSP algorithms of the CT device now ran on a laptop computer with the Mac OS X operating system. The OS X Core Audio component [Apple 2008] enabled low latencies between voltage input, computation, and output, which other mainstream operating systems (e.g. versions of Linux and Windows) could not match. To further lower the latency of voltage I/O, the Motu 828mkII audio signals interface was replaced by a Motu UltraLite mk3.

**2.2.6.2** *Combined I/O: obtaining a roundtrip latency for haptics* The latency between sensor input and actuator output is an important property for haptic devices. For prototype 4, it was understood from [Sinclair and Wanderley 2007] that in general, such latency should not exceed 1 ms, if the haptic experience of e.g. touching a rigid, solid surface in space is to be induced adequately. The CT device latency was measured using the roundtrip latency from actuator output to sensor input. This included determining the separate latencies added in different sections of the I/O chain. It was found that most latency was added by the electromagnet circuit, and by the stage from voltage input to voltage output (including digital computation). The electromagnet circuit was tested using pulsed digital output, with latency measurements timed on 99% of a target maximum amplitude difference being reached in recorded magnetic field strength.

In prototype 2, the electromagnet circuit had added 33 ms, to an overall roundtrip latency of 42.5 ms. By prototype 4, due to the improvements already discussed in Sections 2.2.5.4 and 2.2.5.5, electromagnet circuit latency was reduced to 1.0 ms. DSP computation was moved to a 192 KHz sample rate, since this setting coincided with the lowest possible latency for voltage I/O provided by the Motu UltraLite mk3 device. This resulted in a new total roundtrip latency of 4.0 ms, with 1.6 ms added by the input stages, and 2.4 ms by the output stages.

This represented a tenfold reduction in latency, but was still short of the target 1 ms. The logical next step to reduce the roundtrip latency toward this target would have been to re-implement all voltage I/O and computation in a dedicated, embedded device. However, that would also have meant introducing separate DSP algorithms for auditory and somatosensory perception running on separate devices, and written in separate programming languages. This would have introduced a serious practical obstacle to achieving the goal of complete integration of computed touch and computed sound (see Section 1.6.9). Also, in practice, the obtained latency was found to be already quite usable for inducing aspects of haptic perception. Therefore, it was decided to proceed, tentatively, with 4.0 ms as the roundtrip latency obtained for haptics.

**2.2.7 Research goals attained in the resulting transducer** In the preceding subsections of Section 2.2, we have described the development of a series of prototypes. How, in summary, has the resulting transducer technology implemented the computed fingertip touch model?

**2.2.7.1** *Computed fingertip touch: qualitative aspects* The CT device implements the computed fingertip touch model in such a way as to attain the majority of chapter goals identified in Section 1.6.9: Via the flat, closed, and rigid contact surface of the keystone component attached to the fingertip, the fingerpad becomes both the source and target of transduction, during unidirectional fingertip movement orthogonal to a surface. Input from human motor activity is obtained via the sensor component, which tracks the vertical distance between fingerpad and device surface. Output to somatosensory perception is provided via the electromagnet component, which simultaneously applies orthogonal forces to the fingertip (see Section 2.2.1).

**2.2.7.2** *Computed fingertip touch: quantitative aspects* Obtaining a fixed and linear distance input range was not trivial (see Sections 2.2.4.2, 2.2.5.2, and 2.2.4.3). The resolution of distance input was set to 0.2 mm (see Section 2.2.4.4). This seemed a usable value when tracking voluntary orthogonal fingertip movements. The distance input range that could then be obtained extended to 35.0 mm above the device surface (see Section 2.2.5.3). This seemed acceptable, when compared to the distance ranges in historical and prehistorical examples of unidirectional fingertip movement orthogonal to a surface. The sampling rate obtained for distance input was 4000 Hz (see Section 2.2.5.3).

Obtaining a fixed and linear magnetic field strength range controlled via the electromagnet also was not trivial (see Section 2.2.4.5). Obtaining accurate force output then further required rigid attachment of the contact surface to the fingerpad. On the other hand, the attachment of keystone to fingertip also had to remain comfortable during use (see Section 2.2.5.4). To obtain a rigid yet comfortable attachment, the final keystone design enabled a precisely adjustable, personal fit (see Section 2.2.6.1).

To choose between different candidate keystone-electromagnet pairs, force output range turned out to be not a sufficient criterion. Instead, a measure labeled MPSR was used, which preserved the orthogonality of fingertip movement during force output. Potential multi-finger practicality also was a decisive criterion (see Sections 2.2.4.6 and 2.2.5.4).

The resulting force output had an amplitude resolution of at least 2195 steps (see Section 2.2.4.7). This was regarded as sufficient, as single-step changes in force amplitude no longer seemed noticeable. The temporal resolution of the force output enabled wave output at frequencies across the 0-1000 Hz range (see Sections 2.2.4.8 and 2.2.5.5). This was also regarded as sufficient, since accurate wave output was now possible across the frequency ranges of the various human mechanoreceptor types known to be involved in fingertip vibration perception.

Finally, the latency that was obtained between distance input and force output was 4.0 ms (see Section 2.2.6.2). Further pursuit of a lower latency would have introduced serious practical obstacles against achieving the goal of completely integrated computed touch and computed sound (see Section 1.6.9). And, while not optimal, the obtained latency was found to be usable for inducing aspects of haptic perception.

**2.2.7.3** *Computed fingertip touch: novel aspects* The CT device differs both in its intended purpose and in its actual I/O from the haptic transducers that form its technological context (see Section 2.2.2).

First, where other devices implicitly happen to fit the computed fingertip touch model identified in Section 1.5.2, the CT device was made to fit the model. Obtaining computed fingertip touch was a goal in itself (see Section 1.5.1).

Second, where other devices implicitly provide support for unidirectional fingertip movement orthogonal to a surface, by supporting general fingertip movement within a 3D volume, the CT device targets this type of movement specifically and explicitly. Where other devices provide implicit support by extending from the piano-type key, the CT device instead attempts to fit a general anatomical movement lying at the root of piano-type keys and other devices (see Section 1.2).

Third, the CT device enables transducer I/O that is specific to those flexing movements of the human finger that are independent, precise, and directly controlled by the motor cortex (see Section 2.2.5.1).

Fourth, where other devices only enable the controlled application of upward forces to the fingertip, the CT device also enables this for downward forces. This implies that a greater range of causal relationships between human actions and changes in heard musical sound may be implemented (see Sections 2.2.2 and 1.5.1.1).

Fifth, unlike other fingerpad transducers, the CT device avoids the use of any connected mechanical parts moving relative to the target anatomical site (see Section 2.2.1). This supports precise output to human somatosensory perception, and has enabled implementing accurate mechanical wave output across the frequency range involved in human fingertip vibration perception (see Sections 2.2.4.7, 2.2.4.8, and 2.2.5.5).

## 2.3  Fingerpad-parallel force output:
## The kinetic surface friction transducer (KSFT)

**2.3.1  Operating principles**    The second transducer technology presented in this chapter, like the first, will have the human fingerpad as its source and target location, via a flat, closed, and rigid contact surface. For this general transducer type, fingerpad-parallel force output can be regarded as the other main case, beside fingerpad-orthogonal force output, for inducing aspects of somatosensory perception (see Section 1.6.2.3). Therefore, we will now explore implementation of the computed fingertip touch model in this second direction.

An existing example of instrumental control of musical sound during which fingerpad-parallel forces are applied via a flat contact surface, is scratching: Here, the fingerpad is pressed down on a vinyl record placed on a turntable, after which it can perform rotational movements in a plane (see Section 1.2.3). The applied forces have a large enough range to co-determine the movement of fingertip control actions. Generalizing from this, the second transducer technology will support fingertip movement parallel to a surface.

Fingerpad-parallel force output during surface-parallel fingertip movements can be implemented by having the fingerpad press down on a rigid object, which it then pushes around on an underlying, flat surface. If this surface contains a ferromagnetic layer, and the movable object contains an electromagnet, regulating the vertically directed attraction between the two can be used to produce a varying, horizontally directed friction, which is transferred to the fingerpad surface.

To put the minimum possible kinetic surface friction close to zero, smooth layers may be used where the movable object and underlying surface connect mechanically. Optionally, a permanently magnetic layer might be added to the surface, so that besides regulated vertical attraction, rejection would also be possible between electromagnet and surface, to further extend the friction output range toward zero.

Finally, to obtain input from fingertip movement, the optical sensor of a standard computer mouse [Agilent Technologies 2001] may be built into the movable component, too, to track its displacement across the magnetic surface.

The resulting device – consisting of magnetic surface, movable electromagnet, and surface displacement sensor – was named *kinetic surface friction transducer (KSFT)*. Its movable part, pushed around by the fingerpad, was named *puck*, a term from ice hockey for a manipulandum that is continuously moved about on a surface with low baseline friction.

A limitation of the proposed setup is that it provides no support for passive touch (see Section 1.3.3.4): the output forces can be applied only during (attempted) fingertip movement. Also, during active touch, the magnitude of the frictional force may be controlled, but not its direction, which will always be opposed to the direction of fingertip movement.

A potential advantage of the proposed setup is that it may yield only two connected mechanical parts moving relative to the target anatomical site: the magnetic surface, and any wiring connected to the electromagnet coil. Although having no such parts would have been better (see Section 2.2.1), the proposed setup presents one or two potentially problematic mechanical connections, rather than many.

An advantage of the proposed setup is that using optical mouse sensor technology means using very cheap, off-the-shelf hardware. A hurdle, however, may be that displacement input is not algorithmically represented in physical units: The mouse pointer GUI screen coordinates usually accessed in software typically are unitless, affected by intermediate velocity transformations, and artificially limited to on-screen pixel display positions.

**2.3.2 Technological context**   By definition, the technological context of the proposed KSFT device will consist of other devices that also can provide fingerpad-parallel force output during surface-parallel fingertip movement (see Section 2.3.1). Here, the force output range should be large enough to co-determine the movement of fingertip control actions: This enables implementing a greater range of causal relationships between human actions and changes in heard musical sound (see Section 2.2.2). Given their smaller force output ranges, the T-PaD [Winfield et al. 2007] and TeslaTouch [Bau et al. 2010] surface friction display technologies – which can deliver subtle cues to the fingerpad – here are not considered part of the immediate technological context.

In [Akamatsu and Sato 1994], a computer mouse extended with friction output was presented, using a built-in electromagnet and a ferromagnetic mouse pad. The device had a limited force output range, controlled with 1-bit amplitude resolution.

Although similar in components to the proposed KSFT device, this friction mouse, like computer mice in general, was gripped and pushed around by the hand as a whole, as opposed to use of the fingerpads only. This meant that the fingerpads were not necessarily in contact with the device surface all the time, and when they were, they would not necessarily receive force output in a parallel direction.

Another major difference is that when using a graphical mouse pointer, GUI screen coordinates are used to represent displacement input. This means that by default, instead of providing a linear representation, displacement input will be affected by intermediate velocity transformations. Also, displacement input will not be accurate, as it is clipped to on-screen pixel display positions. Therefore, this input can be used to add somatosensory cues to the visual output and motor input of GUIs; it cannot be used, however, to induce aspects of fingertip surface texture perception that are based on high-resolution surface friction patterns.

The same two differences, of the transducer being hand-pushed rather than fingerpad-pushed, and of surface displacement input being non-linear and non-accurate, also apply to the computer mice extended with surface-parallel force output in [Akamatsu and MacKenzie 1996], [Münch and Dillmann 1997], [Dennerlein et al. 2000], and [Schneider et al. 2004].

On the other hand, a transducer technology that *is* pushed by the fingerpad and that *does* provide linear and accurate displacement input, is the Pantograph of [Hayward et al. 2004] [Hayward et al. 1994]. Its displacement input has a resolution of 0.01 mm. Also, importantly, this device provides fingerpad-parallel force output in any of the horizontal planar directions, within a working area of 1.6 dm$^2$. This is done using a flat contact surface, actuated by multiple motors via mechanical linkages.

Here, it seemed that the proposed KSFT device might still differentiate, for the friction case, based on two other points. First, by providing output to somatosensory perception that is not based on the use of a number of motors and linkages, but on just one or two connected mechanical parts still moving relative to the fingerpad (see Section 2.3.1). Second, by using cheap off-the-shelf optical mouse sensor technology to obtain displacement input.

**2.3.3 Prototype 1**     **2.3.3.1** *Transducer components*     In the first prototype, the magnetic surface was made of a ferromagnetic layer covered by a paper top layer. The displacement sensor used for the puck was that of a Trust optical computer mouse, providing digital input over a wired USB connection. The electromagnet used for the puck was of the same type used in CT prototype 1 (see Section 2.2.3.1). It was rigidly attached to the proximal side of the mouse base (see Figure 2.9). This was done in a configuration putting both poles of the electromagnet near the magnetic surface, resting on it close together, so as to obtain a larger force output range than possible via a single pole. The poles were covered by a layer of PVC tape where they connected mechanically to the underlying paper surface. Digital output controlling coil current was implemented using the same setup as in CT prototype 1.

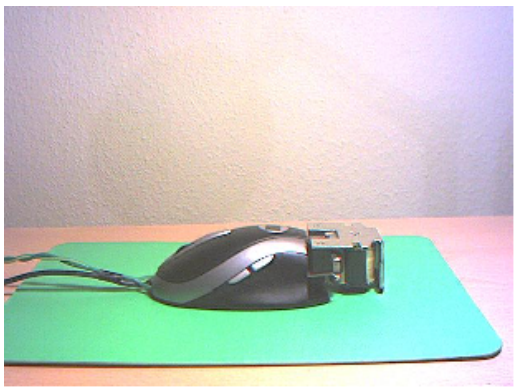

**Figure 2.9** *KSFT prototype 1.*

**2.3.3.2** *Displacement input*    The contact surface used during fingerpad I/O was located more distally than the electromagnet, at the top of the partially remaining housing of the optical mouse. Here too there was a PVC top layer (see Figure 2.9). During translational fingertip movements parallel to the magnetic surface, the accuracy of displacement input was negatively affected by planar rotation of the base of the puck around the relatively heavy electromagnet core. This problem was especially pronounced during sideways fingertip movements, in the ulnar-radial direction. The digital representation of this displacement input was not developed beyond the temporary use of GUI screen coordinates.

**2.3.3.3** *Friction output*    The weight of the puck was 240 g. This ought to be less, it seemed, both to lower the baseline kinetic surface friction, and to thereby make using the device less tiring. Still, using prototype 1, two important properties of the proposed KSFT device could be tested for with positive results: First, it was found that fingerpad-parallel force output had been obtained that had a large enough range to co-determine the movement of fingertip control actions: These could be slowed down, or even halted. Second, it was found that this friction output, already when programmed to respond to the displacement input of Section 2.3.3.2, could be used to induce aspects of fingertip surface texture perception.

**2.3.4  Prototype 2    2.3.4.1** *Transducer components*    In the second prototype (see Figure 2.10), the magnetic surface again contained a ferromagnetic layer. Underneath it, a layer of rubber-like plastic now ensured rigid desktop attachment of the magnetic surface during operation. On top, a thin plastic coating applied to thin paper provided a smoother surface for puck movement, adding less perceived noise.

The displacement sensor inside the puck was that of a Microsoft Arc (v0140) optical computer mouse, providing digital input over a wireless USB connection.

The electromagnet inside the puck was built from scratch, and similar to the electromagnet used in CT prototype 3 (see Section 2.2.5), except in having a smaller core, coil, and bridge height. The standard sliding pads of the Arc mouse were attached to the bottom of the electromagnet, mechanically connecting it to the top layer of the magnetic surface during fingertip movement. Digital output controlling coil current was implemented using the same setup as in CT prototype 4 (see Section 2.2.6).

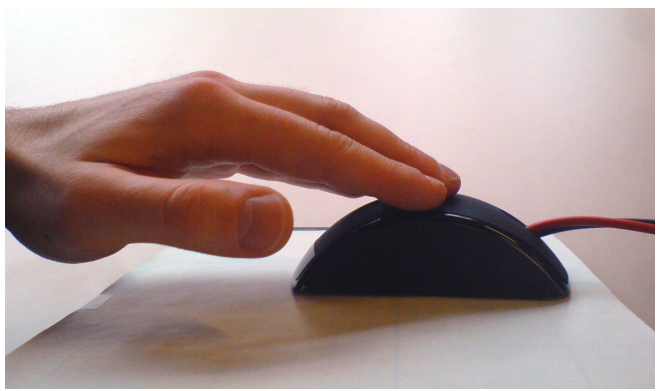

**Figure 2.10** *KSFT prototype 2.*

**2.3.4.2** *Displacement input: obtaining sufficient spatial resolution for fingertip surface texture perception* For the displacement input of prototype 2, the digital values provided by the optical mouse hardware over USB were accessed directly. After a series of measurements for both the $x$ and $y$ surface axes of the displacement sensor, it turned out that these unitless values could be converted to mm simply by applying multiplication with a fixed value. In this way, linear displacement input was obtained.

Direct access to the USB connection also improved the spatial accuracy of displacement input, as bypassing the GUI meant that updates no longer were clipped to zero to maintain on-screen graphical pointer display. Also, this time, the puck was built around the electromagnet, with the fingerpad contact surface (see Figure 2.10) placed more directly above the magnetic poles. This significantly reduced planar rotation of the puck base during translational fingertip movements, improving spatial accuracy.

Unfortunately, it was found that there continuously occurred significant noise peaks in both the $x$ and $y$ digital amplitude series produced by the Microsoft Arc mouse. These were verified not to represent catch-ups relating to measurement timing irregularities. To counter these errors, a filtering algorithm was written which applied a series of empirically determined heuristics, which were based on the typical positive acceleration in absolute displacement caused by noise peaks; their typical maximum duration; the typical minimum time interval between them; and the last probably correct displacement reading available. Figure 2.11 shows an example where this filter processes displacement input during fingertip movement, which is oscillating with increasing speed in the ulnar-radial direction.

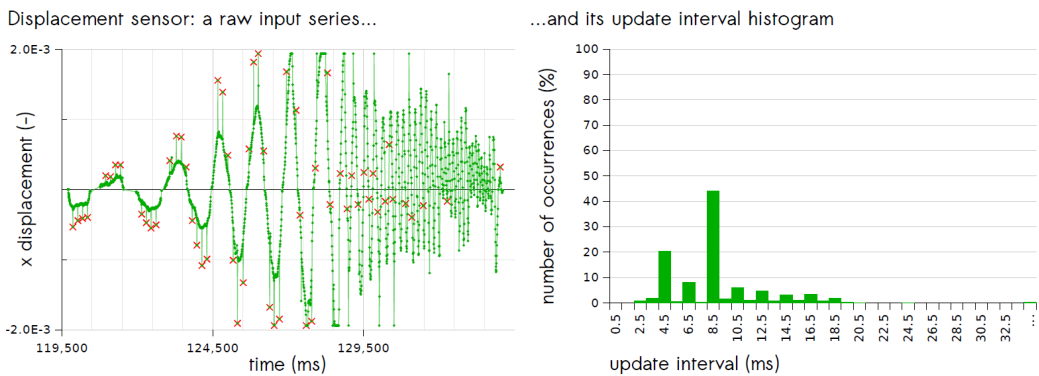

**Figure 2.11** *Displacement input.* **Left, shown in green:** *fingertip movement oscillating with increasing speed in the ulnar-radial direction, tracked over time along the x axis of the displacement sensor. Near t = 128 s, it can be seen how the speed of fingertip oscillation crosses the maximum displacement ceiling of the optical mouse sensor. Beyond this point, displacement input becomes inaccurate.* **Left, shown in red:** *each × marks a noise peak, as detected in real time by the custom noise suppression algorithm.* **Right:** *histogram of the update intervals during the fingertip movement input series.*

Another issue was jitter, also illustrated in Figure 2.11 by a typical histogram of update intervals. The variation over time in update intervals was found to be consistent with possibly precisely clocked displacement updates generated in the mouse hardware, then being picked up by the host operating system using a differently clocked event queue, with considerable jitter introduced along the way. Precise and regular timing could have been restored, by extending the filtering algorithm with buffering functionality. This was not done, however, as it would also have added at least another 8 ms to the already high latency between input and output (see Section 2.3.4.4 below).

The resulting displacement input was characterized by a sampling rate of 125 Hz, subject to jitter between -6 ms and +12 ms. During continuous use, errors in positional accuracy typically would accumulate at a rate of up to 2 cm / min. At the same time, the input was found to be sensitive to displacements down to 0.02 mm. As discussed in Section 1.3.3.5, aspects of surface roughness perception via the human skin are induced by details separated by 0.125 mm or more [LaMotte and Srinivasan 1991]. Therefore, the spatial resolution of displacement input by the KSFT device was now regarded as sufficient to support fingertip surface texture perception.

**2.3.4.3** *Friction output: obtaining a fixed and linear magnetic field strength range*
Puck weight was decreased to 170 g, lowering the baseline kinetic surface friction, and making operating the device less tiring. Use of the new digital output, coil circuit, and electromagnet improved the amplitude resolution and temporal resolution of friction output. A remaining issue, found for output features of short duration, was that resonation within the mechanical structure between electromagnet and fingerpad contact surface would also lead to orthogonally applied forces. Due to this, when reducing the duration of output features from 10 ms to 1 ms, their perception no longer could be verified to be due to parallel force output. Still, importantly, friction output could now be controlled using a fixed and linear magnetic field strength range.

**2.3.4.4** *Combined I/O: latency* The KSFT device latency, from displacement input to friction output, was measured using the roundtrip latency from actuator output to sensor input. This included determining the separate latencies added in different sections of the I/O chain. The output stages from computation to actuation were found to add 2.4 ms. In a series of 10 measurements, the input stages from sensing to computation were found to add between 12.3 ms and 25.7 ms. The resulting average roundtrip latency was 20.5 ms. This meant that prototype 2 was not yet usable for inducing aspects of haptic perception (see Section 2.2.6.2). The prototype could be used, however, to induce aspects of fingertip surface texture perception during active touch.

**2.3.5  Research goals attained in the resulting transducer** In the preceding subsections of Section 2.3, we have described the development of a series of prototypes. Due to a lack of time and resources, this development was cut short, especially regarding the pursuit of a higher displacement input sampling rate; of a lower latency between displacement input and friction output; of a smaller, lighter, and more rigid puck; and of a possibly lower baseline friction based on integrating a permanent magnet layer in the magnetic surface (see Section 2.3.1). Still: How, in summary, has the resulting transducer technology implemented the computed fingertip touch model?

**2.3.5.1** *Computed fingertip touch: qualitative aspects* The KSFT device implements the computed fingertip touch model in such a way as to attain the remaining chapter goals identified in Section 1.6.9: Via the flat, closed, and rigid contact surface of the puck component pressed onto by the fingertip, the fingerpad is both source and target of transduction, during surface-parallel fingertip movement. Input from human motor activity is obtained via the sensor component, tracking fingerpad displacement parallel to the magnetic surface. Output to somatosensory perception is provided via the electromagnet component, applying parallel, frictional forces to the fingerpad during its movement (see Section 2.3.1).

**2.3.5.2** *Computed fingertip touch: quantitative aspects* For the KSFT device displacement input, a spatial resolution of 0.02 mm was obtained (see Section 2.3.4.2). Based on the distances involved in surface roughness perception, this was regarded as sufficient for inducing aspects of fingertip surface texture perception. Furthermore, a spatial range of tens of cm, in both planar directions, and an average sampling rate of 125 Hz were obtained.

For the KSFT device friction output, a range large enough to co-determine the movement of fingertip control actions was obtained (see Section 2.3.3.3), via magnetic field strength output with a fixed and linear range (see Section 2.3.4.3). For the shortest effective features in friction output, a duration between 10 and 1 ms was obtained.

Between the KSFT device displacement input and friction output, an average latency of 20.5 ms was obtained (see Section 2.3.4.4). This limited the use of I/O to inducing aspects of fingertip surface texture perception during active touch.

**2.3.5.3** *Computed fingertip touch: novel aspects*    The technological context of the KSFT device is formed by other devices that also can provide fingerpad-parallel force output during surface-parallel fingertip movement. Here, force output ranges must be large enough, so that automaton output may co-determine fingertip movement (see Section 2.3.2). The KSFT device then differs from similar devices in a number of ways.

First, where other devices implicitly happen to fit the computed fingertip touch model of Section 1.5.2, the KSFT device was made to fit the model. Obtaining computed fingertip touch was a goal in itself (see Section 1.5.1).

Second, the KSFT device, being fingerpad-pushed rather than hand-pushed, differs from computer mice extended with surface-parallel force output in that it targets fingerpad transduction. Here, the KSFT device also differs by providing displacement input that is sufficiently linear and accurate to be used for inducing aspects of fingertip surface texture perception that are based on high-resolution surface friction patterns (see Section 2.3.2).

Third, the KSFT device differs from the Pantograph, for the friction case, by providing output to somatosensory perception that is not based on the use of a number of motors and linkages, but rather, on the use of only one or two remaining connected mechanical parts still moving relative to the fingerpad. Also, the KSFT device uses cheap off-the-shelf optical mouse sensor technology to obtain displacement input (see Section 2.3.2).

# 3. New systems for computed fingertip touch

**CHAPTER SUMMARY**

In this chapter, we present two systems for computed fingertip touch: one expanding from the cyclotactor (CT) device, and another expanding from the kinetic surface friction transducer (KSFT) device.

The CT system offers a fingerpad distance input with a range of 35.0 mm, a resolution of 0.2 mm, and a sampling rate of 4000 Hz. Simultaneously, it offers a fingerpad force output with a range over distance that is shown in Figure 3.2, a resolution of ± 0.003 N, and accurate wave output up to 1000 Hz. The latency between the input and output is 4.0 ms.

The CT system provides excellent support for real-time instrumental control of musical sound; moreover, its output covers the frequency ranges involved in fingertip vibration perception; and its I/O is capable of inducing aspects of haptic perception.

The KSFT system offers a fingerpad displacement input along two planar directions, with a range of tens of cm, a resolution of 0.02 mm, and an average sampling rate of 125 Hz. Simultaneously, it offers a fingerpad kinetic friction output with a range of 0.14-1.43 N, and with a temporal resolution between 1 and 10 ms. The average latency between input and output here is 20.5 ms.

The KSFT system supports inducing aspects of fingertip surface texture perception during active touch. For fingertip movement at lower speeds and accelerations, I/O can cover the spatial range of surface roughness perception.

The I/O of both systems can be programmed via classes implemented in the SuperCollider language. Using this language also enabled complete integration with computed sound: Input from motor activity, output to somatosensory perception, and output to auditory perception are easily combined within a single written algorithm. Finally, both systems are cheaply mass-producible.

De Jong S, 2010a Presenting the cyclotactor project. In *Proceedings of the 2010 international conference on Tangible, Embedded, and embodied Interaction* (ACM, January 24-27 2010, MIT, Cambridge, MA, USA) 319-320.

De Jong S, 2010b Kinetic surface friction rendering for interactive sonification: an initial exploration. In *Proceedings of the 2010 international workshop on Interactive Sonification* (April 7 2010, Stockholm, Sweden) 105-108.

De Jong S, 2010e The cyclotactor. Best Demonstration Award. *The 2010 EuroHaptics international conference* (July 8-10 2010, Amsterdam, the Netherlands).

# 3.1 Introduction

In this chapter, we will pursue the second set of goals identified in Section 1.6.9. This means expanding the new transducer technologies of Chapter 2 to new systems for computed fingertip touch – which should algorithmically represent transducer state using physical units; integrate computed sound; support real-time instrumental control of musical sound; and be powerful while cheaply mass-producible.

In Section 3.2, we will pursue these goals for the cyclotactor (CT) device; in Section 3.3, for the kinetic surface friction transducer (KSFT) device. The resulting CT system won Best Demonstration Award at the international EuroHaptics conference of 2010 [De Jong 2010e].

# 3.2 The CT system

**3.2.1 Technical capabilities** **3.2.1.1** *Distance input in mm*   The way in which accurate and linear fingerpad-to-surface distance input was obtained for the CT device corresponded to obtaining an algorithmic representation of transducer input in terms of millimeters (see Sections 2.2.4.3 and 2.2.5.3). This representation had a range of 35.0 mm, a resolution of 0.2 mm, and a sampling rate of 4000 Hz (see Section 2.2.7.2).

**3.2.1.2** *Force output in N*   For transducer output, the physical unit chosen for algorithmic representation was the force, in newtons, applied orthogonally to the human fingerpad via the keystone permanent magnet. After obtaining a final design of keystone and electromagnet which included accurate and linear magnetic field strength output (see Section 2.2), more work was needed to obtain a force output in N.

With the keystone permanent magnet aligned straight above the electromagnet core, measurements were made to determine how applied orthogonal force varied over coil current and distance above device surface. This was done using a device custom-built for the purpose, which allowed precisions of 0.005 A, 0.01 mm, and 0.01 N.

Part of the resulting measurements are shown in Figure 3.1. At the back, it can be seen how a fixed current of 10 A results in an upward force, which diminishes over distance. This decrease follows a curve, which is also represented by the green coloring becoming ever lighter over distance. At the front, it can be seen how a fixed current of 0 A results in a downward force, which also diminishes over distance, represented by red coloring becoming ever lighter.

Between the front and back of Figure 3.1, however, it can be seen how the force curves for fixed currents between 0 and 10 A may first increase, then decrease over distance. Also, the coloring shows how here, the same fixed current may result in a downward as well as an upward force, depending on the distance. Such a perceptually different result illustrates how it would be a pitfall to regard the volts or ampères of the coil circuit – although they too describe transducer state – as sufficient physical units to algorithmically represent transducer output.

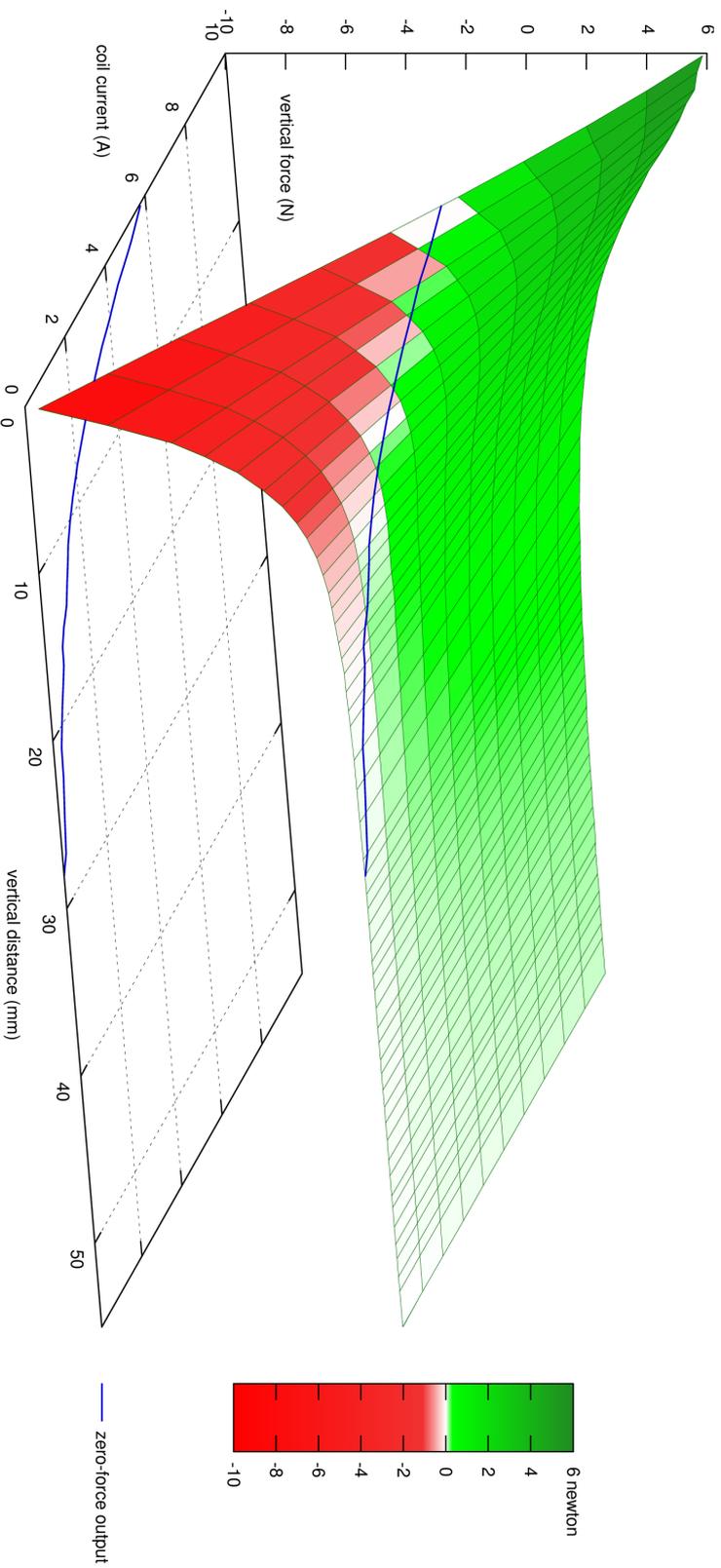

**Figure 3.1** *The CT system: measurements underlying fingerpad-orthogonal force output in newtons.*

The blue curve in Figure 3.1 highlights where a force of 0 N is applied. This curve has been duplicated on the graph floor, to show more precisely the current required over distance. Only after using this characteristic, to counteract magnetization of the electromagnet core by the keystone, did it become possible to have the CT device appear like an ordinary, inactive surface.

To enable this, and more, a DSP algorithm was written which took a requested force in N as input, and then computed the required coil current based on the distance input and interpolation tables derived from the measurements just discussed. Using this algorithm to control the coil circuit, a force output in N was implemented. The output range, varying over distance, is shown in Figure 3.2.

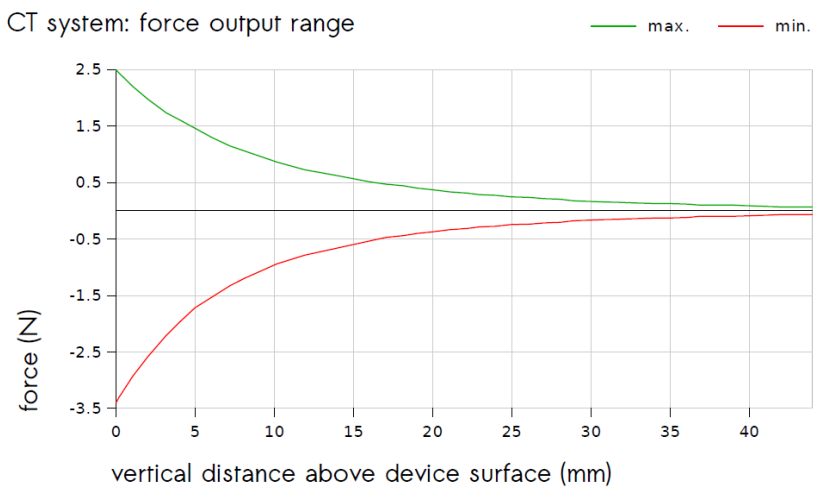

**Figure 3.2**  *The CT system: force output range over distance.*

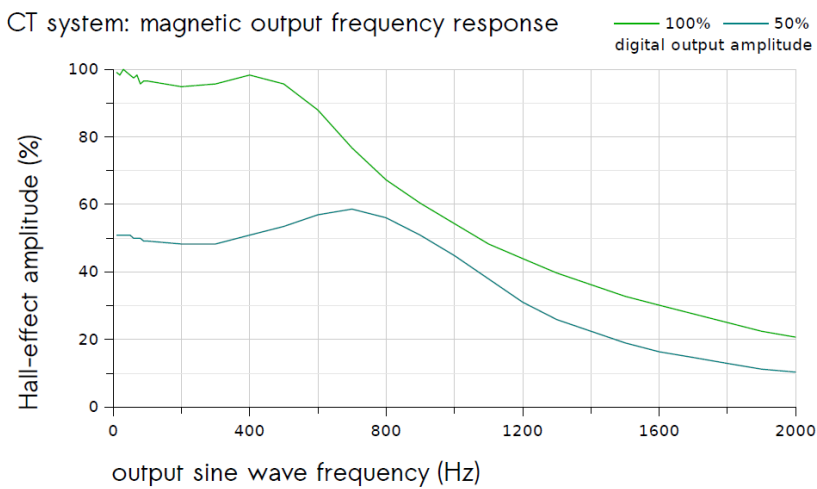

**Figure 3.3**  *The CT system: magnetic output frequency response.*

Based on the force output range at device surface, and the amplitude resolution discussed in Section 2.2.4.7, the amplitude precision of force output was estimated at 0.003 N. The temporal resolution when controlling magnetic field strength to produce force output allowed accurate wave output across the 0-1000 Hz range (see Section 2.2.5.5). The attenuation of this wave output, measured over frequency, is shown in Figure 3.3. As can be seen, over the 0-400 Hz range – important for inducing aspects of fingertip vibration perception – attenuation is linear and, with a dip of 5%, approaches a flat response.

**3.2.1.3** *I/O latency in ms*   The latency achieved between the above distance input and force output was 4.0 ms (see Section 2.2.7.2). This enabled inducing aspects of haptic perception.

**3.2.1.4** *Integration with computed sound*   During transducer development, described in Section 2.2, the software components of the CT device were made in the Max/MSP programming environment. However, after completion of prototype 4, when creating algorithms for the CT device that also incorporated computed sound, I/O failed. This was found to be caused by DSP computations being dropped in Max/MSP.

All software components were then ported to the SuperCollider programming language [McCartney 2002], which did not present this problem. Also, the SuperCollider programming libraries supported the implementation of all types of computed musical sound discussed in Section 1.4, in any combination algorithmically possible. Additional classes were written to provide access to the CT I/O discussed in the Sections above. After this, algorithms could be written in SuperCollider which, apart from output to auditory perception, also included input from motor activity, and output to somatosensory perception. An example of this will be given in Section 3.2.2 below.

**3.2.1.5** *Real-time instrumental control of musical sound*   When making music in real time, for human control to be satisfying, the latency between sensor input and audio output should be at most 10 ms, according to [Wright 2002]. To this end, voltage I/O in the CT system was implemented using the Motu UltraLite mk3 interface of prototype 4, controlled using OS X Core Audio (see Section 2.2.6.1). This resulted in a latency from distance input to (headphone) audio output of 3 ms, well below the stated limit.

**3.2.2 Programming interface**   As discussed above, the software components of the CT system were re-implemented in the SuperCollider language. In addition to this, a class library was written which provided DSP primitives usable for both computed sound and computed touch. On the one hand, this library served as a wrapper providing access to a range of synthesis primitives via a single uniform syntax. On the other hand, its implementation added precise and accurate control over amplitude and phase, important for forms of parametrized waveform generation.

In Figure 3.4, a code example using the implemented CT I/O class is shown. This example also illustrates the integration, within a single algorithm, of input from human

```
c = CyclotactorIO.new;
(
  {
    a =
      SynthDef ( "example",
        {
          var outputHeight_mm = 10,
              cycleHeight_mm = outputHeight_mm / 2,
              zPos_mm = In.ar (c.zPos_mm_continuous_bus),
              zPosRamp = (zPos_mm / cycleHeight_mm) % 1,
              zPosBipolarRamp = (zPosRamp * 2) - 1,
              zPosBipolarRamp_N = zPosBipolarRamp * 0.20,
              zWithinRange = (zPos_mm <= outputHeight_mm);

          Out.ar // audio left and right
          ( [10,11],
            SinOsc.ar
            ( freq: LinExp.ar (zPosRamp, 0, 1, 20, 8000),
              mul: zWithinRange * 1e-02
            )
          );

          Out.ar // upward/downward force
          ( c.zForce_N_bus,
            zWithinRange * zPosBipolarRamp_N
          );

        }).play (c.generalSynthesis_group);

  }.value;
)
```

**Figure 3.4** *The CT system: code example. Highlighted in blue: regular audio output. Highlighted in green, from top to bottom: distance input in mm, force output in N.*

motor activity, output to human auditory perception, and output to human somatosensory perception.

**3.2.3 Cost and mass-producibility**    The hardware components of the CT system can be divided into an off-the-shelf personal computer; an off-the-shelf electronic signals interface; and the custom transducer electronics. The personal computer and signals interface used were both mass-produced, costing € 400 and € 500, respectively (in 2014; for other details, see Section 2.2.6.1). All of the custom transducer electronics are mass-producible as well (see Sections 2.2.4.1, 2.2.5.1, and 2.2.6.1). However, especially the cost of the coil circuit was an open question.

After CT prototype 4 had been completed, Arno van Amersfoort of the Electronics Department at the Leiden Institute of Physics [ELD 2014] designed and built a miniaturized version of the coil circuit, using newer and cheaper components. Testing this circuit then confirmed that it indeed matched the original circuit in terms of the parameters of force output and I/O latency discussed in the preceding Sections. The new circuit also meant that a desktop form factor for the custom transducer electronics now would need a volume at most the size of a pizza box. The unit cost of the new coil circuit, like the unit cost of the other transducer electronics taken together, would be less than € 400.

Therefore, in summary, the hardware components of the CT system are mass-producible, at a cost (excluding the host laptop computer) below € 1300.

**3.2.4 Research goals attained in the resulting system**    In the preceding subsections of Section 3.2, we have presented the CT system for computed fingertip touch. How, in summary, have the chapter goals identified in Section 1.6.9 been achieved in this system?

The first goal was to algorithmically represent transducer state using physical units. In the CT system, this is done using a distance input in mm, a force output in N, and a known latency in ms between the two (see Sections 3.2.1.1 to 3.2.1.3). The programming interface for writing algorithms that use this I/O has been discussed in Section 3.2.2.

The second goal was to integrate computed sound. In the CT system, this has been done through use of the SuperCollider programming language (see Section 3.2.1.4). The code example given in Section 3.2.2 illustrated the resulting, complete integration.

The third goal was to support real-time instrumental control of musical sound. In the CT system, the latency between input from motor activity and output to auditory perception is more than sufficiently small to achieve this goal (see Section 3.2.1.5).

The fourth and final goal was to realize a powerful, yet cheaply mass-producible system.

Here, we understand implementations of computed fingertip touch to be more powerful, if characteristics such as automaton processing speed, memory size, and transducer fidelity enable inducing a wider range of perceptual phenomena (see Sections 1.6.7 and 1.4.4). The CT system, then, can be regarded as powerful because its performance enables satisfactory real-time instrumental control of computed musical sound (see Section 3.2.1.5); because its force output covers the frequency ranges involved in fingertip vibration perception (see Section 3.2.1.2); and because its I/O latency enables inducing aspects of haptic perception (see Section 3.2.1.3).

Finally, we understand a system for computed touch to be cheaply mass-producible if large-scale production is possible at a cost comparable to that of common, widely used devices for personal computing. As discussed in Section 3.2.3, the CT system, excluding the host personal computer, is mass-producible at a unit cost below that of a mid-range laptop computer.

# 3.3 The KSFT system

**3.3.1 Technical capabilities** **3.3.1.1** *Displacement input in mm* The way in which surface-parallel fingerpad displacement tracking was implemented in the KSFT device corresponded to obtaining an algorithmic representation of transducer input in millimeters (see Section 2.3.4.2). Here, the values due to individual sensor updates were accumulated over time, resulting in a displacement input representing relative position rather than velocity. This representation had a range of tens of cm in both planar directions, a resolution of 0.02 mm, and an average sampling rate of 125 Hz (see Section 2.3.5.2).

Because the sampling rate of displacement input was much lower than the rates for friction and audio output, its use could lead to perceivable quantization artefacts in the output to somatosensory and auditory perception. To counter this where necessary, a variant of the displacement input was implemented, providing an approximate reconstruction of the displacement trajectory at a higher temporal resolution. This was done using linear interpolation over the average update interval. Cubic spline and sinc interpolation were considered, also, but as these required more data points than just the current and previous displacement update, they were not used, to avoid adding extra latency. For the same reason, a mechanism to resynchronize with the sensor hardware update clock was left out, as well (see Section 2.3.4.2).

**3.3.1.2** *Friction output in N* For the algorithmic representation of transducer output, the physical unit chosen was the kinetic friction, in newtons, applied to the moving puck. Via the puck contact surface, this corresponded to the force applied, in a parallel direction, to the moving human fingerpad. Measurements were made to determine how kinetic friction varied over coil current. This was done using a device custom-built for the purpose, which allowed precisions of 0.005 A and 0.01 N. The resulting measurements are shown in Figure 3.5.

Using these measurements, a DSP algorithm was written which implemented a kinetic friction output in N. The range obtained for this output was 0.14-1.43 N. This meant that the KSFT system covered the 0.15-0.42 N friction range recommended in [Crommentuijn and Hermes 2010] for haptic devices operated by the fingers and wrist. For the shortest effective features in friction output, a duration between 10 ms and 1 ms was obtained (see Section 2.3.5.2).

**3.3.1.3** *I/O latency in ms* The average latency achieved between the above displacement input and friction output was 20.5 ms (see Section 2.3.5.2). This did not enable inducing aspects of haptic perception, but did enable inducing aspects of fingertip surface texture perception during active touch.

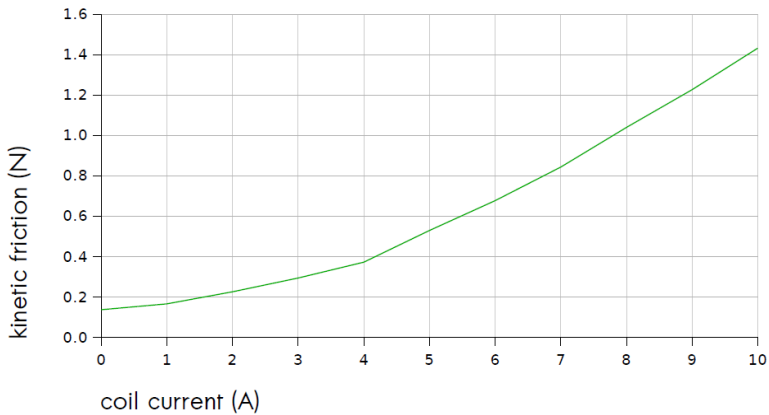

**Figure 3.5** *The KSFT system: measurements underlying fingerpad-parallel force output in newtons.*

**3.3.1.4** *Integration with computed sound* The software components of the KSFT device, like those of the CT device, were ported to the SuperCollider programming language. Classes were written to provide access to the KSFT I/O discussed in the preceding Sections. After this, algorithms could be written in SuperCollider which, apart from output to auditory perception, also included input from motor activity, and output to somatosensory perception. An example of this will be given in Section 3.3.2 below.

**3.3.1.5** *Real-time instrumental control of musical sound* Due to the input stages of the KSFT device, the average latency between displacement input and (headphone) audio output was 19.5 ms (see also Section 2.3.4.4). This was well above the 10 ms limit already discussed above, and when making fast changes to heard musical sound, the delay could become noticeable and unsatisfactory.

**3.3.2 Programming interface** As discussed above, the software components of the KSFT system were implemented in the SuperCollider language. Here too, the custom class library providing DSP primitives for both computed sound and computed touch could be used when writing algorithms.

In Figure 3.6, a code example using the implemented KSFT I/O class is shown. This example also illustrates the integration, within a single algorithm, of input from human motor activity, output to human auditory perception, and output to human somatosensory perception.

**3.3.3 Cost and mass-producibility** The KSFT system uses the same off-the-shelf personal computer and off-the-shelf electronic signals interface as the CT system (see above). Also, it uses the same coil circuit as part of its custom transducer electronics. This circuit here controls a different electromagnet, however, while input is now obtained via an optical mouse displacement sensor. This does not raise costs, since the electromagnet is made in the same way as the one in the CT system; while the optical mouse hardware costs only € 40 (in 2014; for other details, see Section 2.3.4.1). Therefore, the hardware components of the KSFT system are mass-producible, at a cost (excluding the host laptop computer) again below € 1300.

**3.3.4 Research goals attained in the resulting system**    In the preceding subsections of Section 3.3, we have presented the KSFT system for computed fingertip touch. How, in summary, have the chapter goals identified in Section 1.6.9 been achieved in this system?

```
t = TactileIO.new;
(
  {
    a =
      SynthDef ( "example",
      {
        var cycleWidth_mm = 10,
            xPosRamp = ( In.ar ( t.xPos_mm_bus ) / cycleWidth_mm ) % 1,
            yScalingFactor = LinLin.ar
                             ( In.ar ( t.yPos_mm_bus ),
                               50, 90, 0, 1
                             );

        Out.ar // audio left and right
        ( [10,11],
          SinOsc.ar
          ( freq: LinExp.ar (xPosRamp, 0, 1, 20, 8000),
            mul: LinExp.ar (yScalingFactor, 0, 1, 1e-02, 5e-04)
          )
        );

        Out.ar // kinetic friction
        ( t.kFric_N_bus,
          LinLin.ar
          ( xPosRamp,
            0,     1,
            0.14, LinLin.ar (yScalingFactor, 0, 1, 1.40, 0.14)
          )
        );

      }).play (t.generalSynthesis_group);

  }.value;
)
```

**Figure 3.6**  *The KSFT system: code example. Highlighted in blue: regular audio output. Highlighted in green, from top to bottom: x displacement input in mm, y displacement input in mm, and kinetic friction output in N.*

The first goal was to algorithmically represent transducer state using physical units. In the KSFT system, this is done using two displacement inputs in mm, a friction output in N, and a known latency in ms inbetween (see Sections 3.3.1.1 to 3.3.1.3). The programming interface for writing algorithms that use this I/O has been discussed in Section 3.3.2.

The second goal was to integrate computed sound. In the KSFT system, this has been done through use of the SuperCollider programming language (see Section 3.3.1.4). The code example given in Section 3.3.2 illustrated the resulting, complete integration.

The third goal was to support real-time instrumental control of musical sound. In the KSFT system, the latency between input from motor activity and output to auditory perception is only small enough to achieve this goal for slowly-made changes to heard musical sound (see Section 3.3.1.5).

The fourth and final goal was to realize a powerful, yet cheaply mass-producible system. Here, we understand the relative powerfulness of computed fingertip touch implementations in the same way as already discussed in Section 3.2.4. The KSFT system, then, can be regarded as powerful because, for fingertip movement at lower speeds and accelerations, its displacement input and friction output can cover the spatial range of surface roughness perception.

More precisely, in Section 2.3.4.2, we discussed how aspects of surface roughness perception are induced by details separated by 0.125 mm or more, and we also discussed how the obtained displacement input resolution of 0.02 mm seemed sufficient to support this. Now, when computing friction output to induce aspects of surface roughness, due to the average I/O latency of 20.5 ms, there will be a difference between the displacement speed on which computation is based, and the actual displacement speed during output. This error will be smaller, however, during fingertip movements with lower accelerations.

Also, as the shortest duration for perceptually effective features in friction output was found to be between 10 ms and 1 ms, the KSFT system can render apparent details of size 0.125 mm and larger as long as the absolute fingertip displacement speed stays below 12.5-125 mm/s. (In addition to the positional displacement inputs, the KSFT system was extended with displacement speed inputs, derived from the same sensor data, and yielding ranges from -334 mm/s to +334 mm/s, with a 2.6 mm/s resolution.)

Finally, we again understand a system for computed touch to be cheaply mass-producible, if large-scale production is possible at a cost comparable to that of common, widely used devices for personal computing. As discussed in Section 3.3.3, the KSFT system, excluding the host personal computer, is mass-producible at a unit cost below that of a mid-range laptop computer.

# 4. New forms
# of instrumental control

CHAPTER SUMMARY

In this chapter, we demonstrate that the KSFT system supports active touch; and that the CT system supports passive touch, active touch, and manipulation.

To reflect these fundamental types of touch, we define *computed passive touch*, *computed active touch*, and *computed manipulation*. We develop working definitions for these subtypes of computed touch that differentiate based on the presence and dynamics of a perceptually induced exterospecific component.

Simultaneously, we demonstrate computational liberation, by using it to implement new forms of instrumental control.

First, we pursue such new forms of instrumental control indirectly, by using computed fingertip touch to *display the state of the sound-generating process*, at a higher level of detail than provided by existing technologies: The idea is that such display may better inform, and thereby alter, fingertip control actions.

In Section 4.2.2, we use computed passive touch to enable display of granular synthesis at the level of individual grains. Using fingerpad-orthogonal force pulses, it seems display can occur based on presence, duration, amplitude, and vibrational content – and, significantly, using a timescale identical to that of sonic grains.

In Section 4.3.2, we use computed active touch, during instrumental control similar to turntable scratching, to enable touch display that is more specific to the stored sound fragment being played back. This is done using fingerpad-parallel friction output, with millisecond resolution.

In Section 4.3.3, we use computed active touch during surface-orthogonal fingertip movements to expand touch display outside of the moment of percussive impact.

Finally, we pursue new forms of instrumental control directly: using computed touch to *implement new types of fingertip control action*. In Section 4.5.2, we use horizontally applied output forces during horizontally directed fingertip movements to enable *pushing against a virtual surface bump*. In Section 4.5.3, we use computed touch to enable *fingertip tensing during force wave output* – and show this can be used *simultaneously* with control based on surface-orthogonal fingertip movement.

De Jong S, 2008 The cyclotactor: Towards a tactile platform for musical interaction. In *Proceedings of the 2008 international conference on New Interfaces for Musical Expression* (NIME08, June 5-7 2008, Genova, Italy) 370-371.

De Jong S, 2011 Making grains tangible: microtouch for microsound. In *Proceedings of the 2011 international conference on New Interfaces for Musical Expression* (NIME11) 326-328.

# 4.1 Introduction

In this chapter, we will pursue the third and final set of goals identified in Section 1.6.9. That is, to realize support for new forms of instrumental control of musical sound, using the systems for computed fingertip touch introduced in Chapter 3.

In Sections 4.2 to 4.4, we will discuss implemented examples demonstrating support for new forms of passive touch, active touch, and manipulation. This will be preceded by explicit definitions of *computed passive touch,* of *computed active touch,* and of *computed manipulation*. In each of the examples, computed touch will be used to display the state of the sound-generating process. The underlying idea is that such display can be used to inform human fingertip control actions.

Then, in Section 4.5, we will discuss implemented examples of computational liberation. Here, computed touch will be used to implement new types of fingertip control action.

### 4.1.1 Approach when discussing examples of instrumental control of musical sound
When discussing the implemented examples, the same five-part structure will be used. First, an introduction gives context and motivation for the form of touch implemented in the example. Then, three subsequent sections give a technical description of the causal relationships set up in the example.

This is done first for the sound-generating process. Here, supposing that an example would not involve computed touch, but instead, say, a piano key, this is where we would describe the strings used, and how these, if hit at a specific location, would become subject to specific mechanical waves, resulting via the further mechanical structure of the piano in sound waves traversing the air.

Second, the computed touch of the example is described. In the hypothetical piano key example, we would here instead describe the "mechanical touch": the structure and mechanism of the piano key, and its dynamics during operation.

Third, the causal relationships that enable instrumental control of musical sound are described. In the piano key example, this is where we would describe how lever action of the key activates a hammer mechanism, striking the strings.

After the causal relationships that define the example have been described in the middle three sections, the final section evaluates the results of these relationships from the perspective of human action and perception.

# 4.2 Implemented examples of computed passive touch

### 4.2.1 Computed passive touch    We can formulate a working definition of *computed passive touch* by drawing directly from the definitions of passive touch in Section 1.3.3.4, of computed fingertip touch in Section 1.5.2, and of computed touch in Section

2.1.1. Specifically, below, we will understand computed passive touch as the phenomenon where, for some target anatomical site, four key ingredients are in place: First, a humanly programmable, Turing-complete automaton is present, causally determining the physical state of one or more output transducers. Second, changes over time in output transducer state then influence transduction by somatosensory receptors at the target anatomical site. Third, this then induces aspects of human somatosensory perception, involving the central nervous system. Fourth, all of the above occurs in the absence of human motor activity causing movement of the target anatomical site.

## 4.2.2 Example: isotemporal display of grain generation using the CT system
**4.2.2.1** *Introduction*    In Section 1.3.3.4, we mentioned the fingertip resting on the surface of a piano key as an example of induced aspects of passive touch potentially influencing future control actions. An extension of this is the general idea that passive touch may be used to inform future control actions by having it display the state of the sound-generating process. Here, we will pursue this idea for granular synthesis, as a form of computed sound (see Section 1.4.2.6). More specifically, we will investigate computed passive touch as a means for displaying processes of grain generation [De Jong 2011].

Existing, widely used types of general-purpose controller hardware, such as buttons, sliders, knobs, pads, and keys (see Section 1.2.3) enable control actions to, as well as some display of grain-generating processes. These devices may be used to initiate, modulate, and terminate such processes in real time. In general, then, stages in the operation of these devices can be set up to usefully coincide with stages of grain generation. For example, pressing and releasing a piano-type key may be used to initiate and terminate grain generation, with simultaneously moving a slider used to alter additional real-time parameters.

However, display providing higher levels of detail than this can be envisioned. The generation of each granular sound fragment will happen according to a set of explicitly defined parameters. Usually, the values for these parameters are separately determined for each grain, at the moment of its instantiation, to then remain fixed for the rest of its duration. It follows that display realized at the highest level of detail would also represent the states of individual grains.

As an initial exploration of this possibility, we will attempt to implement display of individual grains via computed passive touch. This will be done on the CT system exclusively, as the KSFT system does not support passive touch (see Section 2.3.1). The required absence of human motor activity here means that the fingertip will be held still. Output will then apply non-vibratory and vibratory force pulses, constructed in a one-to-one relationship with sonic grains.

**4.2.2.2** *The sound-generating process*    The audio signals interface used for voltage I/O by the CT system (see Section 2.2.6.1) here also was used to provide monophonic audio output to a pair of headphones. This also meant that all A/D and D/A conversions for voltage I/O occurred at the same, constant $c_{\text{sampling rate}}$ (Hz). Consequently, for algorithms computing signals, time $t \in \mathbb{N}$ increased in discrete, equal steps, with $t = 0$ the first time step of algorithm execution.

A grain envelope signal then was computed as a block pulse, with a duty cycle of $c_{\text{grain duration}}$ seconds, varied between algorithm runs. Over time, the amplitude series of this signal was characterized by:

$$s_{\text{grain envelope}}[t] = \begin{cases} 1, & (t \bmod c_{\text{sampling rate}}) < (c_{\text{grain duration}} \times c_{\text{sampling rate}}) \\ 0, & (t \bmod c_{\text{sampling rate}}) \geq (c_{\text{grain duration}} \times c_{\text{sampling rate}}) \end{cases}$$

This meant that the interval between successive grain event onsets was precisely 1 second. The grain envelope signal then controlled audio output by modulating a sine wave signal:

$$s_{\text{audio sine}}[t] = \sin\left((t / c_{\text{sampling rate}} \times 2\pi \times c_{\text{audio sine frequency}}) - \tfrac{1}{2}\pi\right)$$

$$o_{\text{audio}}[t] = s_{\text{grain envelope}}[t] \times s_{\text{audio sine}}[t]$$

Here, a relatively high $c_{\text{audio sine frequency}} = 4000$ Hz was used, to perceptually retain pitch more easily when smaller values were used for $c_{\text{grain duration}}$. The initial phase of $-\tfrac{1}{2}\pi$ meant that $s_{\text{audio sine}}[t]$ would start out at the minimum -1 amplitude. While this did not give a gradual transition from zero amplitude into the sonic grain event, it did allow re-use of the algorithmic primitive used for computed touch (see below).

Finally, the latency of the audio output was adjusted, so that temporally, the sonic grain events would coincide as much as possible with their counterparts in the force output, described next.

**4.2.2.3** *Computed touch*     The grain envelope signal determining audio output was also used to construct unipolar, fingerpad-orthogonal force pulses. The resulting "force grains" were in a one-to-one relationship with sonic grains:

$$s_{\text{non-vibratory force grain}}[t] = s_{\text{grain envelope}}[t]$$

As the name of the above signal implies, alternative, vibratory force pulses were generated, too. These, in their construction, were analogous to the sonic grains:

$$s_{\text{force sine}}[t] = \sin\left((t / c_{\text{sampling rate}} \times 2\pi \times c_{\text{force sine frequency}}) - \tfrac{1}{2}\pi\right)$$

$$s_{\text{vibratory force grain}}[t] = s_{\text{grain envelope}}[t] \times (s_{\text{force sine}}[t] + 1) / 2$$

Here, like in Section 2.2.4.8, using a fixed $c_{\text{force sine frequency}} = 250$ Hz was understood to provide a good match for human vibration sensitivity at the fingertip. With $s_{\text{force sine}}[t]$ being scaled to a unipolar signal, its initial phase of $-\tfrac{1}{2}\pi$ meant that the amplitude transition into the vibratory grain event would not be abrupt, but gradual. Then, with $s_{\text{force grain}}[t]$ equal to either $s_{\text{vibratory force grain}}[t]$ or $s_{\text{non-vibratory force grain}}[t]$, force output in N was computed as:

$$o_{\text{force}}[t] = c_{\text{force bias}} + (s_{\text{force grain}}[t] \times c_{\text{force scaling}})$$

This with $c_{\text{force bias}} = 0.14$ N, and $c_{\text{force scaling}}$ (N) varied between algorithm runs.

**4.2.2.4** *Instrumental control of musical sound*  The heard musical sound induced by the above algorithm can be described as individual pure-sine constant-pitch grains, in a steady tempo, separated by silence. As discussed in the introduction, the question motivating the example was whether detailed display of the state of the sound-generating process would be possible via computed passive touch. Therefore, no additional causal relationships implementing instrumental control were set up.

**4.2.2.5** *Evaluation*  The algorithm described in Sections 4.2.2.2 and 4.2.2.3 was informally tested for a large number of parameter configurations [De Jong 2011]. Three algorithmic parameters were varied between runs, as follows:

- $c_{\text{grain duration}} = 100, 50, 10,$ or 1 ms;

- $s_{\text{force grain}}[t] = s_{\text{non-vibratory force grain}}[t]$, or $s_{\text{vibratory force grain}}[t]$;

- $c_{\text{force scaling}} = 0.86, 0.58,$ or 0.29 N.

The $c_{\text{force scaling}}$ alternatives resulted in a [0.14, 1.00], [0.14, 0.72], or [0.14, 0.43] N range for $o_{\text{force}}[t]$, respectively. The 1 ms duration was not tested for vibratory force pulses, as it could not contain at least one full vibration cycle (of duration 4 ms). Otherwise, for all possible combinations of the above options, output was applied to the fingertip, with and without simultaneous headphone audio output, and observations were noted down.

What seemed clearly perceivable was, first, the presence or absence of force pulses; second, differences in their durations; and third, for a given duration, differences in amplitude. Also, at 100 ms and 50 ms durations, vibratory force pulses indeed seemed to give an impression of vibration, different from the sensation produced by a non-vibratory force pulse of the same duration. This changed at 10 ms, however, to a pulsed sensation which seemed less distinct.

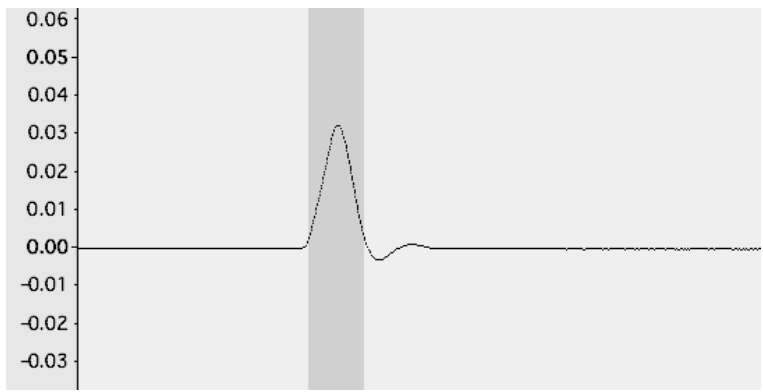

**Figure 4.1**  *Linear magnetic field strength recorded during the application of a perceived force pulse by the CT system. The dark grey bar spans a 1.0 ms duration.*

This seems to indicate that force pulses applied to the fingerpad can be used to display information, based on their presence, duration, amplitude, and – for longer durations – the presence or absence of vibration. In the case of granular synthesis, this can happen "isotemporally": force pulses may be generated simultaneously with sonic grains, using a timescale for their onsets and durations that is identical to that of individual sonic grains (see Section 1.4.2.6, and Figure 4.1). Therefore, it seems that for granular synthesis, computed fingertip touch may enable display of the state of the sound-generating process at higher levels of detail than currently offered by existing controller hardware.

Also, the above demonstrated that the CT system supports passive touch.

## 4.3  Implemented examples of computed active touch

**4.3.1  Computed active touch**    We can formulate a working definition of *computed active touch* by drawing directly from the definitions of active touch in Section 1.3.3.5, of computed fingertip touch in Section 1.5.2, and of computed touch in Section 2.1.1. Specifically, below, we will understand computed active touch as the phenomenon where, for some target anatomical site, again three of the key ingredients mentioned in Section 4.2.1 are in place: First, a humanly programmable, Turing-complete automaton is present, causally determining the physical state of one or more output transducers. Second, changes over time in output transducer state then influence transduction by somatosensory receptors at the target anatomical site. Third, this then induces aspects of human somatosensory perception, involving the central nervous system.

Now, however, all of the above occurs in the *presence* of human motor activity causing movement of the target anatomical site. Furthermore, aspects of this motor activity causally determine the physical state of one or more input transducers. Changes over time in input transducer state then influence the state of the Turing-complete automaton.

Because of this, causal relationships implemented via the Turing-complete automaton may induce sensations that include exterospecific components. These are then produced in such a way as not to be subject to change.

**4.3.2  Example: isospatial display of stored sound fragment playback using the KSFT system**    **4.3.2.1** *Introduction*    In Section 1.2.3, we discussed the electromechanical playback of stored sound fragments using gramophone record players, or turntables. We also discussed how electronic playback followed, using digital samplers. We then discussed sampling independently of specific hardware, in Section 1.4.2.3, as a form of computed sound (see Figure 1.10).

The human control actions that came into use with turntables included *scratching*. Here, the fingerpad is pressed down on a flat contact surface, after which surface-parallel fingertip movements are performed. During this, fingerpad-parallel forces are applied via the contact surface (see Section 2.3.1). The causal relationship enabling instrumental control of musical sound, here, is that for the given stored sound fragment, spatial fingertip position is made to correspond to temporal playback position.

Like the examples in Section 4.2.2.1, this setup enables display, via touch, of the state of the sound-generating process. But again, we may envision touch display that might better inform control actions, by providing higher levels of detail: As the signal variations in the record groove do not result in felt forces, touch display during scratching is not specific to the actual stored sound fragment being played back.

Therefore, below, we will explore display that *is* more specific to the stored sound fragment. We will use the KSFT system to implement friction output during active touch that reflects ongoing stored sound fragment playback with millisecond resolution [De Jong 2010b].

**4.3.2.2** *The sound-generating process*    As before, we use a time $t \in \mathbb{N}$, based on a constant $c_{\text{sampling rate}}$ (Hz). The sound fragment, then, was represented by a series of amplitude values within a [-1, 1] range, stored as a one-dimensional array in a memory buffer $\vec{b}_{\text{stored sound fragment}}$. Playback was characterized by:

$$o_{\text{audio}}[t] \ = \ f_{\text{index and interpolate}}\big(s_{\text{sample index}}[t], \ \vec{b}_{\text{stored sound fragment}}\big)$$

Here, $f_{\text{index and interpolate}}$ calculates the amplitude value for the specified index, which may be a non-integer number: cubic interpolation is performed to enable intermediate values. A delay of 1.0 ms was applied to the resulting signal, to optimally synchronize audio output with related friction output (see Sections 3.3.1.3 and 3.3.1.5).

**4.3.2.3** *Computed touch*    Friction output displayed stored sound fragment playback:

$$s_{\mu}[t] \ = \ \frac{1}{n} \sum_{i=0}^{i=n-1} \big| o_{\text{audio}}[t-i] \big|$$

This with $n \ = \ c_{\text{sampling rate}} \times 0.001$ and from $t \ \geq \ n-1$, so that $s_{\mu}[t]$ tracked the average absolute amplitude over the most recent millisecond of audio output. Then, to better match the perceived loudness of audio output, friction output was not computed directly from the average absolute amplitude, but via:

$$o_{\text{friction}}[t] \ = \ f_{\text{dB to N mapping}}\big(20 \times \log_{10}\big(\frac{s_{\mu}[t]}{c_{\text{ref}}}\big)\big)$$

This was done with reference amplitude $c_{\text{ref}} = 1$, so that the maximum absolute amplitude corresponded to 0 dB input to the $f_{\text{dB to N mapping}}$ function, which clipped and linearly mapped the range of [-30, 0] dB to [0.14, 0.5] N.

**4.3.2.4** *Instrumental control of musical sound*    The causal relationship enabling instrumental control was:

$$s_{\text{sample index}}[t] \;=\; i_{\text{x displacement}}[t] \;\times\; c_{\text{samples per mm}}$$

Here, $i_{\text{x displacement}}[t]$ (mm) is the displacement input for the $x$ planar direction of the KSFT system (see Section 3.3.1.1). The constant $c_{\text{samples per mm}}$ was varied between algorithm runs.

**4.3.2.5** *Evaluation*    When using appropriate values for $c_{\text{samples per mm}}$, the algorithm described in Sections 4.3.2.2 to 4.3.2.4 enabled instrumental control similar to turntable scratching, with sideways fingertip movements controlling variable-pitch playback of stored sound fragments. The algorithm was tested in a pilot experiment with 5 volunteer test subjects [De Jong 2010b]. The subjects were presented with a stored sound fragment consisting of silence, followed by 240 cycles of a sinusoid, followed by more silence. Sine cycles were identical, of maximum amplitude, and had a length of 100 samples each. The test subjects were asked to explore the stored fragment, by using sideways fingertip movement at different speeds. This was done repeatedly, for the following parameter configurations:

• $c_{\text{samples per mm}}$ = 4000, 500, or 8 samples/mm.

During exploration, the above $c_{\text{samples per mm}}$ options resulted in frequency ranges for audio output of [104, 13360] Hz, [13, 1670] Hz, or [0.2, 26.7] Hz, respectively. Each of the three options was presented twice: first without, then with friction output enabled in software.

For $c_{\text{samples per mm}}$ = 4000 samples/mm, and for $c_{\text{samples per mm}}$ = 500 samples/mm during fast movements, audio playback of the sine wave fragment coincided with a heightened friction level, computed from the averaged signal intensity (see Section 4.3.2.3). Here, test subjects clearly felt that sound and touch were related, both resulting from movement in or over the same specific surface area.

The algorithm was publicly demonstrated at the 2010 international workshop on Interactive Sonification in Stockholm, Sweden. There, participants also could record the stored sound fragment live via a microphone, to then explore the result as in the experiment (see also Figure 4.2).

The results of the pilot experiment indicate that computed friction output can be used to add touch display during scratching that is more specific to the stored sound fragment being played back. This can happen "isospatially": with temporally coinciding features in sound and touch being perceived as belonging to the same spatial location. This suggests a potential use of computed friction as an aid to spatial orientation before or during fingertip control actions.

Also, the above demonstrated that the KSFT system supports active touch.

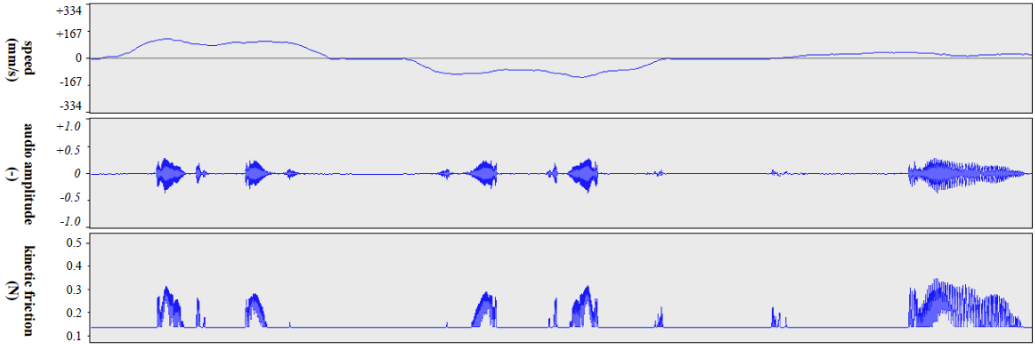

**Figure 4.2** *Movement speed input, audio output, and friction output during 5 seconds of instrumental control of a stored sound fragment. In the left section of the figure, a signal feature is first played back in one direction; then, in the middle section, this is done in the opposite direction. Finally, with fingertip movement having returned to somewhat before the initial position, in the rightmost section part of the feature is again played back, but now using a slower movement.*

### 4.3.3 Example: isotemporal display of triggered waveform synthesis using the CT device

**4.3.3.1** *Introduction*    In Section 1.2.3, we discussed pressure-sensitive pads operated by the fingertips (see Figure 1.4). Here, by varying the applied downward contact pressure over time, input may be provided to the control of computed sound, continuously. However, when rhythmic musical sound is made, pads may rather be operated much like the membranes of acoustic drums: using percussive hand and finger movements that result in contact pressure being applied only during a brief moment of impact. As in Sections 4.2.2.1 and 4.3.2.1, we may then envision touch display, of the state of the sound-generating process, which provides a higher level of detail: by having this display occur not just during the moment of impact.

Below, we will explore such display, implemented using prototype 1 of the CT device (see Section 2.2.3) [De Jong 2006]. Here, percussive movements triggered a fixed audio and touch response, generated using parametrized waveform synthesis (see Section 1.4.2.2). As in Section 4.2.2, the goal here was to provide "isotemporal" display: The touch response is transduced during the same time period as the audio response.

**4.3.3.2** *The sound-generating process*    As before, we use a time $t \in \mathbb{N}$, based on a constant $c_{\text{sampling rate}}$ (Hz). Audio output, then, was zero, or silent, except when a trigger signal $s_{\text{trigger}}[t]$ would activate a fixed audio response of $c_{\text{response duration}}$ seconds. More precisely, supposing that at some time $t_{\text{activation}}$, with no ongoing audio response, $s_{\text{trigger}}[t_{\text{activation}}] = 1$ while $s_{\text{trigger}}[t_{\text{activation}} - 1] = 0$; then for $i \in \mathbb{N}$, $0 \leq i < (c_{\text{response duration}} \times c_{\text{sampling rate}})$:

$$o_{\text{audio}}[t_{\text{activation}} + i] \ = \ s_{\text{audio response}}[i]$$

Here, $s_{\text{audio response}}[i]$ was a fixed amplitude series, generated algorithmically using parametrized waveform synthesis: An ADSR envelope primitive (see Section 1.4.2.2) controlled the frequency of a sine wave amplitude series over time. Here, a brief Attack-Decay peak to maximum frequency was followed by a prolonged Sustain period at a low frequency, which was then finally cut off by a brief Release phase, coinciding with a return to zero amplitude.

**4.3.3.3** *Computed touch*    Like the audio output, touch output was based on a fixed response of $c_{\text{response duration}}$ seconds, identically activated by $s_{\text{trigger}}[t]$, so that from some time $t_{\text{activation}}$, for $i \in \mathbb{N}$, $0 \leq i < (c_{\text{response duration}} \times c_{\text{sampling rate}})$ :

$$s_{\text{to touch output}}[t_{\text{activation}} + i] \; = \; s_{\text{touch response}}[i]$$

Again, $s_{\text{touch response}}[i]$ was a fixed amplitude series, generated algorithmically using parametrized waveform synthesis. Here, the amplitude of a unipolar triangle wave series over time decreased from maximum to zero, while being subtracted from a level bias at maximum amplitude. The resulting signal was passed on to the prototype 1 output transducer, which mapped the range from minimum to maximum amplitude onto the range from minimum (downward) to maximum (upward) force currently possible for the given keystone distance (see Section 2.2.3.3 and Figure 4.3).

**4.3.3.4** *Instrumental control of musical sound*    The causal relationship enabling instrumental control was:

$$s_{\text{trigger}}[t] \; = \; \begin{cases} 1, & i_{\text{z distance}}[t] \; \leq \; c_{\text{distance threshold}} \\ 0, & i_{\text{z distance}}[t] \; > \; c_{\text{distance threshold}} \end{cases}$$

Here, $i_{\text{z distance}}[t]$ is the distance input, not yet in mm, of the prototype 1 CT device (see Section 2.2.3.2). Given the still low accuracy of the $i_{\text{z distance}}[t]$ range, the constant $c_{\text{distance threshold}}$ remained fixed across algorithm runs.

**4.3.3.5** *Evaluation*    The algorithm described in Sections 4.3.3.2 to 4.3.3.4 enabled instrumental control via downward percussive movements toward the device surface, made with the keystone held between the fingertips. Before the device surface could be reached, this would trigger a sound initially similar to a drum kick, then quickly transforming into a low-pitched bass tone. Simultaneously, a sudden upward force would create a sensation that seemed akin to coming into contact with some object, while causing a reverberation.

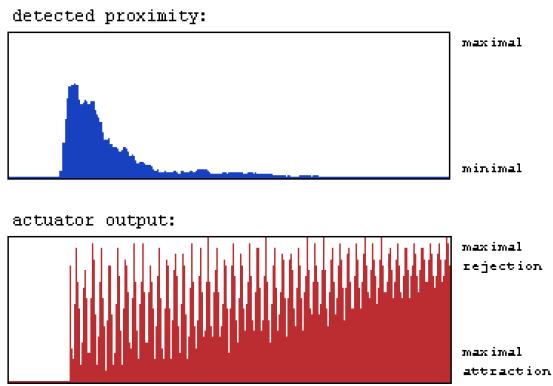

**Figure 4.3**    *Executing isotemporal display of triggered waveform synthesis: Simultaneous I/O by prototype 1 of the CT device during a percussive movement toward the device surface. Time increases to the right, 100 samples/s.*

Figure 4.3 shows I/O by the prototype 1 CT device during such a percussive movement. Spikes in the actuator output can be seen reflected in the proximity input. As shown, the reverberation would gradually die away, together with the sound being produced. In this way, isotemporal display of the sound-generating process was realized – occurring not just during the moment of percussive impact.

The algorithm was successfully tried out by the visitors of an exhibition featuring the device, and was later publicly demonstrated at the 2006 international conference on New Interfaces for Musical Expression in Paris, France.

The above also demonstrated that, from prototype 1 onward, the CT system has supported active touch.

# 4.4  Implemented examples of computed manipulation

**4.4.1  Computed manipulation**    We can formulate a working definition of *computed manipulation* by drawing directly from the definitions of manipulation in Section 1.3.3.5, and of computed active touch in Section 4.3.1. Specifically, below, we will understand computed manipulation as the phenomenon identical to computed active touch except for the following: Now, the causal relationships implemented via the Turing-complete automaton are set up in such a way, that an exterospecific component *is* made subject to change over time. Also, importantly, this is done as a consequence of specific motor activity.

**4.4.2  Example: isotemporal display using a dynamic virtual wall on the CT system**
**4.4.2.1** *Introduction*    The purpose of the following example is to demonstrate that the CT system supports computed manipulation. As in the earlier examples of Sections 4.2.2, 4.3.2 and 4.3.3, here too, computed touch is used to display the state of the sound-generating process.

To induce an exterospecific component, we implemented a version of the *virtual wall*: the classic primitive often used as a benchmark for the performance of haptic technologies. Our implementation is based on the basic knowledge that mechanical *stiffness* and *damping* are involved in virtual walls [Colgate et al. 1993]. These two factors directly defined the computation of forces applied to the fingertip.

**4.4.2.2** *The sound-generating process*    As in Section 4.3.3.2, the sound-generating process was characterized by a fixed audio response, generated algorithmically using parametrized waveform synthesis. Now, however, the content of this audio response was a fixed-frequency sine wave, amplitude-modulated by an increasing-then-decreasing envelope function. This response was activated by a trigger signal $s_{\text{audio trigger}}[t]$.

### 4.4.2.3 *Computed touch*

Before implementing this example, a fingertip speed input $i_{\text{z speed}}[t]$ (mm/s) was added to the CT system, for use alongside its existing $i_{\text{z distance}}[t]$ (mm) distance input. Computed touch was then defined by the following set of amplitude series:

$$s_{\text{spring}}[t] \;=\; \frac{\left| i_{\text{z distance}}[t] - c_{\text{z limit}} \right|}{c_{\text{z limit}}} \times c_{\text{max. spring force}}$$

$$s_{\text{dashpot}}[t] \;=\; i_{\text{z speed}}[t] \times c_{\text{z damping}}$$

$$s_{\text{fingertip within range}}[t] \;=\; \begin{cases} 1, & i_{\text{z distance}}[t] \;\leq\; c_{\text{z limit}} \\ 0, & i_{\text{z distance}}[t] \;>\; c_{\text{z limit}} \end{cases}$$

$$s_{\text{virtual wall}}[t] \;=\; \left( s_{\text{spring}}[t] + s_{\text{dashpot}}[t] \right) \times s_{\text{fingertip within range}}[t]$$

$$o_{\text{force}}[t] \;=\; s_{\text{virtual wall}}[t] \times s_{\text{wall on/off}}[t]$$

This with $s_{\text{wall on/off}}[t]$ a signal that should be either 0 or 1; $c_{\text{z limit}}$ = 5.0 mm; $c_{\text{max. spring force}}$ = 2.0 N; and $c_{\text{z damping}}$ = -3.0 × 10$^{-3}$ N/(mm/s).

### 4.4.2.4 *Instrumental control of musical sound*

Central to instrumental control was a function implementing a Schmitt trigger:

$$s_{\text{Schmitt trigger}}[t] \;=\; f_{\text{Schmitt trigger}} \left( s_{\text{Schmitt trigger}}[t-1], \; i_{\text{z distance}}[t], \; c_{\text{high threshold}}, \; c_{\text{low threshold}} \right)$$

This function would compute 1 if fingertip distance was above $c_{\text{high threshold}}$ = 30.0 mm. It would then continue to do so, until fingertip distance would drop below $c_{\text{low threshold}}$ = 0.2 mm. From then onward, its output would remain 0 until fingertip distance again would rise above $c_{\text{high threshold}}$. The causal relationship that enabled instrumental control was then:

$$s_{\text{audio trigger}}[t] \;=\; 1 - s_{\text{Schmitt trigger}}[t]$$

A second causal relationship, enabling touch display of the state of the sound-generating process, was:

$$s_{\text{wall on/off}}[t] \;=\; s_{\text{Schmitt trigger}}[t]$$

### 4.4.2.5 *Evaluation*

The measurements in Figure 4.4 show 4.5 seconds of I/O by the algorithm described above. The four recorded measures share the same time axis, and are, from top to bottom: vertical fingertip position, in mm; vertical fingertip speed, in mm/s; vertical force output to the fingertip, in N; and to-headphone audio output, as a spectrogram. The occurrence of four consciously executed fingertip actions has been indicated along the time axis.

• First, there is a downward tap against the virtual wall, clearly visible in the fingertip position profile, and also in the simultaneous, peaking output force profile computed during the downward impact and upward retreat. Here, audio output remains silent.

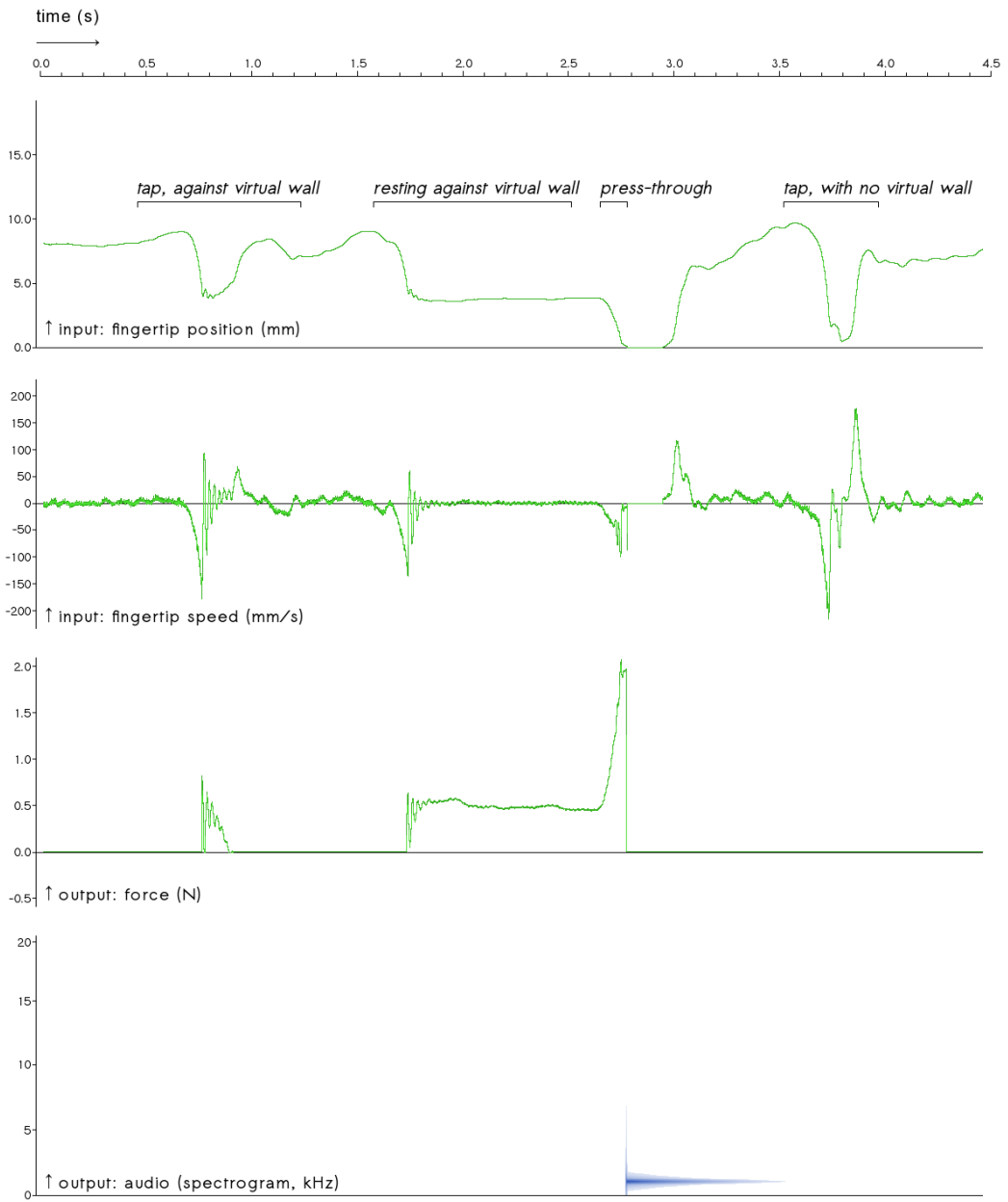

**Figure 4.4** *Computed manipulation: dynamic virtual wall I/O on the CT system. For a detailed description, see the text.*

• Next, the fingertip is made to rest against the virtual wall, which is visible in the drop in fingertip position and its subsequent flattening out over time. Here, after the initial impact, the computed output force can be seen to change to a pattern of relatively small fluctuations around a positive bias.

• Third, the fingertip can then be seen to press through the virtual wall, while countering a markedly increasing output force. This then triggers a fixed-pitch sonic response that breaks the silence and coincides with a steep drop to 0 N output force.

• Fourth and last, another downward tap occurs, again clearly reflected in the fingertip position profile. This time, however, it occurs in the absence of a virtual wall, with the computed output force remaining at 0 N throughout.

Not recorded in Figure 4.4 anymore, but evident from the description of the algorithm (see Section 4.2.4) is that the fingertip must now be raised above 30.0 mm before the virtual wall will reappear – and the sonic response again can be triggered.

So, in this way, the virtual wall *displays the state of the sound-generating process:* Pressing through it triggers a pitched sound, while probing its presence or absence by tapping movements displays whether triggering a sound is currently ("isotemporally") possible.

Finally, we have now illustrated how the above algorithm, running on the CT system, *implements computed manipulation:* Its virtual wall corresponds to an exterospecific component in touch perception, which also is made dynamic: It will disappear completely in response to the specific motor activity of pressing-through.

# 4.5  Implemented examples of computational liberation

### 4.5.1  Demonstrating computational liberation     We understand *computational liberation*, in the context of instrumental control of musical sound, as the gradual minimization of constraints on implementable causal relationships between human actions and changes in heard musical sound, through the development of transducers combined with Turing-complete automata (see Section 1.5.1.2). An endpoint for computational liberation would be the theoretical capability of *completely computed instrumental control of musical sound* (see Section 1.5.1.1).

From this, it follows that computational liberation may be demonstrated by using it to implement new forms of instrumental control. Here, our goal is to provide such demonstration for the systems of Chapters 2 and 3 implementing the computed fingertip touch model (see Section 1.5.2.1).

So far, in the implemented examples of this chapter, computed touch has been used to display the state of the sound-generating process. The idea here has been that this may inform and thereby alter fingertip control actions, indirectly enabling new forms of instrumental control.

In the sections now following, our goal will be to demonstrate computational liberation more directly: Computed touch will now be used to implement new types of fingertip control action. Thereby, it will directly enable new causal relationships between human actions and changes in heard musical sound.

### 4.5.2 Example: pushing against virtual bumps using the KSFT system

**4.5.2.1** *Further pilot experiment results*  In Sections 4.3.2.1 to 4.3.2.4, we introduced and described an algorithm running on the KSFT system for the isospatial display, via fingertip touch, of stored sound fragment playback. In Section 4.3.2.5, results of a pilot experiment featuring the algorithm were discussed. Apart from the results already discussed, additional results were obtained at $c_{\text{samples per mm}}$ = 8 samples/mm. Here, the repeating kinetic friction pattern the test subjects were presented with had a periodic length of 6.25 mm, and described a symmetric curve, which is evident in the example output shown in Figure 4.5.

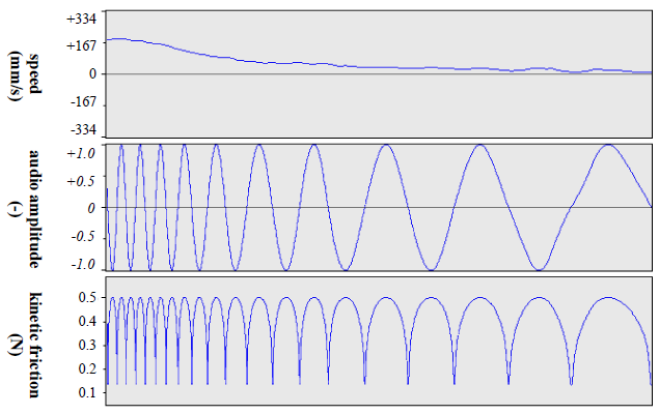

**Figure 4.5** *Movement speed input, audio output, and friction output, during 2 seconds of fingertip movement across a computed, repeating kinetic friction pattern, with a periodic length of 6.25 mm.*

The above friction output, during active touch, resulted in test subjects mostly reporting the perception of regular bumps on a surface [De Jong 2010b]. It seemed remarkable that, based on horizontally applied output forces during horizontally directed fingertip movements, test subjects would report surface variation in a vertical direction.

This was reminiscent, however, of how horizontal-only forces were used to create the sensation of vertical bumps in [Robles-De-La-Torre and Hayward 2001], verified for higher spatial frequencies for the device presented in [Ogawa and Shimojo 2006].

**4.5.2.2** *A new type of fingertip control action*  We understand the above results to demonstrate a new type of fingertip control action, based on two observables: the movement over time of the fingertip, and the resulting perception.

Here, we will discuss newness relative to the single-fingertip control actions widely used with touch pads and touch screens (see Section 1.2.4). In these control actions, too, the fingerpad contacts a flat, closed, and rigid transducer surface, often during surface-parallel fingertip movements.

For example, consider a touch pad control action making changes to heard musical sound if the fingertip moves across some fixed location on the device surface. This control action could then be re-implemented using the KSFT system – but now with the fixed location also made the center of an instance of the pilot experiment friction pattern.

We may then regard the result of this as a *different* type of fingertip control action: In general, within the range set for the pilot experiment, friction output was capable of variably slowing down fingertip movements (see also Section 2.3.2). Specifically, output of the friction pattern, too, results in different patterns of fingertip displacement over time than occur when no variable friction output is applied.

This then also implies a *new* type of fingertip control action: The fingertip displacement patterns will be unlike those of all other implementable touch pad control actions, too, simply due to the general absence of controlled, variable surface friction.

Additionally, given the reported sensation of fingertip movement across regular bumps on a surface, the proposed control action will also be *different* and *new* in terms of the induced perception.

Then, both in terms of fingertip movement and perception, the pilot experiment has yielded a new type of fingertip control action: pushing against or across a virtual surface bump. This directly enables the construction of new causal relationships between human actions and changes in heard musical sound. Thereby, it demonstrates computational liberation.

### 4.5.3 Example: multiplexing a rigidity DOF on orthogonal fingertip movement using the CT device

**4.5.3.1** *Introduction*    In Section 1.5.1.1, we discussed the importance of a fundamental question: *What forms of instrumental control are possible to implement?* Here, a relevant sub-question is: *What human control actions are possible?* In the context of unidirectional fingertip movement orthogonal to a surface (see Section 1.2.5), the human actions that are possible often will be ones that produce downward or upward fingertip movement. However, human actions *suppressing* downward or upward fingertip movement exist, too. Therefore, here, our immediate research question will be: *Can computed touch be used to enable instrumental control that is based on finger rigidity?*

And then, also: *Can such control occur simultaneously with control based on down- and upward fingertip movement?* This is relevant, as a positive answer here could point to ways of obtaining more control possibilities per finger, for simultaneous change to the sound-generating process. A positive answer also may serve to clarify and demonstrate the difference between what we here refer to as "movement-producing" and "movement-suppressing" actions.

To answer the two-part research question, prototype 2 of the CT device (see Section 2.2.4) was used to test an approach based on first generating a wave pattern in fingerpad-orthogonal force output, to then multiplex a rigidity degree-of-freedom (DOF) on the orthogonal fingertip movement input [De Jong 2008]. Here, a form of computed sound – again parametrized waveform synthesis (see Section 1.4.2.2) – was implemented so as to also demonstrate actual control of a sound-generating process.

**4.5.3.2** *The sound-generating process*    As before, computed sound was based on a constant $c_{\text{sampling rate}}$ (Hz), with time $t \in \mathbb{N}$ (see Section 4.2.2.2). Over time, the amplitude series of audio output was characterized by:

$$o_{\text{audio}}[t] = f_{\text{audio band noise}}(t, s_{\text{center frequency}}[t], s_{\text{bandwidth}}[t])$$

Here, $f_{\text{audio band noise}}$ computed a noise signal, characterizable as a sum of sine waves diminishing in amplitude below and above a given center frequency. This $s_{\text{center frequency}}[t]$ was variable across a [100, 24000] Hz range. How quickly sine wave amplitudes would diminish over frequency was determined by the bandwidth parameter $s_{\text{bandwidth}}[t]$, variable across a [20, 320] unitless range. Variation across both ranges seemed independently perceivable in the resulting heard sound.

**4.5.3.3** *Computed touch*    Touch output via the prototype 2 CT device was controlled by a sine wave amplitude series:

$$s_{\text{to touch output}}[t] = \sin(t \,/\, c_{\text{sampling rate}} \times 2\pi \times c_{\text{force sine frequency}})$$

Here, the frequency used was fixed at $c_{\text{force sine frequency}} = 10$ Hz. The output transducer then mapped the $s_{\text{to touch output}}[t]$ amplitude range to the full magnetic field strength range available (see Sections 2.3.3.3, 2.2.4.5, and 2.2.4.6). Effectively, this mapped to the full range from minimum (downward) to maximum (upward) force currently possible for the given keystone distance. However, in a final adjustment, the digital amplitude range of $s_{\text{to touch output}}[t]$ was reduced by 10%, to avoid collapses in magnetic field strength output (see Section 2.2.4.8).

**4.5.3.4** *Instrumental control of musical sound*    The prototype 2 CT device provided a linear distance input $i_{\text{z distance}}[t]$ (see Sections 2.2.4.2 to 2.2.4.4). From it, a "nearness" signal was computed, using a moving average:

$$s_{\text{nearness}}[t] = \frac{1}{n} \sum_{i=0}^{i=n-1} i_{\text{z distance}}[t-i]$$

Here, $t \geq n-1$ and $n = c_{\text{sampling rate}} \,/\, c_{\text{force sine frequency}}$, so as to average out variation in distance input due to the force output wave described in Section 4.5.3.3.

Then, the absolute deviation of distance input $i_{z\,distance}$ from the $s_{nearness}$ average was used, also averaged over the most recent output wave cycle, to compute a "rigidity" signal that tracked how much the fingertip attached to the keystone dampened vibration amplitude:

$$s_{rigidity}[t] \; = \; f_{distance\,compensation}\left(s_{nearness}[t], \; \frac{1}{n} \sum_{j=0}^{j=n-1} \mid i_{z\,distance}[t-j] - s_{nearness}[t-j] \mid\right)$$

This with $n$ as before, and now $t \geq 2\,n - 2$. Here, the $f_{distance\,compensation}$ function compensated for the amplitude of the output force wave itself already varying over fingertip distance. This compensation was based on measurements, made both while keeping a finger maximally rigid and while keeping it maximally loose, of the uncompensated rigidity signal across the $s_{nearness}$ range.

The remaining causal relationships which then enabled instrumental control can be roughly characterized by:

$$s_{center\,frequency}[t] \; \approx \; f_{map\,to\,Hz}\left(s_{nearness}[t]\right)$$

$$s_{bandwidth}[t] \; \approx \; f_{map\,to\,bandwidth\,range}\left(s_{rigidity}[t]\right)$$

Here, $f_{map\,to\,Hz}$ mapped the $s_{nearness}$ range to [100, 24000] Hz, and $f_{map\,to\,bandwidth\,range}$ mapped the $s_{rigidity}$ range to [20, 320] (see Section 4.5.3.2).

**4.5.3.5** *Evaluation*     When executed, the algorithm described above in Sections 4.5.3.2 to 4.5.3.4 produced a continuous sensation of the fingertip being shaken up and down, rapidly and regularly. During this, performing slow up- or downward fingertip movements (see Figure 4.6, left side) would lower or raise the pitch of a continuously heard sound similar to white noise. Simultaneously, performing muscle contractions tensing the finger (see Figure 4.6, right side) would change the timbre of the noise [De Jong 2008].

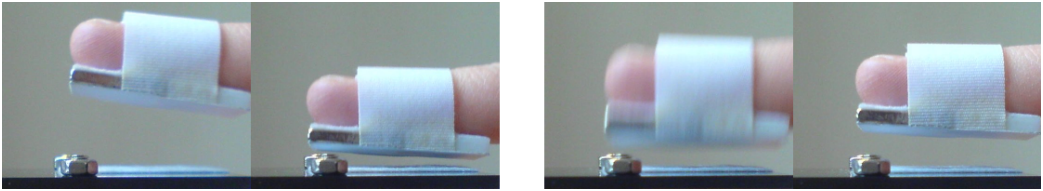

**Figure 4.6** *To the left: fingertip up/down movement, varying $s_{nearness}$. To the right: fingertip tensing, varying $s_{rigidity}$.*

This can be seen in Figure 4.7, where I/O of the algorithm is shown during a slow downward, then upward fingertip movement, during which the fingertip also was tensed and relaxed twice. Correspondingly, two $s_{rigidity}$ peaks can be seen below a slower overall $s_{nearness}$ movement, with both parameters reflected in the sonogram of the audio output. The algorithm was publicly demonstrated at the 2008 international conference on New Interfaces for Musical Expression in Genova, Italy.

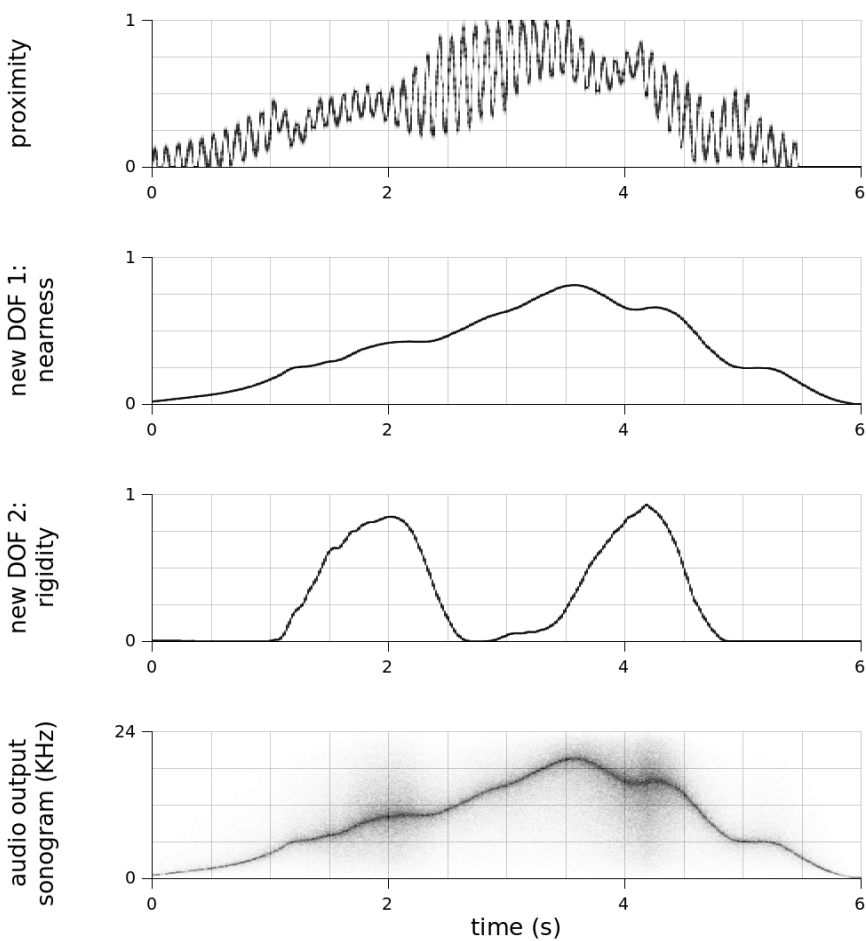

**Figure 4.7** *Simultaneous I/O of the algorithm, recorded during 6 seconds of instrumental control.* **Top**: $i_{z\,distance}$ *is shown here as "proximity": using a linear but normalized and inverted scale, where a larger value means a smaller z distance.* **Upper middle**: *the derived $s_{nearness}$ parameter, also shown using a normalized and inverted scale.* **Lower middle**: *the derived $s_{rigidity}$ parameter, shown using a normalized scale.* **Bottom**: *sonogram of the audio output, shown using a linear frequency scale.*

The above demonstrates a positive answer to the research question posed in Section 4.5.3.1: Computed touch can be used to enable instrumental control that is based on finger rigidity; and this control may occur simultaneously with control based on down- and upward fingertip movement.

Finally, the above also shows – based on both the movement over time of the fingertip, and on the described resulting perception – how the algorithm has yielded a new type of fingertip control action: fingertip tensing during force wave output. This directly enables the construction of new causal relationships between human actions and changes in heard musical sound. Thereby, it demonstrates computational liberation.

# 5. Excursion I: One-press control for pressure-sensitive computer keyboards

CHAPTER SUMMARY

This chapter presents *one-press control*, a fingertip input technique for pressure-sensitive keyboards, based on the detection and classification of pressing movements on the already held-down key. To seamlessly integrate the added control input with existing practices on ordinary computer keyboards, the redefined notion of *virtual modifier keys* is introduced.

A number of application examples are given, especially to point out a potential for simplifying existing user interactions by replacing modifier key combinations with single key presses.

Also, a new class of interaction scenarios employing the technique is proposed, based on a user interaction model named *"What You Touch Is What You Get (WYTIWYG)"*. Here, the proposed fingertip input technique is used to navigate interaction options, get full previews of potential outcomes, and then to either commit to one outcome or abort altogether – all in the course of one key press/release cycle.

The results of user testing indicate some remaining implementation issues, which application of the computed fingertip touch model may help to address. The results also indicate that effective one-press control can be learned within about a quarter of an hour of hands-on operating practice time.

De Jong S, Jillissen J, Kirkali D, De Rooij A, Schraffenberger H, Terpstra A, 2010c One-press control: A tactile input method for pressure-sensitive computer keyboards. In *Extended Abstracts CHI 2010* (ACM, April 10-15 2010, Atlanta, Georgia, USA) 4261-4266.

↑  *Requested clarification:* The burden of executing this research project was shared among the authors, with all major choices made together in an open and democratic process. The concepts and algorithms that were decided upon for publication were by the first author, who also had writing duties. Further information is in the Acknowledgments.

# 5.1 Introduction: Pressure-sensitive computer keyboard hardware

In Section 1.2.3 of Chapter 1, we compared the manual operation of computer keyboards to that of piano-type keyboards, in terms of basic similarities and differences. The differences included serial versus simultaneous keypresses as the norm during use; a compact versus a wide general layout; and different fingertip manipulation of individual keys. The latter meant that with a computer key, after overcoming the initial counterpressure of its surface, further subtleties in movement and applied pressure are not picked up or used for control – something which does happen during the press/release cycles of piano-type keys.

This relative limitation, however, does not need to persist. In [Dietz et al. 2009], a pressure-sensitive computer keyboard was introduced, based on low-cost membrane technology suitable for mass production. This device, in its look and feel exactly like an ordinary keyboard, can independently report the forces applied to those keys that are being depressed. This per-key sensing, and its continuous measurement of pressure while a key is held down, are important additional capabilities compared to the work in progress reported in [Iwasaki et al. 2009]. (There, the built-in accelerometer of a laptop was used to estimate typing pressure while striking keys on the laptop's keyboard.) Basically, the technology proposed in [Dietz et al. 2009] can be said to extend the amplitude resolution of continuous keydown pressure on computer keyboards from 1 bit to more than that.

In this chapter, we will study general opportunities for human-machine control arising where this type of pressure-sensitive computer keyboard hardware is available [De Jong et al. 2010c]. In [Dietz et al. 2009], already a number of example techniques were suggested: mapping a depressed key's force level directly to some control parameter; use as a low-resolution, multi-touch sensor for gesture recognition; and measuring the force level when striking keys, e.g. to control font size of textual output. Next, in Section 5.2, we will first present a different approach, resulting in a new class of keyboard events. Then, in Section 5.3, we will present a new user interaction model, to exploit this new class of events. This is evaluated in Section 5.4 based on user testing. Reflection in Section 5.5 then concludes the chapter.

# 5.2 A new class of keyboard events: One-press control

**5.2.1 Force peaks and virtual modifier keys**    Since the new class of keyboard events we now introduce enables the control of multiple different events during a single key press/release cycle, we call it *one-press control*. This is implemented as a software layer, placed between regular applications and the raw input from the pressure-sensitive keyboard (see Figure 5.1).

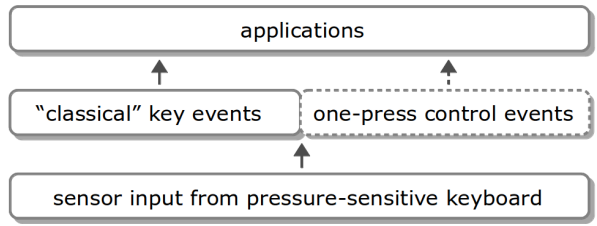

**Figure 5.1** *The proposed software layer.*

Basic functionality can be summarized as follows: "Classical" key depress and release events are transparently recreated and passed on. However, if a key is depressed and held down relatively *softly* for some timeout period (similar to that for typematic character repetition), this does *not* lead to a classical key depress event. Instead, the software will now track the relevant force sensor's input in order to look for subsequent and possibly repeated *pressing movements* on the already held-down key. Figure 5.2 shows a typical response of one force sensor, indicating a usable input range between 0.6 and 3.0 Newton. This overlaps with the 0 - 3 Newton range given in [Mizobuchi et al. 2005] as a reasonable choice (in terms of performance and comfort) for the pen-based control of user interfaces.

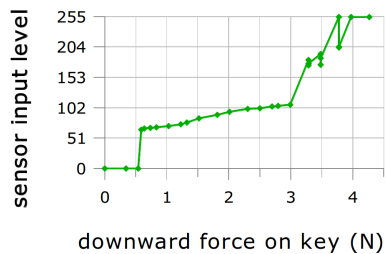

**Figure 5.2** *Typical response of one key's force sensor.*

Using the force sensor's input, pressing movements are then detected as discrete events, and labeled `[mediumRepeat]` or `[hardRepeat]` according to their intensity (see Figure 5.3). This may be compared to other research where pressing events of varying amplitude have also been extracted from touch input employing fixed flat contact surfaces. One example of this is the "Quick release" pressure technique described in [Ramos et al. 2004] and used (in the context of touchscreen mobile devices) in [Brewster and Hughes 2008]. An important difference, however, is that in the current situation the finger is not necessarily lifted from the surface after a pressing event, and detection is not triggered by this. Another example of prior research is that in [Rekimoto and Schwesig 2006], where finger pressure input on a touchpad initiates and terminates layered depress states as it crosses force thresholds. In the current approach, force thresholds are also used to classify amplitude, but pressing events are now regarded as atomic in nature, and detected based on the first-order derivative of force input.

After detection, the label assigned to an input force peak is passed on to applications by attaching it to the source key in question as a *virtual modifier key*. For example, in addition to a classical event like `[alt][F1]`, an application may now also see and respond to a `[mediumRepeat]` or `[hardRepeat][F1]`.

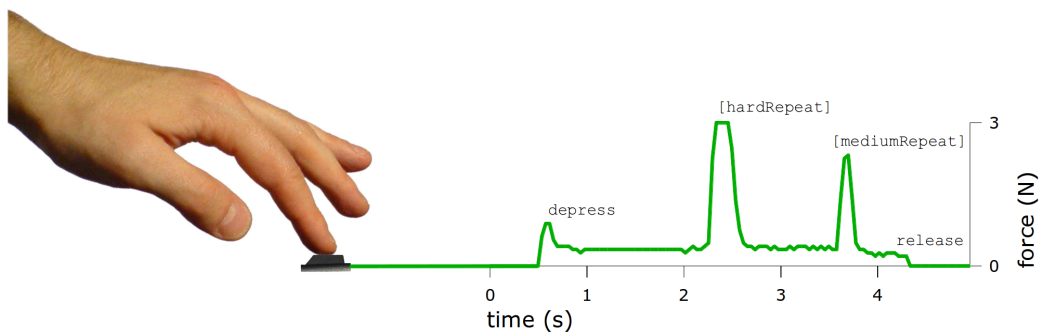

**Figure 5.3** *Force peak extraction after an initial soft depress. (Sensor data sampled after some conditioning.)*

**5.2.2 Some potential advantages**  The approach just described presents a number of potential advantages:

• Augment, rather than replace: keyboard interaction can be extended while leaving traditional mechanisms in place.

• The required skill of making pressing movements may be more familiar to keyboard users than, say, having to control steady force levels to provide input.

• While more control possibilities may become directly accessible, "bailing out" of any new-type interactions can be made as easy and intuitive as a timely key release.

• Using virtual modifier keys, developers can easily plug new-type key events into existing applications. For example, by employing a `[hardRepeat][del]` just like `[shift][del]`, to permanently delete a file; or by employing a `[hardRepeat][a]` just like `[shift][a]`, to type "A" instead of "a".

• Awkward modifier key combinations may be replaced by single key presses, simplifying interaction. For example, by using a `[hardRepeat][F4]` instead of `[alt][F4]` to close a program window; or by using a `[mediumRepeat][tab]` instead of `[alt][tab]` to switch to a next window.

# 5.3  A new user interaction model: WYTIWYG

**5.3.1 What You Touch Is What You Get**  In order to further illustrate and assess the potential applicability of one-press control, a number of demos implementing example scenarios of its use have been created. Behind each scenario is the overall goal of navigating interaction options in a user interface. This is done following a user interaction model based on two design principles:

• For touch input: Let the user navigate interaction options and express confidence in them through fingertip force.

• For visual output: Give previews of possible interaction outcomes, in a way matching actual output as closely as possible.

We named this approach *"What You Touch Is What You Get (WYTIWYG)"*, analogous to the well-known "What You See Is What You Get (WYSIWYG)" paradigm. The approach is similar in nature to the "previewable user interfaces" presented in [Rekimoto et al. 2003]. There, in the instances where preview of an option is activated by *touching* a key, and commitment by *pressing* it, fingertip force can also be said to express confidence. Furthermore, both approaches aim to replace trial-and-error interactions (based on the execution of do-undo actions) by more explorative interactions, where there is room for doubt and variable levels of confidence in the execution of available interaction options.

Still, an important difference in the current approach is its explicit emphasis on providing a preview which, while still flagged as such, in fact allows the user inspection at the same level of detail as actual output. This is intended to maximize a preview's usefulness for informing subsequent actions in exploratory processes of human-machine interaction.

**5.3.2  A typical scenario**      Since typing is the main application of computer keyboards, implemented example scenarios have focused on this aspect of keyboard input. In a typical scenario in the context of web searching, illustrated in Figure 5.4a, a hypothetically extended version of Google Suggest [Google 2010] is controlled via the `[space]` key:

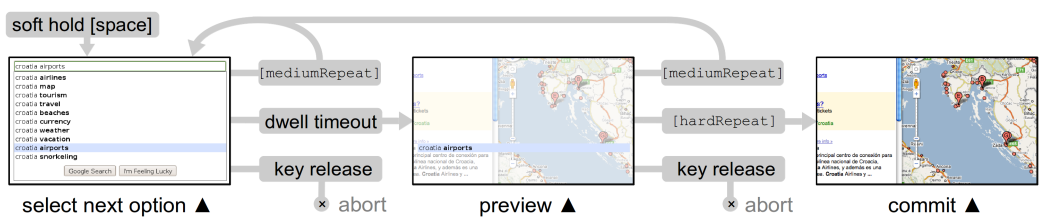

**Figure 5.4a**   *An implemented WYTIWYG example scenario: one-press control of Google Suggest using the `[space]` key.*

• While composing a search query, the user softly depresses and holds `[space]`. After a short timeout, this activates entry into a dropdown menu containing possibly relevant alternatives for text input.

• Now, repeated `[mediumRepeat]` presses cycle through the available items (also expressing a tentative confidence in them). Dwelling on an option activates a preview of the associated search results. The preview is distinguished from actual output only by having a reduced visual contrast, and by having the selected menu item still visible inside it.

• At any point, the user may "bail out" from the presented alternatives, by simply releasing `[space]` again (expressing zero confidence). Alternatively, a firm final `[hardRepeat]` (expressing maximum confidence) commits to the currently selected option.

Many more examples of this type of user interaction are imaginable. A few, also implemented as example scenarios, are illustrated in Figure 5.4b.

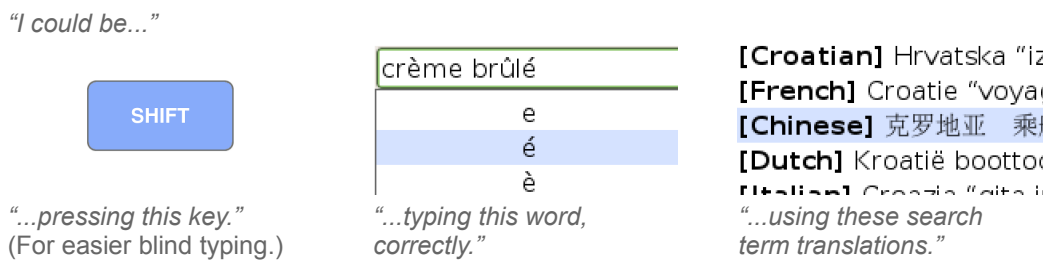

**Figure 5.4b** *More implemented WYTIWYG example scenarios, from left to right increasing in exploratory sophistication.*

## 5.4 Evaluation

In order to assess the learnability of the one-press control fingertip technique, and to identify weak spots in its implementation, an experimental evaluation was performed. This was done using the example scenario described in Section 5.3.2, as it utilizes all aspects of the technique, and places them in a realistic user interface context.

7 volunteer test subjects took part in the experiment: 3 females and 4 males, aged between 23 and 40 years old. All were new to the keyboard and input technique. None reported having any visual, manual or other impairments relevant to everyday computer use. All rated themselves between "somewhat" and "very" familiar with computer keyboard use.

**5.4.1 Procedure** Each test subject would sit at a table with a Microsoft experimental pressure-sensitive keyboard placed within easy reach. A Macbook laptop screen was used for visual output, showing a simulated browser window. First, a short introduction to the fingertip input method would be given. Then, subjects familiarized themselves with the task to be evaluated, in 4 cumulative practice stages: (1) Performing `[mediumRepeat]` presses on one specific key to navigate to arbitrary menu items (10 items total). (2) Holding the key steady when on an option, until the associated preview would appear. (3) Performing `[hardRepeat]` presses after such preview activations, to commit to the related search result. (4) Doing this for a fixed target: menu item #8.

Before each practice stage, the subject would be shown how to handle the key, and the correct visual feedback to expect. When making mistakes while practicing with the key, the test subject would receive verbal feedback indicating what was going wrong and how it could be corrected. This would continue until the test subject indicated being comfortable with the technique so far. Test subject key operating practice time was explicitly delimited, and recorded at the end of each stage.

After completing practice stage 4, test subjects were asked to execute its task 10 times in a row to the best of their ability. The first 10 subsequent attempts were then logged and automatically classified as successes or failures. Any key press/release cycle not navigating to the correct option, activating its preview, and committing by a `[hardRepeat]` would be classified as a failure. (This included, among many other things, quick aborted keypresses.)

**5.4.2 Results** The main results regarding learnability are shown in Figure 5.5. As is plotted, the total operating practice times of participants ranged from 6:30 to 19:06 min:sec. Final task scores ranged from 5 to 9, with participants scoring 7.4 perfect executions on average. All subjects rated final task easiness between "neutral" and "easy", except for the subject having both the lowest score and the longest practice time, who rated it "hard". It should be noted that this participant did log 9 perfect executions in a row. However, only 4 of these were still part of the first 10 attempts counted.

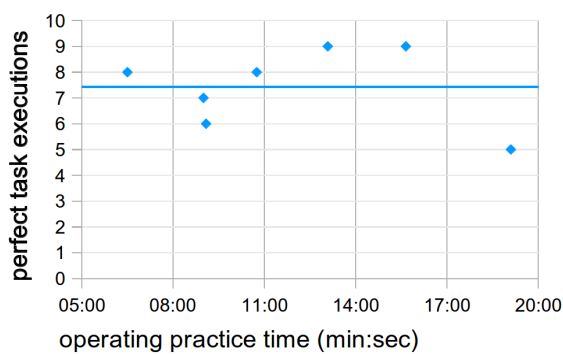

**Figure 5.5** *Test subjects' performance in terms of operating practice time (horizontal, min:sec) and perfect task executions (vertical).*

Analyzing the trial logs, it was possible to classify all failed attempts (18 out of 70) into 4 categories, which are shown in Figure 5.6. The three main causes for failure turned out to be two different types of mix-ups between `[hardRepeat]` and `[mediumRepeat]` presses, and unintended key releases. The latter seemed the dominant issue, especially since 5 test subjects noted avoiding premature key releases as one of the most difficult aspects of the final task.

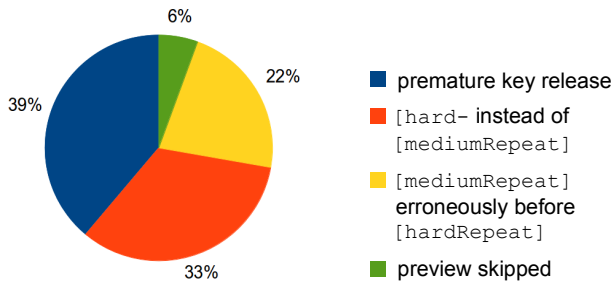

**Figure 5.6** *Classification of failed task executions.*

## 5.5 Reflection

Recapitulating, in this chapter, we have introduced one-press control, a fingertip input technique for pressure-sensitive keyboards, based on the detection and classification of pressing movements on the already held-down key. One-press control can be integrated seamlessly with existing practices on ordinary computer keyboards, by using the concept of virtual modifier keys. Existing user interactions can then be simplified, by replacing modifier key combinations with single key presses using the new fingertip input technique. New user interactions, of a more exploratory nature, can be implemented following the WYTIWYG model. Here, one-press control is used to navigate interaction options, get full previews of potential outcomes, and then to commit to one option, or abort altogether – all in the course of one key press/release cycle.

Using a specific example scenario of this type, user testing of the fingertip input technique was performed, in order to assess its learnability and to identify weak spots in its prototype implementation. The results indicate that it is possible for people to learn the technique within about a quarter of an hour of hands-on operating practice time, to then execute it with a reasonable degree of perfection. Still, a number of issues were identified as well, most prominently the occurrence of unintended key releases. Therefore, any future evaluations comparing one-press control to established input techniques should be preceded by additional work on the force event detection algorithm, to first lessen or remove this issue.

Finally, one currently inherent disadvantage of the approach is that the repeated, variable-intensity pressing movements lack specific tactile feedback confirming their execution. Here, it could help to follow the computed fingertip touch model discussed in Chapter 1, Section 1.5: Computed output to human somatosensory perception involving the fingertips may be used to construct and test suitable candidates for tactile feedback. In a future version of the keyboard technology, this tactile feedback might then be incorporated cheaply e.g. by using piezo actuators such as in [Rekimoto and Schwesig 2006], or by falling back on non-computed forms of touch, such as fixed mechanical implements.

# 6. Excursion II: The computed retinal afterimage

CHAPTER SUMMARY

In this chapter, we develop an automated computational technique for displaying 2D shapes in the retinal afterimage.

The retinal afterimage is the familiar effect in the human visual system where the ongoing perception of light is influenced by the preceding exposure to it. Historically, the retinal afterimage has been used in various techniques for the visual arts, associated with Neo-impressionism, Op art, Stan Brakhage, and the Bauhaus school.

When considering techniques for the visual arts in general, the history of computer graphics shows that incorporating stages of automated computation can offer a fundamental advantage to visual artists: a control over perceived visual complexity that is otherwise unattainable.

This motivates the question whether the retinal afterimage, too, can be induced by the output from automated processes of computation. We pursue this general question for the representative case where input should specify 2D shape.

This raises a fundamental problem: How can we ensure that shape recognition by the viewer actually is due to the retinal afterimage, and is not due to normal viewing of the stimuli, which also occurs?

First, we define a general approach using visual fixation, a rasterization method, and image sequences. For this context, we then formulate a naive but formal model of induced afterimage intensity. Then, based on the model, we develop a series of rule sets for automatically computing the image sequences that serve as stimuli.

The rule sets implement different formally defined strategies toward shape display exclusive to the retinal afterimage. In *ambiguous* rule sets, the screen intensities that are used can each lead to multiple or all of the target afterimage intensities used. In *scrambling* rule sets, visual grouping underlying shape perception is actively subverted during normal viewing. In *hybrid* rule sets, both strategies are combined.

Two rule sets representative of the approach were tested in a pilot experiment with five subjects. When using five-letter word shapes as input, the result for both rule sets was that none of the participants recognized the shapes in the separate images of the sequence used, while all of them did so in the induced afterimage effect. This seems to indicate that the automated computational technique proposed here can be used to display shape specifically in the retinal afterimage.

In the Electronic Appendix, the following is provided: video examples referred to in the text; the image sequences used in the pilot experiment; and software implementing the approach, in source format.

# 6.1 Introduction

**6.1.1 Automated computation enables human control over perceived visual complexity in the arts** Due to the historical development of computer graphics, ever more aspects of human experience that are based on visual perception can be induced by the output from automated processes of computation. This includes, for example, the apparent presence and geometry of flat, two-dimensional (2D) shapes, as well as their colors and movements. It also includes the apparent presence and spatial geometry of three-dimensional (3D) objects, and their positions and movements.

Visual artists routinely use this, for example when operating (or writing) software based on standards like OpenGL [Woo et al. 1997], e.g. to create the 2D or 3D visuals for installations or computer games. Similarly, other standards based on automated computation like Pixar's RenderMan [Upstill 1989] are routinely used to create the images for 3D animated movies.

In such routine use, the visual artist provides input specifying desired visual results to an algorithm. The algorithm then typically executes automatically on one or more electronic digital computers. Ultimately, viewers are exposed to the output from the algorithm, via additional stages of physical transduction culminating in some type of visual display technology. These final stages typically function in such a way as to ensure that given digital output will produce similarly perceived visual effects in different individual viewers.

In many other respects, the subjective experiences of the viewers may differ, of course: Different viewers may feel quite differently about some visualized object A, all agreeing, however, that it spatially appears in front of some other object B. We will here refer to aspects of visual perception of this type – induced similarly across viewers in general, and by the output from automated processes of computation – using the term "computed aspects of visual perception".

The use of computed aspects of visual perception in the visual arts then offers a fundamental advantage, apparent when considering real-life examples. For example, during most scenes of a typical, cinematically released 3D animated movie, the viewer, at any given moment, will be presented with the simultaneous visualization of many different 3D objects. Each of these will be characterized by a range of visual properties that may be regarded as individual to it. Also, each object may be seen as visually relating to the other objects that are present, in any of a number of ways. Moreover, over time, all of this is subject to change. For the experiences it induces, the movie relies on these types of visual complexity, offered perceptually to the viewer. Now suppose, that the visual artists who created the movie would have had to rely on manual techniques to keep track of all visual components, and to realize their display. (By manual techniques, we mean techniques not incorporating any stages of automated computation.) The artists simply would not have been able to create the movie: The personal labor required by unautomated processes achieving similar results would have been so time-consuming, as to make completion of the work (i.e., within the lifetimes of the artists) impossible.

This illustrates a more general point, that is not restricted to a specific aspect of visual perception, or to a specific type of artwork: Given a type of visual effect, by using techniques for its production that are partly automated, a visual artist may create works based on the effect that offer a controlled visual complexity which would otherwise be unrealizable.

Here, we will consider the question whether this fundamental advantage might also be extended to the type of visual effect that is often called the *retinal afterimage*.

**6.1.2  The retinal afterimage**    The retinal afterimage is the familiar effect in the human visual system where the ongoing perception of light is influenced by the preceding exposure to it.

Inside the eye, images of the outside world are projected onto the retina, a layer containing light-sensitive cells. Of these, the "rods" respond to night-time light levels, while the "cones" respond to daytime light levels (see e.g. [Angel 2000]). Usually, cone cells are present in three types, each responding to activity within a different frequency range of the incoming light. This trichromatic response, for the frequency ranges associated with red, green, and blue light, enables human color perception. Afferent neural connections link the rod and cone cells of the retina to the rest of the central nervous system (see e.g. [Kalat 2004]). Further processing of color signals is thought to occur according to the opponent-process model: neural connections signal the red versus green; blue versus yellow; and bright versus dark intensities of incoming light (see e.g. [Rathus 2012]). The cells involved in this continuously adapt to the light levels they are exposed to, and this underlies a range of effects in human visual perception.

One such effect is the well-known phenomenon of "peripheral fading". When looking at an unchanging scene, in which some central point is surrounded by low-contrast shapes, the latter may seem to disappear completely into the background, after staring at the central point for a while [Troxler 1804]. In fact, if there were no eye movement at all, the whole scene would vanish from perception, as the retinal cells would completely adapt to the stable image projected in such a situation, and cease to respond [Martinez-Conde et al. 2004]. This means that for normal vision, eye movement must always be present: even while staring, there will be involuntary, small saccades. (These can be observed directly using Figure 6.1, which also induces a retinal afterimage.) However, for areas where low-contrast shapes are projected onto the retina, small eye movements will not vary detected light levels much, and adaptation, and the consequent fade from perception, can still occur.

Like peripheral fading, the retinal afterimage is thought to occur because of adaptation. When staring at a red figure on a white background, the retinal cells under the influence of the figure shape will adapt to its incoming light, with cells responding to blue and green light adapting to relatively lower levels of intensity. If, after a while, vision is then suddenly directed to a completely white area, the retinal cells adapted to the figure shape will respond to light levels which have increased in the blue and green frequency ranges, while staying constant in the red frequency range. This can then result in the repeated perception of the figure shape, but in a color complementary to red. This demonstrates the classic afterimage effect, which is thought to happen according to the opponent-process model of human color perception [Hering 1964] [Hurvich and Jameson 1957] [Rathus 2012].

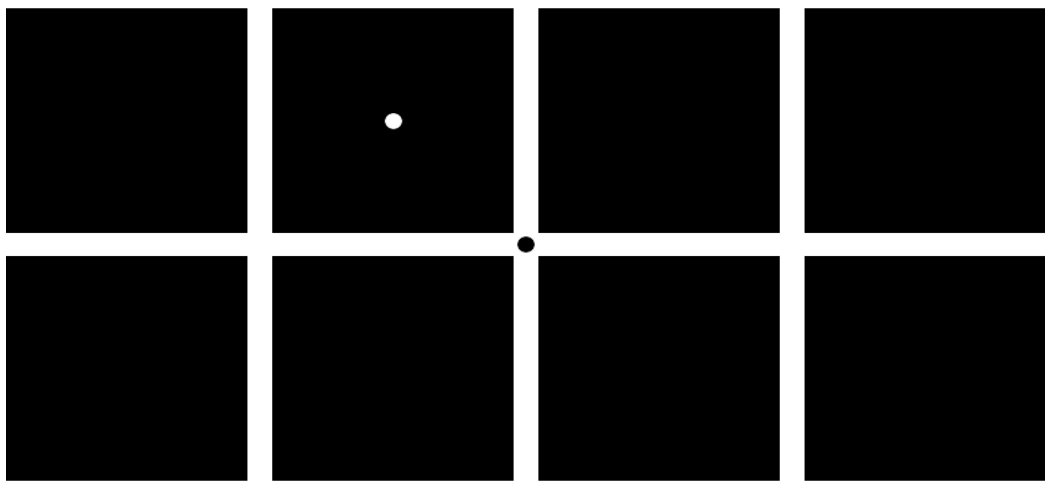

**Figure 6.1** *The retinal afterimage, and the presence of small involuntary eye movements. Please look at the black dot for one minute; then look at the white dot. Even while staring, the afterimage of the crossing lines will seem to move around. This is due to small eye movements. (Image reproduced from [Martinez-Conde et al. 2004] [Verheijen 1961].)*

**6.1.3 Use of the retinal afterimage in techniques for the visual arts** Historically, the retinal afterimage has been consciously used in various visual techniques, associated with different groups of visual artists. These groups include the artists of the *Neo-impressionism* and *Op art* movements, and visual artists part of, or influenced by, the *Bauhaus* school.

In [Chevreul 1839], translated in [Chevreul 1855], the 19th-century scientist Michel Eugène Chevreul discussed different forms of contrast between colors, and their occurrence and use in the arts. This included the retinal afterimage, discussed as "successive contrast" and "mixed contrast". Decades later, the painter Georges Seurat became the seminal figure in the movement of Neo-impressionism, and also for the related techniques of divisionism and pointillism. Seurat based his work on the color theories publicized by Chevreul and other scientists of his time [Poplawski 2003] [Gardner and Kleiner 2010]. Neo-impressionism, with the goal of maximizing the brilliance of color, rejected mixing paint on the palette, and instead relied on mixing colors during the process of viewing [Signac 1899]. This was done using the technique of divisionism: paint is applied dot by dot, with adjacent dots colored according to pairs of complementary colors. In recent teaching materials, it is assumed that when viewing painted areas containing such pairs of dots, a given color may be perceived as brighter or more intense due to the retinal afterimages induced by its complementary color [Scholastic Inc. 2008].

Starting in the second half of the 20th century, the Optical art or Op art movement [Houston and Hickey 2007] [Museum of Modern Art and Seitz 1965] also emphasized the process of viewing in art, making heavy use of illusory effects in the human visual system. This included the retinal afterimage, e.g. in works by the painters Bridget Riley [Sylvester and De Sausmarez 2012], Richard Anuszkiewicz [Madden et al. 2010], and Larry Poons [Morgan 2007]. More generally, the retinal afterimage is part of the techniques associated with Op art [Parola 1969].

Painters, in the traditional sense, have not been the only visual artists to consciously make use of the retinal afterimage. The influential experimental film maker Stan Brakhage, for example, felt at one point that afterimage colors were the only true colors [Brakhage 1967]. Brakhage produced some of his work by directly painting on the successive frames of analog film strips, then shown using a movie projector.

Josef Albers was one of the teachers at the original Bauhaus in Germany. In the course of the 20th century, his work, both as an artist and as an educator, became very influential in visual art and design (see e.g. [Chilvers 2009]). In his teaching, Albers presented the retinal afterimage as a fundamental aspect of human color perception, generally to be taken into account when using color in visual art and design [Albers 2006].

**6.1.4 The computed retinal afterimage: 2D shape**  Historically, the visual artists and movements discussed in Section 6.1.3 regarded the retinal afterimage as a significant effect, studied it, and incorporated it into the visual techniques they used. To us, this motivates the question whether techniques for producing the retinal afterimage, like those for producing other visual effects, could be partly automated. This could increase the scope of the personal labor of visual artists interested in using the retinal afterimage: By using partially automated techniques, controlled visual complexity could be arrived at in less time, both for preliminary studies and for finished works – just as is already possible for other aspects of human visual perception, e.g. those discussed in Section 6.1.1.

Like other computed aspects of visual perception, forms of the computed retinal afterimage would include a specification stage, followed by automated computation, output display, and viewer perception. A fundamental choice before implementing such stages is what aspects of the retinal afterimage to make subject to specification by the visual artist. Candidates for this include color, textural effects, and shape.

Color specification would enable the artist to automatically induce specific recognizable afterimage colors, similar to those identified by Chevreul and Albers. Textural specifications would enable the artist to automatically induce specific recognizable patterned afterimage effects across areas of display, similar to effects obtained by divisionists using dot patterns. Shape specification would enable the artist to automatically induce specific recognizable 2D shapes in the afterimage. In Op art, afterimage shape has often been used to present abstract forms, while figurative use has occurred in countless demonstrations of the retinal afterimage, from [Goethe 1795] to [Jenkins and Wiseman 2009].

Here, we will aim to implement the computed retinal afterimage based on shape specification. This because shape seems to be a basic aspect of afterimage perception, suitable for demonstrating feasibility of the computed retinal afterimage in general.

**6.1.5 The problem of shape recognition outside the afterimage**   Our current goal can be described as finding a method for automatically inducing specific recognizable 2D shapes in the retinal afterimage. However, if the output display stage of such a method will present the viewer with images, there is a problem to be aware of, clarified by Figures 6.2a and 6.2b.

Both of these figures can be used to induce a retinal afterimage. Before this is done, a pattern of bird shapes can be recognized in Figure 6.2a, and the image of a face in Figure 6.2b. When trying out Figure 6.2a, due to its peculiar symmetry, the afterimage will contain what seems like an almost identical, shifted copy of the bird shapes already seen. When trying out Figure 6.2b, the afterimage will briefly show the positive of a face, having a much less obvious likeness to its negative predecessor. This is reflected in the greater element of surprise associated with seeing the second afterimage.

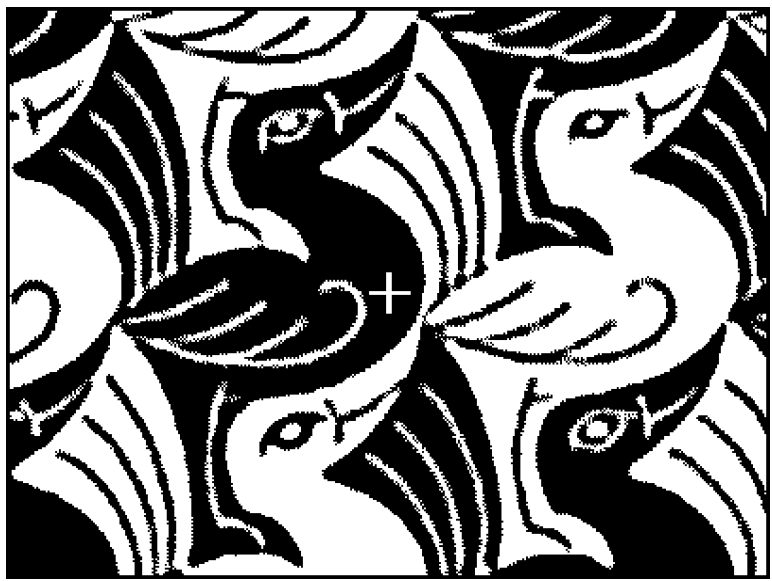

**Figure 6.2a**   *Example afterimage shapes. First, please enlarge this image, and increase its visual contrast, as much as is possible and comfortable. Then, from close by, focus on the crosshair in the middle for about one minute. After this, close your eyes. Briefly, an afterimage will appear. Subsequent blinking may bring it back, as it becomes less and less distinct. (Image: detail from a work by M.C. Escher.)*

These examples demonstrate how, to a varying extent, the contents of the retinal afterimage may also be recognized while looking in an ordinary fashion at the imagery used to induce the effect. When considering a method for automatically inducing shapes in the retinal afterimage, this poses a problem: How can we be sure that recognition of shapes by the viewer is actually due to the afterimage effect, and not due to ordinary viewing – clearly also a possibility?

Here, to avoid such false positives, we will require explicitly negative results for shape recognition outside of the afterimage effect. Furthermore, we will limit our automated approach to the greyscale case, and not use different perceived light intensities for different types of retinal color receptors. Also, we will induce the retinal afterimage using the minimum amount of images: one *bias image* for retinal adaptation, followed by one *trigger image* triggering the effect. (In Figures 6.2a and 6.2b, these images correspond to the image looked at initially, and then closing one's eyes, respectively.) Having two such greyscale images produce an afterimage showing the specified 2D shape, while this shape is not recognizable in either of the images separately, then corresponds to visualization exclusive to the retinal afterimage.

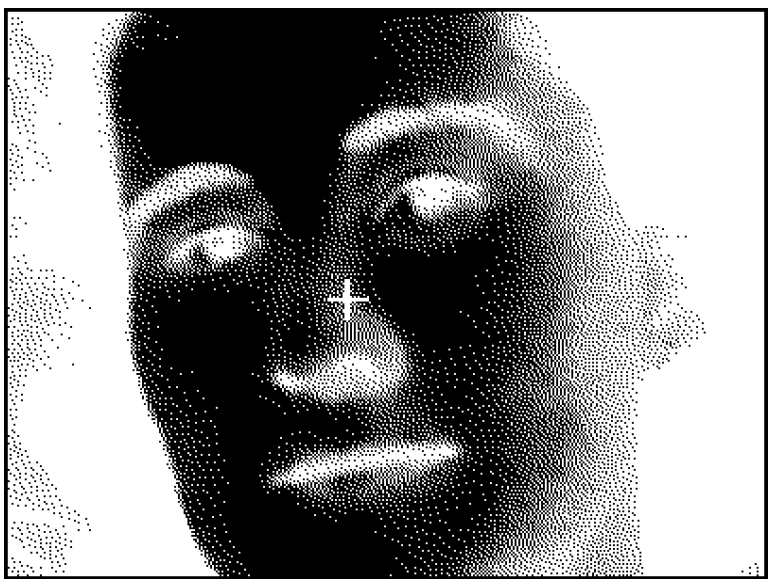

**Figure 6.2b**   *Another example afterimage shape. (See Figure 6.2a for viewing directions.)*

## 6.2  Computing shapes in the retinal afterimage

Below, the construction of a method for computing 2D shapes in the retinal afterimage will be discussed. This will be illustrated by figures in the text, and by video examples which can be found in the Electronic Appendix[1]. All development was done using the TFT LCD display of a Packard Bell R3450 laptop computer, which has a resolution of 98 dpi. This display was used during daytime (set to maximum brightness), in an indoors setting otherwise without artificial lighting. Image sequences were viewed from a distance of approximately 60 cm, chosen as a typical and comfortable viewing distance. Bias images were displayed for 20 seconds, and immediately followed by their trigger images.

---

1   The Electronic Appendix can be retrieved via *http://staas.home.xs4all.nl* .

In the Electronic Appendix, software implementing the method can also be found, allowing direct experimentation with various aspects of the method. These include input patterns, visualization parameters, and rule sets, discussed below. It may be necessary to use the software to recreate video examples, for playback on other display devices than the one used here: Differences in color gamut between various display types and technologies can be considerable, and may initially prohibit an effective reproduction of the greyscale intensities used. The solution in such cases would be to repeat construction, as it is described in the text, for the specific display in question.

**6.2.1 Controlling induced afterimage intensity: visual fixation**    We will use visual fixation in order to cause predictable and distinct retinal adaptation in the viewer. To enable visual fixation, we will use a crosshair: the type of shape already used in Figures 6.2a and 6.2b. The crosshair that will be used here is shown in Figure 6.3, and consists of two hairlines of one pixel wide, intersecting in a precisely defined area. The hairlines are 40 pixels long, and colored green in order to stand out in the greyscale imagery they will be used with. Thick black edges have been added in order to ensure that the crosshair's center will be easy to focus on, regardless of the contents of the surrounding image. Since we are using an electronic display, this surrounding image can be replaced seamlessly, so that triggering the afterimage effect no longer requires eye movement by the viewer.

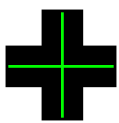

**Figure 6.3** *The crosshair for visual fixation.*

During a typical image sequence, the viewer will be asked to focus constantly on the crosshair's center, which in practice will probably mean constantly correcting for small fixational errors. This can be facilitated by supporting the head with both hands while the elbows are resting on some surface, as suggested in [Verheijen 1961].

**6.2.2 Simultaneously inducing different afterimage intensities: a rasterization method**    To display 2D shapes, the retinal afterimage will have to create the simultaneous perception of different light intensities at different locations. To do this, we will simply divide the bias and trigger images into square areas of $m \times m$ pixels. Through visual fixation, each square in a bias image will correspond to a square in the following trigger image, creating a separate sequence of greyscale intensities. We will use a total image size of 800 × 600 pixels, which allows full-size playback on many different types of displays.

If we try this out for $m = 50$ however, using a chequered black and white pattern (see Figure 6.4, to the left) followed by a similar image where black is replaced by a light grey (of 90% intensity), we find that unintended lighter and darker shades appear along the square edges in the trigger image. This can be seen in the first sequence of Video Example I.

A possible explanation is that we do not succeed in projecting the shapes in the afterimage precisely over those in the trigger image. This is not surprising when recalling the afterimages produced by Figures 6.2a and 6.2b, which were not exact inversions, but blurred variations of the original images. In addition, the already blurred pattern of retinal adaptation cannot be placed exactly and steadily over the trigger image, because of the small eye movements illustrated in Figure 6.1.

We may, however, try to alleviate the effects of this, by blurring the edges in the trigger image. The idea here is that the middle sections of squares, which do overlap, produce the desired afterimage intensities; while the borders in between those intensities, although unstable, can be made into smooth transitions, by having the light differences resulting from small eye movement be more gradual.

The contents of a pixel matrix, such as a trigger image, can be blurred by convolving them with the contents of another matrix, which specifies how each pixel's greyscale value is to be recomputed as a weighted sum of the values of itself and its neighbours. We will use $n \times n$ convolution matrices, with all elements equal to $1 / n^2$, so that in general the replacement of the value $p_{x, y}$ of a pixel at location $(x, y)$ will be the mean

$$\frac{1}{n^2} \sum_{i=0}^{n-1} \sum_{j=0}^{n-1} p_{x-\lfloor n/2 \rfloor + i, \ y-\lfloor n/2 \rfloor + j} \ .$$

(Here, pixels lying outside the pixel matrix are assigned the value of the nearest pixel inside the matrix.) For small $n$, this results in bands of $n$ - 1 pixels wide between the square areas, linearly traversing the difference in greyscale intensity. For odd $n$, the middle of such a band will be at the location of the original sharp edge (see Figure 6.4 for an example). Now, as $n$ increases from $n = 1$, the perception of unintended shades along the square edges in the trigger image decreases; around $n = 13$ it may seem largely gone, and replaced by an impression of squares of even intensity, with fuzzy borders between them. This can be seen in the second sequence of Video Example I, where the trigger image has been convolved for $n = 13$.

This sequence also shows another effect, however: squares in the afterimage may sometimes seem to join their neighbours, resulting in areas of even intensity which break the regular chequering pattern. We might have anticipated that different sequences of greyscale intensities could be used to create similar afterimage intensities, but these irregularities seem surprising in that they show that apparently similar causes – the regular structures in the bias and trigger images – do not always lead to similar results in the afterimage. In [Wade 1978], a range of potential factors influencing the complete or partial disappearance and reappearance of structured afterimages is examined. In [Lou 2001], the influence of selective attention in this is highlighted and studied, including an effect of filling-in of enclosed regions. The irregularities that can be observed in the chequering pattern might provide an example of related effects.

Still, it seems possible that these squares with fuzzy borders could be used to construct the display of 2D shapes in the retinal afterimage. For this, it would be desirable to increase shape resolution by decreasing square size $m$. A minimum for this would be $m = n$, since below this value, convolution would not leave the original greyscale intensities of trigger image squares present. Going down from $m = 50$, there initially ($m = 38$, $m = 32$, $m = 25$) is the impression of fuzzy squares as described before – imperfect, but apparently similarly so. For lower values ($m = 22$, $m = 19$, $m = 13$) the squares give an increasingly unstable impression, distorted by large diagonal patterns. The third sequence of Video Example I again shows the chequered sequence, now with square size decreased to $m = 25$ (and $n = 13$ as before).

Above, we have determined $n$ and $m$ only tentatively, and the software in the Electronic Appendix allows free experimentation with both parameters.

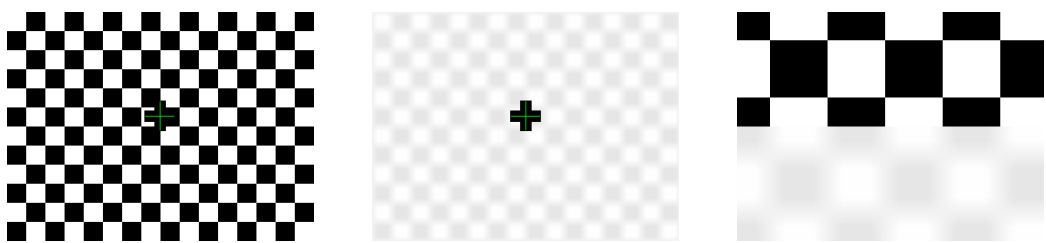

**Figure 6.4** *Example sequence. Left: middle section, surrounding the crosshair, of a bias image. Middle: the corresponding section of a trigger image. Right: a more detailed comparison, illustrating convolution of the trigger image ($n = 13$).*

**6.2.3 A naive but formal model of induced afterimage intensity**    In the previous section, the retinal afterimage of Figure 6.4 and Video Example I seemed to give an overall impression according with the discussion of the afterimage in Section 6.1.2: The light grey squares preceded by black ones appeared to light up, which can be explained by retinal adaptation to lower light levels. In order to realize a general method for 2D shape display, we would like to explore these effects of adaptation in a formal way. We will do this by making a number of naive assumptions, giving us a simple model to work with – explicitly not intended as a model for the retinal afterimage in any general sense. Factors influencing afterimage color perception which will not be explicitly taken into account, here, e.g. include post-adaptation contour alignment [Daw 1962] and the presence of induced contrast, both during and after adaptation [Anstis et al. 1978]. This will be further discussed in Section 6.4.

Before formulating assumptions, we have to make explicit a distinction between *bias* and *trigger intensity* on the one hand, and *afterimage intensity* on the other. The former will mean actual light levels produced by screen pixels, corresponding to greyscale values stored in display memory. The latter, a perceived intensity, will indicate a scale between dark and light based on the subjective impression created in the viewer.

The first naive assumption we will use, then, is that using one particular sequence of bias and trigger intensities will normally result in the perception of one particular afterimage intensity. Denoting the set of possible pixel display intensities as a finite subset $I \subset [0, 1]$ (where 0 means black and 1 means white), we model afterimage intensity as a real-valued function $f_a : I \times I \rightarrow R$ of pixel display intensities (while assuming environment light, exposure times and retinal sensitivity to be constant).

Then, if bias intensity $b$ and trigger intensity $t$ both have the same value for a sequence (as e.g. for the white squares in Figure 6.4), we denote the resulting afterimage intensity by this value, and assume that changing the bias intensity would cause it to change as well:
$$b = t \quad \Leftrightarrow \quad f_a(b, t) = t$$
From the discussion of retinal adaptation in Section 6.1.2, we would expect such a change in bias intensity to be associated with either a decrease or an increase in afterimage intensity (as e.g. for the light grey squares in Figure 6.4):
$$b > t \quad \Leftrightarrow \quad f_a(b, t) < t$$
$$b < t \quad \Leftrightarrow \quad f_a(b, t) > t$$
Our final assumption (which will be exemplified by Figure 6.5 in Section 6.2.4.1) is that one particular trigger intensity will give a darker impression in the afterimage if and only if there has been retinal adaptation to a lighter bias intensity:
$$b_1 > b_2 \quad \Leftrightarrow \quad f_a(b_1, t) < f_a(b_2, t) \tag{1}$$

**6.2.4 Visualization exclusive to the retinal afterimage: rule sets**  Having made these assumptions, suppose now we would want to produce afterimages using some set $A = \{a_1, a_2, \dots, a_n\} \subset R$ of $n > 1$ afterimage intensities, with $a_1 < a_2 < \dots < a_n$. There would probably be various ways to arrive at such a set, but in any case, we would have to select pairs of bias and trigger intensities with which to produce it. We define an afterimage *rule set* for producing $A$ as a partial function $f_r : I \times I \rightarrow A$ which is surjective, so that it produces all of $A$:
$$\forall a \in A \quad \exists\, b, t \in I \quad : \quad f_r(b, t) = a$$
and which respects the afterimage function $f_a$:
$$\forall a \in A \quad : \quad f_r(b, t) = a \quad \Rightarrow \quad f_a(b, t) = a \tag{2}$$
Now, given a rule set and a particular pattern of afterimage intensities to be produced, we can determine the contents for corresponding bias and trigger images by applying the rule set to the pattern. In the next sections, we will define various types of rule sets (accompanied by concrete examples), with properties in favor of shape display exclusive to the afterimage. For this, it is important to mention first that rule sets will be used non-deterministically: If more than one $(b, t)$ pair may produce a given $a = f_r(b, t)$, one of them will be chosen at random.

**6.2.4.1** *Ambiguous rule sets*    First, we will focus on the subgoal that a pattern of afterimage intensities should not be recognizable in the trigger image used to produce it. A certain way to achieve this would be to use a rule set in which any trigger intensity can lead to any afterimage intensity: We could then freely choose our pattern of trigger intensities, regardless of the afterimage that is to be produced. We say a rule set $f_r$ is *trigger-ambiguous* if:

$$f_r\,(b,\,t) = a \quad \Rightarrow \quad \forall a' \in A \quad \exists\,b' \in I \;:\; f_r\,(b',\,t) = a'$$

The simplest examples of such rule sets would use a single trigger intensity to produce the minimum two afterimage intensities. Consider for example the rule set $f_1$, shown in the diagram of Figure 6.5. (In the diagram, **B** and **T** indicate the subsets of **I** containing $f_1$'s bias and trigger intensities, respectively. These subsets are formally defined below.) Rule set $f_1$ is based on a dark grey trigger intensity, which is either darkened by a preceding white intensity, giving $a_1 = f_1$ (1, 0.25), or lit up by a preceding black intensity, giving $a_2 = f_1$ (0, 0.25). This will not allow a choice in trigger intensity, but the ambiguity requirement still guarantees that shapes in the afterimage will not be recognizable in the always evenly grey trigger images.

This is illustrated by the image sequence depicted in Figure 6.5, and demonstrated in Video Example II, which is the result of applying $f_1$ to an example afterimage pattern consisting of a symbol and a simple regular pattern. This sequence also shows how, by generating negative bias images and neutral trigger images, rule set $f_1$ implements the classic afterimage effect.

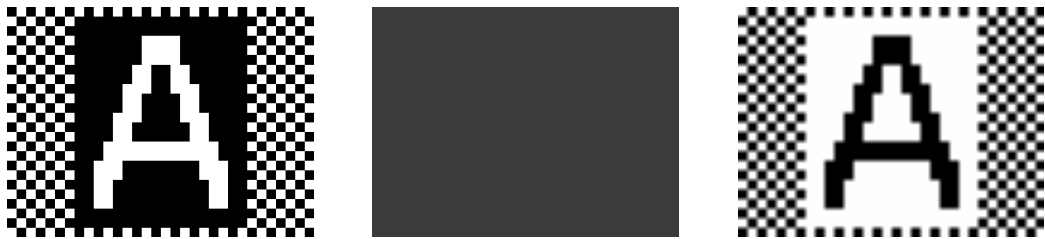

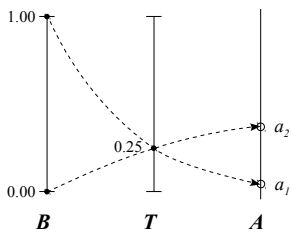

**Figure 6.5** *Trigger ambiguity. Left: diagram of rule set $f_1$, with arrows linking the bias and trigger intensities that are used to their target afterimage intensities. Above, left: the bias image from an image sequence generated by rule set $f_1$. Above, middle: the evenly grey trigger image. Above, right: the target afterimage pattern, with $a_1$ shown as black, and $a_2$ shown as white.*

Since we would like shapes in the afterimage to go unrecognized in the bias image also, we introduce a second concept, analogous to trigger ambiguity. We say a rule set $f_r$ is *bias-ambiguous* if:

$$f_r\,(b,\,t) = a \quad \Rightarrow \quad \forall a' \in A \quad \exists\,t' \in I \;:\; f_r\,(b,\,t') = a'$$

Our example for this type of rule set will introduce the simultaneous use of different bias-trigger intensity pairs to produce the same afterimage intensity. Starting point here will be the fact that black and white bias intensities can lead to two distinct afterimage intensities by simply remaining constant. This corresponds to a rule set $f_2$, shown in the diagram of Figure 6.6, of which the first half is given by $a_1 = f_2(0, 0)$ and $a_2 = f_2(1, 1)$. To satisfy the ambiguity requirement, the second half will then have to allow each bias intensity to lead to both of the afterimage intensities.

In the case of black, for example, we have to find a $t$ for which $a_2 = f_2(0, t)$. This can be done by varying $t$ and judging how well the combination of $f_2(0, t)$ and $f_2(1, 1)$, applied to an all-$a_2$ afterimage pattern, succeeds in creating an even impression in the afterimage. Three ranges will then be distinguishable for $t$: a high range where $f_2(0, t)$ will seem lighter than $f_2(1, 1)$; a low range where the opposite is true; and a range inbetween, where the afterimage intensities will seem similar, but still may flicker or otherwise remain visually separate. The most stable results in this middle range seemed to be around $t = 0.87$, so that we will use $a_2 \approx f_2(0, 0.87)$.

For the white case, a similar procedure can be followed, using an all-$a_1$ afterimage pattern. This resulted in an $a_1 \approx f_2(1, 0.15)$.

Applying our now-complete rule set $f_2$ to the same afterimage pattern that we have used before has resulted in the image sequence shown in Figure 6.6, and demonstrated in Video Example III. Here, for each instance of an afterimage intensity, the choice between applicable bias-trigger intensity pairs has been arbitrary, so that the bias image has become a random pattern of black-and-white, which is reflected in slight distortions in the trigger image.

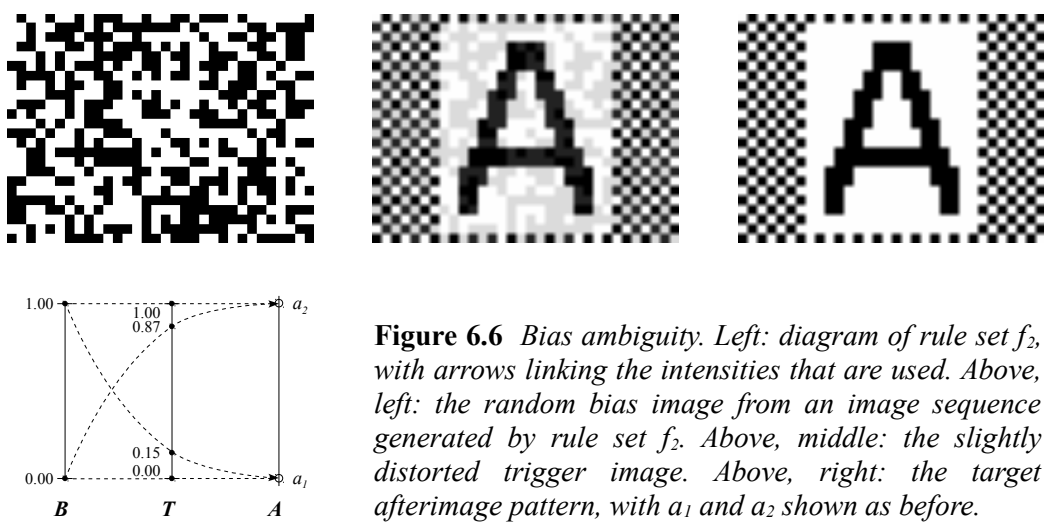

**Figure 6.6** *Bias ambiguity. Left: diagram of rule set $f_2$, with arrows linking the intensities that are used. Above, left: the random bias image from an image sequence generated by rule set $f_2$. Above, middle: the slightly distorted trigger image. Above, right: the target afterimage pattern, with $a_1$ and $a_2$ shown as before.*

Both of the previous examples have shown in practice how an ambiguity requirement guarantees the unrecognizability of afterimage shapes, in either the bias image or the trigger image. Neither example did however conceal the afterimage pattern in the image not subject to ambiguity. We would like to have a rule set $f_r$ which is *fully ambiguous:* both bias-ambiguous and trigger-ambiguous. Unfortunately, such a rule set cannot exist, which we can prove within our framework.

**Proof.** Suppose a rule set $f_r$ is bias-ambiguous.

Define the sets $\boldsymbol{B}$ and $\boldsymbol{T}$ of $f_r$'s bias and trigger intensities:

$\boldsymbol{B} = \{\, b \mid b \in \boldsymbol{I} \wedge (\exists\, t \in \boldsymbol{I}, a \in \boldsymbol{A} : f_r\,(b, t) = a)\,\}$

$\boldsymbol{T} = \{\, t \mid t \in \boldsymbol{I} \wedge (\exists\, b \in \boldsymbol{I}, a \in \boldsymbol{A} : f_r\,(b, t) = a)\,\}$

Then choose

$b_{max} \in \boldsymbol{B}$ so that $\forall b \in \boldsymbol{B} : b \leq b_{max}$

$a_{max} \in \boldsymbol{A}$ so that $\forall a \in \boldsymbol{A} : a \leq a_{max}$

$a_{min} \in \boldsymbol{A}$ so that $\forall a \in \boldsymbol{A} : a \geq a_{min}$.

There must exist a $t' \in \boldsymbol{T}$ with $f_r\,(b_{max}, t') = a_{max}$ (bias ambiguity).

However, for this $t'$, trigger ambiguity will not hold, because

$\neg\, \exists\, b' \in \boldsymbol{B} : f_r\,(b', t') = a_{min}$.

> *Proof.*  Suppose $\exists\, b' \in \boldsymbol{B} : f_r\,(b', t') = a_{min}$.
> Then $f_a\,(b', t') < f_a\,(b_{max}, t')$ by Prop. (2),
> since $n > 1$ guarantees $a_{min} < a_{max}$.
> It then follows by Prop. (1) that $b' > b_{max}$,
> which is impossible by definition.

Therefore $f_r$ cannot be both bias- and trigger-ambiguous. ∎

This does not have to mean that ambiguity is completely useless to our purposes however: we may still realize a decrease in recognizability in both the bias and the trigger image by informally relaxing requirements to the level of a *partial ambiguity*, where it suffices that each bias or trigger intensity can lead to more than one of the afterimage intensities.

This is demonstrated by rule set $f_3$, which again will use black and white as its bias intensities, but this time to produce three afterimage intensities. The construction of $f_3$, shown in the diagram of Figure 6.7, starts by searching for a $t_1$ and $t_2$ for which $f_a\,(0, t_1)$ $= f_a\,(1, t_2)$, and using the result as the middle afterimage intensity $a_2$. This may be done starting out from a medium grey trigger intensity, as in $t_1 = 0.5 - d$ and $t_2 = 0.5 + d$, with $d$ increasing from 0. Varying $d$, we can then search for matching afterimage intensities in a process similar to that described for rule set $f_2$. This seemed to give the most stable impression around $d = 0.13$, so that $a_2 \approx f_3\,(0, 0.37)$ and $a_2 \approx f_3\,(1, 0.63)$.

We quickly obtain additional brighter and darker afterimage intensities, in a way that satisfies partial ambiguity, by choosing $a_3 = f_3\,(0, 0.63)$ and $a_1 = f_3\,(1, 0.37)$.

The rule set defined above will create image sequences in which the parts of the target afterimage pattern using $a_1$ or $a_3$ will be unambiguously repeated in the bias and trigger images, using separate greyscale intensities. However, the recognition of parts will be hampered, because the remaining image space, which leads to $a_2$, uses the same greyscale values randomly. This is demonstrated in Video Example IV and illustrated in Figure 6.7, using the example target afterimage pattern, which this time has been adapted to use three intensities.

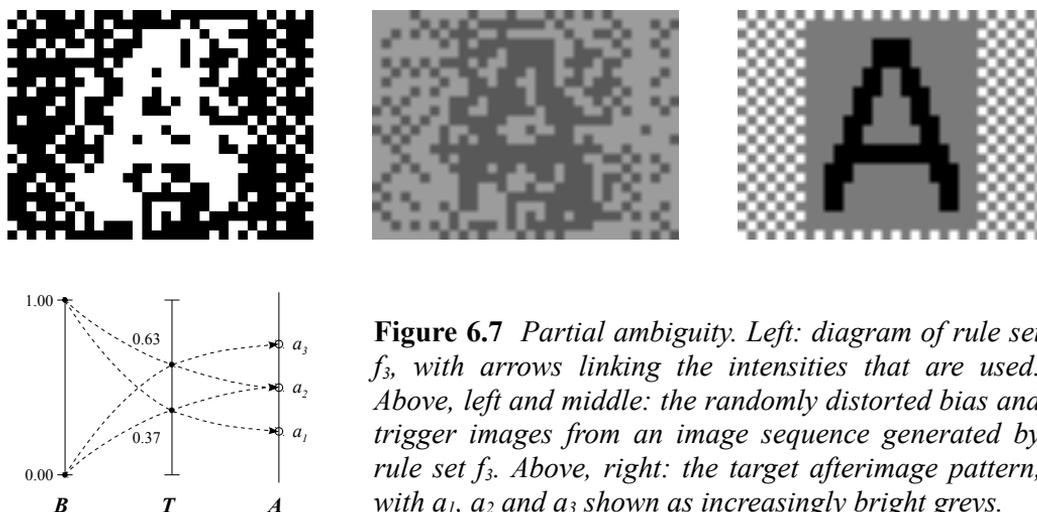

**Figure 6.7** *Partial ambiguity. Left: diagram of rule set $f_3$, with arrows linking the intensities that are used. Above, left and middle: the randomly distorted bias and trigger images from an image sequence generated by rule set $f_3$. Above, right: the target afterimage pattern, with $a_1$, $a_2$ and $a_3$ shown as increasingly bright greys.*

**6.2.4.2** *Scrambling rule sets*    A basic property of human visual perception is that adjacent areas of a similar shade tend to be grouped together and perceived as a shape. We will now introduce another approach to defining rule sets, using this tendency in a subversive manner. As before, the goal is to have shapes recognized in the afterimage not be recognized by normal viewing of the image sequence producing the afterimage.

First, we will need a tool to look at how rule sets reorder intensities, when comparing the bias images they generate to the target afterimage patterns. Given a rule set $f_r$, we can enumerate the set $\boldsymbol{B}$ of its bias intensities (defined in Section 6.2.4.1) according to $b_1 < ... < b_{|\boldsymbol{B}|}$ – just as we have done for the set $\boldsymbol{A}$ of afterimage intensities from the outset. We then define $f_r$'s *mapping scheme* as a tuple of $|\boldsymbol{B}|$ subsets from the set $\{1, ... , |\boldsymbol{A}|\}$, where the $i$-th subset consists of all $j$ for which $\exists\, t \in \boldsymbol{I} : f_r\,(b_i, t) = a_j$.

This means that, reading a mapping scheme from left to right, we find for each bias intensity, from dark to light, the ranks of the afterimage intensities to which it is linked. For example, the mapping scheme for rule set $f_1$ is given by $(\{2\}, \{1\})$, meaning that firstly, its darkest bias intensity is mapped to its lightest afterimage intensity; and that secondly, its lightest bias intensity is mapped to its darkest afterimage intensity (see the diagram of Figure 6.5). As another example, the mapping scheme for rule set $f_2$ is $(\{1, 2\}, \{1, 2\})$; just as it will be for any other bias-ambiguous rule set which uses two bias and two afterimage intensities (see the diagram of Figure 6.6).

Now, suppose we have a target afterimage pattern displaying a shape on a background, with each in one intensity. Both will be repeated in the bias image, in bias intensities according to the rule set used. Arbitrarily choosing one of the bias intensities present for the shape, suppose that it is nearer in brightness to one of the bias intensities present for the background than to all other intensities also present for the shape. Where adjacent, these two bias intensities will tend to visually group together, distorting the perception of the original shape. A rule set $f_r$ will have this property for all possible combinations of uniformly tinted shapes if:

$$f_r(b, t) = a \quad \Rightarrow \quad \forall a' \in A, a' \neq a : \exists b', t' \in I :$$
$$[\ f_r(b', t') = a' \ \wedge \ \forall b'', t'' \in I, b'' \neq b :$$
$$f_r(b'', t'') = a \quad \Rightarrow \quad |\ b - b'\ | < |\ b - b''\ |\ ] \quad (3)$$

However, given a rule set $f_r$, it may be more intuitive to look at its mapping scheme: if in it, multiple occurrences of the same afterimage intensity are always separated by occurrences of all other afterimage intensities lying strictly inbetween, the above property will hold. Consider for example the mapping schemes ({1}, {2}, {1}) and ({1}, {2}, {1}, {2}). (For an example illustrating the use of ({1}, {2}, {1}, {2}), please preview how the rule set shown in the diagram of Figure 6.9 links its trigger intensities to its afterimage intensities.) Both of the above mapping schemes correspond to rule sets satisfying the property, and both have a scrambling effect on the shapes of the afterimage, as is illustrated in Figure 6.8. However, the effect of the first mapping scheme is crippled by the fact that it has afterimage intensity $a_2$ preceded by bias intensity $b_2$ only, which greatly aids shape recognition. We still need to make explicit that a rule set $f_r$ should produce each afterimage intensity using multiple bias intensities:

$$f_r(b, t) = a \quad \Rightarrow \quad \exists b', t' \in I, b' \neq b : f_r(b', t') = a \quad (4)$$

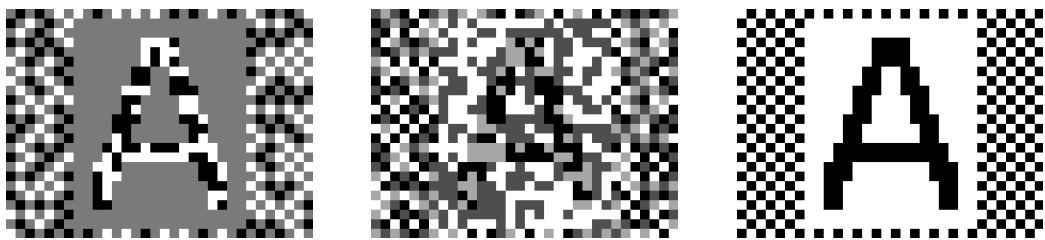

**Figure 6.8** *Scrambling effects. Right: the target afterimage pattern. Left: a corresponding bias image using the mapping scheme ({1}, {2}, {1}). Middle: a stronger result using mapping scheme ({1}, {2}, {1}, {2}).*

Now, we say a rule set $f_r$ is *bias-scrambling* if it has both properties defined in Props. (3) and (4). Also, by repeating the argument for trigger images instead of bias images, we get analogous definitions for trigger intensity mapping schemes and *trigger-scrambling* rule sets.

Example rule set $f_4$ will be bias-scrambling as well as trigger-scrambling, using four bias and four trigger intensities. Its construction starts by combining the outer bias intensities, black and white, with the inner trigger intensities, two greys both near medium grey, but still separately distinguishable (this is shown in the diagram of Figure 6.9). More precisely, we choose $a_1 = f_4 (1, 0.52)$ and $a_2 = f_4 (0, 0.48)$, so that each trigger intensity will produce an afterimage intensity on the opposite side of an intermediate grey. We then finish construction by looking for pairs of identical bias and trigger intensities which also produce $a_1$ and $a_2$. This can be done in a process similar to that described for rule set $f_2$ in Section 6.2.4.1, using all-$a_1$ and all-$a_2$ afterimage patterns. The pairs producing the most stable impressions seemed to be $a_1 \approx f_4$ (0.39, 0.39) and $a_2 \approx f_4$ (0.62, 0.62).

This means that both 0.39 and 0.62 will play double roles in $f_4$: on the one hand, as the outer trigger intensities, resulting in the trigger intensity mapping scheme ({1}, {2}, {1}, {2}); and on the other hand, as the inner bias intensities, resulting in the bias intensity mapping scheme ({2}, {1}, {2}, {1}).

The result of this is demonstrated in Video Example V, of which the image sequence is shown in Figure 6.9. We can see that the trigger image has a weakened scrambling effect, due to the relative closeness of its middle two intensities: an outer trigger intensity will visually group almost equally well with both of these intensities. Increasing the difference in brightness between the middle trigger intensities involves a trade-off however, as it means decreasing the difference between afterimage intensities, thus reducing contrast of the afterimage.

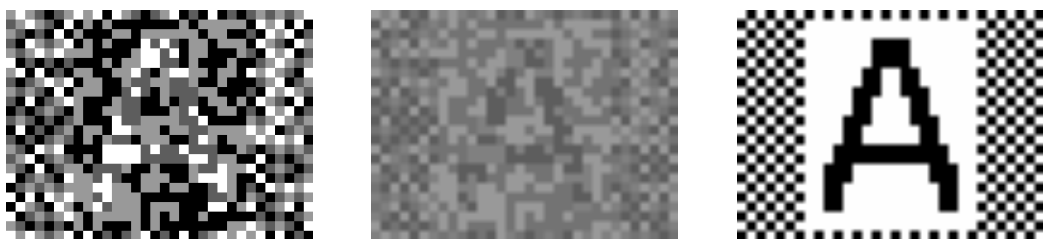

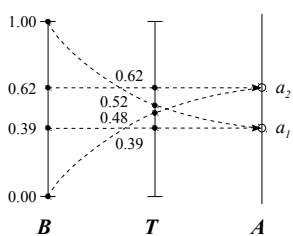

**Figure 6.9** *Bias-scrambling, trigger-scrambling. Left: diagram of rule set $f_4$, with arrows linking the intensities that are used. Above, left and middle: the scrambled bias and trigger images from an image sequence generated by rule set $f_4$. Above, right: the target afterimage pattern, with $a_1$ shown as black and $a_2$ shown as white.*

**6.2.4.3** *Hybrid rule sets* When comparing the ambiguity requirement to the scrambling requirement, it is obvious that the former is stronger in that it really guarantees unrecognizability of the afterimage shape, although this is limited to either the bias image or the trigger image; while the latter, although we have seen it used in both images simultaneously, cannot rule out recognizability. It therefore becomes worthwhile to look at ways of combining both approaches in a single rule set, and we will now discuss two examples of this.

Our first example, rule set $f_5$, will be bias-scrambling and trigger-ambiguous, using four bias and two trigger intensities (shown in the diagram of Figure 6.10). It was found that, when starting out similarly to rule set $f_4$ of Section 6.2.4.2, but with the trigger intensities near medium grey somewhat further apart, as in $a_1 = f_5 (1, 0.57)$, $a_2 = f_5 (0, 0.43)$, matching pairs of identical bias and trigger intensities were given by $a_1 \approx f_5 (0.43, 0.43)$ and $a_2 \approx f_5 (0.57, 0.57)$. This results in a rule set with trigger ambiguity and a bias intensity mapping scheme given by ({2}, {1}, {2}, {1}), just as in rule set $f_4$. An image sequence created by this rule set is demonstrated in Video Example VI, and depicted in Figure 6.10.

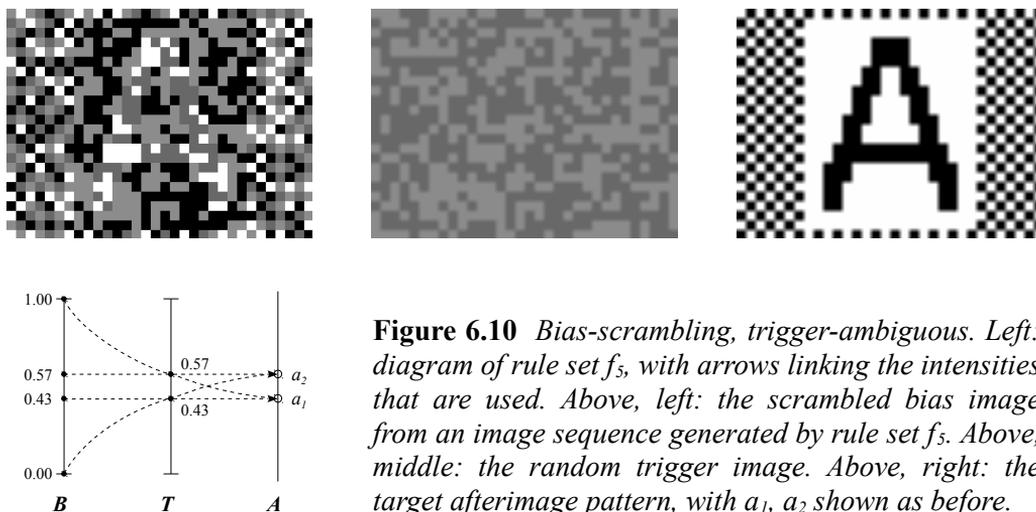

**Figure 6.10** *Bias-scrambling, trigger-ambiguous. Left: diagram of rule set $f_5$, with arrows linking the intensities that are used. Above, left: the scrambled bias image from an image sequence generated by rule set $f_5$. Above, middle: the random trigger image. Above, right: the target afterimage pattern, with $a_1$, $a_2$ shown as before.*

As it turns out, in the afterimages produced by rule set $f_5$, shape and background are hard to distinguish, seeming of an almost even grey. The afterimage intensities $a_1$ and $a_2$ apparently lie too close to eachother to get a good contrast. Instead of trying to improve the current rule set, we will move on to the next, however: It will swap the ambiguity and scrambling requirements, while making sure from the outset that we will get an afterimage contrast similar to that of rule set $f_4$.

Example rule set $f_6$ will be bias-ambiguous and trigger-scrambling, using two bias and four trigger intensities. The first part of its construction is the same as for rule set $f_4$, choosing $a_1 = f_6 (1, 0.52)$ and $a_2 = f_6 (0, 0.48)$ (shown in the diagram of Figure 6.11). We continue by searching for a $t_1$ with $a_1 \approx f_6 (0, t_1)$ and a $t_4$ with $a_2 \approx f_6 (1, t_4)$. Here, testing with all-$a_1$ and all-$a_2$ patterns seemed to give the most stable results for $a_1 \approx f_6 (0, 0.25)$ and for $a_2 \approx f_6 (1, 0.74)$. This results in a bias-ambiguous rule set, with its trigger intensity mapping scheme given by ({1}, {2}, {1}, {2}), again as in rule set $f_4$. Figure 6.11 shows an image sequence generated by the rule set, which is demonstrated in Video Example VII.

Although the afterimage intensities of rule set $f_6$ give a better contrast than those of $f_5$, its scrambling suffers from the same weakening effect mentioned for rule set $f_4$, only more so, as the outer trigger intensities now lie further apart. The scrambling effect created by rule set $f_5$ will probably be better, as its middle bias intensities lie relatively further apart.

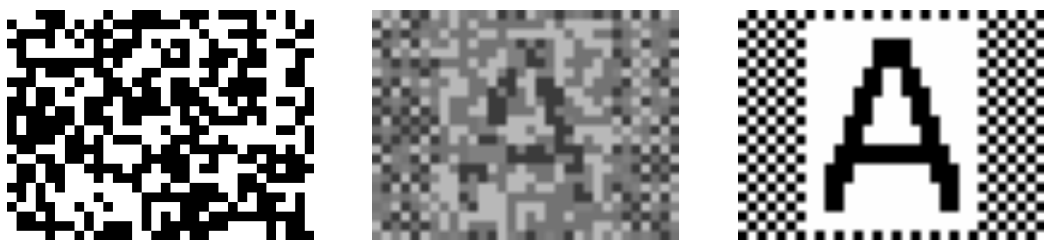

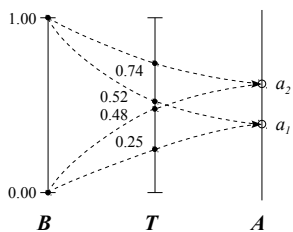

**Figure 6.11** *Bias-ambiguous, trigger-scrambling. Left: diagram of rule set $f_6$, with arrows linking the intensities that are used. Above, left: the random bias image from an image sequence generated by rule set $f_6$. Above, middle: the scrambled trigger image. Above, right: the target afterimage pattern, with $a_1$, $a_2$ shown as before.*

**6.2.4.4** *Multi-trigger sequences*    Bias-ambiguous rule sets have the property that, once a random bias image is chosen, it may still lead to any possible afterimage pattern, depending on the contents of the trigger image. Using a rule set which is both bias-ambiguous and trigger-scrambling, this can be tried out in practice: Unlike in the similar case for trigger ambiguity, it may be possible to combine a strong bias impression with multiple different trigger images, thereby inducing multiple different afterimage patterns in a single run.

In Video Example VIII, this idea has been implemented by replacing a single trigger image with two images shown in quick succession. The two corresponding target afterimage patterns show the words "hello" and "world", respectively. (See Figure 6.12.) Here, bias duration has been extended from 20 to 30 seconds, to improve contrast in the second afterimage. Extending duration further seemed to lead to an uncomfortable viewing experience with unstable afterimages. Choosing the duration of the first trigger image also involved a trade-off, between smaller values giving little time for recognition of the first afterimage, and greater values deteriorating the quality of the second. This resulted in a tentative choice for 1.5 seconds. As before, the software in the Electronic Appendix allows free experimentation.

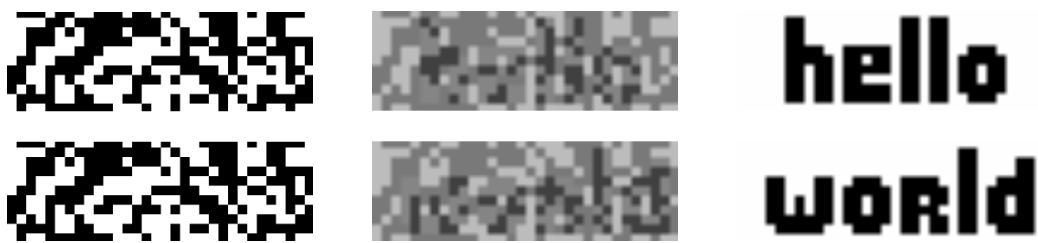

**Figure 6.12** *Multiple trigger images for a "hello world" example. A sequence generated by rule set $f_6$. Left: the middle section of the bias image (shown twice). Middle: the corresponding sections of the first and second trigger images. Right: the target afterimage patterns ($a_1$, $a_2$ as before).*

# 6.3  Evaluation: pilot experiment

**6.3.1  Overview**    We tested the approach described in Section 6.2, in order to verify that it indeed can be used to automatically display specified 2D shapes in the retinal afterimage. This included verifying that ordinary viewing of the images used does not result in recognition of the specified shapes. To this end, a pilot experiment was conducted, with its participants naive to the visualization method and without prior knowledge of the target afterimage patterns used.

In the experiment, the parameters for visual fixation and rasterization were set to the tentative defaults identified in Sections 6.2.1 and 6.2.2. Then, to determine representative rule sets for use in the experiment, first trigger-ambiguous rule set $f_1$ and bias-ambiguous rule set $f_2$ (see Section 6.2.4.1) were dropped from consideration, since both trivially allow shape recognition in the image not subject to ambiguity. Of the remaining rule sets, bias-ambiguous, trigger-scrambling rule set $f_6$ (see Section 6.2.4.3) seemed a better candidate than bias-scrambling, trigger-scrambling rule set $f_4$, since $f_6$'s bias ambiguity guarantees unrecognizability, where $f_4$'s bias scrambling does not. Rule set $f_6$ also seemed a better candidate than bias-scrambling, trigger-ambiguous rule set $f_5$, since $f_6$ was designed to give a stronger afterimage contrast than $f_5$. Rule set $f_6$ did not seem a necessarily better candidate than the remaining, partially ambiguous rule set $f_3$ (see Section 6.2.4.1), however. Therefore, both $f_6$ and $f_3$ were chosen to represent the approach in the experiment.

Regarding target afterimage patterns, a choice was made to use word shapes, on an even background, as quickly recognizable everyday shapes, presented without distraction. The 2D word shapes were then placed centered, near the crosshair. This reflected the fact that in regular vision, the recognition of letter strings as words happens more quickly when the letters are placed in the center, rather than in the periphery, of visual fixation [Lee et al. 2003].

When using a line thickness of 1 square unit (see Section 6.2.2), it turned out that individual letters in the afterimage sometimes would appear so distorted as to be unreadable. For this reason, mostly, a line thickness of 2 square units was used.

In order to have naive test subjects get used to the visualization method, an introductory sequence, serving as a dry-run, was included. Preliminary testing of rule sets $f_3$ and $f_6$ using two image sequences based on 3-letter word shapes showed improving afterimage recognition results when extended with two sequences based on 5-letter word shapes. This resulted in the choice to use two sequences based on 5-letter word shapes, preceded by two sequences based on 3-letter word shapes for the pilot experiment.

**6.3.2  Procedure**    The experiment was performed separately with 5 unpaid volunteers between 20 and 24 years old. All test subjects were new to the visualization method. Three of them were female, with two having normal vision, and one being somewhat far-sighted. Of the two males, one had normal vision, and one was near-sighted (with correction). None of the participants was dyslexic, and all were both willing and able to fully concentrate on the images and image sequences presented to them.

Conditions for the experiment were as described in Section 6.2: Image sequences were viewed on the TFT LCD display of a Packard Bell R3450 laptop computer (resolution 98 dpi), from a distance of approximately 60 cm. The experiment was performed during daytime, in an indoors setting, with indirect daylight and without artificial lighting. When images were shown separately, they were displayed for 20 seconds; when they were shown in sequence, the bias image was displayed for 20 seconds and immediately followed by the trigger image, displayed for 5 seconds. The specific images and image sequences used in the experiment can be found and played back in the Electronic Appendix, which is accessible using any standard web browser.

The experiment started with the chequered, third sequence of Video Example I (see also Figure 6.4), as a dry-run to get used to the visualization method. After test subjects were seated in front of the screen, they were shown the bias image, then asked whether they had recognized a pattern in it, and if so, what pattern. This was repeated for the trigger image. The test subjects were then told they would now see the first image switch to the second, and were asked to concentrate on the moment of the switch. Instructions were given to constantly focus on the crosshair's center throughout the sequence, while resting one's head on both hands, with the elbows placed on armrests. After the sequence had completed, the same question asked for the separate images was asked again. The attention of the test subjects was then pointed to how the light grey fuzzy squares seemed to light up in the sequence, in which they were preceded by black squares. The sequence was then repeated a few times until the subjects indicated being accustomed to this effect.

After this, four sequences were presented using the same procedure, only now the subjects wrote down their answers in a form, and did not receive feedback anymore. They could however indicate uncertainty about their response, in which case a sequence was repeated until a final answer was settled upon. The first sequence to be presented was generated by rule set $f_3$, based on a 3-letter pattern spelling "red". The second sequence was generated by rule set $f_6$, based on a 3-letter pattern spelling "low". After these were completed, the participants had the opportunity to write down any general remarks, and there was a short break. Then, the third sequence was shown, generated by rule set $f_3$, based on a 5-letter pattern spelling "light". This was followed by the fourth sequence, generated by rule set $f_6$, based on a 5-letter pattern spelling "hello". Finally, there was another opportunity to write down general remarks.

**6.3.3 Results**     For all test subjects and all image sequences, none of the target afterimage patterns were ever recognized in the separate images. For the scrambled images, no responses came near word shape recognition, apart from one remark for the fourth sequence about "dark letter-like shapes in the center". For the partially ambiguous images, what came nearest recognition were remarks, sometimes made for the third sequence, about the separate images having a distinctly brighter or darker region in the center. For the ambiguous (i.e. random) images, the responses, although sometimes richly imaginative (e.g. "a hurricane in some kind of frozen circular motion") most often were a simple "no".

As expected from preliminary testing, the results for shape recognition in the afterimage were mixed for the 3-letter sequences. Participants often expressed doubt, and often needed multiple iterations to settle on their answers. In the first sequence, "red" was recognized 3 out of 5 times, and in the second sequence, "low" was recognized 2 out of 5 times.

For the 5-letter sequences, results improved considerably. All test subjects recognized the target words, in both sequences, mostly immediately: In the third sequence, "light" was once reported after four runs (at approximately 50 instead of 60 cm viewing distance); in all other cases, it was reported on the first run. In the fourth sequence, with a different participant being the exception, "hello" was once reported after two runs (and written down with the first letter capitalized); in all other cases, it was reported on the first run. In Figure 6.13, the experimental results for word shape recognition outside and within the afterimage effect are summarized.

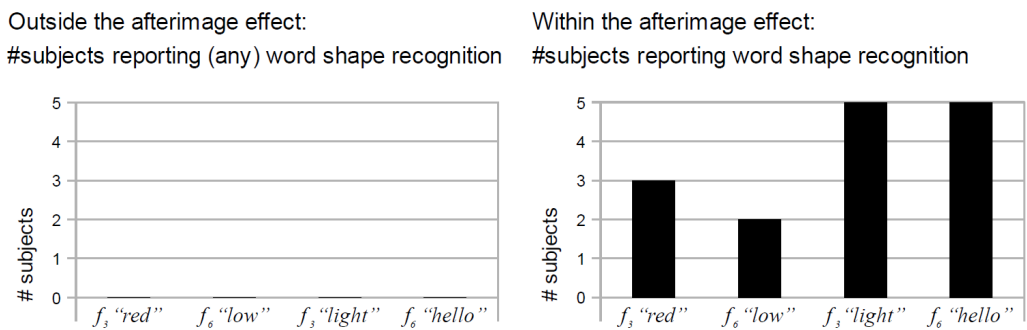

**Figure 6.13** *Results of the pilot experiment: word shape recognition outside and within the retinal afterimage.*

## 6.4 Conclusion

In the pilot experiment, rule sets $f_3$ and $f_6$ were each used to generate an image sequence based on a 5-letter target afterimage word shape specified in square units. Showing these image sequences to 5 test subjects resulted in none of them recognizing the 5-letter word shapes in the separate bias and trigger images, and all of them doing so in the afterimage effect (see Figure 6.13).

As expected, a similar procedure for 3-letter word shapes showed weaker performance, with word shapes being less often recognized successfully within the afterimage effect. One possible explanation for this, is that the afterimage display of 3-letter word shapes might be less robust than that of 5-letter word shapes in the face of the type of irregularities first identified in Section 6.2.2. If this is the case, it could be a reason for increasing the sophistication of the rasterization method, and its explicitly naive formal model of induced afterimage intensities. This could be informed by the examination of a range of potential factors influencing the complete or partial disappearance and reappearance of structured afterimages in [Wade 1978]; and also by the study of the influence of selective attention in this, including an effect of filling-in of enclosed regions, in [Lou 2001]; as well as by the study of the influence of post-adaptation contour alignment, where an afterimage may appear with much more stability and intensity if the geometry of contours still present for a previously induced afterimage aligns retinally with the geometry of contours present in the image being viewed, as in [Daw 1962]; and also by the study of how post-adaptation contours influence, and indeed can be used to control, the way in which color filling-in and spreading effects occur in the retinal afterimage [Van Lier et al. 2009].

In general, and also when developing partly automated techniques for other aspects of the retinal afterimage than 2D shape, e.g. textural effects and color (see Section 6.1.4), another important factor to be explicitly taken into account is the presence of induced simultaneous contrast in colors, both during and after adaptation, and the influence this has on the appearance of colors and greys in the retinal afterimage [Anstis et al. 1978]. Also more generally, the use of display technology with integrated eye tracking might support more natural eye movement during retinal adaptation; and the latter could be made to occur faster, by using darker surroundings and brighter stimuli than we have done here.

In any case, the results of the pilot experiment indicate that we have developed an automated method which, for a subset of all specifiable, recognizable 2D shapes, can successfully induce these *specifically* via the retinal afterimage. Thereby, and in this sense, our method realizes a form of the computed retinal afterimage (see Section 6.1). This also demonstrates feasibility of the computed retinal afterimage in general. Like other visual effects (see Section 6.1.1) the retinal afterimage, too, can be a computed aspect of human visual perception. This gives visual artists interested in the retinal afterimage the advantage of being able to create controlled visual complexity that uses the effect, in less time (see Section 6.1.4).

In the Electronic Appendix (see Section 6.2), we have included a software toolkit allowing complete reproduction of and free experimentation with the approach. The toolkit, released in source format under a public license, has been implemented for the Max programming environment widely used by contemporary artists [Puckette 2002].

# 7. Conclusion

# 7.1 Observations and a possible theory

Like other natural phenomena, human music making can be described as a causal chain, consisting of components that are each subject to measurement and empirical investigation. Therefore, as a working definition, in Section 1.1 we proposed *instrumental control of musical sound:* the phenomenon where human actions make changes to a sound-generating process, resulting in heard sound which induces musical experiences within the brain.

To create a technique for making music, then, is to set up a causal relationship, between aspects of human action and changes in heard musical sound. The development of different techniques may then be motivated by a single, overarching question: How can the instrumental control of musical sound be improved? As discussed in Section 1.5, answering this question requires also answering another, fundamental question: *What forms of instrumental control of musical sound are possible to implement?*

Parts of the answer to this question are already present in the many techniques for making music that have been developed from prehistory onward. These techniques initially involved the human body only, or made use of mechanical technologies. More recently, electromechanical technologies have also come into development and use (see Section 1.2). However, it is only when *computational* technologies are given the central role of causally linking human action to heard musical sound, that a unique advantage appears: Unlike earlier technologies, Turing-complete automata combined with transducers explicitly minimize the constraints on implementable causations (see Sections 1.4.4 and 1.5).

There is evidence for this theoretical advantage resulting in the practical implementation of new forms of instrumental control of musical sound: The combination of electronic digital computer and electric loudspeaker, introduced in the 1950s, has enabled the development of a wide variety of new types of sound-generating processes, which have come into wide use. In Section 1.4, an overview of these types was given, based on how instrumental control moved away from direct manipulation of the wave table.

Considering this evidence, the following question in particular seems potentially rewarding: How can we systematically *extend* the scope of computational technologies, from the sound-generating process, to the other components of the causal chain? In Section 1.4.4, we first formulated a *computed sound* model, to describe the fundamentals which enabled the wide variety of implementable sound-generating processes. Here, a Turing-complete automaton must become causally linked, via a transducer, to human auditory perception. Then, in Section 1.5.1.1, we generalized this to the notion of *completely computed instrumental control of musical sound*. Here, the Turing-complete automaton can track, represent and induce all relevant aspects of human action and perception. This seems to approach a general capability for implementing *all* perceivably different causal relationships between human actions and changes in heard musical sound.

Realizing a system capable of completely computed instrumental control seems hard. We can work toward this goal, however, by progressively developing transducers combined with Turing-complete automata. We call this process the *computational liberation* of instrumental control, as it will *gradually minimize constraints* on implementable causal relationships (see Section 1.5.1.2). We view computational liberation then as a de facto ongoing historical process, exemplified in the development of computed sound.

What area to consider next for computational liberation? Fingertip use is extremely important to the instrumental control of musical sound (see Section 1.2.5). Therefore, it makes sense to give priority to the area of fingertip touch.

To describe the prerequisites for computational liberation that are specific to fingertip use, we formulated a model for *computed fingertip touch* (see Section 1.5.2.1). Here, a Turing-complete automaton may be causally linked, via transducers, both to human somatosensory perception and to human motor activity involving the fingertips. The model is explicitly rooted in existing knowledge on fingertip movement and touch, including its anatomy, physiology, and neural processes (see Section 1.3 and Appendix A). This allows us to explicitly distinguish between different fundamental subtypes of touch. In Chapter 4, this resulted in working definitions for *computed passive touch*, *computed active touch*, and *computed manipulation*. These working definitions differentiate based on the presence and dynamics of a perceptually induced exterospecific component.

Within the area of fingertip use, a specific type of movement deserves additional priority. After considering fingertip use throughout time and across musical instruments, we identified *unidirectional fingertip movement orthogonal to a surface* as a widespread, common component (see Section 1.2). Here, fingertip movement approximates a single path of movement, at right angles with a surface, and extending across at most a few centimeters. Technologies implementing the computed fingertip touch model should support this type of fingertip movement with priority, because it offers advantages in terms of control precision, effort, speed, and multiplication (see Section 1.2.5). This motivated a number of choices directing transducer development: to have the human fingerpad as transducer source and target; to create transducers based on flat, closed, and rigid contact surfaces; and to provide orthogonal as well as parallel force output to the fingerpad (see Section 1.6.2).

Finally, for reasons discussed in Sections 1.4.4 and 1.6.7, it is crucial that technologies implementing the computed fingertip touch model are developed in such a way as to be mass-producible, cheap, and powerful.

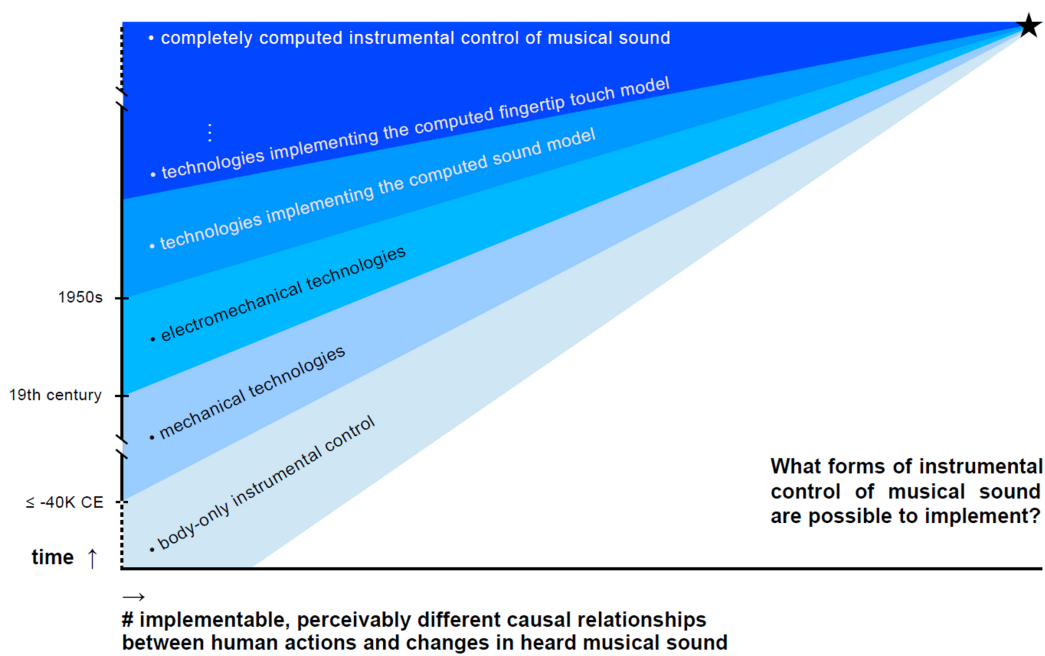

Figure 7.1 of the graph:

- completely computed instrumental control of musical sound
- technologies implementing the computed fingertip touch model
- technologies implementing the computed sound model
- electromechanical technologies
- mechanical technologies
- body-only instrumental control

1950s

19th century

≤ -40K CE

time ↑

→
# implementable, perceivably different causal relationships
between human actions and changes in heard musical sound

What forms of instrumental control of musical sound are possible to implement?

**Figure 7.1** *Overview. What forms of instrumental control of musical sound are possible to implement?*

• *Along the vertical axis: time. Along the horizontal axis: the number of implementable and perceivably different causal relationships between human actions and changes in heard musical sound.*

• *The rising edges then indicate the increasing contributions, over time, by body-only instrumental control and by different types of technology. The gradual shift in background hue at each edge is meant to indicate that these different contributions combine, to enable the overall total of implementable causal relationships.*

• *As is shown, body-only instrumental control has contributed from the earliest times. Mechanical technologies have contributed from at least 40 millennia BCE; while electromechanical technologies have done so, roughly, from the 19th century onward.*

• *From the 1950s onward, the labels are shown in white, to highlight the turn to a process of computational liberation: Here, the use of computational technologies enables an explicit minimization of constraints on implementable causal relationships.*

• *Zooming in, contributing technologies here include those implementing the computed sound model; the computed fingertip touch model; and other, as yet unspecified models.*

• *The star in the upper right of the graph marks a hypothetical point in the future where this development arrives at a theoretical endpoint: the completely computed instrumental control of musical sound.*

## 7.2 Hypothesis

Based on the above, we formulate the following *Hypothesis:*

> H
>
> Developing technologies that newly fit the computed fingertip touch model enables the implementation of new forms of instrumental control of musical sound.

Here, the implementation of new forms of instrumental control should result after some *limited* amount of time and effort. This makes it possible to design experiments disproving the hypothesis, ensuring its falsifiability.

## 7.3 Testable predictions

In Chapters 2 and 3, we developed the *cyclotactor (CT)* system, which provides fingerpad-orthogonal force output while tracking surface-orthogonal fingertip movement. This system newly implemented the computed fingertip touch model. Then, by deduction, the Hypothesis yields the following *Prediction 1:*

> P 1
>
> The CT system developed in Chapters 2 and 3 will enable the implementation of new forms of instrumental control of musical sound.

In Chapters 2 and 3, we also developed the *kinetic surface friction transducer (KSFT)* system, which provides fingerpad-parallel force output while tracking surface-parallel fingertip movement. This system, too, newly implemented the computed fingertip touch model. Then, by deduction, the Hypothesis also yields the following *Prediction 2:*

> P 2
>
> The KSFT system developed in Chapters 2 and 3 will enable the implementation of new forms of instrumental control of musical sound.

## 7.4 Experimental results

The first part of the experimental results is about technology: How do the CT and KSFT systems each indeed newly implement the computed fingertip touch model? Tables 7.2a and 7.2b, below, first describe the main qualitative and quantitative aspects of both technologies. Their resulting capabilities, in terms of human action and perception, are then summarized in Table 7.2c. Here, the phrase "complete integration with computed sound", used in Table 7.2a, means that input from human motor activity, output to somatosensory perception, and output to auditory perception are easily combined within a single written algorithm. Table 7.3 then concludes this part, by listing novel aspects of each technology. An important property here is *to avoid the use of connected mechanical parts moving relative to the target anatomical site*. We emphasize this as a general principle for transducer construction, with the potential benefit of enabling more precise output to somatosensory perception.

| qualitative aspect | system | |
| --- | --- | --- |
| based on the use of a flat, closed, and rigid contact surface | CT | KSFT |
| provides *fingerpad-orthogonal* force output, tracks *surface-orthogonal* fingertip movement | CT | |
| provides *fingerpad-parallel* force output and tracks *surface-parallel* fingertip movement | | KSFT |
| I/O made programmable in physical units; via SuperCollider classes | CT | KSFT |
| complete integration with computed sound | CT | KSFT |
| precisely adjustable, personal fit, for more accurate & comfortable I/O | CT | |
| cheaply mass-producible | CT | KSFT |

**Table 7.2a** *Qualitative aspects of the technologies developed in Chapters 2 and 3.*

| quantitative aspect | | CT system | KSFT system |
| --- | --- | --- | --- |
| input | spatial resolution | 0.2 mm | 0.02 mm |
| | spatial range | 35.0 mm | hundreds of $cm^2$ |
| | temporal resolution | 4000 Hz | 125 Hz, average |
| output | force resolution | ± 0.003 N | N/A |
| | force range | bipolar, varying over distance: see Chapter 3, Figure 3.2 | 0.14-1.43 N, kinetic friction |
| | temporal resolution | accurate wave output up to 1000 Hz | features between 1-10 ms |
| I/O | latency | 4.0 ms | 20.5 ms, average |

**Table 7.2b** *Quantitative aspects of the technologies developed in Chapters 2 and 3.*

| resulting capability | system | |
| --- | --- | --- |
| force output can co-determine the movement of fingertip control actions | CT | KSFT |
| I/O can induce aspects of haptic perception | CT | |
| accurate mechanical wave output across the frequency ranges involved in fingertip vibration perception | CT | |
| excellent support for real-time instrumental control of musical sound | CT | |
| I/O can induce high-resolution aspects of fingertip surface texture perception during active touch | | KSFT |

**Table 7.2c** *Resulting capabilities, in terms of human action and perception, of the technologies developed in Chapters 2 and 3.*

| novel aspect | system |
|---|---|
| explicit and specific support for unidirectional fingertip movement orthogonal to a surface | CT |
| transducer *completely avoids* the use of connected mechanical parts moving relative to the target anatomical site | CT |
| I/O specific to those flexing movements of the human finger that are independent, precise, and directly controlled by the motor cortex | CT |
| transducer *partially avoids* the use of connected mechanical parts moving relative to the target anatomical site | KSFT |
| inducing high-resolution aspects of fingertip surface texture perception using cheap, off-the-shelf optical mouse sensor input | KSFT |

**Table 7.3** *Novel aspects of the technologies developed in Chapters 2 and 3.*

The final part of the experimental results is about forms of instrumental control: What new forms have been implemented using both systems? This is summarized in Table 7.4. In its first examples, computed fingertip touch was used to display the state of the sound-generating process – at a higher level of detail than provided by existing technologies. This to better inform, and thereby alter, fingertip control actions. In the final examples, new forms of instrumental control were pursued more directly: Here, computed touch was used to implement new types of fingertip control action.

| see | type | system | key points |
|---|---|---|---|
| 4.2.2 | passive touch display | CT | • display of granular synthesis at the level of individual grains<br>• via presence, duration, amplitude, and vibrational content of fingerpad-orthogonal force pulses<br>• using a timescale identical to that of sonic grains |
| 4.3.2 | active touch display | KSFT | • during actions similar to turntable scratching: display more specific to the stored sound fragment<br>• via fingerpad-parallel friction, millisecond resolution |
| 4.3.3 | active touch display | CT | • during surface-orthogonal percussive fingertip movements<br>• touch display expanded outside moment of impact |
| 4.5.2 | control action | KSFT | • pushing against a virtual surface bump<br>• using horizontally applied output forces, during horizontally directed fingertip movements |
| 4.5.3 | control action | CT | • fingertip tensing during force wave output<br>• can be used simultaneously with control based on surface-orthogonal fingertip movement |

**Table 7.4** *New forms of fingertip instrumental control presented in Chapter 4.*

# 7.5 Discussion [1]

We will now reflect on the main experimental results, presented above, after first discussing the results of the research excursions of Chapters 5 and 6.

The first research excursion followed the phenomenon of unidirectional fingertip movement orthogonal to a surface elsewhere. Chapter 5 presented *one-press control*, a fingertip input technique for pressure-sensitive computer keyboards, based on the detection and classification of pressing movements on the already held-down key. We showed how this new technique can be seamlessly integrated with existing practices on ordinary computer keyboards, and how it can be used to simplify existing user interactions by replacing modifier key combinations with single key presses. In general, the proposed technique can be used to navigate GUI interaction options, to get full previews of potential outcomes, and then to either commit to one outcome or abort altogether – all in the course of one key press/release cycle. The results of user testing indicated that effective one-press control can be learned within about a quarter of an hour of hands-on operating practice time.

The second research excursion followed the idea of using computation to induce aspects of human perception elsewhere. In Chapter 6, we first considered how the use of techniques incorporating stages of automated computation offers visual artists a control over perceived visual complexity that is otherwise unattainable. This motivated the question whether computed output also can induce 2D shape perception in the *retinal afterimage:* the familiar effect in the human visual system where the ongoing perception of light is influenced by the preceding exposure to it. A fundamental problem here is how to exclude the possibility that shape recognition is caused by normal viewing of the stimuli, which may occur simultaneously. To solve this, we developed a rasterization method, a model of the afterimage intensities it induces, and then a series of candidate formal strategies for concrete rule sets computing stimuli. The Electronic Appendix to the chapter provides video examples, the image sequences used in a pilot experiment, and also software implementing the approach, in source format. The results of the pilot experiment testing the approach confirmed shape display specific to the retinal afterimage. This result also demonstrated feasibility of the *computed retinal afterimage* in general.

Finally, the overview of the main experimental results, given in Section 7.4, provides a summary of how the research goals for the body of this thesis (see Section 1.6.9) have been achieved. Moreover, the results of Tables 7.2a to 7.4 confirm both Prediction 1 and Prediction 2. This supports the Hypothesis: Developing technologies that newly fit the computed fingertip touch model enables the implementation of new forms of instrumental control of musical sound. At the start of this chapter, we presented an articulated, empirical view on what human music making is, and on how this relates to computation. The experimental evidence which we obtained seems to indicate that this view can be used as a tool, to generate models, hypotheses and technologies that enable an ever more complete answer to the fundamental question as to what forms of instrumental control of musical sound are possible to implement.

---

1 Some contributions discussed in this chapter are also briefly summarized in Appendix C.

## 7.6 The future: computed instruments [2]

**7.6.1 Algorithmic primitives for computed touch**   Reflecting on how the examples of computed fingertip touch of Chapter 4 were implemented, a general issue becomes apparent: The algorithmic representations that are used, in reflecting transducer I/O, are still quite removed from describing the aspects of human action that are involved. To bridge this gap, we may develop a set of parametrized algorithms that implement:

- forms of computed passive touch (see Section 4.2.1);

- forms of computed active touch (see Section 4.3.1).

Here, the parameters offered could still be defined using physical units. However, instead of directly describing transducer state, they would enable control that is more specific to the patterns of passive touch and exterospecific components being induced in the user.

Parametrized algorithms like this can then be used as programming primitives, to build the more complex algorithms that implement new forms of instrumental control. Here, we can apply lessons on the design of complex algorithms from the field of Operating Systems: "As a general rule, having a small number of orthogonal elements that can be combined in many ways leads to a small, simple, and elegant system." [Tanenbaum 2001]. Interpreting this, our computed touch primitives should be:

- conceptually simple;

- orthogonally combinable;

- expressive.

Here, by "expressive" we mean that the computed touch primitives, already when present in few different types, should enable the implementation of many perceivably different forms of touch.

**7.6.2 Computed manipulanda**   The algorithmic primitives that implement forms of computed active touch can then be used to implement forms of computed manipulation (see Section 4.4.1): by varying their parameters, over time, in response to specific transducer input. A *computed manipulandum* is then an algorithm that can:

- induce an exterospecific component in human touch perception;

- change this component, over time, in response to specific motor activity.

Computed manipulanda, running on a given system for computed touch, may well be complemented by forms of passive touch display and active touch display, executing simultaneously.

---

2   Continuing from Chapter 4, proposed future work more immediately includes the pursuit of new forms of fingertip touch display, and of new fingertip control actions. This includes the touch display of sonic grains, and making the same time-varying content both heard and felt during control. This Section discusses a broader context for such future work.

**7.6.3  Target properties for implemented control actions**    As discussed in Sections 1.3.3.3 and 1.3.4.2, during control actions, the activation of learned motor programs enables faster movement execution, with a reduced claim on attention and consciousness once movement has been initiated. This enables skilled and virtuose playing: Over a given time period, more changes may be made, intentionally and successfully, to the sound-generating process.

Then, effectively, the resulting musical sound will be a choice from a wider range of possibilities. In this sense, motor programs enable more powerful real-time control over the induced musical experience – e.g. during improvisation.

Also, in the player, the use of motor programs could allow attention and consciousness to become more occupied with the musical experiences that are the object of instrumental control.

Given the above fundamental advantages, it seems that the default for control actions, also when implemented using computed active touch or computed manipulation, should be that they are suited to execution by learned motor programs. As discussed in Section 1.3.3.3, this implies that, after some learning phase:

  • movement execution does not need visual feedback anymore;

  • movement execution occupies attention and consciousness less, or not at all.

**7.6.4  The role of the computed image**    If, when making music, it is not known what the possible control actions are, it will not be possible to exercise control in the sense of reliably producing outcomes according to intentions. However, unlike predecessor technologies, the use of computed touch implies that the set of possible control actions may become highly dynamic over time. This raises a question: How does the user know what control actions currently are possible?

The answer may be: via some form of real-time display. One way of providing this seems inherently present: The user may probe using active touch. However, this will take some time, during which actual control cannot be exercised. An added *visual* display might well convey the relevant information more immediately. For the user to know what control actions are possible, active touch probing might then be necessary only during an initial learning phase, after which visual recognition would suffice.

This then justifies further refining the view discussed in Section 7.1 by adding a *computed image* model, to characterize suitable technologies for visual display. Here, a Turing-complete automaton will be causally linked, via a transducer, to human visual perception. Then, if we select a transducer technology that can induce aspects of visual perception that include spatial display at the scale of control action movements, the computed image may provide:

  • symbolic display enabling cognition of possible control actions;

  • to-scale spatial display of the induced exterospecific components.

**7.6.5 Computed instruments**    Hearing, touch, and vision are main areas of human action and perception via which traditional musical instruments enable forms of instrumental control of musical sound. Therefore, if an algorithm enables forms of instrumental control via computed sound (see Section 1.4.4), computed touch (see Section 2.1.1), and computed image (see above), it does not seem exaggerated to call it a *computed instrument*.

Naturally, computed instruments may combine computed touch, computed image, and computed sound in any way that is algorithmically possible. Certain specific types of causal relationships can be expected, however. Changes to heard musical sound will be enabled by causalities from computed touch to computed sound. Especially when, within the overall algorithm, this involves degrees of freedom controlling scalar parameters of audio synthesis, this will relate to the concept of *mapping* [Hunt et al. 2000] [Hunt and Wanderley 2002]. Also, as discussed in Section 7.6.4, cognition of possible control actions may be enabled by causalities from computed touch to computed image. This will relate to the concept of *affordance* [Gibson 1977] [Norman 1988].

A possibly interesting question here seems whether, beyond the execution of movement during control actions, all other aspects of touch and vision, too, can be made to disappear from consciousness, so that it may engage more purely in making changes to the induced musical experience.

The considerations presented in Sections 7.6.1 to 7.6.5 have guided work on a concrete system that can run computed instruments. With computed touch based on the CT system, at the time of writing, the work on its hard- and software is nearing completion.

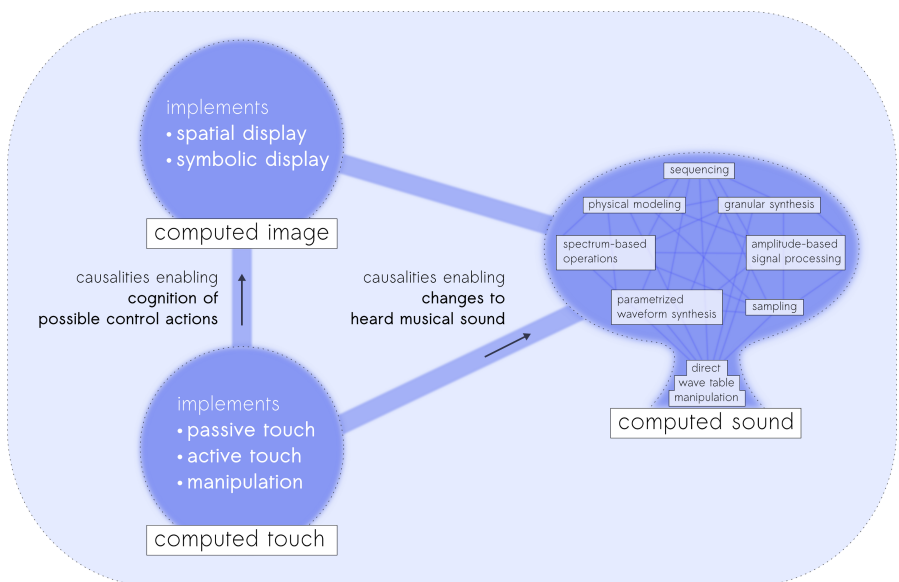

**Figure 7.5** *A view on computed instrumental control of musical sound.*

# Appendix A: Functional anatomy, physiology, and neural processes of fingertip movement and touch

**A-1  Bones involved in fingertip movement**     Here, we will first list the bones involved in fingertip movement, for both arms, from shoulder to fingertip. An overview of these bones is shown in Figure A-1. Attached nearest to the torso, or most proximal, is the humerus [Zygote Media 2012]. This bone lies within the upper arm. The next two bones lying more distal, or further away in their attachment to the torso, are the ulna and the radius. These lie roughly parallel, within the forearm, both having their proximal ends attached to the distal end of the humerus. Near the hand, the distal end of the radius lies on the side of the thumb, while that of the ulna lies on the side of the little finger. Here, both radius and ulna attach to the bones of the wrist, or carpus.

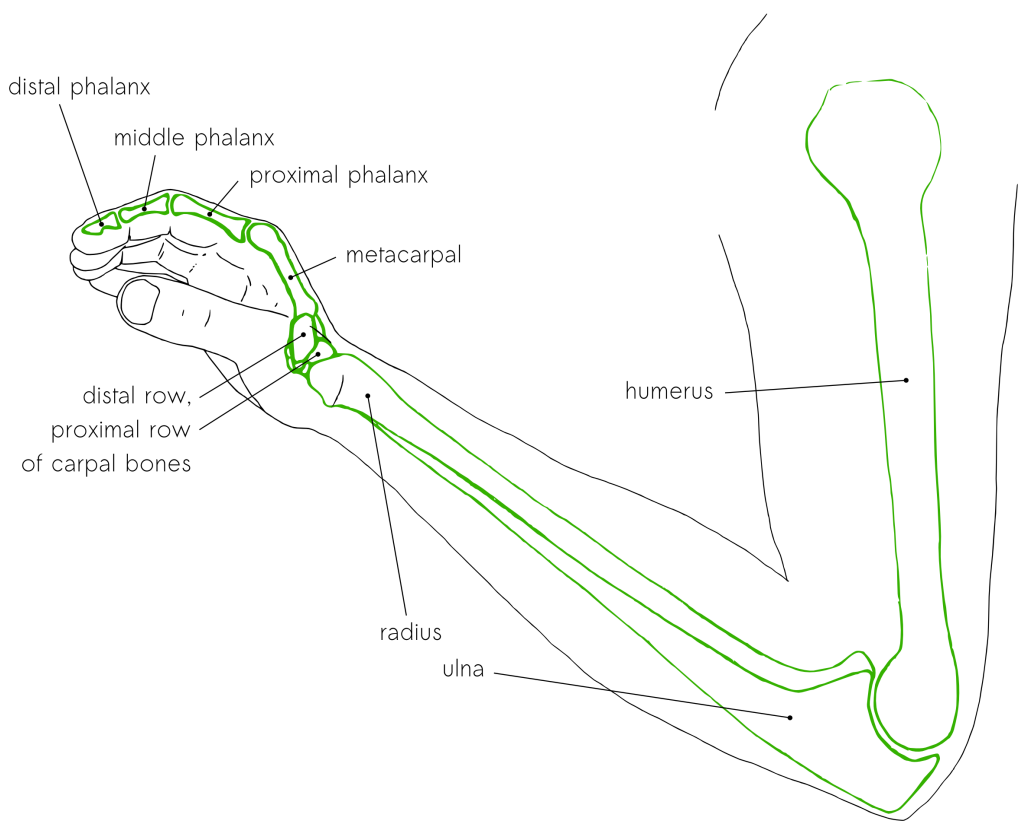

**Figure A-1**  *Bones involved in index fingertip movement. Here, shown for the right hand and arm.*

There are eight carpal bones, fitting together within the wrist. Unlike the other bones discussed here, the carpal bones do not have an elongated shape. If the other bones may seem like sticks, the carpal bones will seem more like pebbles. They can be divided into two rows, one lying proximal, one distal [Tubiana et al. 1996]. The proximal row of carpal bones, from the radial to the ulnar side, consists of the scaphoid, the lunate, the triquetrum, and the pisiform. The distal row of carpal bones, from the radial to the ulnar side, consists of the trapezium, the trapezoid, the capitate, and the hamate [Sobotta et al. 1994].

Beyond the carpal bones are the five metacarpal bones, with their proximal ends attached to the distal row of the carpus. They extend further distally, lying next to eachother across the breadth of the hand. Each metacarpal can be considered the first bone of a separate chain of bones, continuing distally, with each chain corresponding to one of the five digits of the hand. For each chain, the next more distally attached bone is a proximal phalanx. The proximal phalanges lie within the respective segments of the thumb and fingers which are first seen to clearly stick out from the rest of the hand. Before the end of each chain, the fingers have a middle phalanx within their next more distal segment. Such a phalanx is not present in the thumb. Each chain is ended by a distal phalanx, forming the skeletal basis for each fingertip.

**A-2  Joint movements causing fingertip movement**    How the bones listed above are movable relative to eachother determines fingertip movement during instrumental control of musical sound. To indicate specific types of externally observable movement that result from the various articulations, specific anatomical terms can be used. Paired with the names of the joints involved, these terms can then be used to unambiguously indicate main types of movement which may occur. Here, in this way, we will list contributing joint movements, from shoulder to fingertip. A visual overview is given in Figure 1.7.

At the shoulder joint, the humerus is moved relative to the shoulder blade in three types of rotation [Sobotta et al. 1994]. By moving the upper arm, these rotations also move the forearm and hand. For the purpose of describing each rotation, we will assume here that the upper arm initially hangs down along the torso. Then, in *shoulder anteversion/retroversion*, the humerus rotates forward and backward, respectively, swinging the upper arm forward and backward relative to the torso. In *shoulder adduction/abduction*, the upper arm is swung sideways relative to the torso: toward it, and away from it, respectively. Finally, in *shoulder endorotation/exorotation*, the upper arm is rotated around its longitudinal axis, inward toward the front of the torso, or in the opposite, outward direction, respectively. Other movements of the shoulder, such as shrugging, result from displacements of the shoulder blade and clavicle. Such displacements also occur during the three rotations just mentioned [Ten Donkelaar et al. 2007]. However, we will not further discuss these movements here, considering them of, and not relative to, the torso.

At the elbow joint, the ulna and radius together flex and extend relative to the humerus [Sobotta et al. 1994]. This *elbow flexion/extension* moves the forearm with the hand in a hinge-like motion relative to the upper arm.

Within the forearm, the radius, lying along the ulna, can rotate around it, forming the radioulnar joint. The resulting movement of *forearm pronation/supination* will turn the entire hand, since it is attached to the radius at the wrist. Rotation in the direction turning the palm upward is called supination, rotation turning the palm downward, pronation [Sobotta et al. 1994].

In general, the joint movements at the wrist are the combined result of individual articulations involving the various carpal bones, driven by the excursion of muscles originating within the arm [Sobotta et al. 1994]. Our overview of this, here, and also of the more distal joint movements, below, will be based on information scattered throughout [Tubiana et al. 1996], unless indicated otherwise. In *wrist radial/ulnar deviation*, the hand is turned sideways toward either the radius or the ulna. This involves articulations between the carpal bones inside the wrist, as well as between the proximal row of carpal bones and the radius and ulna. The result is an overall joint movement which can rotate the hand across ± 60 °. In *wrist palmar/dorsal flexion*, either the palm of the hand or its back (dorsum) is flexed toward both the radius and the ulna. The muscles causing this insert, via tendons, at various locations on and near the proximal ends of the metacarpals. In the resulting flexion of the metacarpals relative to both radius and ulna, the index and middle metacarpals are fixed rigidly to bones inside the distal carpal row, so that these all move together as a single unit. This forms the basis for an overall joint movement rotating the hand across ± 160 °.

Other metacarpals however can move independently. On the radial side, the thumb metacarpal can move in multiple ways relative to the carpal bones and the other metacarpals. However, as our focus is on movement of the tips of the fingers, we will not discuss these movements here, only noting that the specific articulations underlying them enable the thumb to become opposed to, and be brought into contact with, any of the fingertips of the same hand. On the other, ulnar side, the little finger metacarpal can be separately flexed, so as to move its palmar side toward the forearm and also rotate it toward the thumb. This movement of *little finger opposition* is based on the contraction of a muscle which originates at the carpus and inserts on the ulnar side of the little finger metacarpal [Ten Donkelaar et al. 2007].

Between the metacarpal bone of each finger and its proximal phalanx lies a metacarpophalangeal (MP) joint. On the outside, the location of the MP joint is visible as the knuckle at the base of the finger (see Figure A-1). On the inside, three tendons and three muscles enter each finger here. This includes two flexor tendons entering on the palmar side, and one extensor tendon entering on the dorsal side. All of these tendons are attached to muscles which lie outside of the hand, within the arm. Also entering the finger, on its radial and on its ulnar side, respectively, are two interosseous muscles, originating between the metacarpal bones of the hand. An exception to this is the little finger, which only has an interosseous muscle entering on its radial side; on the ulnar side, it has the "abductor digiti minimi" muscle instead. Finally, a lumbrical ("worm-like") muscle also enters on the radial side of each finger. It is exceptional, in that it originates not on a bone, but on a tendon. More precisely, the lumbrical muscle originates on a section, lying within the palm, of one of the two flexor tendons also entering the finger [Ten Donkelaar et al. 2007]. Together, these three muscles and three tendons continue distally into the finger, inserting in various combinations at various locations of its bony and fibrous skeleton. Based on a number of interacting and non-trivial mechanisms inside the finger, the coordinated contraction of the muscles involved results in another four types of externally observable joint movement.

In *MP joint flexion/extension*, the proximal phalanx is flexed toward or extended away from the palmar side of the metacarpal. In what, for clarity's sake, we will call *MP joint radial/ulnar deviation*, the proximal phalanx is turned sideways, within a more limited angular range, flexing toward either the ulna or the radius. (Conventionally, this movement is often called abduction/adduction, and defined as rotation away from or toward the middle finger. However, it seems the movement also exists for the middle finger itself.) The MP joint movements, by moving the proximal phalanx, move the finger as a whole. Continuing more distally, between the proximal phalanx of each finger and its middle phalanx lies a proximal interphalangeal (PIP) joint. Here, in *PIP joint flexion/extension*, the middle phalanx is flexed toward or extended away from the palmar side of the proximal phalanx, in a hinge-like motion moving the distal part of the finger from the middle phalanx onward. Finally, between the middle phalanx of each finger and its distal phalanx lies a distal interphalangeal (DIP) joint. In *DIP joint flexion/extension*, the distal phalanx is flexed toward or extended away from the palmar side of the middle phalanx, in a hinge-like motion, moving the fingertip.

### A-3  Some general properties of the joint movements causing fingertip movement

So far, our overview of fingertip movement as the result of combined joint movements may seem to suggest a model not unlike that of, say, a segmented robot arm. However, such an impression could easily lead to a number of implicit assumptions which in fact would not be true. To avoid this, we will now highlight some general properties of the joint movements discussed above.

*Joint movements often do not produce straight movement trajectories.* For example, during MP joint flexion/extension, other rotation than that strictly flexing the proximal phalanx toward the metacarpal is occurring simultaneously. This is due to asymmetries in the components of the bony and fibrous skeleton at the MP joint, as well as other details of the anatomy at this location. Similarly, the rotation during MP joint radial/ulnar deviation is not strictly sideways in the directions indicated, and for similar reasons.

*Joint movements often do not happen around fixed axes of rotation.* For example, during forearm pronation/supination, when viewed in cross section, the rotation of the radius bone can be described as happening around an axis which lies somewhere inside the ulna bone. However, the ulna itself is simultaneously being rotated, and displaced along a curve, and the location of the main rotational axis will vary during a single movement, as well as between subsequent movements. At the wrist, palmar/dorsal flexion and radial/ulnar deviation, too, are examples of joint movements not restricted to a fixed geometric axis.

*Joint movements often are not replicated identically across joints.* For example, the proximal phalanges of the fingers have an ulnar inclination in their attachment to the metacarpals. This ulnar inclination varies in amount across the fingers of one hand, thereby varying the orientation of movements relative to the metacarpal. More generally, although MP joint flexion/extension, MP joint radial/ulnar deviation, PIP joint flexion/extension, and DIP joint flexion/extension are replicated across the fingers, this also includes individual variations in the basic underlying anatomy, resulting in differences between the same movements at different fingers.

*Joint movements often do not happen fully orthogonal to eachother.* Different joint movements may very well occur simultaneously, but, due to details of the underlying anatomy, the range of one movement may depend on the current state of another. For example, elbow extension is limited by wrist palmar flexion. Wrist palmar flexion facilitates, however, wrist ulnar deviation; while wrist dorsal flexion facilitates wrist radial deviation. More distally, MP joint flexion limits MP joint radial/ulnar deviation, and during general flexion of the finger joints, the middle and distal phalanges can apply the greatest forces if there is dorsal flexion and ulnar deviation at the wrist.

Lack of orthogonality may also mean that movements become mirrored across joints. For example, some of the extensor tendons combining to enter at the finger MP joints are attached to eachother across the back of the hand; while some of the flexor tendons of the middle, ring and little fingers attach to the same muscle. As a result of such underlying anatomy, movement at the joints of one finger may often induce simultaneous movement, in a similar direction, in adjacent fingers. Perhaps the strongest example of mirroring, however, are PIP and DIP joint flexion/extension near the fingertip. At least partly due to the anatomy of the fibrous skeleton at these joints [Ten Donkelaar et al. 2007], PIP and DIP joint flexion/extension typically happen simultaneously and in the same direction.

**A-4 Somatosensory receptors for fingertip movement** Distributed across various anatomical areas, there exist various types of somatosensory receptors that are essential to normal human fingertip movement. We will discuss these below, with Table 1.8 providing a summary.

*Muscle spindles* are located within the muscles and transduce muscle length, as well as the speed of muscle length changes. Here, the somatosensory neuron is attached to a spindle made of 3 to 10 thin muscle fibers that stretch along and respond when the nearby ordinary muscle fibers are stretched. During contraction of the ordinary muscle fibers, specialized motor neurons contract the spindle fibers too, ensuring that they keep stretching along with the ordinary muscle fibers. The action potential characteristics of different types of muscle spindles together transmit the properties just mentioned [Wolters and Groenewegen 2004].

*Golgi tendon organs* are located between a tendon and the muscle it attaches to, and transduce muscle tension and changes in muscle tension. They adapt slowly, and may respond when muscle spindles do not, in cases where changes in muscle tension do not produce changes in muscle excursion. On the other hand, a relaxed muscle that is being stretched may change in length while not changing the tension it applies, resulting in muscle spindles that respond while the Golgi tendon organs do not [Wolters and Groenewegen 2004]. There are receptors much like Golgi tendon organs present in tissues within and around the joints, transducing events related to joint movement.

*Ruffini mechanoreceptors* are located in tissues within and around the joints, and also deeper in the skin below the epidermis. They transduce mechanical deformation, and mechanical vibration. They respond best to vibration waves with frequencies in the 15-400 Hz range [Goldstein 2002]. In the skin, they adapt slowly, and their sensory units have relatively large receptive fields. There, they are often called slowly adapting type 2 (SA2) cutaneous receptors [Goldstein 2002]. In the joints, however, Ruffini mechanoreceptors adapt rapidly [Wolters and Groenewegen 2004].

*Vater-Pacini mechanoreceptors* are located in tissues within and around the joints; deeper within the skin below the epidermis; at the surface of tendons and fascia (fibrous connective tissues, e.g. around muscles); in the walls of blood vessels; and in the periosteum (an outer membrane of bones, for long bones absent at the joints) [Kahle 2001] [Wolters and Groenewegen 2004]. They transduce changes in mechanical deformation, and mechanical vibration. Vater-Pacini mechanoreceptors respond best to vibration waves with frequencies in the 10-500 Hz range [Goldstein 2002]. Within the skin, they adapt rapidly, and their sensory units have relatively large receptive fields. Here, they are often called rapidly adapting type 2 (RA2) cutaneous receptors [Goldstein 2002]. In the joints, however, Vater-Pacini mechanoreceptors adapt slowly [Wolters and Groenewegen 2004].

*Merkel (SA1) mechanoreceptors* are located within the skin near the boundary between epidermis and dermis. They transduce mechanical deformation, and low-frequency mechanical vibration. Merkel (SA1) mechanoreceptors respond best to vibration waves with frequencies in the 0.3-3 Hz range [Goldstein 2002]. Adapting slowly, and with their sensory units having relatively small receptive fields, they are often called slowly adapting type 1 (SA1) cutaneous receptors [Goldstein 2002].

*Meissner (RA1) mechanoreceptors* are also located within the skin near the boundary between epidermis and dermis. They transduce (light) mechanical deformation, and low-frequency mechanical vibration. Meissner (RA1) mechanoreceptors respond best to vibration waves with frequencies in the 3-40 Hz range [Goldstein 2002]. Adapting rapidly, and with their sensory units having relatively small receptive fields, they are often called rapidly adapting type 1 (RA1) cutaneous receptors [Goldstein 2002].

*Nociceptors* are located throughout the skin, and throughout various other tissues of the hand and arm. They transduce mechanical, thermal and chemical events associated with the occurrence of tissue damage [Wolters and Groenewegen 2004] [Carlson 1998]. Nociceptors adapt slowly, and in most cases are presumed to be free nerve endings [Wolters and Groenewegen 2004].

*Thermoreceptors* are located within the skin, and transduce changes in temperature. They are widely considered to be types of free nerve endings which increase their firing rates on changes in temperature as small as 0.2 °C, but rapidly adapt to more constant temperature levels [Wolters and Groenewegen 2004]. Upward and downward changes in temperature are transduced by different types of thermoreceptors. Receptors for cooling respond in the 20-45 °C range, and best around 30 °C. Receptors for warming respond in the 30-48 °C range, and best around 44 °C. (The internal body temperature of a human typically is around 37 °C) [Goldstein 2002]. Warmth receptors lie deeper in the tissue of the skin than cold receptors [Carlson 1998].

So far, we have not discussed *Krause receptors* [Kahle 2001] or *Golgi-Mazzoni receptors* [Tubiana et al. 1996]. This has been due to problems in obtaining clear information on the separate nature of these related receptor types. Also, we have not discussed cutaneous somatosensory receptors that are specific to hairy skin, since the skin surrounding the fingertip is glabrous.

**A-5 Types of passive touch and their underlying somatosensory transduction**
Here, we will list a number of roughly indicated types of passive touch, and discuss for each case which of the types of somatosensory receptor discussed in Section A-4 are assumed [Kahle 2001] to cause sensation via transduction.

First, there exist various *skin pressure sensations*. Sensations of pressure being applied, or released, may involve the Vater-Pacini (RA2) mechanoreceptors [Goldstein 2002]. Sensations of slowly being pushed against, being lightly tapped, and detailed pressure changes over time (such as those of raised-dot patterns moving across the skin) may involve the Merkel (SA1) mechanoreceptors. Meissner (RA1) mechanoreceptors are also involved in (light) skin pressure sensations. A measure of the size and density of the receptive fields underlying these sensations is the *two-point threshold*: the smallest distance between two needles placed on the skin for which they are still perceived separately. Across the human body, the two-point threshold is lowest at the fingertips, where it is ± 2 mm. For comparison, the two-point threshold is ± 4 mm at more proximal parts of the finger, and ± 8 mm at the palm [Wolters and Groenewegen 2004].

The Ruffini (SA2) mechanoreceptors are involved in various *skin stretch sensations* [Goldstein 2002].

Furthermore, the Ruffini (SA2), Vater-Pacini (RA2), Merkel (SA1) and Meissner (RA1) mechanoreceptors may be involved in various *skin vibration sensations*, based on the overlapping frequency ranges describing their sensitivity to mechanical waves (see Table 1.8). Given the presence of Ruffini and Vater-Pacini mechanoreceptors in tissues within and around the joints, and the additional presence of Vater-Pacini mechanoreceptors at the surface of the tendons and fascia, in the walls of blood vessels, and in the periosteum, it seems appropriate to mention the possibility of vibration sensations originating from other tissues involved in fingertip movement, too.

The *proprioceptive sensations* are defined as those which convey a sense of position of the limbs [Goldstein 2002]. Such sensations may be based on transduction of muscle length by the muscle spindles [Wolters and Groenewegen 2004], but may also involve the skin mechanoreceptors, e.g. as when the dorsal skin of the hand becomes stretched due to finger flexion [Tubiana et al. 1996].

The *kinesthetic sensations* are defined as those which convey a sense of movement of the limbs [Goldstein 2002]. As the fingertip is being moved passively, sensations of the presence, speed and direction of its movement may involve muscle spindles within the muscles [Wolters and Groenewegen 2004], Golgi tendon organs between the muscles and tendons [Carlson 1998], and within and around the joints, the receptors similar to Golgi tendon organs, as well as Ruffini and Vater-Pacini mechanoreceptors [Wolters and Groenewegen 2004].

Transduction by the nociceptors of various mechanical, thermal, chemical and electrical events associated with the infliction of tissue damage may result in unpleasant *pain sensations* [Carlson 1998] [Kahle 2001] [Wolters and Groenewegen 2004].

Finally, various *temperature sensations* may result from transduction by the warmth and cold thermoreceptors within the skin. Typically, sensations of warmth and cold indicate changes in skin temperature, and soon become replaced by a sense of neutrality when no further changes are detected. The same temperature may then be experienced at once as both warm and cold at different skin sites, if these previously have adapted to different temperature levels. However, at the boundaries of the temperature ranges transduced by the thermoreceptors, the nature of temperature sensations shifts from relative to absolute [Carlson 1998].

# Appendix B: Samenvatting

Dit proefschrift [1] gaat over de unieke en centrale rol die berekening te spelen heeft in de ontwikkeling van hoe mensen muziekmaken.

Net als andere natuurlijke fenomenen kan menselijk muziekmaken beschreven worden als een causale keten, bestaande uit componenten die elk vatbaar zijn voor meting en empirisch onderzoek. Daarom stellen we, als een werkdefinitie, in Sectie 1.1 *instrumentele controle van muzikaal geluid* voor: het fenomeen waar menselijke acties veranderingen aanbrengen aan een geluidgenererend proces, resulterend in gehoord geluid dat muzikale ervaringen in de hersenen opwekt.

Een techniek voor muziekmaken creëren is dan een causale relatie opzetten, tussen aspecten van menselijke actie en veranderingen in gehoord muzikaal geluid. De ontwikkeling van verschillende technieken kan dan gemotiveerd worden door een enkele, overkoepelende vraag: Hoe kan de instrumentele controle van muzikaal geluid verbeterd worden? Zoals wordt besproken in Sectie 1.5 vergt het beantwoorden van deze vraag het óók beantwoorden van een andere, fundamentele vraag: *Welke vormen van instrumentele controle van muzikaal geluid zijn implementeerbaar?*

Delen van het antwoord op deze vraag zijn al aanwezig in de vele technieken voor muziekmaken die vanaf de prehistorie zijn ontwikkeld. Initieel maakten deze technieken gebruik van louter het menselijk lichaam, of van mechanische technologieën. Meer recent zijn electromechanische technologieën ook in ontwikkeling en gebruik gekomen (zie Sectie 1.2). Het is echter pas wanneer *computationele* technologieën de centrale rol gegeven wordt om menselijke actie causaal te verbinden aan gehoord muzikaal geluid, dat een uniek voordeel verschijnt: In tegenstelling tot eerdere technologieën minimalizeren Turing-complete automaten gecombineerd met transducers expliciet de beperkingen op implementeerbare causaties (zie Sectie 1.4.4 en 1.5).

Er is bewijslast die toont dat dit theoretische voordeel resulteert in de praktische implementatie van nieuwe vormen van instrumentele controle van muzikaal geluid: De combinatie van electronische digitale computer en electrische luidspreker, geïntroduceerd in de jaren 1950, heeft de ontwikkeling van een brede variëteit aan nieuwe typen geluidgenererende processen mogelijk gemaakt, die ook breed in gebruik zijn geraakt. In Sectie 1.4 wordt een overzicht van deze typen gegeven, gebaseerd op hoe instrumentele controle wegbewoog van directe manipulatie van de golftabel.

Gegeven deze bewijslast lijkt specifiek de volgende vraag potentieel vruchtbaar: Hoe kunnen we het bereik van computationele technologieën systematisch *uitbreiden,* vanaf het geluidgenererend proces, naar de andere componenten van de causale keten? In Sectie 1.4.4 formuleren we eerst een model van *berekend geluid*, om de fundamentele factoren te beschrijven die de brede variëteit aan implementeerbare geluidgenererende processen mogelijk hebben gemaakt. Dit omvat een Turing-complete automaat die causaal verbonden moet worden, via een transducer, aan menselijke auditorische perceptie. Daarna, in Sectie 1.5.1.1, generaliseren we dit tot de notie van *compleet berekende instrumentele controle van muzikaal geluid*. Dit omvat een Turing-complete automaat die in staat is alle relevante aspecten van menselijke actie en perceptie te volgen, te representeren, en op te wekken. Dit lijkt een algemene capaciteit te benaderen: die tot het implementeren van *alle* waarneembaar verschillende causale relaties tussen menselijke acties en veranderingen in gehoord muzikaal geluid.

---

1 "Berekende vingertop-aanraking voor de instrumentele controle van muzikaal geluid met een excursie naar het berekende retinale nabeeld".

Een systeem bouwen dat in staat is tot compleet berekende instrumentele controle lijkt moeilijk. We kunnen echter naar dit doel toewerken, door een progressie van transducers gecombineerd met Turing-complete automaten te ontwikkelen. We noemen dit proces de *computationele bevrijding* van instrumentele controle, omdat het *gradueel beperkingen* op implementeerbare causale relaties *zal minimalizeren* (zie Sectie 1.5.1.2). We zien computationele bevrijding dan als een de facto voortgaand historisch proces, met als duidelijk voorbeeld de ontwikkeling van berekend geluid.

Welk gebied vervolgens te overwegen voor computationele bevrijding? Vingertop-gebruik is extreem belangrijk voor de instrumentele controle van muzikaal geluid (zie Sectie 1.2.5). Daardoor heeft het zin om het gebied van vingertop-aanraking prioriteit te geven.

Om de specifiek bij vingertop-gebruik horende vereisten voor computationele bevrijding te beschrijven, formuleren we een model voor *berekende vingertop-aanraking* (zie Sectie 1.5.2.1). Dit omvat een Turing-complete automaat die via transducers causaal verbonden kan zijn met zowel menselijke somatosensorische perceptie als met menselijke motor-activiteit inzake de vingertoppen. Het model is expliciet geworteld in bestaande kennis over vingertop-beweging en -aanraking, inclusief anatomie, fysiologie, en neurale processen (zie Sectie 1.3 en Appendix A). Dit staat ons toe expliciet onderscheid te maken tussen verschillende fundamentele subtypen van aanraking. In Hoofdstuk 4 resulteert dit in werkdefinities voor *berekende passieve aanraking, berekende actieve aanraking,* en *berekende manipulatie.* Deze werkdefinities verschillen onderling op basis van de aanwezigheid en dynamiek van een perceptueel opgewekte exterospecifieke component.

Binnen het gebied van vingertop-gebruik verdient een specifiek type beweging extra prioriteit: Na het overwegen van vingertop-gebruik over de tijd en over muziekinstrumenten, identificeren we in Sectie 1.2 *unidirectionele vingertop-beweging loodrecht op een oppervlak* als een wijdverbreid aanwezige component. Vingertop-beweging benadert hier een enkel bewegingstraject, dat rechte hoeken maakt met een oppervlak en zich uitstrekt over hoogstens enkele centimeters. Technologieën die het model van berekende vingertop-aanraking implementeren behoren dit type vingertop-beweging met prioriteit te ondersteunen, omdat het voordelen biedt voor de precisie, inspanning, snelheid, en vermenigvuldiging van controle (zie Sectie 1.2.5). Dit

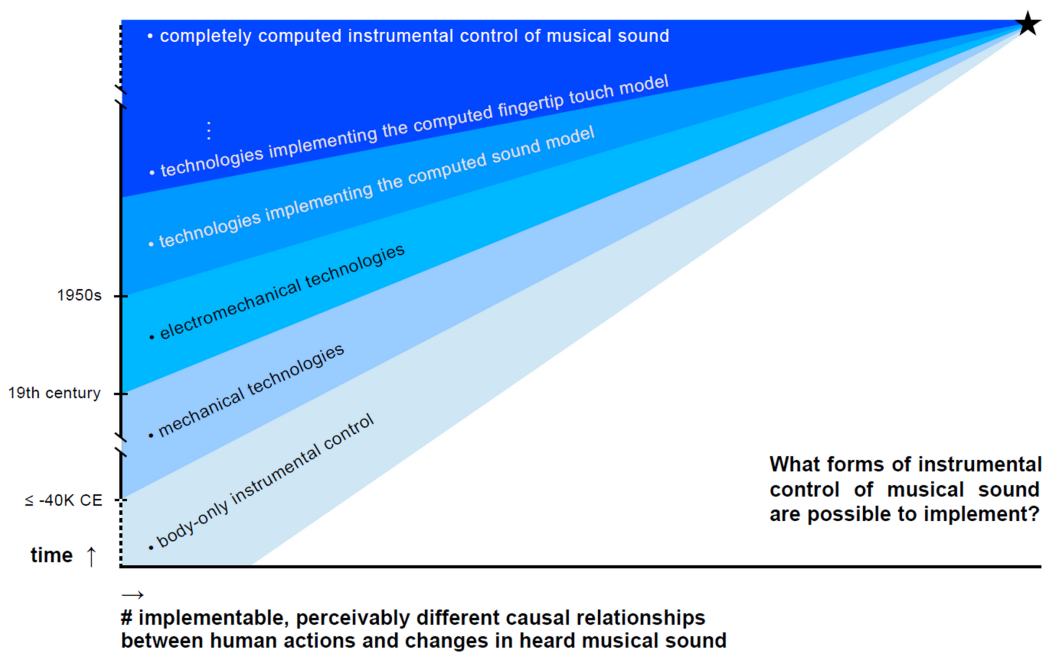

Labels within figure:
• completely computed instrumental control of musical sound
• technologies implementing the computed fingertip touch model
• technologies implementing the computed sound model
• electromechanical technologies
• mechanical technologies
• body-only instrumental control

1950s
19th century
≤ -40K CE
time ↑

What forms of instrumental control of musical sound are possible to implement?

→
# implementable, perceivably different causal relationships between human actions and changes in heard musical sound

**Figuur I**  *Overzicht. Welke vormen van instrumentele controle van muzikaal geluid zijn implementeerbaar?*

• *Langs de verticale as: tijd. Langs de horizontale as: het aantal implementeerbare en waarneembaar verschillende causale relaties tussen menselijke acties en veranderingen in gehoord muzikaal geluid.*

• *De stijgende randen wijzen op over de tijd toenemende bijdragen, door instrumentele controle met louter het menselijk lichaam, en door verschillende typen technologie. De geleidelijke verschuiving in achtergrondtint bij iedere rand is bedoeld om erop te wijzen dat deze verschillende bijdragen combineren, om gezamenlijk het totaal aan implementeerbare causale relaties mogelijk te maken.*

• *Zoals getoond, heeft instrumentele controle met louter het lichaam bijgedragen vanaf de vroegste tijden. Mechanische technologieën hebben bijgedragen vanaf tenminste 40 millennia BCE; terwijl electromechanische technologieën dit, grofweg, gedaan hebben vanaf de 19e eeuw.*

• *Vanaf de jaren 1950 staan labels in wit gedrukt, om zo de draai naar een proces van computationele bevrijding te benadrukken: Het gebruik van computationele technologieën maakt vanaf hier een expliciete minimalizatie van beperkingen op implementeerbare causale relaties mogelijk.*

• *Inzoomend op dit type technologieën zien we technologieën bijdragen die het model van berekend geluid implementeren; dat van berekende vingertop-aanraking; en dan andere, nog niet uitgewerkte modellen.*

• *De ster rechtsboven in de grafiek markeert een hypothetisch punt in de toekomst, waar deze ontwikkeling uitkomt bij een theoretisch eindpunt: de compleet berekende instrumentele controle van muzikaal geluid.*

motiveert een aantal leidende keuzes voor transducer-ontwikkeling: het nemen van het menselijk vingerkussen als transducerbron en -doel; het creëren van transducers gebaseerd op vlakke, gesloten, en rigide contactoppervlakken; en het verschaffen van zowel loodrechte als parallelle krachtuitvoer naar het vingerkussen (zie Sectie 1.6.2).

Tenslotte is het om redenen die besproken worden in Sectie 1.4.4 en 1.6.7 cruciaal dat technologieën die het model van berekende vingertop-aanraking implementeren worden ontwikkeld op zo een manier dat ze massa-produceerbaar, goedkoop, en krachtig zijn.

## Hypothese

Gebaseerd op bovenstaande, formuleren we de volgende *Hypothese:*

> **H**
>
> Het ontwikkelen van technologieën die, op nieuwe wijze, passen binnen het model van berekende vingertop-aanraking maakt de implementatie van nieuwe vormen van instrumentele controle van muzikaal geluid mogelijk.

De implementatie van nieuwe vormen van instrumentele controle behoort hier te resulteren na een *beperkte* hoeveelheid tijd en inspanning. Dit maakt het mogelijk experimenten te ontwerpen die de hypothese ontkrachten, zodat de falsifieerbaarheid ervan gewaarborgd is.

## Testbare voorspellingen

In Hoofdstuk 2 en 3 ontwikkelen we het *cyclotactor (CT)*-systeem, dat krachtuitvoer loodrecht op het vingerkussen geeft tijdens het volgen van vingertop-bewegingen loodrecht op een oppervlak. Dit systeem zal op nieuwe wijze het model van berekende vingertop-aanraking implementeren. De Hypothese levert dan via deductie de volgende *Voorspelling 1* op:

> **V 1**
>
> Het CT-systeem ontwikkeld in Hoofdstuk 2 en 3 zal de implementatie van nieuwe vormen van instrumentele controle van muzikaal geluid mogelijk maken.

In Hoofdstuk 2 en 3 ontwikkelen we ook het *kinetic surface friction transducer (KSFT)*-systeem, dat krachtuitvoer parallel aan het vingerkussen geeft tijdens het volgen van vingertop-bewegingen parallel aan een oppervlak. Ook dit systeem zal op nieuwe wijze het model van berekende vingertop-aanraking implementeren. De Hypothese levert dan via deductie ook de volgende *Voorspelling 2* op:

> **V 2**
>
> Het KSFT-systeem ontwikkeld in Hoofdstuk 2 en 3 zal de implementatie van nieuwe vormen van instrumentele controle van muzikaal geluid mogelijk maken.

# Experimentele resultaten

Het eerste deel van de experimentele resultaten gaat over technologie: Hoe implementeren het CT- en KSFT-systeem ieder inderdaad, op nieuwe wijze, het model van berekende vingertop-aanraking? Tabellen IIa en IIb, beneden, beschrijven eerst de kwalitatieve en kwantitatieve hoofdaspecten van beide technologieën. De resulterende capaciteiten op het gebied van menselijke actie en perceptie zijn daarna samengevat in Tabel IIc. De frase "complete integratie met berekend geluid", die gebruikt wordt in Tabel IIa, betekent hier dat invoer van menselijke motor-activiteit, uitvoer naar somatosensorische perceptie, en uitvoer naar auditorische perceptie gemakkelijk gecombineerd kunnen worden in één enkel geschreven algoritme.

Tabel III sluit het eerste deel af met het opsommen van nieuwe aspecten van beide technologieën. Een belangrijke eigenschap is hier *het vermijden van het gebruik van verbonden mechanische delen die bewegen ten opzichte van het anatomisch doelgebied.* We benadrukken dit als een algemeen principe voor transducerbouw, met als potentieel voordeel het mogelijk maken van meer precieze uitvoer naar somatosensorische perceptie.

Het laatste deel van de experimentele resultaten gaat over vormen van instrumentele controle: Welke nieuwe vormen zijn geïmplementeerd met beide systemen? Dit is samengevat in Tabel IV. In de eerste voorbeelden werd berekende vingertop-aanraking gebruikt om de toestand van het geluidgenererend proces te tonen – in groter detail dan mogelijk via bestaande technologieën. Dit om controle-acties van de vingertop beter te informeren, en daarmee, te veranderen. In de laatste voorbeelden werden nieuwe vormen van instrumentele controle meer direct nagejaagd: Berekende aanraking werd hier gebruikt om nieuwe typen van controle-acties met de vingertop te implementeren.

| kwalitatief aspect | systeem | |
|---|---|---|
| gebaseerd op het gebruik van een vlak, gesloten, en rigide contactoppervlak | CT | KSFT |
| geeft krachtuitvoer *loodrecht op het vingerkussen,* volgt vingertop-beweging *loodrecht op een oppervlak* | CT | |
| geeft krachtuitvoer *parallel aan het vingerkussen,* volgt vingertop-beweging *parallel aan een oppervlak* | | KSFT |
| I/O programmeerbaar gemaakt in natuurkundige eenheden, en dit via klassen in SuperCollider | CT | KSFT |
| complete integratie met berekend geluid | CT | KSFT |
| precies instelbare, persoonlijke pasvorm, voor meer accurate en comfortabele I/O | CT | |
| goedkoop massa-produceerbaar | CT | KSFT |

**Tabel IIa** *Kwalitatieve aspecten van de technologieën ontwikkeld in Hoofdstuk 2 en 3.*

| kwantitatief aspect | | CT-systeem | KSFT-systeem |
|---|---|---|---|
| invoer | spatiële resolutie | 0.2 mm | 0.02 mm |
| | spatieel bereik | 35.0 mm | honderden cm$^2$ |
| | temporele resolutie | 4000 Hz | 125 Hz, gemiddeld |
| uitvoer | krachtresolutie | ± 0.003 N | N/B |
| | krachtbereik | bipolair, variërend over afstand: zie Hoofdstuk 3, Figuur 3.2 | 0.14-1.43 N, kinetische frictie |
| | temporele resolutie | accurate golfuitvoer tot 1000 Hz | elementen van 1-10 ms |
| I/O | latentietijd | 4.0 ms | 20.5 ms, gemiddeld |

**Tabel IIb** *Kwantitatieve aspecten van de technologieën ontwikkeld in Hoofdstuk 2 en 3.*

| resulterende capaciteit | systeem | |
|---|---|---|
| krachtuitvoer kan beweging van vingertop-controle-acties meebepalen | CT | KSFT |
| I/O kan aspecten van haptische perceptie opwekken | CT | |
| accurate uitvoer van mechanische golven over de frequentiebereiken die van belang zijn voor vibratieperceptie via de vingertop | CT | |
| uitstekende ondersteuning voor real-time instrumentele controle van muzikaal geluid | CT | |
| I/O kan tijdens actieve aanraking met hoge resolutie aspecten van vingertop-oppervlaktetextuurperceptie opwekken | | KSFT |

**Tabel IIc** *Resulterende capaciteiten op het gebied van menselijke actie en perceptie van de technologieën ontwikkeld in Hoofdstuk 2 en 3.*

| nieuw aspect | systeem |
|---|---|
| expliciete en specifieke ondersteuning voor unidirectionele vingertop-beweging loodrecht op een oppervlak | CT |
| transducer *vermijdt* het gebruik van verbonden mechanische delen die bewegen ten opzichte van het anatomisch doelgebied *totaal* | CT |
| I/O specifiek voor die vingerbuigbewegingen die onafhankelijk, precies, en direct door de motorische schors aangestuurd zijn | CT |
| transducer *vermijdt* het gebruik van verbonden mechanische delen die bewegen ten opzichte van het anatomisch doelgebied *deels* | KSFT |
| hoge-resolutie-aspecten van vingertop-oppervlaktetextuurperceptie opgewekt met sensorinvoer van een gewone en goedkope optische muis | KSFT |

**Tabel III** *Nieuwe aspecten van de technologieën ontwikkeld in Hoofdstuk 2 en 3.*

| zie | type | systeem | hoofdpunten |
|-----|------|---------|-------------|
| 4.2.2 | display via passieve aanraking | CT | • granulaire synthese tonen op het niveau van individuele grains<br>• via aanwezigheid, duur, amplitude, en vibrationele inhoud van krachtpulsen loodrecht op vingerkussen<br>• met tijdsschaal identiek aan die van sonische grains |
| 4.3.2 | display via actieve aanraking | KSFT | • tijdens op turntable scratching gelijkende acties: specifieker display, van opgeslagen geluidfragment<br>• via vingerkussen-parallelle frictie, met milliseconde-resolutie |
| 4.3.3 | display via actieve aanraking | CT | • tijdens percussieve vingertop-bewegingen loodrecht op een oppervlak<br>• display via aanraking vergroot tot buiten het moment van impact |
| 4.5.2 | controle-actie | KSFT | • duwen tegen een virtuele oppervlaktebult<br>• met louter horizontaal toegepaste krachtuitvoer, tijdens horizontaal gerichte vingertop-bewegingen |
| 4.5.3 | controle-actie | CT | • de vingertop spannen tijdens krachtgolf-uitvoer<br>• simultaan bruikbaar met controle gebaseerd op vingertop-beweging loodrecht op een oppervlak |

**Tabel IV**  *Nieuwe vormen van instrumentele controle via de vingertop, beschreven in Hoofdstuk 4.*

# Discussie

Hieronder zullen we eerst de resultaten van twee onderzoeks-excursies bespreken (voor hun motivatie: zie de Acknowledgments). We sluiten deze samenvatting daarna af met reflectie op de hierboven gepresenteerde experimentele hoofdresultaten. In het eindhoofdstuk gaat Sectie 7.6 hier nog kort op door, met enige speculatie over toekomstige ontwikkelingen gebaseerd op de concepten van *berekend manipulandum* en *berekend instrument*.

De eerste onderzoeks-excursie volgde het fenomeen van unidirectionele vingertop-beweging loodrecht op een oppervlak naar elders. Hoofdstuk 5 presenteert *ééndruks-controle*, een vingertop-invoertechniek voor drukgevoelige computertoetsenborden, gebaseerd op de detectie en classificatie van drukbewegingen op de reeds neergehouden toets. We laten zien hoe deze nieuwe techniek naadloos geïntegreerd kan worden in de bestaande praktijk op gewone computertoetsenborden, en ook hoe de techniek gebruikt kan worden om bestaande gebruikersinteracties te versimpelen: door het vervangen van wijzigingstoetscombinaties door enkelvoudige toetsindrukken. In het algemeen kan de voorgestelde techniek gebruikt worden om te navigeren over interactie-opties binnen een GUI, om daarbij volledige previews te krijgen van mogelijke uitkomsten, en om dan óf te kiezen voor één uitkomst, óf alles af te blazen – allemaal in de loop van een enkele toetsindruk/loslaat-cyclus. De resultaten van gebruikerstests laten zien dat effectieve ééndruks-controle in ongeveer een kwartier oefenen aangeleerd kan worden.

De tweede onderzoeks-excursie volgde het idee berekening te gebruiken om aspecten van menselijke perceptie op te wekken naar elders. In Hoofdstuk 6 overwegen we eerst hoe het gebruik van technieken met stadia van geautomatiseerde berekening visuele kunstenaars een controle over waargenomen visuele complexiteit biedt die anderszins onbereikbaar is. Dit motiveert de vraag of berekende uitvoer ook de waarneming kan opwekken van 2D-vormen in het *retinale nabeeld:* het bekende effect in het menselijk visueel systeem waarbij de voortgaande perceptie van licht beïnvloed wordt door de voorafgaande blootstelling eraan. Een fundamenteel probleem hier is hoe de mogelijkheid uit te sluiten dat vormherkenning wordt veroorzaakt door het normaal zien van de stimuli, wat gelijktijdig kan gebeuren. Om dit op te lossen ontwikkelen we een rasterizatiemethode, een model van de nabeeld-intensiteiten die deze methode opwekt, en dan een serie van formele kandidaat-strategieën voor concrete regelsets die stimuli uitrekenen. De Electronische Appendix bij dit hoofdstuk geeft video-voorbeelden, de beeldreeksen gebruikt in een pilot-experiment, en ook software die de aanpak implementeert, in bronformaat. De resultaten van het pilot-experiment, uitgevoerd om de aanpak te testen, bevestigen vorm-display specifiek in het retinale nabeeld. Dit resultaat demonstreert ook de haalbaarheid van het *berekende retinale nabeeld* in het algemeen.

Eindigend met de hoofdresultaten: De inhoud van Tabellen IIa t/m IV bevestigt zowel Voorspelling 1 als Voorspelling 2. Dit ondersteunt de Hypothese: Het ontwikkelen van technologieën die op nieuwe wijze passen binnen het model van berekende vingertop-aanraking maakt de implementatie van nieuwe vormen van instrumentele controle van muzikaal geluid mogelijk. Aan het begin van deze samenvatting presenteerden we een gearticuleerde, empirische kijk op wat menselijk muziekmaken is, en op hoe dit relateert aan berekening. De experimentele bewijslast die we vergaard hebben lijkt erop te wijzen dat deze kijk gebruikt kan worden als een *gereedschap,* voor het systematisch genereren van modellen, hypothesen, en nieuwe technologieën die een steeds completer antwoord mogelijk maken op de fundamentele vraag welke vormen van instrumentele controle van muzikaal geluid implementeerbaar zijn.

# Appendix C: Stellingen (Propositions)

Attached to the thesis
*"Computed fingertip touch for the instrumental control of musical sound
with an excursion on the computed retinal afterimage"*.

**I.** Computation has a unique and central role to play in the development of how humans make music (Introduction and Chapter 1).

**II.** The question as to what forms of *instrumental control of musical sound* are possible to implement is fundamental to the future development of (digital) musical instruments (Introduction and Chapter 1).

**III.** The technical concept of *computational liberation* demonstrably can be used as a tool to obtain an ever more complete answer to this fundamental question (Chapter 4).

**IV.** When developing transducers and algorithms, fingertip use, and more specifically, *unidirectional fingertip movement orthogonal to a surface* should be prioritized as an area for investigation (Chapters 2 and 3).

**V.** The results presented in [Higham et al. 2012] can be interpreted as evidence for how very ancient the presence of this type of fingertip movement is in the instrumental control of musical sound.

**VI.** The 6 DOF magnetic levitation device designed by Hollis that is used with the human fingerpad by Grieve et al. in [Grieve et al. 2009] is part of a development in haptic transducer technology (also exemplified by the CT system in this Thesis) that can be usefully characterized as *the avoidance of connected mechanical parts moving relative to the target anatomical site*.

**VII.** Technologies such as T-PaD [Winfield et al. 2007] and TeslaTouch [Bau et al. 2010] can deliver subtle touch display during control, but to more completely explore new types of fingertip control actions on frictional surfaces, a larger force output range (as exemplified by the KSFT system in this Thesis) is essential.

**VIII.** The fingertip input of successive iterations of consumer computing devices may well be adopted for the control of musical sound, as for example happens in the Mobile Music Toolkit of [Bryan et al. 2010], but this should never replace developing new transducer technologies from scratch.

**IX.** Repeated and varying pressing movements on the already held-down key of a computer keyboard can be used both to simplify existing user interactions and to implement new ones, that allow the rapid yet detailed navigation of multiple possible interaction outcomes (Chapter 5).

**X.** Automated computational techniques can display shape specifically in the retinal afterimage (Chapter 6).

# References

Agilent Technologies, Inc., 2001 Optical mice and how they work. *Technical report 5988-4554EN* (Santa Clara, CA, USA).

Aimi R, 2007 Percussion instruments using realtime convolution: Physical controllers. In *Proceedings of the 2007 international conference on New Interfaces for Musical Expression* 154-159.

Akamatsu M, Sato S, 1994 A multi-modal mouse with tactile and force feedback. *Journal of Human-Computer Studies, 40* 443-453.

Akamatsu M, MacKenzie I S, 1996 Movement characteristics using a mouse with tactile and force feedback. *Journal of Human-Computer Studies, 45* 483-493.

Albers J, 2006 *Interaction of color – Revised and expanded edition* (New Haven, CT, USA: Yale University Press).

Allegro MicroSystems, Inc., 1997 Hall-effect IC applications guide. *Application Note 27701B* (Worcester, MA, USA).

Angel E, 2000 *Interactive computer graphics – A top-down approach with OpenGL, 2nd edition* (Boston, MA, USA: Addison-Wesley) 53-56.

Anstis S, Rogers B, Henry J, 1978 Interactions between simultaneous contrast and coloured afterimages. *Vision Research, 18* 899-911.

Apple, Inc., 2008 Core Audio. At *https://developer.apple.com/library/mac/documentation/MusicAudio/Conceptual/CoreAudioOverview/Introduction/Introduction.html* (accessed January 2014).

Bau O, Poupyrev I, Israr A, Harrison C, 2010 TeslaTouch: Electrovibration for touch surfaces. In *Proceedings UIST 2010* (ACM) 283-292.

Berdahl E, Niemeyer G, Smith J, 2009 HSP: A simple and effective open-source platform for implementing haptic musical instruments. In *Proceedings of the 2009 international conference on New Interfaces for Musical Expression* 262-263.

Berkelman P, Hollis R, 2000 Lorentz magnetic levitation for haptic interaction: device design, function, and integration with simulated environments. *International Journal of Robotics Research, 9* 644-667.

Bevilacqua F, Rasamimanana N, Fléty E, Lemouton S, Baschet F, 2006 The augmented violin project: Research, composition and performance report. In *Proceedings of the 2006 international conference on New Interfaces for Musical Expression* 402-406.

Biard J R, Pittman G E, 1966 Semiconductor radiant diode. *US patent no. 3,293,513* (December 20 1966).

Bongers B, 2000 Physical interfaces in the electronic arts. Chapter submitted for the book and CDROM *Trends in gestural control of music* (Paris, France: IRCAM).

Bouillot N, Wozniewski M, Settel Z, Cooperstock J R, 2008 A mobile wireless augmented guitar. In *Proceedings of the 2008 international conference on New Interfaces for Musical Expression* 189-192.

Brakhage S, 1967 The stars are beautiful. *Essential Brakhage: Selected writings on filmmaking* (Kingston, NY, USA: McPherson & Company) 134-137.

Braunstein R, 1955 Radiative transitions in semiconductors. *Physical Review, 99(6)* 1892.

Brewster S A, Brown L M, 2004 Tactons: Structured tactile messages for non-visual information display. In *Proceedings of the 5th Australasian User Interface Conference*.

Brewster S, Hughes M, 2008 Pressure-based input for mobile devices. In *Proceedings Volume II HAID 2008* (http://www.auditorysigns.com/haid2008) 10-11.

Bryan N J, Herrera J, Oh J, Wang G, 2010 MoMu: A mobile music toolkit. In *Proceedings of the 2010 international conference on New Interfaces for Musical Expression* 174-177.

Cadoz C, Luciani A, Florens J, 1984 Responsive input devices and sound synthesis by simulation of instrumental mechanisms: The Cordis system. *Computer Music Journal, 8(3)* 60-73.

Cadoz C, Lisowski L, Florens J, 1990 A modular feedback keyboard design. *Computer Music Journal, 14(2)* 47-51.

Cadoz C, Luciani A, Florens J, Castagné N, 2003 Artistic creation and computer interactive multisensory simulation force feedback gesture transducers. In *Proceedings of the 2003 international conference on New Interfaces for Musical Expression* 235-246.

Carlson N R, 1998 *Physiology of behavior* (Upper Saddle River, NJ, USA: Prentice Hall).

Carrillo A P, Bonada J, 2010 The Bowed Tube: A virtual violin. In *Proceedings of the 2010 international conference on New Interfaces for Musical Expression* 229-232.

Chang J, 2005 *Can't stop won't stop: A history of the hip-hop generation* (New York, USA: St. Martin's Press).

Chevreul M E, 1839 *De la loi du contraste simultané des couleurs, et de l'assortiment des objets colorés, considéré d'après cette loi* (Paris, France: Pitois-Levrault).

Chevreul M E, 1855 *The principles of harmony and contrast of colours, and their applications to the arts* (London, UK: Longman, Brown, Green, and Longmans).

Chilvers E I, 2009 Albers, Josef. In *The Oxford dictionary of art and artists (Oxford reference online)* (Oxford, UK: Oxford University Press).

Chowning J, 1973 The synthesis of complex audio spectra by means of frequency modulation. *Journal of the Audio Engineering Society, 21(7)* 526-534.

Colgate J E, Grafing P E, Stanley M C, Schenkel G, 1993 Implementation of stiff virtual walls in force-reflecting interfaces. In *Proceedings of the IEEE 1993 Virtual Reality Annual International Symposium* 202-208.

Conard N J, Malina M, Münzel S C, 2009 New flutes document the earliest musical tradition in southwestern Germany. *Nature, 460* 737-740.

Crevoisier A, Kellum G, 2008 Transforming ordinary surfaces into multi-touch controllers. In *Proceedings of the 2008 international conference on New Interfaces for Musical Expression* 113-116.

Croat J J, Herbst J F, Lee R W, Pinkerton F E, 1984 Pr-Fe and Nd-Fe-based materials: A new class of high-performance permanent magnets. *Journal of Applied Physics, 55* 2078.

Crommentuijn K, Hermes D J, 2010 The effect of Coulomb friction in a haptic interface on positioning performance. In *EuroHaptics 2010, Part II, LNCS 6192* (Berlin, Germany: Springer-Verlag) 398-405.

Davidson P, Han J, 2006 Synthesis and control on large scale multi-touch sensing displays. In *Proceedings of the 2006 international conference on New Interfaces for Musical Expression* 216-219.

Daw N W, 1962 Why after-images are not seen in normal circumstances. *Nature, 196* 1143-1145.

De Jong S, 2006 A tactile closed-loop device for musical interaction. In *Proceedings of the 2006 international conference on New Interfaces for Musical Expression* (NIME06, June 4-8 2006, Paris, France) 79-80.

De Jong S, 2008 The cyclotactor: Towards a tactile platform for musical interaction. In *Proceedings of the 2008 international conference on New Interfaces for Musical Expression* (NIME08, June 5-7 2008, Genova, Italy) 370-371.

De Jong S, 2009 Developing the cyclotactor. In *Proceedings of the 2009 international conference on New Interfaces for Musical Expression* (NIME09, June 3-6 2009, Pittsburgh, PA, USA) 163-164.

De Jong S, 2010a Presenting the cyclotactor project. In *Proceedings of the 2010 international conference on Tangible, Embedded, and embodied Interaction* (ACM, January 24-27 2010, MIT, Cambridge, MA, USA) 319-320.

De Jong S, 2010b Kinetic surface friction rendering for interactive sonification: an initial exploration. In *Proceedings of the 2010 international workshop on Interactive Sonification* (April 7 2010, Stockholm, Sweden) 105-108.

De Jong S, Jillissen J, Kirkali D, De Rooij A, Schraffenberger H, Terpstra A, 2010c One-press control: A tactile input method for pressure-sensitive computer keyboards. In *Extended Abstracts CHI 2010* (ACM, April 10-15 2010, Atlanta, GA, USA) 4261-4266.

De Jong A P A, 2010d Apparatus comprising a base and a finger attachment. *US patent application no. 12/792,432* (June 2 2010) 1-56.

De Jong S, 2010e The cyclotactor. Best Demonstration Award. *The 2010 EuroHaptics international conference* (July 8-10 2010, Amsterdam, the Netherlands).

De Jong S, 2011 Making grains tangible: microtouch for microsound. In *Proceedings of the 2011 international conference on New Interfaces for Musical Expression* (NIME11) 326-328.

De Mestral G, 1955 Velvet type fabric and method of producing the same. *US patent no. 2,717,437* (September 13 1955).

Dennerlein J T, Martin D B, Hasser C, 2000 Force-feedback improves performance for steering and combined steering-targeting tasks. In *CHI Letters, 2(1)* 423-429.

Dietz P H, Eidelson B, Westhues J, Bathiche S, 2009 A practical pressure sensitive computer keyboard. In *Proceedings UIST 2009* (ACM) 55-58.

**D**imitrov S, Alonso M, Serafin S, 2008 Developing block-movement, physical-model based objects for the Reactable. In *Proceedings of the 2008 international conference on New Interfaces for Musical Expression* 211-214.

**D**oornbusch P, 2005 *The music of CSIRAC: Australia's first computer music* (Australia: Common Ground Publishing).

**E**lbert T, Pantev C, Wienbruch C, Rockstroh B, Taub E, 1995 Increased cortical representation of the fingers of the left hand in string players. *Science, 270* 305-307.

**E**LD, 2014 The Electronics Department (ELD) hosted in the Leiden Institute of Physics (LION). At *http://www.eld.leidenuniv.nl/eld/* (accessed January 2014) (Faculty of Math and Sciences, Leiden University, NL).

**E**lektronische Werkplaats of the Royal Conservatoire, 2013 Microlab and Ipsonlab. At *http://kc.koncon.nl/ewp/ewp/development.html* (accessed December 2013) (The Hague, NL).

**E**ssl G, O'Modhrain S, 2006 An enactive approach to the design of new tangible musical instruments. *Organised Sound, 11(3)* 285-296.

**E**ssl G, Müller A, 2010 Designing mobile musical instruments and environments with urMus. In *Proceedings of the 2010 international conference on New Interfaces for Musical Expression* 76-81.

**F**ahrenheit D G, 1724 Experimenta & observationes de congelatione aquae in vacuo factae. *Philosophical Transactions of the Royal Society, 33(381-391)* 78-84.

**F**araday M, 1833 Experimental researches in electricity. Fourth series. *Philosophical Transactions of the Royal Society of London, 123* 507-522.

**F**arwell N, 2006 Adapting the trombone: A suite of electro-acoustic interventions for the piece Rouse. In *Proceedings of the 2006 international conference on New Interfaces for Musical Expression* 358-363.

**F**iebrink R, Wang G, Cook P R, 2007 Don't forget the laptop: Using native input capabilities for expressive musical control. In *Proceedings of the 2007 international conference on New Interfaces for Musical Expression* 164-167.

**F**irestone F A, 1942 Flaw detecting device and measuring instrument. *US patent no. 2,280,226* (April 21 1942).

**F**iste J, 2007 Cello vibrato in lower positions – Analysis of technique. At *http://www.celloprofessor.com/Cello-Vibrato.html* (accessed May 2012).

**F**MD, 2014 The Department of Fine Mechanics (FMD) hosted in the Leiden Institute of Physics (LION). At *http://www.physics.leidenuniv.nl/index.php/fmd-home* (accessed January 2014) (Faculty of Math and Sciences, Leiden University, NL).

**F**reed A, Uitti F M, Zbyszyński M, Wessel D, 2006 Augmenting the cello. In *Proceedings of the 2006 international conference on New Interfaces for Musical Expression* 409-413.

**F**reed A, 2009 Novel and forgotten current-steering techniques for resistive multitouch, duotouch, and polytouch position sensing with pressure. In *Proceedings of the 2009 international conference on New Interfaces for Musical Expression* 230-235.

**G**ardner H, Kleiner F S, 2010 *Gardner's Art through the ages: The Western perspective, volume 2* (Stamford, CT, USA: Cengage Learning).

Geiger G, 2006 Using the touch screen as a controller for portable computer music instruments. In *Proceedings of the 2006 international conference on New Interfaces for Musical Expression* 61-64.

Gibson J J, 1962 Observations on active touch. *Psychological Review, 69(6)* 477-491.

Gibson J J, 1977 The theory of affordances. In *Perceiving, acting, and knowing: Toward an ecological psychology* 67-82.

Gillespie R B, 1996 *Haptic display of systems with changing kinematic constraints: The virtual piano action* (PhD dissertation, Stanford University).

Gillespie B, 2001 Haptics. Chapter from *Music, cognition, and computerized sound – An introduction to psychoacoustics (P. Cook, editor)* (Cambridge, MA, USA; London, UK: The MIT Press).

Goethe J W, 1795/1805 Watercolor with pencil outlines on paper: *Bild eines Mädchens in umgekehrten Farben, gezeichnetes Nachbild in Komplementärfarben* (Weimar, Germany: Klassik Stiftung Weimar).

Goldstein E B, 2002 *Sensation and perception* (Pacific Grove, CA, USA: Wadsworth-Thomson Learning).

Google, Inc., 2010 Google Suggest. Documented at *http://www.google.com/support/websearch/bin/answer.py?hl=en&answer=106230* (accessed Januari 2010).

Gosling P, Noordam B, 2006 *Mastering your PhD: Survival and success in the doctoral years and beyond* (Berlin, Germany: Springer-Verlag).

Grieve T, Sun Y, Hollerbach J M, Mascaro S A, 2009 3-D force control on the human fingerpad using a magnetic levitation device for fingernail imaging calibration. In *Proceedings of the Third Joint EuroHaptics Conference and Symposium on Haptic Interfaces for Virtual Environment and Teleoperator Systems* 411-416.

Grisdale R O, 1941 Resistance material. *US patent no. 2,258,646* (October 14 1941).

Hall E, 1879 On a new action of the magnet on electric currents. *American Journal of Mathematics, 2(3)* 287-292.

Harding R E M, 1933 *The Piano-forte: Its history traced to the Great Exhibition of 1851* (Cambridge, UK: Cambridge University Press).

Hayward V, Choksi J, Lanvin G, Ramstein C, 1994 Design and multi-objective optimization of a linkage for a haptic interface. *Advances in robot kinematics* (Kluwer Academic Publishers) 352-359.

Hayward V, Astley O R, Cruz-Hernandez M, Grant D, Robles-De-La-Torre G, 2004 Haptic interfaces and devices. *Sensor Review, 24(1)* 16-29.

Heinz S, O'Modhrain S, 2010 Designing a shareable musical TUI. In *Proceedings of the 2010 international conference on New Interfaces for Musical Expression* 339-342.

Hering E, 1964 *Outlines of a theory of the light sense* (Cambridge, MA, USA: Harvard University Press).

Hermann T, Bovermann T, Riedenklau E, Ritter H, 2007 Tangible computing for interactive sonification of multivariate data. In *Proceedings of the 2007 international workshop on Interactive Sonification*.

Higham T, Basell L, Jacobi R, Wood R, Ramsey C B, Conard N J, 2012 Testing models for the beginnings of the Aurignacian and the advent of figurative art and music: The radiocarbon chronology of Geißenklösterle. *Journal of Human Evolution, 62(6)* 664-676.

Hiller L, Ruiz P, 1971 Synthesizing musical sounds by solving the wave equation for vibrating objects. *Journal of the Audio Engineering Society, 19* 463-470, 542-551.

Hinton J, 2005 The Lilac Chaser visual illusion. Reported at *http://en.wikipedia.org/wiki/Lilac_chaser* (accessed January 2011).

Hochenbaum J, Vallis O, Diakopoulos D, Murphy J, Kapur A, 2010 Designing expressive musical interfaces for tabletop surfaces. In *Proceedings of the 2010 international conference on New Interfaces for Musical Expression* 315-318.

Houston J, Hickey D, 2007 *Optic nerve: Perceptual art of the 1960s* (London, UK: Merrell Publishers).

Howard D M, Rimell S, Hunt A D, 2003 Force feedback gesture controlled physical modeling synthesis. In *Proceedings of the 2003 international conference on New Interfaces for Musical Expression*.

Hsu W, 2006 Managing gesture and timbre for analysis and instrument control in an interactive environment. In *Proceedings of the 2006 international conference on New Interfaces for Musical Expression* 376-379.

Hudson S E, 2004 Using LED arrays as touch-sensitive input and output devices. In *Proceedings of the 2004 ACM symposium on User Interface Software and Technology* 287-290.

Hunt A D, Wanderley M M, Kirk R, 2000 Towards a model for instrumental mapping in expert musical interaction. In *Proceedings of the 2000 International Computer Music Conference* 209-212.

Hunt A D, Wanderley M M, 2002 Mapping performer parameters to synthesis engines. *Organised Sound, 7:2* 97-108.

Hurvich L M, Jameson D, 1957 An opponent-process theory of color vision. *Psychological Review, 64(6, Part I)* 384-404.

Ishii H, Ullmer B, 1997 Tangible bits: Towards seamless interfaces between people, bits and atoms. In *Proceedings of the 1997 ACM SIGCHI conference on Human Factors in Computing Systems* 234-241.

Iwasaki K, Miyaki T, Rekimoto J, 2009 Expressive typing: A new way to sense typing pressure and its applications. In *Extended Abstracts CHI 2009* (ACM) 4369-4374.

Jenkins R, Wiseman R, 2009 Darwin illusion: Evolution in a blink of the eye. *Perception, 38* 1413-1415.

Jones R, Driessen P, Schloss A, Tzanetakis G, 2009 A force-sensitive surface for intimate control. In *Proceedings of the 2009 international conference on New Interfaces for Musical Expression* 236-241.

Jordà S, 2002 FMOL: Toward user-friendly sophisticated new musical instruments. *Computer Music Journal, 26(3)* 23-29.

**J**ordà S, Alonso M, 2006 Mary had a little scoreTable* or the reacTable* goes melodic. In *Proceedings of the 2006 international conference on New Interfaces for Musical Expression* 208-211.

**J**ordà S, Geiger G, Alonso M, Kaltenbrunner M, 2007 The reacTable: Exploring the synergy between live music performance and tabletop tangible interfaces. In *Proceedings of the 2007 international conference on Tangible, Embedded and Embodied Interaction* 139-146.

**K**ahle W, 2001 *Atlas van de anatomie – Deel 3: Zenuwstelsel en zintuigen* (Baarn, NL: HB Uitgevers).

**K**alat J W, 2004 *Biological psychology, 8th edition* (Belmont, CA, USA: Wadsworth) 146-155.

**K**arplus K, Strong A, 1983 Digital synthesis of plucked-string and drum timbres. *Computer Music Journal, 7(2)* 43-55.

**K**ato Y, Takei T, 1930 Studies on composition, chemical properties and magnetization of zinc ferrite. *Journal of the Japan Mining Society, 539* 244-255.

**K**iefer C, 2010 A malleable interface for sonic exploration. In *Proceedings of the 2010 international conference on New Interfaces for Musical Expression* 291-296.

**K**latzky R L, Lederman S J, 2006 The perceived roughness of resistive virtual textures: I. Rendering by a force-feedback mouse. *ACM Transactions on Applied Perception, 3(1)* 1-14.

**K**vifte T, 2007 *Instruments and the electronic age* (Oslo, Norway: Taragot Sounds).

**L**ähdeoja O, 2008 An approach to instrument augmentation: The electric guitar. In *Proceedings of the 2008 international conference on New Interfaces for Musical Expression* 53-56.

**L**ähdeoja O, 2009 Augmenting chordophones with hybrid percussive sound possibilities. In *Proceedings of the 2009 international conference on New Interfaces for Musical Expression* 102-105.

**L**amotte R H, Srinivasan M A, 1991 Surface microgeometry: Neural encoding and perception. In *Information processing in the somatosensory system (O. Franzen and J. Westman, editors)* (London, UK: Macmillan Press).

**L**ederman S J, Klatzky R L, 1987 Hand movements: A window into haptic object recognition. *Cognitive Psychology, 19* 342-368.

**L**ederman S J, Klatzky R L, 1990 Haptic classification of common objects: Knowledge-driven exploration. *Cognitive Psychology, 22* 421-459.

**L**ederman S J, Klatzky R L, 2009 Haptic perception: a tutorial. *Attention, Perception & Psychophysics, 7* 1439-1459.

**L**ee H-W, Legge G E, Ortiz A, 2003 Is word recognition different in central and peripheral vision? *Vision Research, 43* 2837-2846.

**L**eeuw H, 2009 The Electrumpet, a hybrid electro-acoustic instrument. In *Proceedings of the 2009 international conference on New Interfaces for Musical Expression* 193-198.

**L**evitin D, 2006 *This is your brain on music* (London, UK: Atlantic Books).

**L**inz P, 1997 *An introduction to formal languages and automata* (Sudbury, MA, USA: Jones and Bartlett Publishers).

**L**ippit T, 2006 Turntable music in the digital era: Designing alternative tools for new turntable expression. In *Proceedings of the 2006 international conference on New Interfaces for Musical Expression* 71-74.

**L**osev O V, 1927 Luminous carborundum detector and detection with crystals. *Телеграфия и Телефония без Проводов (Wireless Telegraphy and Telephony), 44* 485-494.

**L**ou L, 2001 Effects of voluntary attention on structured afterimages. *Perception, 30* 1439-1448.

**M**adden D, Spike N, Spike J T, 2010 *Anuszkiewicz: Paintings & sculptures, 1945-2001: Catalogue raisonné* (Firenze, Italy: Edizioni Centro Di).

**M**affei S, 1711 Nuova inventione d'un gravecembalo col forte e piano. *Giornale de' letterati d'Italia, 5* 144-159. Reprinted with translation in Pollens S, 1995 *The early pianoforte* (Cambridge, UK: Cambridge University Press).

**M**agnusson T, 2006 Screen-based musical interfaces as semiotic machines. In *Proceedings of the 2006 international conference on New Interfaces for Musical Expression* 162-167.

**M**ann Y, Lubow J, Freed A, 2009 The Tactus: A tangible, rhythmic grid interface using found-objects. In *Proceedings of the 2009 international conference on New Interfaces for Musical Expression* 86-89.

**M**arshall M T, Wanderley M M, 2006 Vibrotactile feedback in digital musical instruments. In *Proceedings of the 2006 international conference on New Interfaces for Musical Expression*.

**M**artinez-Conde S, Macknik S L, Hubel D H, 2004 The role of fixational eye movements in visual perception. *Nature Reviews Neuroscience, 5(3)* 229-240.

**M**aruyama Y, Terada T, Takegawa Y, Tsukamoto M, 2010 UnitInstrument: Easy configurable musical instruments. In *Proceedings of the 2010 international conference on New Interfaces for Musical Expression* 7-12.

**M**assie T H, Salisbury J K, 1994 The PHANTOM haptic interface: A device for probing virtual objects. In *Proceedings of the 1994 ASME winter annual meeting, symposium on Haptic Interfaces for Virtual Environment and Teleoperator Systems*.

**M**athews M V, Miller J E, Moore F R, Pierce J R, Risset J C, 1969 *The technology of computer music* (Cambridge, MA, USA; London, UK: The MIT Press).

**M**athews M V, 2001 The auditory brain. Chapter from *Music, cognition, and computerized sound – An introduction to psychoacoustics (P. Cook, editor)* (Cambridge, MA, USA; London, UK: The MIT Press).

**M**aupin J T, Vorthmann E A, 1971 Hall effect contactless switch with prebiased Schmitt trigger. *US patent no. 3,596,114* (July 27 1971).

**M**cCartney J, 2002 Rethinking the computer music language: SuperCollider. *Computer Music Journal, 26(4)* 61-68.

**M**cPherson A, Kim Y, 2010 Augmenting the acoustic piano with electromagnetic string actuation and continuous key position sensing. In *Proceedings of the 2010 international conference on New Interfaces for Musical Expression* 217-222.

**M**izobuchi S, Terasaki S, Keski-Jaskari T, Nousiainen J, Ryynanen M, Silfverberg M, 2005 Making an impression: Force-controlled pen input for handheld devices. In *Extended Abstracts CHI 2005* (ACM) 1661-1664.

**M**oore G E, 1965 Cramming more components onto integrated circuits. *Electronics Magazine, 38 (8)* 4-7.

**M**organ A L, 2007 *The Oxford dictionary of American art and artists* (Oxford, UK: Oxford University Press).

**M**ünch S, Dillmann R, 1997 Haptic output in multimodal user interfaces. In *Proceedings of the 1997 international conference on Intelligent User Interfaces* 105-112.

**M**useum of Modern Art, Seitz W C, 1965 *The responsive eye* (New York, USA: Museum of Modern Art).

**N**ishibori Y, Iwai T, 2006 Tenori-on. In *Proceedings of the 2006 international conference on New Interfaces for Musical Expression* 172-175.

**N**ishizawa J, Watanabe Y, 1953 PIN photodiode patent. *Japanese patent no. 2,221,218*.

**N**orman D A, 1988 *The psychology of everyday things* (New York, USA: Basic Books).

**O**boe R, De Poli G, 2002 Multi-instrument virtual keyboard – The MIKEY project. In *Proceedings of the 2002 international conference on New Interfaces for Musical Expression*.

**O**gawa H, Shimojo M, 2006 Development of 2DOF haptic device driven directly by shaft motors. In *Proceedings of the annual conference of the Virtual Reality Society of Japan* 1A4-1.

**O**h J, Herrera J, Bryan N J, Dahl L, Wang G, 2010 Evolving the mobile phone orchestra. In *Proceedings of the 2010 international conference on New Interfaces for Musical Expression* 82-87.

**O**'Modhrain S, 2000 *Playing by feel – Incorporating haptic feedback into computer-based musical instruments* (PhD dissertation, Stanford University).

**P**alacio-Quintin C, 2008 Eight years of practice on the Hyper-Flute: Technological and musical perspectives. In *Proceedings of the 2008 international conference on New Interfaces for Musical Expression* 293-298.

**P**aradiso J A, 2002 The edge of NIME - From cult to conference. In *Proceedings of the 2002 international conference on New Interfaces for Musical Expression*.

**P**arola R, 1969 *Optical art: Theory and practice* (New York, USA: Reinhold Book Corporation).

**P**hilipp H, 2002 Charge transfer sensing. QProx™ white paper. At *http://web.archive.org/web/20020607151810/http://www.qprox.com/background/white_paper.php* (accessed December 2013).

**P**oepel C, Overholt D, 2006 Recent developments in violin-related digital musical instruments. In *Proceedings of the 2006 international conference on New Interfaces for Musical Expression* 390-395.

**P**oplawski P, 2003 *Encyclopedia of literary modernism* (Westport, CT, USA: Greenwood Press).

**P**oupyrev I, Nashida T, Okabe M, 2007 Actuation and tangible user interfaces: The Vaucanson Duck, robots, and shape displays. In *Proceedings of the 2007 international conference on Tangible, Embedded, and embodied Interaction* 205-212.

**P**uckette M, 2002 Max at seventeen. *Computer Music Journal, 26(4)* 31-43.

**P**uckette M, 2007 *The theory and technique of electronic music* (Singapore, Singapore; Hackensack, NJ, USA: World Scientific Publishing).

**R**äisänen J, 2008 Sormina – a new virtual and tangible instrument. In *Proceedings of the 2008 international conference on New Interfaces for Musical Expression* 57-60.

**R**amos G, Boulos M, Balakrishnan R, 2004 Pressure widgets. In *Proceedings CHI 2004* (ACM) 487-494.

**R**athus S A, 2012 *Psychology: Concepts and connections, 10th edition* (Stamford, CT, USA: Cengage Learning).

**R**ekimoto J, Ishizawa T, Schwesig C, Oba H, 2003 PreSense: Interaction techniques for finger sensing input devices. In *Proceedings UIST 2003* (ACM) 203-212.

**R**ekimoto J, Schwesig C, 2006 PreSenseII: Bi-directional touch and pressure sensing interactions with tactile feedback. In *Extended Abstracts CHI 2006* (ACM) 1253-1258.

**R**évész G, 1950 *Psychology and art of the blind* (New York, USA: Longmans, Green and Co.).

**R**oads C, 2004 *Microsound* (Cambridge, MA, USA; London, UK: The MIT Press).

**R**oads C, Strawn J, Abbott C, Gordon J, Greenspun P, 1996 *The computer music tutorial* (Cambridge, MA, USA; London, UK: The MIT Press).

**R**obles-De-La-Torre G, Hayward V, 2001 Force can overcome object geometry in the perception of shape through active touch. *Nature, 414* 445-448.

**R**ovan J, Hayward V, 2000 Typology of tactile sounds and their synthesis in gesture-driven computer music performance. In *Trends in gestural control of music* (Paris, France: Editions IRCAM).

**R**undgren T, 2011 Technical Grammy Award: Roger Linn. At *http://www.grammy.com/ news/technical-grammy-award-roger-linn-0* (accessed May 2012).

**S**agawa M, Fujimura S, Togawa N, Yamamoto H, Matsuura Y, 1984 New material for permanent magnets on a base of Nd and Fe. *Journal of Applied Physics, 55* 2083.

**S**chiesser S, Traube C, 2006 On making and playing an electronically-augmented saxophone. In *Proceedings of the 2006 international conference on New Interfaces for Musical Expression* 308-313.

**S**chlei K, 2010 Relationship-based instrument mapping of multi-point data streams using a trackpad interface. In *Proceedings of the 2010 international conference on New Interfaces for Musical Expression* 136-139.

**S**chneider C, Mustufa T, Okamura A M, 2004 A magnetically-actuated friction feedback mouse. In *Proceedings of the 2004 EuroHaptics international conference.*

Scholastic Inc., 2008 Georges Seurat – working with color. In *Scholastic Art, 39(1) Teacher's Edition* 1-4.

Shive J N, 1953 Semiconductor photoelectric device. *US patent no. 2,641,713* (June 9 1953).

Signac P, 1899 *D'Eugène Delacroix au néo-impressionnisme* (Paris, France: Éditions de la Revue Blanche).

Sinclair S, Wanderley M M, 2007 Defining a control standard for easily integrating haptic virtual environments with existing audio/visual systems. In *Proceedings of the 2007 international conference on New Interfaces for Musical Expression*.

Smith J, 1987 Waveguide filter tutorial. In *Proceedings of the 1987 International Computer Music Conference* 9-16.

Smith W, 1873 The action of light on selenium. *Journal of the Society of Telegraph Engineers, 2* 31.

Sobotta J, Putz R, Pabst R, 1994 *Atlas van de menselijke anatomie – Deel 1: Hoofd, hals, bovenste extremiteit* (Houten, NL: Bohn Stafleu Van Loghum).

Sokolov S Y, 1929 On the problem of the propagation of ultrasonic oscillations in various bodies. *Elektrische Nachrichten-Technik, 6* 454-461.

Sturgeon W, 1825 Improved electro magnetic apparatus. *Transactions of the Royal Society of Arts, Manufactures, & Commerce (London), 43* 37-52.

Sylvester D, De Sausmarez M, 2012 *Bridget Riley: Works 1960-1966* (London, UK: Ridinghouse).

Takegawa Y, Tsukamoto M, Terada T, Nishio S, 2007 Mobile Clavier: New music keyboard for flexible key transpose. In *Proceedings of the 2007 international conference on New Interfaces for Musical Expression* 82-87.

Takegawa Y, Terada T, Tsukamoto M, 2008 UnitKeyboard: An easily configurable compact clavier. In *Proceedings of the 2008 international conference on New Interfaces for Musical Expression* 289-292.

Tanenbaum A S, 2001 Operating system design. Chapter from *Modern operating systems, 2nd edition* (Upper Saddle River, NJ, USA: Prentice Hall).

Taylor S, Hook J, 2010 FerroSynth: A ferromagnetic music interface. In *Proceedings of the 2010 international conference on New Interfaces for Musical Expression* 463-466.

Ten Donkelaar H J, Lohman A H M, Moorman A F M, 2007 *Klinische anatomie en embryologie – Deel 2* (Maarssen, NL: Elsevier gezondheidszorg).

The Magnetic Levitation Haptic Consortium, 2009. At *http://maglevhaptics.org* (accessed 2009).

Troxler I P V, 1804 Über das Verschwinden gegebener Gegenstände innerhalb unseres Gesichtskreises. *Ophthalmologische Bibliothek, 2(2)* 1-53.

Truax B, 1986 Real-time granular synthesis with the DMX-1000. In *Proceedings of the 1986 International Computer Music Conference* 138-145.

Tubiana R, Thomine J M, Mackin E, 1996 *Examination of the hand and wrist* (New York, USA: Informa Healthcare USA).

**T**uring A M, 1936 On computable numbers, with an application to the Entscheidungsproblem. In *Proceedings of the London Mathematical Society, 2(42)* 230-265.

**U**pstill S, 1989 *The RenderMan companion: A programmer's guide to realistic computer graphics* (Boston, MA, USA: Addison-Wesley).

**V**anegas R, 2007 The MIDI pick. In *Proceedings of the 2007 international conference on New Interfaces for Musical Expression* 330-333.

**V**an Lier R, Vergeer M, Anstis S, 2009 Filling-in afterimage colors between the lines. *Current Biology, 19(8)* R323-R324.

**V**erheijen F J, 1961 A simple after image method demonstrating the involuntary multidirectional eye movements during fixation. *Opt. Acta (Lond.), 8* 309-312.

**W**ade N J, 1978 Why do patterned afterimages fluctuate in visibility? *Psychological Bulletin, 85(2)* 338-352.

**W**ang G, 2009 Designing Smule's ocarina: The iPhone's magic flute. In *Proceedings of the 2009 international conference on New Interfaces for Musical Expression* 303-307.

**W**essel D, Wright M, 2001 Problems and prospects for intimate musical control of computers. *CHI01 Workshop New Interfaces for Musical Expression (NIME01)*.

**W**essel D, Avizienis R, Freed A, Wright M, 2007 A force sensitive multi-touch array supporting multiple 2-D musical control structures. In *Proceedings of the 2007 international conference on New Interfaces for Musical Expression* 41-45.

**W**ickens C D, 1992 *Engineering psychology and human performance* (New York, USA: HarperCollins Publishers Inc.).

**W**infield L, Glassmire J, Colgate J E, Peshkin M, 2007 T-PaD: Tactile pattern display through variable friction reduction. In *Proceedings of the Second Joint EuroHaptics Conference and Symposium on Haptic Interfaces for Virtual Environment and Teleoperator Systems* 421-426.

**W**ishart T, 1994 *Audible design* (York, UK: Orpheus the Pantomime).

**W**olters E Ch, Groenewegen H J, 2004 *Neurologie – Structuur, functie en dysfunctie van het zenuwstelsel* (Houten, NL: Bohn Stafleu Van Loghum).

**W**oo M, Neider J, Davis T, OpenGL Architecture Review Board, 1997 *OpenGL programming guide* (Boston, MA, USA: Addison-Wesley).

**W**right M, 2002 Problems and prospects for intimate and satisfying sensor-based control of computer sound. In *Proceedings of the symposium on Sensing and Input for Media-centric Systems*.

**X**enakis I, 1960 Elements of stochastic music. *Gravensaner Blätter, 18* 84-105.

**Y**arrow K, Haggard P, Rothwell J C, 2008 Vibrotactile – auditory interactions are post-perceptual. *Perception, 37* 1114-1130.

**Y**erkes K, Shear G, Wright M, 2010 Disky: A DIY rotational interface with inherent dynamics. In *Proceedings of the 2010 international conference on New Interfaces for Musical Expression* 150-155.

**Y**oung D, Nunn P, Vassiliev A, 2006 Composing for Hyperbow: A collaboration between MIT and the Royal Academy of Music. In *Proceedings of the 2006 international conference on New Interfaces for Musical Expression* 396-401.

**Z**adel M, Scavone G, 2006 Different strokes: A prototype software system for laptop performance and improvisation. In *Proceedings of the 2006 international conference on New Interfaces for Musical Expression* 168-171.

**Z**byszyński M, Wright M, Momeni A, Cullen D, 2007 Ten years of tablet musical interfaces at CNMAT. In *Proceedings of the 2007 international conference on New Interfaces for Musical Expression* 100-105.

**Z**oran A, Maes P, 2008 Considering virtual and physical aspects in acoustic guitar design. In *Proceedings of the 2008 international conference on New Interfaces for Musical Expression* 67-70.

**Z**ygote Media Group, Inc., 2012 Zygote Body™. At *http://www.zygotebody.com* (accessed June 2012).

# Photographic credits

All images by Staas de Jong, unless otherwise noted:

• Collage in the Acknowledgments: Includes one photograph kindly made by Hyunjung Kim; and one by anonymous.

• Figure 1.1: Based on a diagram of the human skull and brain by Patrick J. Lynch and C. Carl Jaffe, 2006.

• Figure 1.3: Photograph kindly made available by the Rijksmuseum voor Oudheden.

• Figure 1.4: Includes images in the public domain.

• Figure 1.5: The image of the button accordion is based on a photograph by Nathanaël Carré. The image of the piano is based on a photograph by Pianomap.

• Figure 6.1: Image reproduced from Martinez-Conde et al., 2004.

• Figure 6.2a: Detail from a work by M.C. Escher.

• Figure 6.2b: Based on a detail from the wedding photograph of Roos van Engelen, who first tried to show me the retinal afterimage, early 1940s.

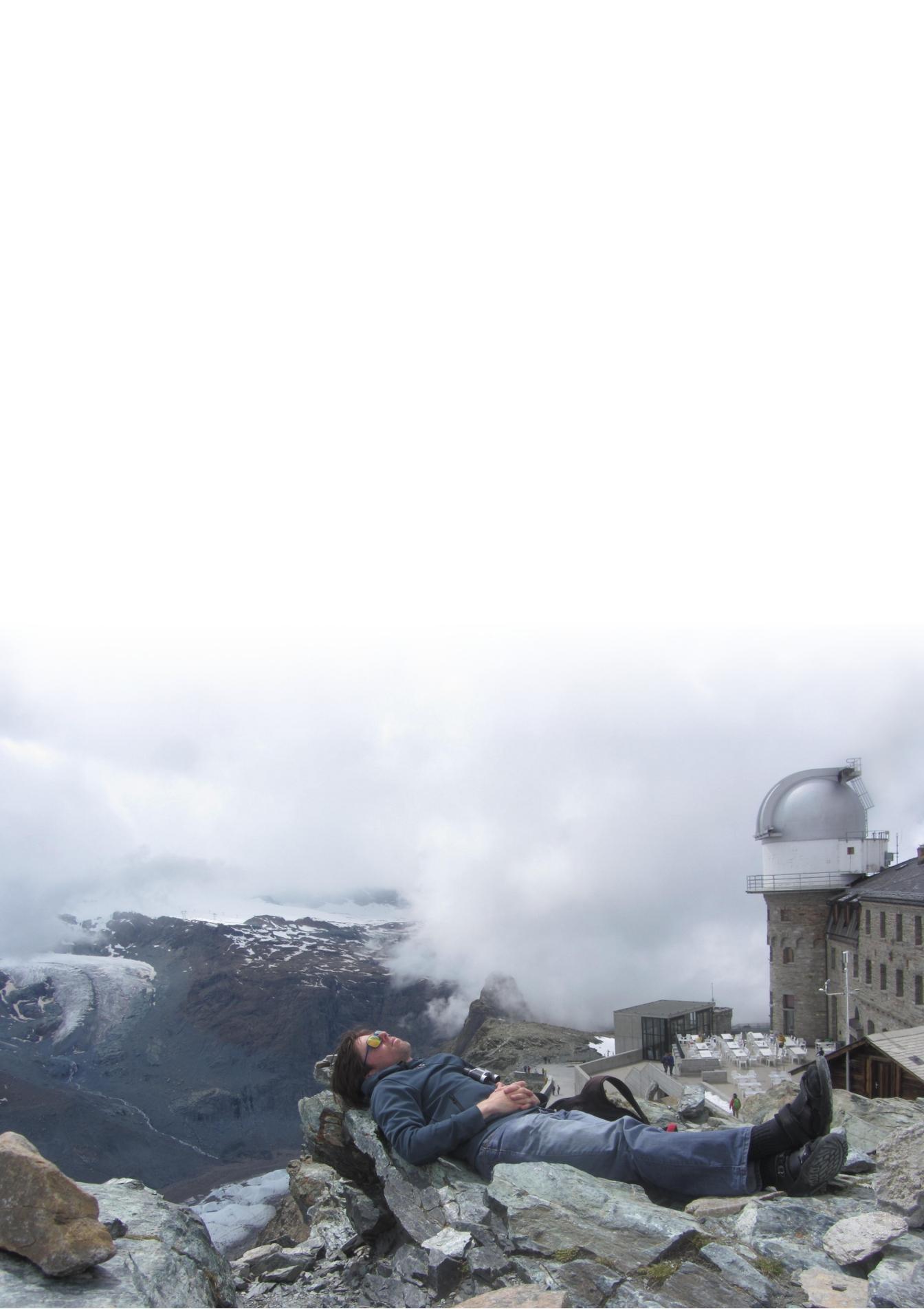